\documentclass[journal,onecolumn]{IEEEtran}
\usepackage[T1]{fontenc}% optional T1 font encoding
\usepackage{xcolor}
\usepackage{longtable,multirow,rotating}
\usepackage{tabu}
\usepackage{tabularx}
\usepackage{booktabs}
\usepackage{array}
\usepackage{comment}

\usepackage{graphicx}
\usepackage{xcolor}
\usepackage{caption}
\usepackage{subcaption}
\usepackage{textcomp}
\usepackage{gensymb}
\usepackage{float}
\usepackage{tabularx}

\usepackage{cite} %to cite multiple references []-[]
\usepackage{varioref}
\usepackage{hyperref}
\usepackage[acronym,nopostdot,automake]{glossaries}
\usepackage[noabbrev]{cleveref}
\newcolumntype{P}[1]{>{\centering\arraybackslash}p{#1}}
\newcolumntype{M}[1]{>{\centering\arraybackslash}m{#1}}
\usepackage{enumitem}

\makeglossaries 
\newacronym{jaxa}{JAXA}{Japan Aerospace Exploration Agency}
\newacronym{hap}{HAP}{High Altitude Platform}
\newacronym{haps}{HAPS}{High Altitude Platform Station}
\newacronym{laps}{LAPS}{Low Altitude Platform Station}
\newacronym{uav}{UAV}{Unmanned Aerial Vehicle}
\newacronym{bs}{BS}{Base Station}
\newacronym{uxbs}{UxBS}{UAVs BS}
\newacronym{urllc}{URLLC}{Ultra Reliable Low Latency Communication}
\newacronym{5g}{5G}{Fifth Generation}
\newacronym{b5g}{B5G}{Beyond 5G}
\newacronym{6g}{6G}{Sixth Generation}
\newacronym{4g}{4G}{Fourth Generation}
\newacronym{3g}{3G}{Third Generation}
\newacronym{2g}{2G}{Second Generation}
\newacronym{leo}{LEO}{Low Earth Orbit}
\newacronym{meo}{MEO}{Medium Earth Orbit}
\newacronym{geo}{GEO}{Geostationary Earth Orbit}
\newacronym{hetnet}{HetNet}{Heterogeneous Network}
\newacronym{vhetnet}{VHetNet}{Vertical Heterogeneous Network}
\newacronym{capex}{CAPEX}{Capital Expense}
\newacronym{mimo}{MIMO}{Multiple-Input Multiple-Output}
\newacronym{siso}{SISO}{Single-Input Single-Output}
\newacronym{sinr}{SINR}{Signal-to-Interference-plus-Noise Ratio}
\newacronym{snr}{SNR}{Signal-to-Noise Ratio}
\newacronym{sir}{SIR}{Signal to Interference Ratio}
\newacronym{sar}{SAR}{Synthetic Aperture Radar}
\newacronym{itu}{ITU}{International Telecommunications Union}
\newacronym{rss}{RSS}{Reconfigurable Smart Surface}
\newacronym{rssi}{RSSI}{Received Signal Strength Indicator}
\newacronym{ai}{AI}{Artificial Intelligence}
\newacronym{ftn}{FTN}{Faster-Than-Nyquist}
\newacronym{sefdm}{SEFDM}{Spectrally-Efficient Frequency Division Multiplexing}
\newacronym{fbmc}{FBMC}{Filter Bank Multicarrier}
\newacronym{noma}{NOMA}{Non-Orthogonal Multiple Access}
\newacronym{smbs}{SMBS}{Super Macro Base Station}
\newacronym{swap}{SWAP}{Size, Weight, and Power}
\newacronym{iot}{IoT}{Internet  of  Things}
\newacronym{ici}{ICI}{Inter-carrier Interference}
\newacronym{lte}{LTE}{Long-Term Evolution}
\newacronym{fso}{FSO}{Free Space Optical}
\newacronym{mmWave}{mmWave}{millimeter Wave}
\newacronym{rf}{RF}{Radio Frequency}
\newacronym{los}{LOS}{Line-of-Sight}
\newacronym{nlos}{NLOS}{Non Line-of-Sight}
\newacronym{ue}{UE}{User Equipment}
\newacronym{its}{ITS}{Intelligent Transportation System}
\newacronym{cav}{CAV}{Connected Autonomous Vehicle}
\newacronym{v2x}{V2X}{Vehicle to Everything}
\newacronym{rr}{RR}{Radio Regulation}
\newacronym{itu-r}{ITU-R}{ITU Radiocommunication Sector}
\newacronym{isi}{ISI}{Inter-symbolic Interference}
\newacronym{icao}{ICAO}{International Civil Aviation Organization}
\newacronym{un}{UN}{United Nations}
\newacronym{imt}{IMT}{International Mobile Telecommunications}
\newacronym{halo}{HALO}{High Altitude Long Operation Network }
\newacronym{wrc}{WRC}{World Radiocommunication Conference}
\newacronym{eirp}{EIRP}{Equivalent  Isotropic  Radiated  Power}
\newacronym{pfd}{PFD}{Power Flux Density}
\newacronym{ism}{ISM}{Industrial, Scientific and Medical }
\newacronym{erast}{ERAST}{Environmental  Research  Aircraft  and  Sensor Technology}
\newacronym{spf}{SPF}{Stratospheric  Platform}
\newacronym{nal}{NAL}{National  Aerospace  Laboratory}
\newacronym{jaea}{JAEA}{Japan  Aerospace  Exploration  Agency}
\newacronym{utias}{UTIAS)}{University of Toronto Institute for Aerospace Studies}
\newacronym{crc}{CRC}{Communications Research Centre Canada }
\newacronym{rs}{RS}{Relay Station}
\newacronym{mems}{MEMS}{Micro-Electro-Mechanical System}
\newacronym{thz}{THz}{Terahertz}
\newacronym{ghz}{GHz}{Gigahertz}
\newacronym{dlr}{DLR}{German Aerospace Centre}
\newacronym{gbsb}{GBSB}{Geometry-Based Single-Bounce}
\newacronym{sos}{SoS}{Sum-of-Sinusoids}
\newacronym{qos}{QoS}{Quality of Service}
\newacronym{ber}{BER}{Bit Error Rate}
\newacronym{phy}{PHY}{Physical}
\newacronym{ofdm}{OFDM}{Orthogonal Frequency Division Multiplexing}
\newacronym{ofdma}{OFDMA}{Orthogonal Frequency Division Multiple Access}
\newacronym{mcs}{MCS}{ Modulation  Coding  Scheme}
\newacronym{cdma}{CDMA}{Code Division Multiple Access}
\newacronym{wcdma}{WCDMA}{Wide-band CDMA}
\newacronym{ds-cdma}{DS-CDMA}{Direct Sequence CDMA}
\newacronym{bet}{BET}{Best Effort Traffic}
\newacronym{ora}{ORA}{Optimal Resource Allocation}
\newacronym{ora-gf}{ORA-GF}{Optimum  Resource  Allocation  subject  to  Global  Fairness}
\newacronym{ora-lf}{ORA-LF}{Optimum  Resource  Allocation  subject  to  Local  Fairness}
\newacronym{sra-gf}{SRA-GF}{Simplified  Resource  Allocation  subject  to  Global  Fairness}
\newacronym{cac}{CAC}{Call Admission  Control}
\newacronym{mbms}{MBMS}{Multicast Broadcast Multimedia Services}
\newacronym{mec}{MEC}{Mobile Edge Computing}
\newacronym{svm}{SVM}{Support Vector Machine}
\newacronym{fl}{FL}{Federated Learning}
\newacronym{son}{SON}{Self-Organizing Networks}
\newacronym{nfp}{NFP}{Network Flying Platform}
\newacronym{ll}{LL}{Low Layer}
\newacronym{nr}{NR}{New Radio}
\newacronym{cir}{CIR}{Carrier-to-Interference Ratio}
\newacronym{cinr}{CINR}{Carrier-to-Interference Plus Noise Ratio}
\newacronym{jt-comp}{JT-CoMP}{Joint Transmission Co-ordinated  Multipoint  }
\newacronym{pso}{PSO}{ Particle  Swarm  Optimization}
\newacronym{fdma}{FDMA}{Frequency Division Multiple Access}
\newacronym{tdma}{TDMA}{Time Division Multiple Access}
\newacronym{sc-fdma}{SC-FDMA}{Single Carrier Frequency Division Multiple Access}
\newacronym{parp}{PAPR}{Peak-to-Average-Power  Ratio  }
\newacronym{ifft}{IFFT}{Inverse-Fast-Fourier Transform} 
\newacronym{cp}{CP}{Cyclic  Prefix}
\newacronym{rrc}{rRC}{Root Raised Cosine}
\newacronym{mlse}{MLSE}{Maximum Likelihood Sequence Estimation}
\newacronym{bpsk}{BPSK}{Binary  Phase  Shift  Keying}
\newacronym{bask}{BASK}{Binary  Amplitude  Shift  Keying}
\newacronym{qpsk}{QPSK}{Quadrature Phase  Shift  Keying}
\newacronym{qam}{QAM}{Quadrature Amplitude  Modulation}
\newacronym{oma}{OMA}{Orthogonal  Multiple  Access}
\newacronym{sdma}{SDMA}{Spatial  Division Multiple Access}
\newacronym{rrm}{RRM}{Radio Resource Management}
\newacronym{ml}{ML}{Machine Learning}
\newacronym{dnn}{DNN}{Deep Neural Network}
\newacronym{rnn}{RNN}{Recurrent Neural Network}
\newacronym{mmtc}{mMTC}{massive Machine-Type Communication}
\newacronym{qoe}{QoE}{Quality of Experience}
\newacronym{tdm}{TDM}{Time Division Multiplexing}
\newacronym{ip}{IP}{Internet Protocol}
\newacronym{ns}{NS}{Network Slicing }
\newacronym{sdn}{SDN}{Software Defined Network}
\newacronym{nfv}{NFV}{Network Function Virtualization}
\newacronym{sd-abn}{SD-ABN}{Software Defined Airborne Backbone Network Architecture}
\newacronym{sll}{SLL}{Side-Lobe Level}
\newacronym{dof}{DoF}{Degree-of-Freedom}
\newacronym{vr}{VR}{Virtual Reality}
\newacronym{ar}{AR}{Augmented Reality}
\newacronym{clpso}{CLPSO}{Comprehensive  Learning  Particle  Swarm  Optimizer}
\newacronym{ioe}{IoE}{Internet of Everything}
\newacronym{mcu}{MCU}{Micro  Controller  Unit}
\newacronym{gpu}{GPU}{Graphical Processing  Unit}
\newacronym{wifi}{WiFi}{Wireless Fidelity}
\newacronym{3d}{3D}{Three Dimensional}
\newacronym{rl}{RL}{Reinforcement Learning}
\newacronym{si}{SI}{Swarm Intelligence}
\newacronym{ntn}{NTN}{Non-Terrestrial Network} 
\newacronym{3gpp}{3GPP}{Third Generation Partnership Project}
\newacronym{tr}{TR}{Technical Report} 
\newacronym{embb}{eMBB}{enhanced Mobile Broadband}
\ifCLASSINFOpdf

\else

\fi

\begin{document}

\title{A Vision and Framework for the High Altitude Platform Station (HAPS) Networks of the Future}

\author{Gunes Karabulut~Kurt,~\IEEEmembership{Senior Member,~IEEE,}
        Mohammad G. Khoshkholgh,~\IEEEmembership{Member, ~IEEE,}
        Safwan Alfattani,~\IEEEmembership{Student Member, ~IEEE,}
        Ahmed Ibrahim,~\IEEEmembership{Member, ~IEEE,}
        Tasneem S. J. Darwish,~\IEEEmembership{Senior Member, ~IEEE,}
        Md Sahabul Alam,~\IEEEmembership{Member, ~IEEE,}
        Halim Yanikomeroglu,~\IEEEmembership{Fellow, ~IEEE,}
        and~Abbas Yongacoglu,~\IEEEmembership{Life Member,~IEEE}% <-this % stops a space
\thanks{Gunes Karabulut Kurt, Mohammad G. Khoshkholgh, Ahmed Ibrahim, Tasneem S. J. Darwish, Md Sahabul Alam, and H. Yanikomeroglu are with the Department 
of Systems and Computer Engineering, Carleton University, Ottawa, 
Canada. Emails: guneskurt@sce.carleton.ca, m.g.khoshkholgh@gmail.com, 
ahmedibrahim@sce.carleton.ca, tasneemdarwish@sce.carleton.ca, 
sahabulalam@sce.carleton.ca, halim@sce.carleton.ca. Gunes Karabulut Kurt is 
also with the Department of Electronics and Communications 
Engineering, Istanbul Technical University, Istanbul, Turkey.}
\thanks{Safwan Alfattani and Abbas Yongacoglu are with the School of Electrical Engineering and Computer Science, University of Ottawa, Ottawa, ON, Canada. Emails: yongac@uottawa.ca, salfa043@uottawa.ca.}% <-this % stops a space
\thanks{Manuscript received July 23, 2020, \textcolor{black}{revised on Jan. 19, 2021}.}}

\markboth{A Vision and Framework for the HAPS Networks}{Shell \MakeLowercase{\textit{et al.}}: Bare Demo of IEEEtran.cls for IEEE Journals}

\maketitle

\begin{abstract}

\textcolor{black}{
%Conventional deployments of \acrfull{haps} are restricted for under-served areas that are sparsely populated with limited infrastructure, however their full potential remains undiscovered.
A \acrfull{haps} is a network node that operates in the stratosphere at an \textcolor{black}{of} altitude around 20 km and is instrumental for providing communication services. \textcolor{black}{Precipitated by} technological innovations in the areas of autonomous avionics, array antennas, solar panel efficiency \textcolor{black}{levels, and battery energy densities, and} fueled by flourishing industry ecosystems, \textcolor{black}{the HAPS has emerged} as an indispensable component of  next\textcolor{black}{-}generations of wireless networks. 
In this article, we provide a vision and framework for the HAPS networks of the future supported by a comprehensive and state-of-the-art literature \textcolor{black}{review}.
%In this article, we aim at providing a comprehensive, state-of-the-art literature review, while delivering a vision supported with a framework for their future applications.
We highlight the \textcolor{black}{unrealized} potential of \acrshort{haps} systems and elaborate on their unique ability to serve metropolitan
areas. The latest  advancements and promising technologies in the \acrshort{haps} energy and payload systems  are discussed.  
The integration of the emerging  \acrfull{rss} technology in the communications payload of \acrshort{haps} systems for providing a cost-effective deployment is proposed. 
%We demonstrate the synergy between \acrshort{haps} and the recent technological advances in the use of \acrfull{rss} in the communications payload of \acrshort{haps} systems. 
A detailed overview of the radio resource management in \acrshort{haps} systems is presented along with  synergistic physical layer techniques, including \acrfull{ftn} signaling. %, \acrfull{sefdm}, \acrfull{fbmc}, and \acrfull{noma}. 
Numerous aspects of  handoff management in \acrshort{haps} systems are \textcolor{black}{described}. The notable contributions of \acrfull{ai} in \acrshort{haps}, including machine learning in the design, topology management, handoff, and resource allocation aspects are emphasized. \textcolor{black}{The extensive overview of the literature we provide is crucial} for substantiating our vision that depicts the expected deployment opportunities and challenges in the next 10 years (next-generation networks), as well as in the subsequent 10 years (next-next-generation networks). 
%We set forth various use-cases of \acrshort{haps} with mounted macro base stations, targeting mega-cities, and elaborate on the \acrshort{haps} mega-constellation as a robust approach  for providing ubiquitous coverage, high throughput, low delay  communication links as well as enabling smart services and applications. 
%Finally, the challenges and open issues are categorized according to the next 10 and 20 years, and many examples of each group along with tentative solutions and the possible road-maps are presented.
}

\end{abstract}

% Note that keywords are not normally used for peerreview  papers.

\IEEEpeerreviewmaketitle

\glssetwidest{HAPS-SMBS}
\hspace{2cm}\printglossary[style=alttree,type=\acronymtype,title=Abbreviations,nogroupskip, nonumberlist]
%\printglossary[type=\acronymtype]

\newpage 

\section{Introduction}

\begin{figure}[!tbp]
	\begin{center}
	\centering
\includegraphics[width=1\columnwidth]{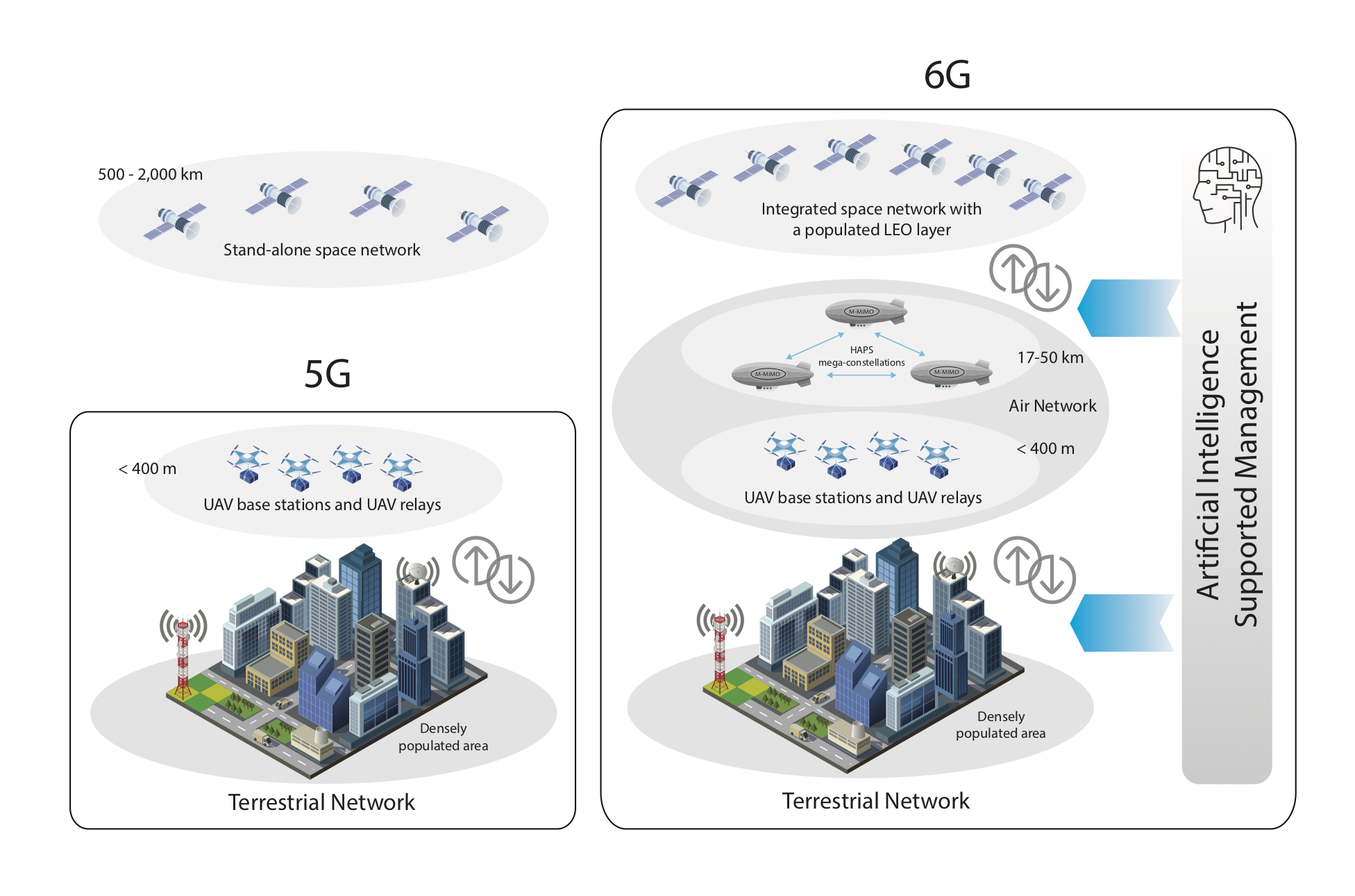}
\caption{\textcolor{black}{An overview \textcolor{black}{of} the transition from 5G to 6G. A fully integrated \acrfull{vhetnet} is envisioned in 6G.}}
\label{fig_6G}
\end{center}
\end{figure}

In the state-of-the-art   \acrfull{6g} network architecture,  \textcolor{black}{a vision of} a three-layer  \acrfull{vhetnet} is under discussion. This vision is consistent with the \acrfull{3gpp} activities regarding  \acrfull{ntn}, as defined in  \acrfull{tr} 38.811 \cite{T338811}\footnote{In this document, the terminology as defined by the \acrfull{itu} is used \cite{ITU-RR}.}. The three layers \textcolor{black}{consist of a } satellites (space) network, aerial network, and  terrestrial network \cite{cao2018airborne}\textcolor{black}{, as shown in Fig. \ref{fig_6G}.} \textcolor{black}{A} \acrfull{haps} is an integral component \textcolor{black}{in the} realization of the vision of \acrshort{vhetnet}s.

\textcolor{black}{A \acrshort{haps} is a network node that operates in the stratosphere at an altitude \textcolor{black}{of} around 20 km. Due to the unique properties of the stratosphere, a \acrshort{haps} can stay at \textcolor{black}{a} quasi-stationary position \textcolor{black}{and contribute significantly to the goal of ubiquitous connectivity}. %Since it is located above the strong wind currents in the atmosphere that take place between 10 and 15 km at mid-latitudes, \acrshort{haps} can serve as a quasi-stationary network element that can provide significant benefits to the ubiquitous connectivity goal.
}  \textcolor{black}{ \acrshort{haps}\textcolor{black}{-}related research activities can be traced back to 1990s \textcolor{black}{through} numerous research perspectives \cite{tozer2001high}. \textcolor{black}{This research showed promise for the advancement of ubiquitous connectivity, %to which \textcolor{red}{the recently terminated  Google Loon project have contributed 
\textcolor{black}{with a particular}  focus on rural areas and disaster relief applications. In what follows, we go a step further by showing how the use of HAPS systems can catalyze advanced mobile wireless communication services with ultra-wide coverage and high capacity.}
%In a nutshell, the prior art, including the deployments such as the Google Loon project, has targeted rural areas and disaster relief applications. %Despite the fact that \acrshort{haps} systems can bring many additional advantages to the telecommunication industry, almost all the \acrshort{haps}  projects are mainly targeting rural connectivity. Motivated by both the techologial advancements in the aeronoutical domaib abd developments  and the industry achievements in the \acrshort{haps} system, the investment in the \acrshort{haps} field has brought back some attentions.
%However, the use of \acrshort{haps} as a new promising platform can catalyze advanced mobile wireless communication services with ultra-wide coverage and high capacity.  
}
 %Eventually, HAPS has become a more viable network element both due to the evolution of the communication networks and also due to the advances in solar panel efficiency, battery energy density, lightweight composite materials, autonomous avionics, and antennas. With these advancements, in practice, the potential benefits of a HAPS can be substantially more than the conventional gains.}

\textcolor{black}{Recently,  \acrshort{haps} has been discussed as a viable aerial network component due to the evolution in communications technologies and  advances in solar panel efficiency, lightweight composite materials, autonomous avionics, and antennas. As  cost\textcolor{black}{s are} time-dependent and more cost-effective technologies and materials are emerging,  the use of \acrshort{haps} systems will become more economically feasible in the future networks.} %5\textcolor{brown}{Nonetheless, with the current technology and costs, it may still be more economical to use \acrshort{haps} than installing a ground-based  network for extending the coverage/increasing capacity in the suburban and remote areas \cite{arum2020review}.} 
\textcolor{black}{With the development of advanced materials and the realization of necessary technological leaps, it is expected that  new enablers will  gradually materialize in coming years. These research trends have resulted in \acrshort{haps} being actively considered \textcolor{black}{to be} a feasible technology for the future of wireless communication networks. Although the choice of energy source was considered as a fundamental issue in \acrshort{haps} \textcolor{black}{research}, solar power coupled with energy storage has been regarded as the primary means of providing energy for \acrshort{haps} \textcolor{black}{systems} since they have large surfaces suitable to accommodate solar panel films \cite{qiu2019air}.} \textcolor{black}{Moreover,} because of its low-delay characteristics in comparison with the emerging satellite networks, a \acrshort{haps} can provide wireless services directly to the users of  terrestrial networks \cite{hoshino2019study}.

 \textcolor{black}{\textcolor{black}{With ongoing disruption in wireless communication designs (e.g., data-driven designs) and emerging use cases (e.g., on-demand distributed machine learning platforms and data centers), \acrshort{haps} systems have become more appealing for their potential benefits.}
\textcolor{black}{From this perspective, a stand-alone balloon providing Internet access to a remote area is a limited example of what is possible}.  The era of portable data-centers, intelligent signal boosters,  flying macro base stations, and machine-learning \textcolor{black}{platforms capable of intelligent decision-making for a massive number} cargo drones and flying cars \textcolor{black}{has arrived}. In effect, we envision the future with a massive constellation of \acrshort{haps}s,  termed as  \textcolor{black}{a}   \textit{\acrshort{haps} mega-constellation}\footnote{
	According to ITU recommendations, a  \acrshort{haps}  should have a wide footprint of about 500 km in radius \cite{lTURF1500}. \textcolor{black}{A} network of few multiple \acrshort{haps} can extend the coverage to serve a whole country. For example, \textcolor{black}{a HAPS constellation of 18 nodes is estimated to be sufficient to cover all of Greece, including all of its islands} \cite{milas2003interference}, \textcolor{black}{and} a constellation of 16 \acrshort{haps} \textcolor{black}{nodes} \textcolor{black}{are held to be sufficient to cover Japan} \cite{miura2001wireless}.} (analogous to \textcolor{black}{a} satellite mega-constellation), enabling high capacity network access, computation offloading, and data analytics tools, to millions of users/devices not only in suburban areas but also in dense urban areas, as shown in Fig. \ref{fig_6G}.}
 \textcolor{black}{As depicted in this figure, \textcolor{black}{we can summarize our proposed framework} as follows: \begin{enumerate}
    \item The \acrshort{haps} layer, \textcolor{black}{performing} as a large-scale intelligent %reflector,
    \textcolor{black}{entity,} enables fast, reliable, and efficient long-distance communication between  satellites, bypassing the need for the installation of millions of \textcolor{black}{ground and offshore relay stations} \cite{Handley2019mega}. It can also function as a distributed data-center for recording the orbital path\textcolor{black}{s} of satellites, monitoring conjunction alerts, and calculating the probability of \textcolor{black}{a} collision between satellites. The availability of such information in \textcolor{black}{a timely} manner \textcolor{black}{for different}  satellite companies is vital for the preservation of the functionalities of the satellite mega-constellations. \textcolor{black}{In addition}, satellites help the \acrshort{haps} layer in improving the handoff performance. 
\item The \acrshort{haps} layer is responsible \textcolor{black}{for managing} the mobility of \textcolor{black}{a} swarm of \acrfull{uav} \textcolor{black}{by} providing edge intelligence, offloading  heavy computations, and handling large-scale sensing and monitoring, which are useful for cargo delivery and  monitoring systems. The communication platform is expected to smoothly handle diverse communication requirements\textcolor{black}{,} such as \acrfull{urllc} and \acrfull{embb} communications. 
\item The \acrshort{haps} layer provides fast Internet access and wireless communication services, such as IoT and distributed machine learning, to urban, suburban, and remote areas, reducing the reliance on  terrestrial and satellite networks.\end{enumerate} }

\textcolor{black}{\textcolor{black}{Due to these capabilities}, we envision that the use of \acrshort{haps} systems can be a remedy to the architectural problems that will be encountered as the use of aerial components  \textcolor{black}{increases in} wireless networks. The \textcolor{black}{use} of \acrshort{haps} systems as new wireless access platforms for  future wireless communication \textcolor{black}{systems has great potential, and its associated promises are detailed in this paper.}}

\subsection{Survey and Overview Articles on HAPS}
\textcolor{black}{Several overview articles have been published on the use of \acrshort{haps} \textcolor{black}{systems} as  communication platform\textcolor{black}{s}. \cite{karapantazis2005broadband} provide\textcolor{black}{d} a summary of the essential technical aspects of \acrshort{haps} \textcolor{black}{systems,} including possible architectures \textcolor{black}{and} cell formations  as of 2005. \textcolor{black}{A summary of \textcolor{black}{current} and potential applications as well as  past field trials along with open technical issues \textcolor{black}{was given} in \cite{widiawan2007high}.} The commercial and research project  deployments of \acrshort{haps} systems as of 2009 \textcolor{black}{was} provided in \cite{gavan2009concepts}.  The literature on the use of optical communications in \acrshort{haps} platforms \textcolor{black}{was} presented in \cite{fidler2010optical}.   The authors of \cite{aubineau2010itu} provide\textcolor{black}{d} an overview of  \acrshort{haps} related activities at past \acrfull{wrc} of the \acrfull{itu}, until 2010. \textcolor{black}{\cite{mohammed2011role}, published in 2011, present\textcolor{black}{ed} an overview of possible architectures for using \acrshort{haps} \textcolor{black}{systems} to provide global connectivity.}   \cite{d2016high} provide\textcolor{black}{d} a survey of the technological changes as of 2016.  %Potential and the existing applications, a summary of the field trials, and open technical issues are enumerated in \cite{widiawan2007high}.
{The books} \cite{ aragon2008high, grace2011broadband} demonstrate\textcolor{black}{d} the use of \acrshort{haps} nodes \textcolor{black}{as either a stand-alone or complementary part of a terrestrial network}. \cite{cianca2005integrated} describe\textcolor{black}{d} the integration of the satellite systems and  \acrshort{haps} \textcolor{black}{systems}, paving the way for hybrid terrestrial-satellite communication system\textcolor{black}{s}. Findings from experimental studies from the HeliNet project \textcolor{black}{were} presented in \cite{thornton2001broadband}.  \textcolor{black}{The work in }\cite{wang2014high} demonstrate\textcolor{black}{d} that \acrshort{haps} accommodates a suitable system for wide-area synthetic aperture radar (SAR) imaging in a microwave remote sensing. } 
\textcolor{black}{ %Wireless power transfer to provide power to \acrshort{haps} systems is investigated in \cite{gavan2010microwave}. 
An overview \textcolor{black}{of} the channel models for \acrshort{haps} \textcolor{black}{and} satellite systems for both \acrfull{siso} and \acrfull{mimo} antenna systems \textcolor{black}{was given} in \cite{arapoglou2011land}. The book \cite{kanatas2016radio} provide\textcolor{black}{d} an in depth overview of the channel models in \acrshort{haps} and satellites. \textcolor{black}{Although the works described above provided overviews of past and suggested use cases, their main focus was} on  the use of \acrshort{haps} \textcolor{black}{systems} in sparsely populated areas or areas with underdeveloped infrastructure.}

\textcolor{black}{With respect to UAV communications, which mainly cover low to medium altitude platforms, there are many survey and tutorial articles that  extensively \textcolor{black}{overviewed} the related literature\footnote{It is worth \textcolor{black}{noting} that while 3GPP technical reports, e.g., TR 22,829, \textcolor{black}{use} the term ``UAV'' for low-altitude vehicles with altitudes \textcolor{black}{of} roughly up to 150 m, the general literature sometimes uses this term for broader applications \textcolor{black}{that range} from low to medium and occasionally to high altitude platforms \cite{Mozaffari2019SurveyUAV, Zhang2019SurveymmWaveUAV}. In this paper, we stick to the \acrshort{itu} definition\textcolor{black}{,} unless otherwise \textcolor{black}{noted}.}. Among them\textcolor{black}{,}  \cite{Zeng2019proceedingUAV} provide\textcolor{black}{d} a comprehensive tutorial of the subject, review\textcolor{black}{ed}
  recent developments, and highlight\textcolor{black}{ed} important open issues. The authors extensively discuss\textcolor{black}{ed} the problem of joint rate allocation and trajectory design in UAV systems. \textcolor{black}{They also comprehensively overviewed channel modeling in UAV systems, including UAV to ground and \acrshort{bs} to UAV configurations.}  A review of recent development\textcolor{black}{s in}
 UAV communications from a 5G perspective is given in \cite{Mozaffari2019SurveyUAV}. \textcolor{black}{T}he authors also discuss\textcolor{black}{ed} several problems in UAV trajectory design \textcolor{black}{as they pertained to communication requirements and the limited battery life of UAVs} for IoT applications. Also, \cite{Zhang2019SurveymmWaveUAV} comprehensively overview\textcolor{black}{ed} the use of mmWave communications in low-altitude platforms. \cite{Haijun2019SurveyCyberUAV} survey\textcolor{black}{ed} the literature \textcolor{black}{on} UAV\textcolor{black}{s} from a cyber\textcolor{black}{-}physical system perspective by reviewing the three components of communication, computation, and control platforms in a versatile UAV system. However, acknowledging distinctive traits of \acrshort{haps} nodes in comparison to UAVs, these literature is not relevant to the scope and subject of this paper. For examples, important issues in UAV communications, among them the trajectory design with respect to limited energy on board, the management of temporar\textcolor{black}{y}, small-scale, and on-demand service provider\textcolor{black}{s} or computational platform\textcolor{black}{s} for  terrestrial networks, and the control and management of swarms of UAVs have very different scales in \acrshort{haps} systems. Instead of providing communication/computation service to a hand-full of users on demand, we now face \textcolor{black}{the prospect of doing} so for a coverage area of  60 km to 400 km. \textcolor{black}{Additionally}, we now face the problem of preserving the functionality \textcolor{black}{of HAPS systems} for a couple of months and preferably years. }

\textcolor{black}{The literature on \acrshort{haps} have progressed in a limited scale in the period between 2015 to 2018. Yet in 2018, along with possible research directions, a revival on the literature can be observed. A fresh view on the literature is given in the survey paper \cite{arum2020review}. Therein, the authors discuss\textcolor{black}{ed} how the coverage of a \acrshort{haps} can be extended \textcolor{black}{by} several order\textcolor{black}{s}  of magnitude compared to conventional use-cases, e.g., from 60 km coverage radius to about 500 km coverage radius. Different communication techniques\textcolor{black}{,} such as resource allocation, \acrshort{mimo} communications and advanced antenna systems, and handoff \textcolor{black}{were} listed as main enabling factors and many relevant papers \textcolor{black}{were} reviewed accordingly. We should note that \cite{arum2020review} mainly consider\textcolor{black}{ed}  communications issues of \acrshort{haps} systems almost related to 3G/4G technologies, while this work attempts to position \acrshort{haps} systems in the era of 5G and beyond by covering various applications of \acrshort{haps} systems for large-scale communications, intelligent relaying, computation offloading, and distributed machine learning. Furthermore, \cite{arum2020review} often \textcolor{black}{did} not explicitly distinguish between \acrshort{uav}\textcolor{black}{s} and \acrshort{haps} \textcolor{black}{systems} to the exten\textcolor{black}{t} that many developed ideas for the former \textcolor{black}{were} implicitly assumed to be (automatically) transferable to the latter, which, as  discussed above, may not be valid or precise. From a technological viewpoint, the use of \acrshort{haps} in \cite{arum2020review} is mainly restricted for the single-station applications for remote areas and disastrous situations. \textcolor{black}{By contrast, our main goal in this work}}  \textcolor{black}{is to shed light on undiscovered potentials of the \acrshort{haps}, where, in particular, dense-urban areas can greatly benefit from. \textcolor{black}{As the current research literature does not address the use of \acrshort{haps} systems in densely populated metropolitan areas,}  \textcolor{black}{the literature does} not fully reflect the potential of stratospheric platforms.} \textcolor{black}{ For an overview of the surveys and books on \acrshort{haps} systems refer to Table \ref{tab:surveys}. 
% These works consider the use of \acrshort{haps} systems in under served areas. In this work, by focusing on wireless architectural aspects, we extend their usage to densely populated metropolitan areas and provide a fresh new perspective on the deployment of \acrshort{haps} systems as a part of mega-constellations.
}
\begin{table}
\begin{center}
\centering
\caption{\textcolor{black}{An overview of survey papers and books on  \acrshort{haps} systems}}\label{tab:surveys} 
\begin{tabular}{ |M{1.2cm}|M{1cm}|M{2.2cm}|M{10cm}| }
 \hline
\textcolor{black}{Reference}& \textcolor{black}{Year} & %\textcolor{black}{Scope} &
\textcolor{black}{Focus} &  \textcolor{black}{Description}  \\  \hline  \hline
\textcolor{black}{\cite{karapantazis2005broadband}}& \textcolor{black}{2005} & %\textcolor{black}{Survey} &
\textcolor{black}{Wireless architecture} & \textcolor{black}{\begin{itemize}[leftmargin=*] \vspace{-0.3cm} 
 \item Summarizes the technical aspects and potential architectures of HAPS deployment.
 \item The survey is based on old use-cases of HAPS systems.
\end{itemize}
\vskip -0.3cm
} \\ 
 \hline
\textcolor{black}{\cite{widiawan2007high}} & \textcolor{black}{2007} & %\textcolor{black}{Survey} &
\textcolor{black}{Project deployments} & \textcolor{black}{\begin{itemize}[leftmargin=*]\vspace{-0.3cm}
     \item Summarizes the main concepts of HAPS technology, applications and field-trials.
 \end{itemize}\vskip -0.3cm}  \\ \hline
 
 \textcolor{black}{ \cite{aragon2008high}} & \textcolor{black}{2008} & %\textcolor{black}{Book} &
 \textcolor{black}{Wireless architecture (Book)} & \textcolor{black}{\begin{itemize}[leftmargin=*]\vspace{-0.3cm}
     \item Introduces the main concepts for HAPS as an alternative for telecommunications services.
 \end{itemize}\vskip -0.3cm} \\ \hline
 
  \textcolor{black}{\cite{gavan2009concepts}} & \textcolor{black}{2009} & %\textcolor{black}{Survey} &
  \textcolor{black}{Project deployments} & \textcolor{black}{\begin{itemize}[leftmargin=*]\vspace{-0.3cm}
     \item Description of the developments of HAPS projects and main potential applications.
 \end{itemize}\vskip -0.3cm}  \\ \hline
  \textcolor{black}{\cite{fidler2010optical}} & \textcolor{black}{2010} & %\textcolor{black}{Survey} &
  \textcolor{black}{Optical links} & \textcolor{black}{\begin{itemize}[leftmargin=*]\vspace{-0.3cm}
      \item \textcolor{black}{A review of the technologies, studies, and field-trials} for HAPS optical communication links.
  \end{itemize}\vskip -0.3cm} \\ \hline
     \textcolor{black}{\cite{aubineau2010itu}} & \textcolor{black}{2010} & %\textcolor{black}{Survey} &
     \textcolor{black}{Spectrum management} & \textcolor{black}{\begin{itemize}[leftmargin=*]\vspace{-0.3cm}
     \item A review of  technical studies for  HAPS developments in \textcolor{black}{past meetings of the WRC and} the ITU-R recommendations for HAPS systems.
 \end{itemize}\vskip -0.3cm} \\ \hline
 \textcolor{black}{  \cite{mohammed2011role}} & \textcolor{black}{2011} & %\textcolor{black}{Survey} &
 \textcolor{black}{Wireless architecture} & \textcolor{black}{\begin{itemize}[leftmargin=*]\vspace{-0.3cm}
      \item An overview of possible scenarios in which HAPS can be interconnected with terrestrial and satellite networks.
  \end{itemize}\vskip -0.3cm} \\ \hline
  \textcolor{black}{\cite{arapoglou2011land}} & \textcolor{black}{2011} & %\textcolor{black}{Survey} &
  \textcolor{black}{Channel model} & \textcolor{black}{\begin{itemize}[leftmargin=*]\vspace{-0.3cm}
     \item A survey of \textcolor{black}{measurement campaigns} and modeling approaches for HAPS and satellite communication links.
 \end{itemize}\vskip -0.3cm} \\ \hline
 
 \textcolor{black}{ \cite{grace2011broadband}} & \textcolor{black}{2011} & %\textcolor{black}{Book} &
  \textcolor{black}{Wireless architecture and communication links (Book)} & \textcolor{black}{\begin{itemize}[leftmargin=*]\vspace{-0.3cm}
     \item  Describes the basics of HAPS systems and the technological requirements for utilizing HAPS for broadband communications.
\item It also presents a roadmap for HAPS constellations in future networks. 
 \end{itemize}\vskip -0.3cm} \\ \hline

  \textcolor{black}{\cite{d2016high}} & \textcolor{black}{2016} & %\textcolor{black}{Survey} &
  \textcolor{black}{Project deployments} & \textcolor{black}{\begin{itemize}[leftmargin=*]\vspace{-0.3cm}
     \item An \textcolor{black}{overview of the historical development of HAPS technology, including discussions of technological advancements.}
 \end{itemize}\vskip -0.3cm} \\ \hline
   \textcolor{black}{\cite{Mozaffari2019SurveyUAV,Zhang2019SurveymmWaveUAV}} & \textcolor{black}{2019} & %\textcolor{black}{Survey} &
   \textcolor{black}{Wireless architecture and communication links} & \textcolor{black}{\begin{itemize}[leftmargin=*]\vspace{-0.3cm}
     \item These surveys discuss the future applications and challenges of aerial platforms and their use in mmWave communications. 
     \item These studies are mostly focused on low-altitude vehicles (UAVs).
 \end{itemize}\vskip -0.3cm} \\ \hline
 
  \textcolor{black}{\cite{arum2020review}} & \textcolor{black}{2020} & %\textcolor{black}{Survey} &
  \textcolor{black}{Wireless architecture} & \textcolor{black}{\begin{itemize}[leftmargin=*]\vspace{-0.3cm}
     \item \textcolor{black}{Considers the use of HAPS with legacy technologies, 5G, and beyond 5G applications.}
     \item Sparsely populated, under-served areas are considered.
 \end{itemize}\vskip -0.3cm} \\ \hline 
   \hline
  \textcolor{black}{\it This manuscript} & \textcolor{black}{2020}  & \textcolor{black}{Wireless architecture, and use-cases } & \textcolor{black}{\begin{itemize}[leftmargin=*]\vspace{-0.3cm}
     \item  \textcolor{black}{Presents a vision and framework} for  HAPS networks including discussion of  various use-cases. Also, general requirements, design issues, and important  parameters regarding each use-case are elaborated.
     \item The latest advancements in  HAPS  systems as well as the promising technologies and techniques for HAPS communication links  are discussed with a synergetic perspective.
     \item \textcolor{black}{Highlights the potential of} AI/ML to facilitate/empower  the  design,  topology  management,  handoff,  and  resource  allocation  aspects. 
     \item \textcolor{black}{Discusses important challenges and open issues.}
 \end{itemize}\vskip -0.3cm} \\ \hline
\end{tabular}
\end{center}
\end{table}

 \subsection{Contributions and Outline}
 
  \begin{figure}[!tbp]
	\begin{center}
	\centering
\includegraphics[width=1\columnwidth]{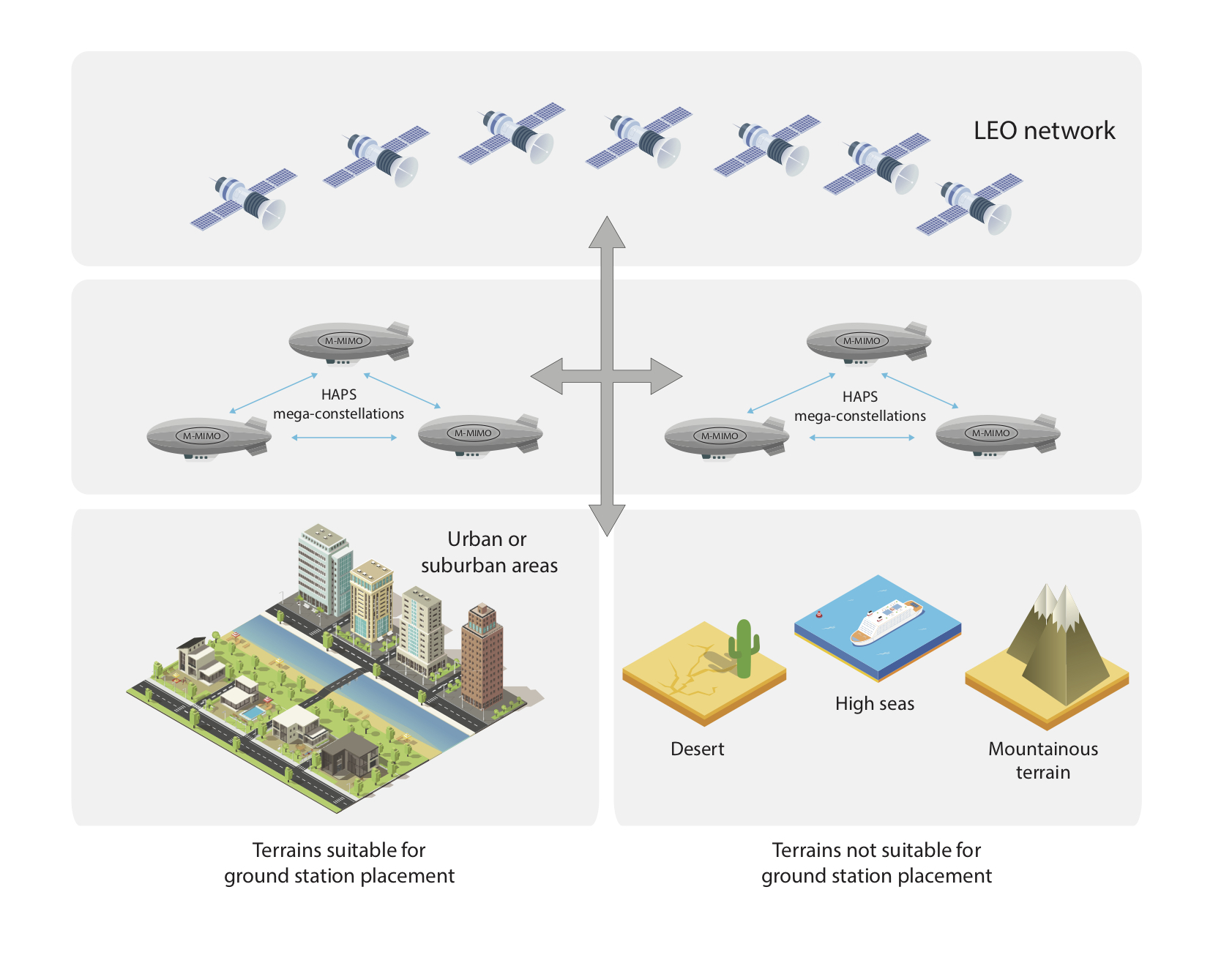}
\caption{\textcolor{black}{The vision of \acrshort{haps} mega-constellations over the next 20 years}, \textcolor{black}{bridging space and terrestrial networks} over densely populated urban centers, providing connectivity and computation even \textcolor{black}{over terrains} that are not suitable for ground network architectures. }
\label{fig_mega}
\end{center}
\end{figure}

 \textcolor{black}{The aim of this article is to \textcolor{black}{present promising research directions and a comprehensive overview of the current literature.} Although the use of \acrshort{haps} \textcolor{black}{systems can extend beyond functioning as an essential networking component (e.g., they can also be used for component wide-area synthetic aperture radar (SAR) imaging in microwave remote sensing \cite{wang2014high}), in this paper we mainly discuss} communication, computation, and networking aspects towards the \acrshort{haps}-mega constellations, as depicted in Fig. \ref{fig_mega}. \textcolor{black}{We distinguish between next-generation and next-next-generation deployment scenarios by looking to the future over the next 10 and 20 years, respectively.} Our main contributions are \textcolor{black}{as follows}:
 \begin{itemize}
     \item \textcolor{black}{\textcolor{black}{Presenting} use-cases of \acrshort{haps} systems and introducing the \textit{\acrshort{haps}-mounted  \acrfull{smbs}} as a promising and cost-effective solution for addressing  traffic demands \textcolor{black}{in the} \acrshort{5g} and \acrshort{6g} era\textcolor{black}{s}. Unlike  conventional macro \acrshort{bs}s, the envisioned \acrshort{haps}-mounted   \acrshort{smbs} not only enhances  coverage and capacity, but   also support\textcolor{black}{s} data acquisition, computing, caching, and processing in a plethora of application domains.}
     %We further elaborate on the HAPSs mega-constellation as a robust approach for providing fast broadband Internet access.
      {\color{black}
     \item \textcolor{black}{Describing recent advancements in \acrshort{haps} energy subsystems} and the latest \textcolor{black}{technological innovations} for communications payload\textcolor{black}{. At the same time, we highlight the evolution of \acrshort{haps} network architecture in accordance with the development of components of \acrshort{haps} systems.}
    %\item Sorting out historically the prominent past HAPS projects up to the most recent ones. The type classifications along with key features of each HAPS project  is briefly discussed. 
    %\item Highlighting the evolution of the HAPS network architecture, and also the  important and recent development in the components of the HAPS system. This includes the recent advancements in the HAPS energy subsystem and the latest technologies introduced for communications payload. % For the energy management subsystem, we describe the developments of the HAPS energy sources along with the most current technology achievements  in both primary and secondary HAPS energy sources. Also, we present efficient methods introduced in the literature for minimizing the HAPS energy consumption and maximizing the collected energy by the energy sources.
 %Regarding the communications payload, after describing the conventional active payloads and their characteristics and requirements, we introduce the promising technology of passive payloads offered by the Reconfigurable Smart Surfaces (RSS). In particular, we briefly describe  the technology features and the recent research and industry activities to investigate and utilize its benefits for communication systems.  Also, we illustrate the integration of RSS technology in HAPS systems and potential use cases as well as the associated benefits brought by such integration.
  \item  Introducing the use of \textit{\acrfull{rss}}  in the communications payload \textcolor{black}{of} \acrshort{haps} systems \textcolor{black}{and discussing their potential use cases and the benefits of such an integration.}}
     \item \textcolor{black}{ Providing a detailed review and discussion of the \acrfull{rrm}  and interference management schemes reported in the \acrshort{haps} literature in the past 20 years. \textcolor{black}{This includes a discussion of power control schemes, such as techniques that take into account mobility,} multicasting, and computational power for edge computing over \acrshort{haps}. Channel/sub-channel allocation and spectrum sharing as well as joint power, sub-channel and time allocations are  discussed. %The placement strategies for \acrshort{haps} where constellations are optimized in possibly multiple aerial network layers are highlighted as well. 
     Antenna and interference management are provided, including cell shape adaptation, coordinated multipoint transmission and platform diversity, massive \acrshort{mimo} for \acrshort{haps}, \textcolor{black}{as well as a discussion of motivations for} virtual massive \acrshort{mimo} over a \acrshort{haps} mega constellation. }
     \item     \textcolor{black}{Proposing suitable waveform designs and multiple access techniques for \acrshort{haps} communication links,  where  potential technologies such as \acrfull{ftn} signaling, \acrfull{sefdm}, \acrfull{fbmc}\textcolor{black}{,} and \acrfull{noma} are also extensively discussed.} %Finally, in Section \ref{RRM_Challanges} we highlight the key challenges and open areas in RRM for HAPs where we talk about the role of AI enivisioned for HAPs' RRM, as well as the considerations that must be accounted for towards 6G.
    \item \textcolor{black}{Addressing mobility management by discussing both inter-\acrshort{haps} and intra-\acrshort{haps} handoff algorithms used in \acrshort{haps} systems and some critical issues that need to be considered in future \acrshort{haps} systems. We also highlight existing techniques in \acrshort{haps} network management and how \acrshort{haps} networks can benefit from the application of  softwarized techniques\textcolor{black}{,} such as network slicing, software defined networks\textcolor{black}{,} and network function virtualization. %Also, the section describes the role of HAPS systems in supporting computing in aerial networks and computational offloading from/to terrestrial networks. In addition, several issues to consider in future HAPS network management and computational support are discussed as well. Section IX explored the current use cases that exploited AI in HAPS systems, and the necessary AI enablers in future HAPS systems are introduced in points.
    }
     \item \textcolor{black}{\textcolor{black}{Describing} the unique role 
     \textcolor{black}{of \acrfull{ai} and \acrfull{ml} in design, topology management,}
     {handoff\textcolor{black}{,} and resource allocation in \acrshort{haps} communication systems}.}
     \item \textcolor{black}{Elaborating on various challenges that the widespread implementation of \acrshort{haps} may encounter in coming years. We categorize the challenges and open issues \textcolor{black}{two groups, next-generation (next 10 years) and next-next-generation (10-20 years), and provide numerous examples of each} group along with tentative solutions and  possible roadmaps. }
 \end{itemize}
 This article is organized as follows. 
 \textcolor{black}{In the next section (Section II), we describe next-generation \acrshort{haps} use-cases. In Section \ref{sec:regulations}, we focus on aviation and spectrum regulations} that aim to harmonize the worldwide usage of \acrshort{haps}. In Section \ref{sec:characteristics}, 
 we describe the main components of \textcolor{black}{a} \acrshort{haps} communication system \textcolor{black}{and} its onboard subsystems, while highlighting  prominent past and recent projects.   The channel models that characterize the performance limits of the \acrshort{haps} nodes are presented in Section \ref{sec:channels}. Section \ref{resource} provides a comprehensive perspective on  radio resource management and interference management of \acrshort{haps} nodes from \textcolor{black}{an} overall network performance perspective. The %mobility 
{handoff} management of \acrshort{haps} nodes, in accordance with the existing terrestrial networks is detailed in Section \ref{sec:handoff}. In Section \ref{sec:management}, \textcolor{black}{we outline the network management perspective}. The indispensable role of AI is %details
{detailed} in Section \ref{sec:AI}.  In Section \ref{sec:open}, in addition to the next-generation networks' needs, the open issues that need to be addressed in the  in the next 20 years, and the associated  challenges,  are listed}. Finally, \textcolor{black}{we present our conclusions} in Section \ref{sec:conclusion}.

\textcolor{black}{\section{Promising Use-Cases for HAPS \textcolor{black}{Systems} in Next-Generation Networks}}\label{sec:HAPS-use-cases20}
%\subsection*{Conventional Use Cases}
%HAPS is a mature research field since late 1990's with many distinct areas of investigation, all been relevant to urban areas including the Google loon project to bring remote parts of the globe online and for disaster monitoring. Some of the conventional applications are:}

%\textcolor{black}{\textbf{HAPS to support sparse populations:}
%Limited by the network coverage, terrestrial communication systems can not provide wireless access services to sparse populations especially in harsh environments like rural, ocean, and mountains with high data rate and reliability. Due to larger footprint, HAPS can accommodate diverse services and applications with different quality of services in such scenarios.}

%\textcolor{black}{\textbf{HAPS act as a back-up to support disaster disrupted infrastructure:}}

%\textcolor{black}{\textbf{HAPS to provide coverage for some known temporary hotspot areas:} Like Olympic Games venues}

\begin{table*}
\centering
\caption{Complementary features of \acrshort{haps} systems compared to \acrfull{leo} satellite}
\label{Table:HAPSveLEO}
\begin{tabular}{|m{4cm}|m{13cm} |}%{|l|l|l|}%{|p{6cm}|p{4cm}|p{6cm}|}
\hline
Advantage & Description\\ 
\hline 
Low altitude deployment with favorable channel conditions &  $\circ$ \acrshort{haps} constellation deployments are expected to be at a low altitude when compared to \acrshort{leo} satellites located from 350 km to 2000 km, leading to a favorable link budget and a high \acrfull{snr} for the downlink providing a coverage advantage. Considering the uplink connectivity, the relatively low path loss enables the use of \acrshort{ue}s as the terminals which have limited transmit power levels, without the need for specialized ground stations.  \\ \hline
Almost \textcolor{black}{stationary} positions &  $\circ$ \acrshort{leo} satellites can cross over continents within several minutes due to their high speeds. As a result, \textcolor{black}{some of the capacity of \acrshort{leo} satellite communication is wasted while they pass over oceans} and underpopulated areas. \textcolor{black}{By contrast, the} relatively stationary position of \acrshort{haps} systems prevents a waste of capacity.  \\ &  $\circ$ \textcolor{black}{The stationary position of the links} avoids the introduction of a significant Doppler shift.
%Due to the stationary of the links, the effect of significant Doppler shift can be avoided.}
%The stationary of the links avoids the introduction of a significant Doppler shift.  
\\ \hline
\textcolor{black}{{Smaller footprint with a large surface volume}} &  $\circ$ \textcolor{black}{A \acrshort{haps} system has a smaller footprint compared to \acrshort{leo} that provisions a higher area throughput.} Due to its large volume,  a \acrshort{haps}  is suitable for \acrshort{mimo} and massive-\acrshort{mimo} deployments. \textcolor{black}{Aided by} multi-antenna arrays, \acrshort{haps} systems can generate highly directional \acrshort{3d} beams with narrow beamwidths that improve the \acrshort{sinr} for all users. \\
&  $\circ$ The larger volume of \acrshort{haps} systems can be equipped with huge solar panels and energy storage systems. Due to advancements in solar panel efficiency and energy-storage, \acrshort{haps} systems \textcolor{black}{can stay airborne for a long period of time with minimal energy consumption.} \\ 
\hline

Reduced round-trip delay & $\circ$ \textcolor{black}{{Due to a lower altitude, a \acrshort{haps} system corresponds}} to a round trip delay of 0.13 to 0.33 ms which makes them a good option for 
\textcolor{black}{low-latency applications, such as \acrshort{urllc}.} Hence, \textcolor{black}{a \acrshort{haps}} constellation-based communication system can overcome the inherent high-latency problem of satellite networks. \\ \hline

Deployment and maintenance advantages & $\circ$ The costs and risks \textcolor{black}{of deployment are lower} in the case of \textcolor{black}{\acrshort{haps} systems compared to \acrshort{leo}s.} Moreover, \acrshort{haps} systems are easier to bring back to earth once they finish their \textcolor{black}{mission, while satellites are not recoverable.} \\ \hline
%mission while satellites may turn to debris. 
 
\end{tabular}
\end{table*} 
%The envisioned HAPS-based wireless
%access architecture is a compelling alternative

%\subsection{Promising Use-cases}
\textcolor{black}{ \acrshort{haps} systems have promising advantages over satellite communications, as summarized in Table \ref{Table:HAPSveLEO}. These \textcolor{black}{advantages will make them} an indispensable component \textcolor{black}{for next-generation} wireless networks.} \textcolor{black}{ Conventional} wireless communication services provisioning using \acrshort{haps} systems are limited to rural and remote areas to provide broadband access as an alternative to terrestrial systems and for \textcolor{black}{disaster relief} \cite{arum2020review}, mainly \textcolor{black}{targeting areas with low user densities}. However,  communication services in urban and suburban areas are heavily concentrated with an ever-increasing demand. \textcolor{black}{The \acrshort{haps}-based wireless access architecture we envision presents a compelling alternative} to terrestrial network densification due to the possibility \textcolor{black}{of using one platform} for multiple applications, as detailed below. Table \ref{Table:summary} summarizes the features \textcolor{black}{of our envisioned \acrshort{haps} systems} over conventional \acrshort{haps} systems.

\begin{table*}
\centering
\caption{Features of the envisioned \acrshort{haps} systems \textcolor{black}{compared to} conventional \acrshort{haps}}
\label{Table:summary}
\begin{tabular}{|p{2.4cm}|p{6cm}|p{8.3cm}|}%{|l|l|l|}%{|p{6cm}|p{4cm}|p{6cm}|}
\hline 
\textcolor{black}{Aspect of comparison} & Conventional \acrshort{haps} & Envisioned \acrshort{haps}\\ 
\hline 
Application scenarios & Rural and remote areas, emergency \textcolor{black}{cases.} & Urban and suburban areas in addition to remote \textcolor{black}{areas.}\\
\hline
\textcolor{black}{Population density} & \textcolor{black}{Applicable only \textcolor{black}{to regions with low user density.}} & \textcolor{black}{Suitable for areas with high user \textcolor{black}{density.}}\\
\hline
\textcolor{black}{Goals} & \textcolor{black}{Extending the coverage of \textcolor{black}{a terrestrial network.} } & \textcolor{black}{Maximizing the achievable capacity to cover a lot of \textcolor{black}{users.}  } \\
& & \textcolor{black}{Guaranteeing low latency for mission-critical \textcolor{black}{applications.}} \\
\hline
\textcolor{black}{Functions} & Providing connectivity for ground \textcolor{black}{users.} & In addition to connectivity, supporting computation, control, and \textcolor{black}{caching.}\\
& & Connecting \acrshort{uav} and satellite mega-constellation \textcolor{black}{nodes.} \\
\hline
 \textcolor{black}{Target use-cases} & \textcolor{black}{Broadband coverage, \textcolor{black}{Internet} access, natural disaster recovery, and environment \textcolor{black}{monitoring.}} & \textcolor{black}{ \acrfull{iot} applications, intelligent transportation systems, high-stake cargo drones, high-capacity \acrfull{ar}/\acrfull{vr} applications, temporary unpredictable events, computation {offloading}, and \textcolor{black}{filling coverage gaps.}}\\
\hline
 \textcolor{black}{Coexistence} & \textcolor{black}{As an alternative to terrestrial \textcolor{black}{networks.}}  & \textcolor{black}{As \textcolor{black}{a complement to} terrestrial} \textcolor{black}{and satellite \textcolor{black}{networks.}}\\
\hline
\textcolor{black}{Network type} & \textcolor{black}{Related to \acrshort{3g}/\acrshort{4g} \textcolor{black}{technologies.}} & \textcolor{black}{Related \textcolor{black}{to the \acrshort{5g}} and beyond \textcolor{black}{era.}} \\
\hline
\textcolor{black}{Deployment} & \textcolor{black}{Single \acrshort{haps} in isolation to provide coverage and \textcolor{black}{capacity.}} & \textcolor{black}{Multiple \acrshort{haps} systems forming a network to provide coverage and \textcolor{black}{capacity.}}  \\
\hline
\end{tabular}
\end{table*}

\subsection{\textcolor{black}{HAPS-Mounted} Super Macro Base Station (HAPS-SMBS)}
 
A macro \acrshort{bs} is a crucial component in wireless access architectures to provide coverage and support capacity. Currently, the concept of network densification through small cell deployments has been widely acknowledged in \acrshort{4g}, \acrfull{lte}, and \acrshort{5g} standards to address the requirements of coverage and capacity in terrestrial networks \cite{bhushan2014network}. However, the communication needs of metropolitan
areas \textcolor{black}{are constantly increasing, and small cell deployments are not up to the task of matching this ever-increasing demand \cite{andrews2016we}.}
%However, the communication needs of metropolitan areas are very high and constantly increasing. Hence, small cell deployments will become insufficient to address the ever-increasing demand, as they may fail to solve the supply and demand matching problem \cite{andrews2016we}. 
Although network coverage and capacity can be improved through \textcolor{black}{the addition} of \acrshort{uav}-mounted \acrshort{bs}s, their \textcolor{black}{\acrfull{swap}} constraints 
\textcolor{black}{limit the lifetime and coverage area of \acrshort{uav} \acrshort{bs}s.}
%Moreover, the coverage and capacity improvement through the use of UAV mounted aerial BSs is a well-studied topic in the literature \cite{}. However, due to their size, weight, and power (SWAP) constraints, the applications of UAV BSs are limited to small coverage area in general. 
Also, the mobility of \acrshort{uav} \acrshort{bs}s introduces a fast on/off restriction, where the \acrshort{bs} needs to be activated/deactivated very rapidly. 

\begin{figure}[!htbp]
	\begin{center}
	\centering
\includegraphics[width=0.9\columnwidth]{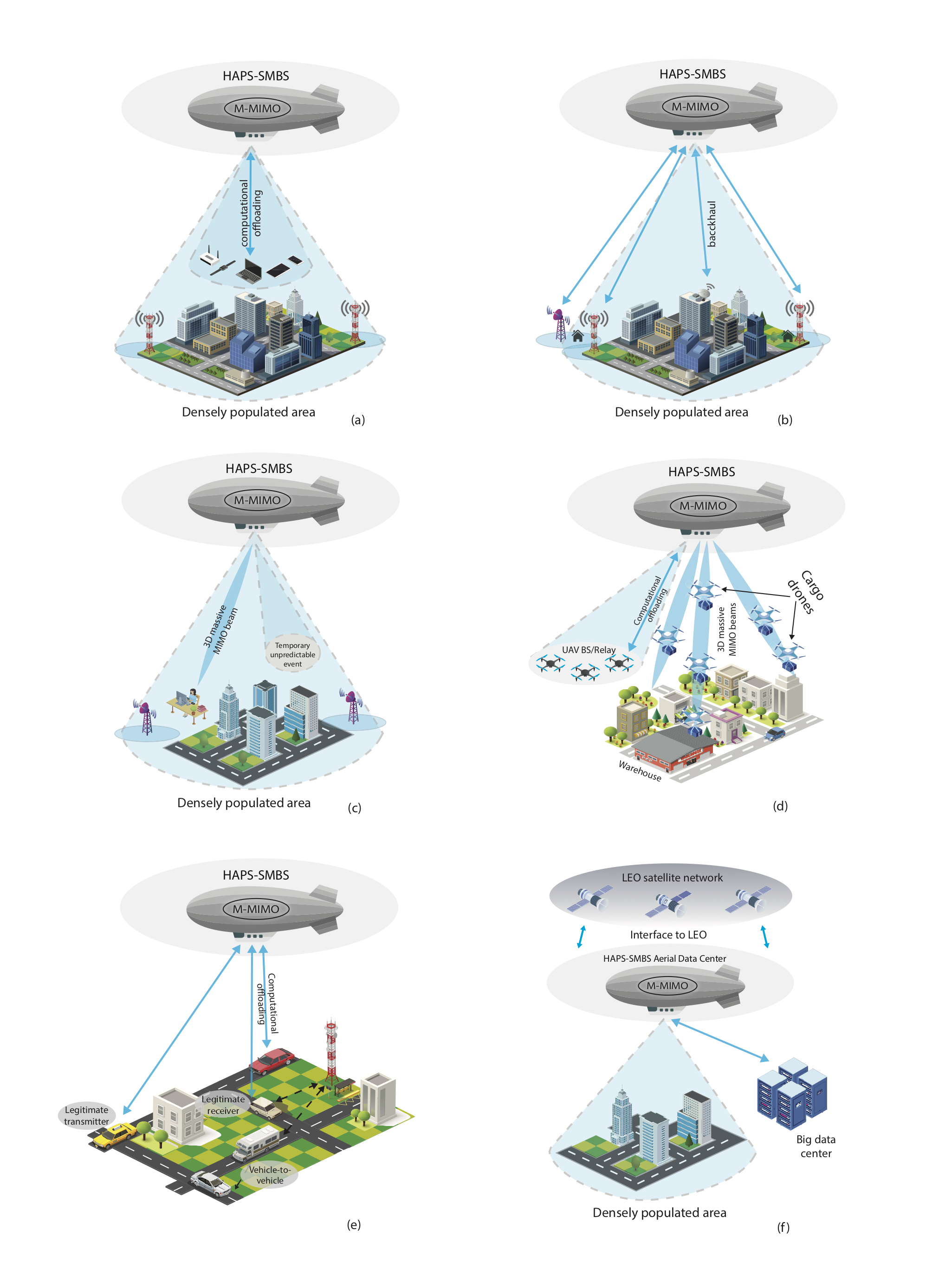}
\caption{(a) \textcolor{black}{A HAPS-SMBS delivering} \acrshort{iot} services. (b) \textcolor{black}{A HAPS-SMBS backhauling} small and isolated \acrshort{bs}s. (c) \textcolor{black}{A HAPS-SMBS covering unplanned} events and \textcolor{black}{filling coverage gaps.} (d) \textcolor{black}{A HAPS-SMBS supporting and managing} aerial networks. (e) \textcolor{black}{A HAPS-SMBS supporting} intelligent transportation systems. (f) \textcolor{black}{A HAPS-SMBS acting as} an interface \textcolor{black}{with \acrshort{leo} satellites} \textcolor{black}{and as an aerial data center.}}
\label{Fig:usecases}
\end{center}
\end{figure}

\textcolor{black}{Compared to \acrshort{uav}s, \acrshort{haps} systems, which are inherently quasi-stationary, have a larger footprint, more computational power, and better \acrshort{los} communication links.} \textcolor{black}{A \acrshort{haps}-mounted \acrshort{smbs} (HAPS-SMBS) can therefore be regarded as a powerful platform to enhance connectivity.} \textcolor{black}{HAPS-SMBS systems, however, are not alternatives} to terrestrial \acrshort{bs}s; \textcolor{black}{instead, they are a complementary} solution for network management and control. The use of HAPS-SMBS systems to support \textcolor{black}{a terrestrial} communication network introduces agility and enables rapid capacity improvement solutions in an intelligent manner to address high and variable traffic demands. \textcolor{black}{With this agility in network design, average user demands can be met with a terrestrial network, and rapidly changing or unpredictable user demands can be addressed by a complementary HAPS-SMBS.} Due to \textcolor{black}{{a} larger volume,} the application of massive \acrshort{mimo} techniques can be exploited \textcolor{black}{in a HAPS-SMBS} to provide improved channel capacity. \textcolor{black}{Also, the use} of multiple coordinated HAPS-SMBS systems, equipped with multi-antenna arrays, can enable further flexibility of the extremely precise beams through a distributed \acrshort{mimo} set-up. \textcolor{black}{The coordinated use of multiple HAPS-SMBS systems is also envisioned for metropolitan areas.} \textcolor{black}{It should be added that,} unlike conventional macro \textcolor{black}{\acrshort{bs}s, HAPS-SMBS systems} not only \textcolor{black}{enhance} coverage and capacity, but also \textcolor{black}{serve as computational platforms.} \textcolor{black}{They function as intelligent frameworks} to enable communication, computation, and caching while exploiting the power of machine learning algorithms. \textcolor{black}{With these features,  the potential benefits of a HAPS-SMBS can be substantially \textcolor{black}{greater than} a conventional macro-\acrshort{bs}. \textcolor{black}{We envision that future HAPS-SMBS architectures will support} data acquisition, computing, caching, and processing in diverse application domains, as exemplified in Fig.~\ref{Fig:usecases}, and detailed below.}
\textcolor{black}{{These potential use cases have been recently presented in \cite{alam2020high}, however their overall general requirements to access the feasibility of deployments have not yet been discussed. In this article, we discuss the requirements in terms of design and technical analysis to attain these use cases with the goal of revealing their full potential. }}

% \begin{figure*}
% \centering
% \begin{subfigure}[b]{.49\linewidth}
% \includegraphics[width=0.8\linewidth]{HAPS_SMBS/HAPS_IoT.pdf}
% \caption{}\label{haps_iot}
% \end{subfigure}
% \begin{subfigure}[b]{.49\linewidth}
% \includegraphics[width=0.8\linewidth]{HAPS_SMBS/HAPS_Backhaul.pdf}
% \caption{}\label{haps_backhaul}
% \end{subfigure}

% \begin{subfigure}[b]{.49\linewidth}
% \includegraphics[width=0.8\linewidth]{HAPS_SMBS/HAPS_uncertain_event_1.pdf}
% \caption{}\label{haps_uncertain}
% \end{subfigure}
% \begin{subfigure}[b]{.49\linewidth}
% \includegraphics[width=0.85\linewidth]{HAPS_SMBS/HAPS_UAV_1.pdf}
% \caption{}\label{haps_UAV}
% \end{subfigure}

% \begin{subfigure}[b]{.49\linewidth}
% \includegraphics[width=0.8\linewidth]{HAPS_SMBS/HAPS_CAV.pdf}
% \caption{}\label{haps_cav}
% \end{subfigure}
% \begin{subfigure}[b]{.49\linewidth}
% \includegraphics[width=0.8\linewidth]{HAPS_SMBS/HAPS_DaTa_Center_1.pdf}
% \caption{}\label{haps_cav}
% \end{subfigure}
% \caption{(a) A HAPS-SMBS to deliver \acrshort{iot} services. (b) HAPS-SMBS for backhauling small and isolated \acrshort{bss}. (c) HAPS-SMBS to cover unplanned events and defeat coverage holes. (d) HAPS-SMBS to support and manage aerial networks. (e) HAPS-SMBS to support intelligent transportation systems. (f) HAPS-SMBS as an interface to \acrshort{leo} satellites and aerial data center.}
% \label{Fig:usecases}
% \end{figure*}

\subsection{Use-Cases for \textcolor{black}{HAPS-SMBS Systems}} \label{sec:USECases}

\subsubsection{{{HAPS-SMBS systems to support IoT services}}}
\textcolor{black}{ 
It is expected that HAPS will play a key role \textcolor{black}{in supporting} diverse \acrfull{iot} applications \cite{sibiya2019reliable,gineste2017narrowband,abu2019performance}. The ever-increasing \textcolor{black}{proliferation of \acrshort{iot} technologies presents} substantial challenges to the research community \textcolor{black}{in terms of addressing connectivity}, reliability, and latency requirements \textcolor{black}{of a massive} number of connected devices. In this context, current \textcolor{black}{infrastructures and methods} of designing wireless access architecture are rather limited and incapable of supporting these highly demanding wireless systems and services.
%In past years, there are several research projects on HAPS, however, they are limited to civil applications only such as disaster monitoring, earth observation, etc. For example, the Google Loon project aims at delivering connectivity to people in unserved and underserved communities around the world by deploying networks through stratospheric balloons acting as cellular BSs. 
The wide footprint of \acrshort{haps} systems  is ideal for providing \textcolor{black}{greater coverage} to a high number of \acrshort{iot} devices each with low-rate links. In addition, \acrshort{iot} devices might be located in areas where there is no terrestrial network coverage (e.g., \textcolor{black}{forests, mountains, oceans, etc.).} \textcolor{black}{HAPS-SMBS systems, therefore, are an} attractive solution to complement terrestrial networks to collect data from \acrshort{iot} devices and provide reliable uplink connections \textcolor{black}{for them} in a seamless, efficient, and cost-effective manner, as shown in Fig. \ref{Fig:usecases}(a).}

\textcolor{black}{To support \acrshort{iot} devices on the ground from HAPS-SMBS, a natural question that \textcolor{black}{will arise will be that of} the required transmission power of such \acrshort{iot} devices to communicate directly \textcolor{black}{with a HAPS-SMBS} located at least 20 km away. This becomes even more accentuated for some particular applications with IoT devices that are expected to function \textcolor{black}{over a time span} of decades \textcolor{black}{without requiring batteries to be recharged or replaced.} Among other things, the required transmission power of devices is proportional to the required received SNR to guarantee the QoS, which is inversely proportional to the transmission rate. As \acrshort{iot} devices transmit data at a very low bit rate when they are on (note that devices may occasionally turn on and stay off for long \textcolor{black}{periods of time}), \acrshort{iot} devices with low data transmission rates are therefore capable of communicating directly with a HAPS using low transmission power.}

\textcolor{black}{\textcolor{black}{Instead} of handling massive machine type communications (mMTC) that we have in terrestrial \textcolor{black}{networks, however, we} need to handle mx-MTC $(x>>1$) \textcolor{black}{in the case of a HAPS-SMBS} due to its large coverage area. This will introduce many unprecedented \textcolor{black}{challenges due to the need to design} efficient multiple access techniques \textcolor{black}{for the simultaneous transmission of a massive} number of devices. System \textcolor{black}{designers} may also need to strike a balance between reducing packet collisions which \textcolor{black}{reduce} the need for frequent packet re-transmission attempts---given the very strict energy consumption limit of some devices and/or the strict delay requirements of some applications---and the reliability requirements of mission-critical applications, which may call for re-transmission of the packets.}

\subsubsection{\textcolor{black}{{HAPS-SMBS systems  for backhauling small and isolated BSs}}}
%Although the concept of small cell base station has been widely acknowledged and studied for extremely high data rate coverage in \acrshort{4g} \acrfull{lte} wireless framework and still perceived as a \acrshort{5g} key enabler, this concept cannot be realized without cost-effective and scalable backhaul solutions to support such network densification through small cell deployments \cite{alzenad2018fso}.  
\textcolor{black}{Although fiber} optic communications remain a superior option for backhaul connectivity,  installing fiber for backhauling small cell \acrshort{bs}s may not be an efficient solution for many environments due to its high-cost \cite{dahrouj2015cost}. A cost-effective backhauling solution is the \textcolor{black}{use of wireless microwave links, and this is already a well-accepted approach}. Also, the \textcolor{black}{combination of the wider channel bandwidths of mmWave bands} and \acrshort{mimo} digital beamforming with high gain advanced antennas \textcolor{black}{makes mmWaves a viable solution} for in-band backhauling \cite{taori2015point}. 

\textcolor{black}{Although the distance \textcolor{black}{between a HAPS and a ground} femto-BS can \textcolor{black}{be between 20 km and 200 km} (depending on the coverage footprint of the HAPS), the communication link is almost \acrshort{los} dominant (with pathloss exponent of around 2) with moderate shadowing/fading fluctuations \textcolor{black}{due to a lack} of scattering. \textcolor{black}{Therefore, a rule of thumb calculation suggests} that a small-BS \textcolor{black}{with a 3D distance of 200 km} from the HAPS could gather almost the same average power gain that it \textcolor{black}{would} receive from a macro-BS but with a distance of 1000 meters. \textcolor{black}{As our vision advocates} the use of \textcolor{black}{HAPS systems} for metropolitan areas, industrial areas, and even mega-cities, the coverage \textcolor{black}{footprint of a HAPS} could be as small as 20-50 km. These figures become more \textcolor{black}{promising, however, for} a femto-cell \textcolor{black}{situated at a distance of 50 km from a HAPS while receiving an} average signal power similar to that \textcolor{black}{of a macro-BS at a distance} of about 100 meters. Furthermore, we should note that as femto-BSs are stationary and \textcolor{black}{HAPS systems are quasi-stationary} the \textcolor{black}{establishment and maintenance of such links is less demanding since beam tracking and beam adjusting are less necessary.} As a result, the extra communication \textcolor{black}{delay imposed by long distances between HAPS systems and femto-BSs} can be \textcolor{black}{compensated for with} occasional events of beam adjustment/establishment (compared to terrestrial networks).}

\textcolor{black}{When compared to terrestrial communications, mmWave communication links from \textcolor{black}{a HAPS} to ground femto-BSs may suffer from rain/cloud absorption losses (proportional to $10^{cr/H}$, where $H$ is the HAPS's altitude, $r$ is the distance between the ground and HAPS, and $c$ (dB/km) is a factor absorbing the rain/cloud effect \cite{Yuan19TropicalLoss}). This extra loss may not be an issue given \textcolor{black}{that a HAPS} can \textcolor{black}{compensate for these negative} effects by allocating higher transmission power and harnessing higher directional antenna gain due to possibility of installing very large \textcolor{black}{3D} antenna arrays. Additionally, as we also mentioned before, the link is not subject to \textcolor{black}{severe} shadowing/fading fluctuations, which could boost the average received power by about 10-20 dB compared to the counterpart links in terrestrial communications.}

\textcolor{black}{On the other hand, the de facto solutions are  mmWave or macro-wave communications from macro-BSs. Macro-wave communications are versatile and thus present the first option. However, it suffers from high signal attenuation given that macro-BSs are currently down-tilted and the fact that inter-cell interference is quite dominant. Hence, if possible mmWave communications combined with massive MIMO communications should be considered in order to manage the inter-cell interference.
Note that to ensure the quality of mmWave communications due to maneuverability of UAV-BSs, the Macro-BS should always stay in the communication view of the UAV-BSs. Such a requirement, nevertheless, could be limiting and complicates the mobility management of the UAV-BS, as instead of optimizing the mobility of UAV-BS merely for the best performance for the ground users or the provisioned service (shortest path for package delivery), one needs to include the backhaul communication requirements into account.
On the other hand, via backhauling from HAPS such issues are basically trivial as the mobility management can become only a matter of the quality of the service that the UAV-BS is providing.  However, one should note that compared to the backhauling for the ground femto-BSs, the matter is more involved as constant beam-tracking is required due to the mobility of UAV-BSs. }

\textcolor{black}{Inspired by recent advances in \acrshort{haps} systems and \acrshort{fso} \cite{alzenad2018fso,kaushal2016optical}, outdoor small cell \acrshort{bs}s can be backhauled  through HAPS-SMBS.} 
\textcolor{black}{Note that FSO links are vulnerable to weather conditions. \textcolor{black}{In the case} of cloudy, rainy, and foggy conditions, the quality of FSO links could substantially deteriorate. Hence, as \textcolor{black}{FSO links are} generally more robust \textcolor{black}{in clear weather} conditions, robust solutions to effectively deal with various weather conditions should be investigated. One straightforward solution might be to boost the robustness of the \acrshort{fso} backhaul links by considering hybrid mmWave \acrfull{rf}/\acrshort{fso} technologies  \cite{alzenad2018fso,kaushal2016optical,alsharoa2019facilitating,pham2015hybrid}, as depicted in Fig. \ref{Fig:usecases}(b). While this solution is feasible and has immediate merits, one should note \textcolor{black}{that since mmWave communication} may have smaller spectral efficiency \textcolor{black}{compared} to \acrshort{fso}, the backhaul data rate \textcolor{black}{between a HAPS-SMBS and a ground station} could be affected. As a result, apart from early detection for automatic switching between technologies, sophisticated resource allocations and 3D \textcolor{black}{beamforming may be necessary.}}

%\textcolor{black}{\textbf{HAPS-SMBS for backhauling outdoor small cell base stations:}
%Although the concept of small cell base station has been widely acknowledged and studied for extremely high data rate coverage in 4G LTE wireless framework and still perceived as a 5G key enabler, this concept has never found widespread application mainly due to the difficulty and cost in backhauling a high number small cell base stations \cite{alzenad2018fso}. There are two fundamental ways of backhauling, wired and wireless backhauling. Although, fiber is always the best option, however, installing fiber for backhauling small cell BSs may not be an efficient solution for many environments due to high cost \cite{dahrouj2015cost}. Hence, a cost-effective backhauling solution lies in wireless through microwave links or FSO links.}

%\textcolor{black}{Motivated by the recent advances in HAPS and FSO, backhauling outdoor small cell BSs can be realized through HAPS-SMBS \cite{alzenad2018fso,kaushal2016optical}, i.e., place the outdoor small cell BSs wherever appropriate without worrying about backhaul; then focus the FSO laser to the HAPS-SMBS for backhaul. On the other hand, the availability of HAPS-SMBS backhaul links can be further enhanced by considering hybrid RF/FSO links \cite{alzenad2018fso,kaushal2016optical,alsharoa2019facilitating}.}

\subsubsection{\textcolor{black}{{HAPS-SMBS systems to cover unplanned user events}}}
\textcolor{black}{ 
In case of unexpected and temporary events which are difficult to predict, such as flash crowds, wireless networks might require additional support to maintain ubiquitous connectivity \cite{bor2016new,mirahsan2017hethetnets}. Such events normally happen in crowded \textcolor{black}{cities and can lead} to network congestion. \acrshort{uav} mounted \acrshort{bs}s \textcolor{black}{have} recently gained \textcolor{black}{much attention for boosting wireless capacity} and offload traffic from congested terrestrial \textcolor{black}{\acrshort{bs}s} during such events \cite{alzenad2018coverage}. Compared to \acrshort{uav} mounted \textcolor{black}{aerial} \acrshort{bs}s, which have \acrshort{swap} constraints, \textcolor{black}{HAPS-SMBS systems provide greater capacity for ground users} due to their large \textcolor{black}{platforms, massive-MIMO capabilities, and higher transmission power.} Hence, the envisioned HAPS-SMBS architecture can \textcolor{black}{address the demands of unpredictable events} by increasing relevance between the distributions of supply and demand, as shown in Fig. \ref{Fig:usecases}(c).}

\textcolor{black}{To cover such temporary unplanned user events, \textcolor{black}{HAPS-SMBS systems} can be used opportunistically. Alternatively, these events can also be covered through over-engineering terrestrial networks. \textcolor{black}{In this case}, the expenses of HAPS-SMBS operations may be compared with the expenses of over-engineering \textcolor{black}{a terrestrial network.} {Despite revenues, providing connectivity to these scenarios is important to avoid serious losses and poses challenging demands such as high data rate.} Nevertheless, as massive-\acrshort{mimo} is among the most disruptive technologies to provide capacity improvement in ground networks, the promise of this technology \textcolor{black}{in HAPS-SMBS systems needs} to be investigated. Also, other capacity improving techniques such as NOMA, mmWave, beamforming, and any combination of them in HAPS scenario need to be revisited. In general, advanced big data solutions are required to predict the occurrence of temporary events (along with some estimations with regards to the volume of produced traffic per geographic area and unit time) in order to properly provide resources including bandwidth, power, and computational capacity.}
%For example, a HAPS-SMBS can provide additional beam to support the instantaneous capacity requirements in densely populated areas.}

%\textcolor{black}{\textbf{HAPS-SMBS to cover temporary unpredictable events:}
%In case of unexpected and temporary events which are hard to predict, for example, traffic jam, emergence communications, demonstration in a university and so on, wireless networks might require additional support to maintain ubiquitous connections \cite{bor2016new,mirahsan2017hethetnets}. The proposed HAPS-SMBS architecture can address this need by increasing relevance between the distributions of supply and the demand. For example, HAPS-SMBS can provide additional beam to support the instantaneous capacity requirements in densely populated areas. Such kinds of flash events normally happen in the cities with possibly a bigger number of gatherings, which leads to network congestion.}

\subsubsection{\textcolor{black}{{HAPS-SMBS systems as aerial data centers}}}
\textcolor{black}{HAPS-SMBS systems will also be able to operate as aerial data centers} to support agile computational offloading. As an example, \acrfull{ar} applications may require high computational capabilities. In this regard, efficient computational offloading will be a necessity \cite{kumar2013survey,wang2017computation}. As HAPS-SMBS systems can have more computational power than user terminals (e.g., \acrshort{uav} nodes or ground users), \textcolor{black}{may be useful for providing} different levels of computational services. Moreover, due to its high position, HAPS-SMBS can provide better coverage with \acrshort{los} links, avoiding the possibility of disconnection while offloading data. \textcolor{black}{As aerial data centers, HAPS-SMBS systems can also provide} a back-up computational facility.

\textcolor{black}{To envision a flying data center, \textcolor{black}{HAPS-SMBS systems} should have enough power from solar \textcolor{black}{panels} to support additional computation. This requires the investigation of how much power \textcolor{black}{a HAPS-SMBS} will require to support additional computation and how much solar power can be harvested. \textcolor{black}{Also, cooling is an important requirement of data centers.} The \textcolor{black}{atmospheric temperature at a HAPS' operating altitude is quite low (on average in the range of} $[-15^oC, -50^oC]$ \cite{NWS}), so we might not need too much energy for cooling as we can use the naturally low temperature around the HAPS. In addition, \textcolor{black}{the size of HAPS-SMBS data center} will be limited by the onboard payload capacity. Moreover, one of the important design issues in data centers is to reduce the response delays. Analyzing data in the sky will reduce response delays and decrease the burden on air-to-ground communication links.}

\subsubsection{\textcolor{black}{HAPS-SMBS systems to fill coverage gaps}}
\textcolor{black}{HAPS-SMBS systems can supplement existing terrestrial networks by filling coverage gaps} in a cost-effective manner. Coverage \textcolor{black}{gaps} are encountered when \textcolor{black}{terrestrial \acrshort{ue}s are faced with an insufficient \acrfull{sinr} from a terrestrial \acrshort{bs} due to a physical obstruction} \cite{bai2014analysis}. Such blockage effects become more severe for mmWave cellular networks and may have a higher negative impact on user associations. 

\textcolor{black}{To handle this problem, a HAPS-SMBS requires to steer a beam \textcolor{black}{in the targeted direction}. When compared to \acrshort{uav} \acrshort{bs}s, the advantage of using HAPS-SMBS systems is \textcolor{black}{their large platform size} and ability to perform \acrshort{3d} beamforming \cite{halbauer20133d} with massive \acrshort{mimo} \textcolor{black}{that allows to create disjoint narrow beams for each user in the \acrshort{3d} space.}
%that enables the creation of separate narrow beams in the three-dimensional space at the same time for different users. 
In addition, a HAPS-SMBS system can provide a permanent service rather than the temporary service of \acrshort{uav} \acrshort{bs}. This use-case is also shown in Fig. \ref{Fig:usecases}(c). Nevertheless, the creation of very narrow beams with higher capacity and accurate beam steering directions \textcolor{black}{{should be investigated for HAPS-SMBS systems to UEs}.} This can be problematic due chiefly to the long \textcolor{black}{distances} between users and HAPS-SMBS \textcolor{black}{systems, as} CSI estimation/feedback may render unaccepted delay and therefore outdated \textcolor{black}{beamforming} solution. As a result, \textcolor{black}{beamforming} and resource allocation needs to be less sensitive to accurate/up-to-date knowledge of the channel. In effect, solutions \textcolor{black}{that rely more on the long-term behavior} of the channel, for example, statistical CSI needs to be developed.}

\subsubsection{\textcolor{black}{HAPS-SMBS systems for supporting and managing aerial networks}}
Enhancing the computational capabilities of \acrshort{uav}s is becoming more important in order to maintain the critical tasks at \acrshort{uav}s. However, due to their \acrshort{swap} constraints, \acrshort{uav}s have limited onboard computational resources   \cite{cao2018mobile,messous2017computation}. \textcolor{black}{{A HAPS-SMBS system can be suited with powerful processors that can enhance the computation power of UAV networks elements with limited resources as a complement of terrestrial networks.}}
%that can provide computational support as a complementary of terrestrial network for limited-resources UAV network elements, to enhance the computation power of UAVs. 
\textcolor{black}{The larger coverage area of a single HAPS-SMBS \textcolor{black}{enables data collection} from large portions of the aerial network which reduces the dependency on terrestrial stations that are already overcrowded in most urban areas. Moreover, the effect of interference would be much higher in ground base stations compared to \textcolor{black}{HAPS-SMBS systems} for such computational \textcolor{black}{offloading} for UAVs.}
%The HAPS-SMBS system can be equipped with powerful processors that can provide computational support \textcolor{black}{as a complementary of terrestrial network} for limited-resources \acrshort{uav} network elements, %to enhance the he to enhance the computation power of \acrshort{uav}s. 
In addition, 
%due to its inherent characteristics of quasi-stationary position and large coverage area, HAPS-SMBS systems can dynamically learn about the network status, resources and topology, from large portions of the \acrshort{uav} network, and using machine learning algorithm, 
using \acrfull{ml} algorithms, %it
HAPS-SMBS can control and manage the \acrshort{uav} network intelligently with minimum dependence on terrestrial-based control, as exemplified in Fig. \ref{Fig:usecases}(d). 

\textcolor{black}{To control and manage UAV networks from \textcolor{black}{HAPS-SMBS systems}, seamless connectivity of UAV nodes with the HAPS-SMBS systems will be guaranteed. \textcolor{black}{HAPS-SMBS systems} should \textcolor{black}{guarantee reliable and wide} connectivity with relatively low latency. In addition, in \textcolor{black}{the near future}, in densely-populated urban areas thousands of cargo-UAVs are expected to be flying around daily. \textcolor{black}{To ensure their safe operation,} a massive amount of data about \textcolor{black}{them will need to be continuously collected} and analyzed. \textcolor{black}{For this}, on-board \textcolor{black}{processors} with enough power and cooling support \textcolor{black}{will be required.} }

\subsubsection{\textcolor{black}{HAPS-SMBS systems for supporting intelligent transportation systems}}
\textcolor{black}{ 
The full-scale introduction of \textcolor{black}{the}  \acrfull{its}/ \acrfull{cav} paradigm will be the most powerful automobile revolution  %the
\textcolor{black}{in history}  \cite{nikitas2019examining}. Recent advances in sensors, high-end computational units, and the introduction of in-car wireless communication capabilities have paved the way for \acrshort{cav} \textcolor{black}{that will enable} unprecedented scenarios for road transportation \cite{talebpour2016influence}. Nowadays, \textcolor{black}{automakers} are spending billions of dollars to \textcolor{black}{promote} the idea that \acrshort{cav}s can greatly reduce \textcolor{black}{road accident rates} and create a safer society. However, such breakthroughs will certainly create new challenges for the design and implementation of \acrshort{cav} infrastructure. For example, \acrshort{cav}s \textcolor{black}{will need to support services such as interacting with drivers, cooperating with other vehicles, offering decision-making support and strategies for traffic control and management.} Also, \acrshort{cav}s \textcolor{black}{will need to recognize their surroundings, plan a route, and control vehicular motion} without any human input. \textcolor{black}{But wide-scale data fusion} and processing are necessary for such \acrshort{cav} applications \cite{zhang2011data}. Interestingly, a HAPS-SMBS can play a key role in providing the ubiquitous coverage \textcolor{black}{for this} \acrshort{its}/\acrshort{cav} paradigm. \textcolor{black}{Since vehicles} may be \textcolor{black}{limited in their computational processing} capabilities, \textcolor{black}{they may need to offload data} \cite{wang2011real,stojmenovic2014fog}. Due to \textcolor{black}{their large coverage areas and significant computational} capabilities, a HAPS-SMBS can be used for data offloading \textcolor{black}{with minimal} communication delays. 
%However, due to vehicles high mobility data offloading is interrupted by frequent handoffs. In addition, the data processing outcome need to be delivered through the \acrshort{bs} which is  accessible by the vehicle. 
%A HAPS-SMBS can provide both the large area coverage and computational capabilities in low communication delays. 
%Thus, a HAPS-SMBS can eliminate the effect of frequent handoffs in vehicular networks. 
%Moreover, a HAPS-SMBS can have the wide view of a vehicular network which is essential for coordinating \acrfull{v2x} communications and optimizing the network performance. 
Moreover, HAPS-SMBS systems can provide coverage in rural and remote areas, which is essential for travailing on highways and using trains, flights, or ships (Fig. \ref{Fig:usecases}(e)).}

\textcolor{black}{\textcolor{black}{To support such operations,} information from vehicle sensor \textcolor{black}{nodes will need to be} forwarded to the HAPS-SMBS \textcolor{black}{that then will either act} as a \textcolor{black}{relay, forwarding the received signal} to a terrestrial \textcolor{black}{gateway, or will process} the received data on board and \textcolor{black}{send it back} to the vehicles with further instructions. This choice requires optimal \textcolor{black}{planning in the distribution of data offloading} and computing services in terrestrial and HAPS networks taking into account the delays of both communication and computation. Some other design issues for this system would be to support high QoS levels (delay, packet error, outage probability) \textcolor{black}{for vehicle-to-HAPS-SMBS telecommunications links} in order to ensure reliable and \textcolor{black}{fast message exchanges and} guarantee transport and safety applications.}

\textcolor{black}{\textcolor{black}{HAPS-SMBS systems} can also provide coverage for cargo drones \textcolor{black}{that are likely to disrupt} the retail industry in the near future. Usually, cargo drones are supported through terrestrial networks \cite{stanczak2018enhanced}. The use of cargo drones is currently being promoted by mega-retailers to carry courier packages. For instance, cargo drones can be used for Amazon's prime air drone delivery \textcolor{black}{service and the} autonomous delivery of emergency drugs \cite{bamburry2015drones}. In this scenario, a large number of cargo drones \textcolor{black}{will constantly be flying and filling the skies, and} hence \acrshort{3d} \textcolor{black}{highways are to be expected in support of} the cargo package distributions using these drones. A single HAPS-SMBS can be used to provide coverage for a high number of cargo drones in major cities.} 

\textcolor{black}{In this use-case, \textcolor{black}{HAPS-SMBS systems} should ensure reliable connectivity and safe \textcolor{black}{operations for cargo-drones} in the airspace, probably \textcolor{black}{involving a combination} of both radio-based \textcolor{black}{and vision-based} solutions. This requires \textcolor{black}{the provision of communications channels of high reliability and low latency} for many cargo-drones in a large geographical areas. Furthermore, as \textcolor{black}{HAPS-SMBS systems} can provide a computational platform for path-planning and navigation \textcolor{black}{in accordance with} supply-chain requirements, sophisticated solutions for massive computational offloading are required. }

\subsubsection{\textcolor{black}{HAPS-SMBS systems for handling LEO satellite handoffs and providing seamless connectivity}}
\textcolor{black}{The  high \textcolor{black}{speeds}  of \acrshort{leo} satellites \textcolor{black}{necessitate} frequent handoffs at terrestrial gateways \cite{sarddar2011handover}, which is undesirable. Fortunately, HAPS-SMBS systems can cover many satellites simultaneously due to its wide upper footprint.} \textcolor{black}{Therefore, to provide seamless \acrshort{leo} satellite connectivity to aerial and terrestrial networks, 
\textcolor{black}{a HAPS-SMBS} system can serve as an interface to manage the handoff in the \acrshort{leo} satellite network, as shown in   Fig. \ref{Fig:usecases}(f). \textcolor{black}{In this scenario, the frequent handoffs} of \acrshort{leo} satellites will be handled by \textcolor{black}{HAPS-SMBS systems}. Also, if ground users are able to communicate with \textcolor{black}{the HAPS-SMBS} interface directly, then there is no need for users to accommodate special devices for communication with \acrshort{leo} satellites.}

\textcolor{black}{There are two types of \textcolor{black}{links} in this system: \textcolor{black}{user-to-HAPS links; and HAPS-to-LEO satellite links.} The link between a user and a HAPS-SMBS can be realized through \textcolor{black}{RF links, whereas FSO links would} be a better choice for HAPS to LEO \textcolor{black}{connections.} The achievable performance improvement of the aforementioned architecture can be realized through link budget analysis. An improvement in the link budget can be translated reducing the transmit power as well as the cost and size of the user terminal. However, to establish reliable and uninterrupted connections \textcolor{black}{between ground/aerial users and \acrshort{leo} satellites through HAPS-SMBS systems,} HAPS-SMBS systems need to learn the mobility patterns of the \acrshort{leo} satellites in order to predict their handoff, then establish a connection to a coming satellite before losing the current connection. In this regard, machine learning approaches will play a significant \textcolor{black}{role in learning these} mobility patterns. \textcolor{black}{It should also be noted that} as new satellite constellations \textcolor{black}{are added to current} satellite communication systems, \acrshort{ml} solutions \textcolor{black}{will need to be flexible enough} to handle continual environmental changes. \textcolor{black}{Alternatively}, one may also require \textcolor{black}{the incorporation of satellite} tracking system/data---gathered and processed in order to predict any possible collisions among satellites---into the model in order to  compensate for the relatively sudden change in the orbital movements of some satellites.}

%As an example, HAPS can efficiently provide vehicular connectivity due to their larger coverage 
%.......................................%

%\textcolor{orange}{\textbf{HAPS-SMBS as an interface to provide seamless communication to LEO satellites:} LEO satellites move in very high speed resulting frequent disconnection and handover at the terrestrial gateways \cite{sarddar2011handover}. HAPS-SMBS can communicate with LEO satellites with less delay and the have wide upper footprint that can cover many satellites simultaneously. Therefore, it is envisioned that a HAPS-SMBS can act as an interface of the satellite network and provide seamless satellite communication to the aerial and terrestrial networks. In this scenario, HAPS-SMBS will handle the frequent handover of LEO satellites, and if users devices can communicate with HAPS-SMBS directly then users do not have to use special devices or ground stations to communicate with satellites. In this regard, supervised machine learning can be utilized by the HAPS-SMBS to learn the mobility patterns of the satellites in order to predict their handover then establish a connection to a coming satellite before losing the current connection. Figure xxx demonstrate the role of a HAPS-SMBS as an interface for the LEO satellite network.}

\section{Regulatory Aspects}\label{sec:regulations}

\textcolor{black}{The regulation in \textcolor{black}{the}
 aerospace industry is crucial for \textcolor{black}{the}
 safe and harmonious operation of \acrshort{haps} supported networks. \acrfull{itu} \acrfull{rr} defines  \acrshort{haps} as a network element that operates between 20 km and 50 km and at a specified, nominal, fixed point relative to the Earth \cite{ITU-RR}. \acrfull{itu-r} F.1569 indicates that there is a local minimum  in the wind speed \textcolor{black}{of}
 around 20 \textcolor{black}{km} to 25 km, targeting to minimize the required propulsion power for keeping the \acrshort{haps} nodes stationary \cite{ITU-1569}. In the recent deployments, \acrshort{haps} have been frequently deployed at 17 \textcolor{black}{km}
 or 18  km \textcolor{black}{altitude}
 \cite{d2016high}. \textcolor{black}{Different countries determine the different maximum altitudes} of controlled airspaces, and a typical value is 20 km \cite{widiawan2007high}. Although at the borderline between the controlled and the uncontrolled airspace, regulations of \acrshort{haps} need to be carefully designed for safe and secure operations, and currently there %is
\textcolor{black}{are} limited studies addressing  \acrshort{haps} safety \cite{yuniarti2018regulatory}.} \textcolor{black}{The recently founded industry consortium\textcolor{black}{,} HAPS Alliance\footnote{https://hapsalliance.org/}, also works \textcolor{black}{in areas of aviation and commercialization}  to build a strong \acrshort{haps} ecosystem.}

\textcolor{black}{The regulation activities are mostly limited by the \acrshort{itu-r} and \acrfull{icao}. \acrshort{itu-r} regulates the spectrum aspects of \acrshort{haps} while \acrshort{icao}, a \acrfull{un} specialized agency,  governs the safety aspects of \acrshort{haps} and  relations with civil aviation activities. The licensing and  control of  airspace \textcolor{black}{lies} within  the \textcolor{black}{jurisdiction} of  national civil aviation authorities, and rules vary from country to country. }

\subsection{Aviation Regulations}

\textcolor{black}{\acrshort{icao} defines two distinct \acrshort{haps} classes\textcolor{black}{:} unmanned free balloons and the unmanned aircraft. Accordingly\textcolor{black}{,} an unmanned free balloon is defined as a non-power driven, unmanned, lighter-than-air aircraft in free flight, whereas \textcolor{black}{an} unmanned aircraft  \textcolor{black}{is defined as an aircraft intended} to operate with no pilot on board \cite{ICAO-UA}. Although the regulatory guidance is still \textcolor{black}{developing}, regulations associated with these two classes have significant differences. \textcolor{black}{The main difference is that balloons} are excluded from real-time management. The regulations are applied according to the specifics of each development. \textcolor{black}{For example,}  \textcolor{black}{the}
 Google Loon Project\textcolor{black}{, terminated in January 2021, was} included in the unmanned free balloon category. Yet due to  the increasing computational capabilities along with an effective propulsion system, even balloons can be managed \textcolor{black}{in %a
real-time} with smart approaches, as noted by \cite{yuniarti2018regulatory}.}

\textcolor{black}{\textcolor{black}{C}urrent aviation regulations are monitored according \textcolor{black}{to} the  rules of the national civil aviation authorities. Yet,  large-scale \acrshort{haps} deployments are envisioned to be conducted by \textcolor{black}{an} international consortium. To catalyze  successful  large-scale \acrshort{haps} deployments, an international set of rules and regulations \textcolor{black}{are needed to control licensing and operations}. Addressing this concern, Liu and Tronchetti  \cite{liu2019regulating} propose\textcolor{black}{d} the categorization of near space, from 18 km  to 100 km\textcolor{black}{,} as exclusive utilization space along with \textcolor{black}{a} corresponding set of rules. This solution may avoid the uncertainty associated with the international legal status of near space.} 

\subsection{Spectrum Regulations}
\begin{figure}[!tbp]
	\begin{center}
	\centering
\includegraphics[width=1.0\columnwidth]{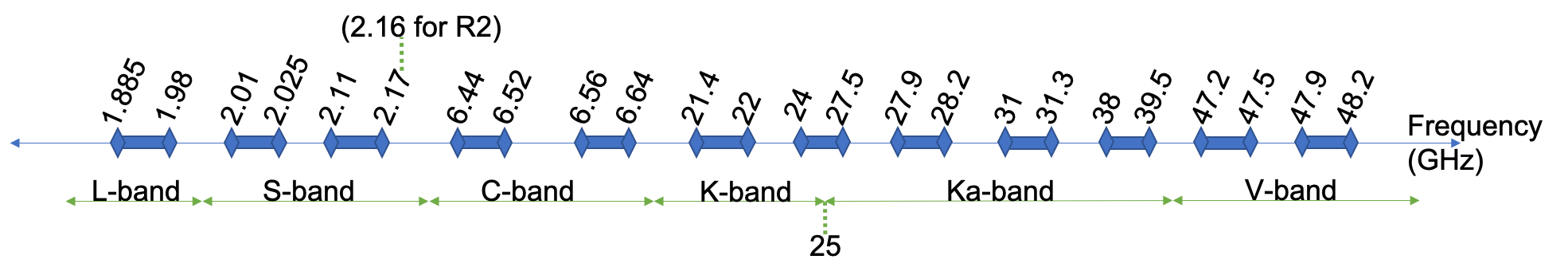}
\caption{\textcolor{black}{An overview of the spectrum bands dedicated for \acrshort{haps}.}}
\label{fig_spectrum}
\end{center}
\end{figure}

\textcolor{black}{\acrshort{itu} has been working \textcolor{black}{to support and integrate} \acrshort{haps} nodes in communication networks since 1997. Based on technical investigations, as reported in recently published reports, including F.2471 \cite{ITU-2471},
F.2472 \cite{ITU-2472} and F.2475 \cite{ITU-2475}, it is concluded that a bandwidth of  396 MHz to 2969 MHz is needed for  ground-to-\acrshort{haps} links. A bandwidth of 324 MHz to 1505 MHz is determined as necessary for the \acrshort{haps}-to-ground links. At \textcolor{black}{the}
World Radiocommunication Conference \textcolor{black}{in} 2019 (WRC-19), which aimed to revise the regulatory framework for \acrshort{haps} and non-geostationary satellite systems, it \textcolor{black}{was} agreed to append the 31 - 31.3 GHz, 38 - 39.5 GHz bands for  \acrshort{haps} usage, in addition to the already dedicated  47.2 – 47.5 GHz and 47.9 – 48.2 GHz bands for worldwide usage. These bands will be used in addition to the previously dedicated International Mobile Telecommunications (IMT) bands in the 2 GHz and  6 GHz bands. \textcolor{black}{Furthermore,} 21.4 - 22 GHz and  24.25 - 27.5 GHz frequency bands can be used by \acrshort{haps} in the fixed services in Region 2, which covers the Americas\textcolor{black}{,} including Greenland, and some of the eastern Pacific Islands.  
The potential of using mmWave bands in \acrshort{haps} networks has been noted in \cite{colella2000halo} \textcolor{black}{and} dates back to 2000, in the High Altitude Long Operation Network (HALO) concept. In addition to the presence  of quite limited ambient interference,  the use of mmWave \textcolor{black}{bands also introduced}  inherent advantages associated with  small antenna sizes and  small array sizes, \textcolor{black}{which} can serve as an advantage, as opposed to the inherent high path loss of these bands. 
An overview of the designated frequency bands %are 
\textcolor{black}{is} provided in Figure \ref{fig_spectrum}. \textcolor{black}{As we can see there,} the frequency bands cover L, S, C, K, Ka, and V bands, some of which also serve other applications. \textcolor{black}{For example,} the L-band and  S-band allocations are also dedicated for terrestrial IMT services. These frequency bands will not only serve disaster relief missions but they will also be used to address the increasing connectivity \textcolor{black}{demands of} end-users by providing commercial broadband services. The \acrshort{itu} regulations also limit the interference of  communication services on  earth observation sensors in radio astronomy stations. WRC-19 \textcolor{black}{provided recommended requirements} for the maximum transmit equivalent isotropic radiated power (EIRP), antenna radiation pattern, power flux density (PFD) limits, the separation distance between radio astronomy station to limit the interference, and the nadir of a \acrshort{haps} platform \cite{ITU-WRC19}.}

% \subsection{\textcolor{black}{Recommendations and Challenges}}
% \textcolor{black}{The spectrum provided by \acrshort{itu} for dedicated \acrshort{haps} usage is quite critical. As another extension unlicensed bands are considered in the trails. Unlicensed bands are specifically designated \textcolor{black}{bands} worldwide that are intended for industrial, scientific and medical (ISM) applications. Although WiFi-based systems prove the successful usage of communication purposes in ISM bands, this is not their main functionality, as the name ISM also implies. In fact, the use of unlicensed ISM bands may have a significant effect on radio astronomy due to imposed electromagnetic interference. This matter is substantiated in Google’s Project Loon tests in Oceania \cite{AMSAT-UK}. Hence the use of these bands in \acrshort{haps} nodes have to be carefully planned in order to protect the radio-astronomy research from unintended interference. To this end, dynamic frequency allocation techniques with cognitive radio capabilities \textcolor{black}{seem} promising to manage interference. }

% \cite{aubineau2010itu}
%........................................%

\section{The HAPS System}\label{sec:characteristics}
{\color{black}
%From TexStudio:
%*A general view of the HAPS system:
In this section, we present \textcolor{black}{a} general view of the \acrshort{haps} communication system along with its onboard subsystems, as illustrated in Fig. \ref{HAPS_system}. In particular, we discuss in detail the recent advancements in both energy and  communications payload subsystems. Moreover, the characteristics of different types of \acrshort{haps} along with the classification and  key features of the most popular \acrshort{haps} projects are highlighted.

  \begin{figure}[!htbp]
	\begin{center}
	\centering
\includegraphics[scale=1]{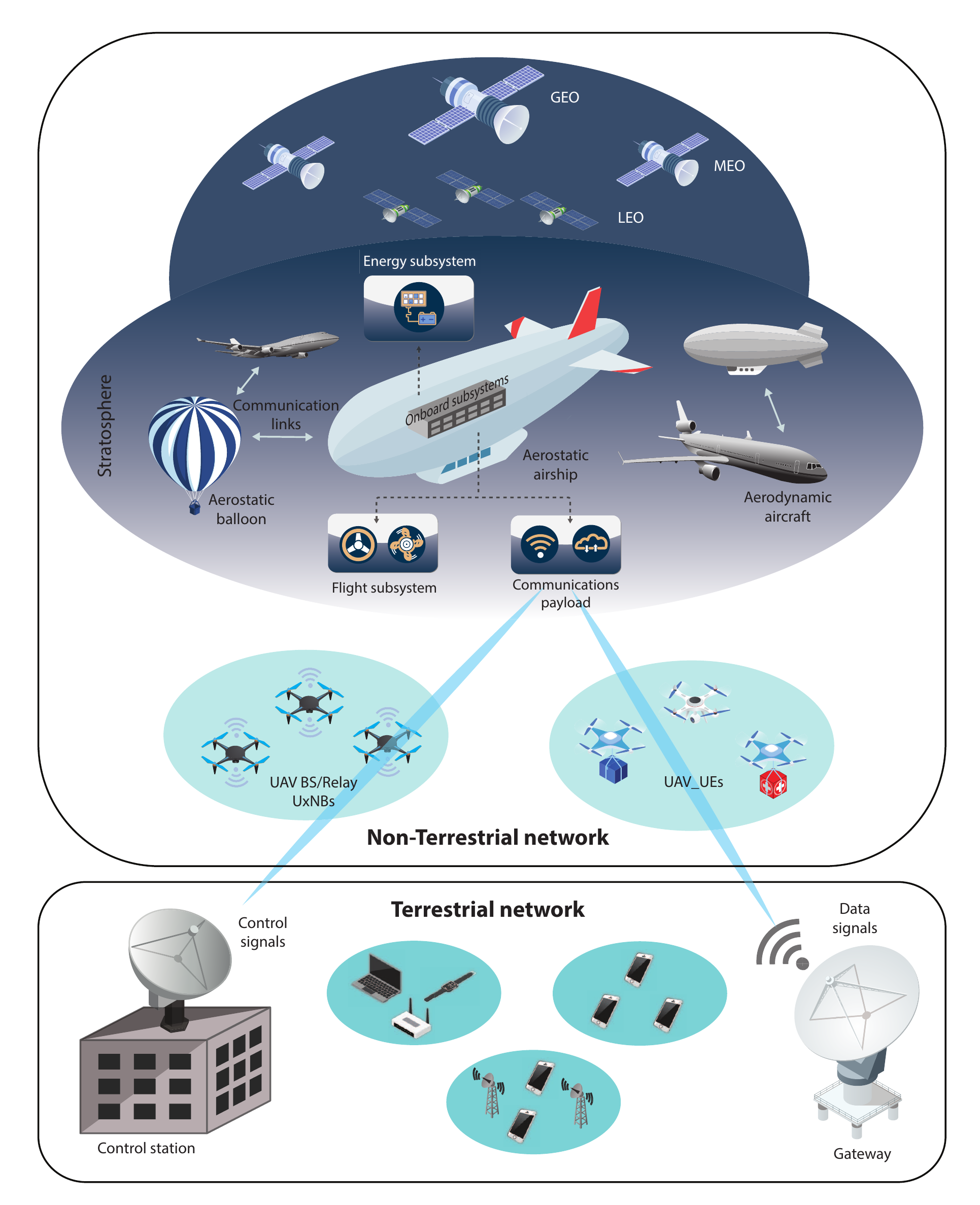}
\caption{\textcolor{black}{A general view of the \acrshort{haps} system and its main components.}}
\label{HAPS_system}
\end{center}
\end{figure}

\subsection{System Components} 
% In order to describe the broad view and the envisioned heterogeneous \acrshort{haps}  communication system, we can  divide it in two parts, aerial part and ground part.
A \acrshort{haps} is located in the stratosphere\footnote{\textcolor{black}{ The lower edge of the stratosphere is about 20 km near the equator. At mid-latitudes, it reaches around 10 km, and at the poles about 7 km. The speed of winds in the stratosphere can exceed those in the troposphere, reaching near 60 m/s in the \textcolor{black}{southern polar vortex.}}}, a layer \textcolor{black}{of} the Earth's atmosphere. This layer has unique properties, which makes it suitable for \acrshort{haps} deployment.
% 	The stratosphere is ranging from 10 to 50 km in altitude, and it  is characterized by  high static stability and relatively mild winds.  
\textcolor{black}{It} is almost free from any weather \textcolor{black}{disturbance, such as lightning or  thunderstorms}. Because of the absence of  clouds in this layer, solar energy can be effectively utilized without atmospheric pollution.
Moreover, \textcolor{black}{the stratosphere} is safe for deployment as \textcolor{black}{it is above the altitude of commercial air traffic.}  Due to these intrinsic features of the stratosphere,  two different types of stratospheric \textcolor{black}{platforms (aerostatic and aerodynamic) can}  be deployed to stay in a quasi-stationary position above the \textcolor{black}{Earth for extended durations,} {\textcolor{black}{as
we will be explain  in the following subsection.}} In general, a \acrshort{haps} communications system \textcolor{black}{consists of two parts: %an aerial
a non-terrestrial  part and a terrestrial part part.}

\subsubsection{Non-terrestrial Part}	\label{sysComp}
\color{black}{
This part includes all the main relative network components in the air \textcolor{black}{or the space} as well as the essential onboard subsystems for \textcolor{black}{an} effective HAPS deployment and successful communication system. Generally, it  consists of two segments:
\begin{itemize} 
\item \textit{Onboard subsystems:} 
They mainly consist of three subsystems; \textit{flight control subsystem}, \textit{energy management subsystem},  and \textit{communications payload subsystem}.
	The goal of the flight control subsystem is to handle the stabilization of the platform, control its mobility, and point it toward the targeted direction. To achieve these,  sensors to measure  altitude and direction of \acrshort{haps}, a computing unit for decision making, and actuators to carry out the desired movement and orientation, are required. Moreover, the flight control unit manages the
interface between the platform and the ground control station. This is performed by the telemetry, tracking and command signals, which reports the health of the platform, and provides an important two-way flow of information between a \acrshort{haps}
and its ground control station \cite{aragon2008high}.
The {energy management subsystem} handles the energy generation and storage process, \textcolor{black}{and it} regulates the energy consumption of other subsystems.  The { communications payload subsystem} is responsible for managing the communications  between the \acrshort{haps} and other entities. Based on the mission of the \acrshort{haps} and the targeted applications, different equipment and technologies can be incorporated in the payload.  Further details of the energy and payload subsystems  will be discussed in the subsections \ref{energy} and \ref{payload}.
                                                                                                  
\item \textit{Non-terrestrial networks:} 
This segment represents all the %aerial 
\textcolor{black}{non-terrestrial communication} nodes \textcolor{black}{in the aerospace domain} that are potentially involved in the \acrshort{haps} communication systems, as depicted in Fig. \ref{HAPS_system}. A HAPS might be connected \textcolor{black}{with other \acrshort{haps} and form a constellation } \cite{dong2016constellation, grace2005improving}, or it could be a part of a network  with different layers of satellites \cite{lin2019robust, cianca2005integrated}.  Moreover,  the \acrshort{haps} layer might be connected \textcolor{black}{with various} types of \acrfull{laps}s, such as UAVs base stations or relays (UxNBs),  \textcolor{black}{or it might serve} a swarm with diverse kinds of UAV users (UAV-UEs) \cite{alzenad2018fso}.

\end{itemize}
\subsubsection{\textcolor{black}{Terrestrial} Part}	
This part represents the ground elements of the \acrshort{haps} communication system. It can be divided into three segments:
\begin{itemize}
	\item \textit{Control station:}
	\textcolor{black}{This} manages the communication operations between  \acrshort{haps} and different types of users. Also, it orchestrates the  communication links and manages the resources between multiple \acrshort{haps} nodes and other  non-terrestrial or terrestrial networks. Moreover, the control station handles the takeoff/landing process, monitors remotely the position of the \acrshort{haps} and controls its direction to maximize the antenna efficiency and enhance the performance.
	
    \item \textit{Communications gateway:}
    It connects the \acrshort{haps} to the core network \textcolor{black}{ through a wired backhaul infrastructure.}
    Depending on the \acrshort{haps} payload and the type of  terrestrial  network, a \acrshort{haps} may either communicate directly with terrestrial users, or the data information \textcolor{black}{may be exchanged} through the communications gateway. \textcolor{black}{The control station and communications gateway} could be either co-located within the same building facilities or have separate locations. They  basically  consist of amplifiers, processing units, and antennas.  \textcolor{black}{Antennas that are typically used} have parabolic dish reflectors to guarantee high directivity gain.

	\item \textit{Terrestrial networks:}
This segment includes all the terrestrial nodes or users  involved in the \acrshort{haps} communication systems. \textcolor{black}{This includes terrestrial BSs and different types of users, such as mobile users and IoT sensors.}
\end{itemize}}}

\subsection{Types of HAPS and Related Projects}
	Generally, \acrshort{haps} nodes can be  classified into manned or unmanned aerial platforms. 
% 	Several manned \acrshort{haps} flights in the stratosphere were made early in the 1930s using balloons. For example, Explorer II gondola carried a two-man crew and reached a record  altitude of 22 km. 
In the 1960s, jet-powered manned \acrshort{haps} were  \textcolor{black}{developed, such as the} as B-57 Canberra and Lockheed F-104 \cite{ehrenfried2014stratonauts}. Most of these manned \acrshort{haps} \textcolor{black}{systems} were considered for meteorology, scientific purposes, or military applications. The Proteus is an  example \textcolor{black}{of a manned} \acrshort{haps} designed for telecommunication usage. 
% It was built by Scaled Composites company for
% 	 NASA's Environmental Research Aircraft and Sensor
% 	Technology (ERAST) project. Its first flight was in 1998, and typically it is flown by two pilots. However, it has the potential to accomplish its missions semi-autonomously or flown remotely from the ground. The Proteus was designed as a %long-duration
% 	 telecommunications relay platform carrying an 8-foot diameter  antenna system for supporting broadband data over major cities. It features an unconventional tandem-wing, twin-boom configuration with two rear-mounted turbofan engines providing power for flight duration up to 18 hours \cite{Protues}. 
However, typical communications applications require prolonged support, and it is difficult for human pilots to fly   \textcolor{black}{for extended durations} in the harsh stratospheric environment.
	Therefore, unmanned  \acrshort{haps} nodes are more popular and preferable for communications. 
	
	 \textcolor{black}{Depending on} the  underlying physical principle that provides the lifting force for the \acrshort{haps}, they are classified as  aerostatic (a.k.a. lighter than air) platforms or aerodynamic  (a.k.a. heavier than air) platforms. While aerostatic platforms \textcolor{black}{make use} of buoyancy to float in the air, aerodynamic platforms use dynamic forces created by the movement through the air \cite{arum2020review,mohammed2011role}. 
	 Aerostatic platforms appear in two shapes, \textcolor{black}{balloons
	 and airships}, and they make use
	 of a lifting gas in an envelope for providing buoyancy to float in the air \cite{mohammed2011role}. 
	 %These gases are mostly hot air, diatomic hydrogen or helium. 
% Although hydrogen is relatively cheap, it may not the best and safest option, since it is highly combustible. Helium is therefore more preferable, since it is  nonflammable and not combustible, if the extra cost is affordable. Other alternatives for lifting gases include methane and ammonia. While methane is used due to its larger molecules, which prevents fast leakage through the envelop, ammonia's main advantage is its relatively high boiling point, which facilitates the altitude control of a \acrshort{haps} \cite{grace2011broadband}.

Balloons are usually unpowered platforms and they can be tethered to easily control their flights.
% However, tethered balloons in the stratosphere have been of little use due to the experiments of tethered balloons using Kevlar cables at an altitude around  20 km in France in the 1980s were totally failed because of gusty winds and air-safety constraints \cite{north2014encyclopedia}. Thus, 
\textcolor{black}{However, tethered}  balloons \textcolor{black}{in the stratosphere have been abandoned due to air-safety constraints, and currently tethered \acrshort{haps}} are mostly restricted to a maximum altitude of 2 km. Google Loon \textcolor{black}{was}  an example of balloons intended for communication proposes.
	 \textcolor{black}{These are made} made from large size sheets of polyethylene equipped with antennas and solar \textcolor{black}{panels, and they can stay in the stratosphere for over 100 days.}
	  Loon's early experiments were conducted in 2011,  and successful WiFi and LTE connections \textcolor{black}{have been} realized through Loon since 2013 \cite{Loon}.
	     
% On the other hand, airships (a.k.a.,  dirigible balloons) are significantly large platforms filled with excessive amounts of  lifting gas and have significantly large payload capacities. Based on the availability of the  internal structural framework, three kinds of airships can be realized;  non-rigid, semi-rigid, and rigid. While non-rigid airships sometimes called  ``blimps", rigid ones are often known as ``zeppelins". 
Airships are  typically powered platforms with propulsion  \textcolor{black}{systems which can stay} in the stratosphere for several months or years \cite{karapantazis2005broadband}.  Although  \textcolor{black}{the immense size} of airships \textcolor{black}{creates dynamic drag during flights} and imposes significant challenges for takeoff and landing, \textcolor{black}{airships offer} great flexibility in terms of payloads  \textcolor{black}{and power generated using solar cells} \cite{mohammed2011role}. 
% An example of airship projects is the Japanese Stratospheric Platform (SPF) Program (a.k.a, SkyNet) 
% %started in 1998 and lasted for 18 months.
% (1998 - 2005) \cite{karapantazis2005broadband,aragon2008high,d2016high}.
%  This project was led by the  National Aerospace Laboratory (NAL), currently the Japan Aerospace Exploration Agency (JAXA). The SkyNet airship had a  semi-rigid hull of ellipsoidal shape with an overall length of nearly 200 m. It is powered by solar cells and regenerative fuel cells and can have a continuous operation for up to 3 years. The objective of SkyNet was to support future telecommunications networks and Earth observation sciences \cite{karapantazis2005broadband,aragon2008high,d2016high}. 
 %Examples:(NAL “SPF” (Stratospheric PlatForm (Japan) ;;;ATG “StratSat” (UK), ...) 
Aerodynamic \acrshort{haps} uses electric motors and propellers as a propulsion system. 
% They rely on the aerodynamic lift %for flying in the air. 
% created by two primary principles, Bernoulli’s Principle and Newton’s Third Law of Physics \cite{karapantazis2005broadband}.
 % ***They use electric motors and propellers as propulsion system.
  %, while solar cells mounted on the wings and stabilizers provide power during the day and charge the onboard fuel cells \cite{karapantazis2005broadband}. %Examples : {.......}
In contrast to aerostatic platforms,  aerodynamic aircraft have limited payload \textcolor{black}{capacities and a higher resistance}
%with limited power (up to 1200 W), and continuous movement of the platform is required, resulting in additional power consumption.
 to strong winds and turbulent conditions \cite{aragon2008high}. 
 Moreover, an aerodynamic \acrshort{haps} has to move forward and circle around the intended area \textcolor{black}{of coverage to maintain its  quasi-stationary position.} Also, they require  \textcolor{black}{large wingspans} (35 to 80 m) for lifting due to the reduced air density \textcolor{black}{at their} operating altitudes. As a result,  the radius of the circular movement will be very large, which requires adjustments in antennas pointing and communications beams. 
%  An example of an early aerodynamic \acrshort{haps} is the HeliNet project (1999-2003).
% %  funded by the $5^{th}$ Framework Programme of the European Union Commission. The project demonstrated the usage of \acrshort{haps} for broadband communications, environmental monitoring,
% %  remote sensing, and localization/navigation applications \cite{aragon2008high,karapantazis2005broadband}.
%  A recent aerodynamic \acrshort{haps}  example is the Japanese project HAPSMobile  launched in 2017.
%  The objective of the project is to provide a reliable network that connects people and things around the world. Its solar-powered aircraft (HAWK30)  has 78 meters wingspan with a high energy density Lithium-ion battery that can tolerate several months of flight duration. The first flight test of HAWK30 was successfully completed in September 2019  \cite{HAPSMobile}. 
 %In 2019, both projects Loon and HAPSMobile signed a long-term strategic relationship for advancing both types of HAPS, aerostatic and aerodynamic, in order to connect more people, places, and things worldwide.

Both   aerostatic and aerodynamic \acrshort{haps}  have their  advantages and disadvantages. The differences are in  deployment costs, coverage areas,  payload capacities, endurance level, positioning control, and  flight duration.
The intended use-case or mission objective plays an important role  \textcolor{black}{in determining} the best \acrshort{haps} option \cite{karapantazis2005broadband, mohammed2011role}. \textcolor{black}{For instance, an aerodynamic \acrshort{haps} might be more preferable for unplanned events or emergency situations due to their reduced deployment costs \cite{aragon2008high}, flexibility in take-off/landing and mobility \textcolor{black}{control. By contrast, an aerostatic \acrshort{haps} might be more appropriate}
 for longer-term \textcolor{black}{use cases, such as} supporting cargo drones, autonomous vehicles, and computation offloading due to their large payload capacities and high energy generation capabilities.} However, station-keeping is more difficult and challenging for aerostatic platforms when \textcolor{black}{ there are strong winds and turbulent conditions.} 

Table \ref{tab:CLASSIFICATION} %enlist the past and recent projects.
\textcolor{black}{lists some  popular past and recent projects along with their classification and key features.}
\textcolor{black}{As we can see, in most cases, each \acrshort{haps} project adopts a certain type of platform,} but perhaps a hybrid type of \acrshort{haps}  that combines the advantages of \textcolor{black}{both aerostatic and aerodynamic types}
 is needed for near-future applications. 
%As a first step, ....
%In 2019,
In this regard, both projects Loon and HAPSMobile signed a long-term strategic relationship in 2019 for advancing both types of aerostatic and aerodynamic \acrshort{haps} systems \cite{Loon}.
\begin{table*}	
\caption{Classification and description of popular HAPS examples}\label{tab:CLASSIFICATION}
\vspace{3pt} \noindent
%\resizebox{\textwidth}{!}{
\begin{tabular}{|p{54pt}|p{46pt}|p{61pt}|p{47pt}|p{42pt}|p{193pt}|}
%\caption{Classification of popular HAPS examples}\label{tab:CLASSIFICATION}\\
	\hline
	\parbox{54pt}{\centering 
		\textbf{Project /Product}
	} & \parbox{46pt}{\centering 
		\textbf{Type}
	} & \parbox{70pt}{\centering 
		\textbf{Company/ Organization}
	} & \parbox{47pt}{\centering 
		\textbf{Country}
	} & \parbox{42pt}{\centering 
		\textbf{Project period}
	} & \parbox{193pt}{\centering 
		\textbf{\textcolor{black}{Description / Important features}}
	} \\
	\hline
	\parbox{54pt}{\centering 
		SHARP
		
		\cite{SHARP}
	} & \parbox{46pt}{\centering 
		Aerodynamic
	} & \parbox{61pt}{\centering 
		Communications Research Centre (CRC)
	} & %\parbox{47pt}{\centering 
	%\textbf{Word-to-LaTeX TRIAL VERSION LIMITATION:}\textit{ A few characters will be randomly misplaced in every paragraph starting from here.}
	
	\parbox{47pt}{\centering 
		Canada
	} & \parbox{42pt}{\centering 
		1980-1987
	} & \parbox{193pt}{%\justify
			\vskip 0.2cm
		$\circ$ \textcolor{black}{It is the first HAPS powered by microwave beams from the ground.}
		\vskip 0.1cm
		
		$\circ$   \textcolor{black}{It was envisioned to operate at an altitude of 21 km providing telecommunications within a diameter of 600 km.}
		
				\vskip 0.1cm
		\textcolor{black}{$\circ$ It demonstrated successful communications for a one-hour flight duration.}
		
		\vskip 0.1cm
		\textcolor{black}{$\circ$ After several successful trial flights, the project was ended because of a large drawdown in the CRC budget.}

		\vskip 0.2cm
		
	} \\
	\hline
	\parbox{54pt}{\centering 
		Pathfinder, Centurion
		\& Helios

		\cite{AV}
	} & \parbox{46pt}{\centering 
		Aerodynamic
	} & \parbox{61pt}{\centering 
AeroVironment for
 NASA \acrfull{erast}
	} & \parbox{47pt}{\centering 
		United States
	} & \parbox{42pt}{\centering 
		1994-2003
	} & \parbox{193pt}{%\raggedright 
	\vskip 0.2cm
		\textcolor{black}{$\circ$ The aim of the project was to develop the technologies of solar aerodynamic HAPS.}   
		
		$\circ$ \textcolor{black}{In 2002, Pathfinder Plus demonstrated the world’s first HAPS \textcolor{black}{at an altitude of 20 km, from which it provided} high-definition TV (HDTV) signals, 3G mobile voice, video and data, and high speed internet connectivity.}  \vskip 0.2cm
	} \\
	\hline
	\parbox{54pt}{\centering 
		(SkyNet)
		
		\cite{karapantazis2005broadband,aragon2008high,d2016high}
	} & \parbox{46pt}{\centering 
		Aerostatic- (Airship)
	} & \parbox{61pt}{\centering 
		(\acrfull{nal})
		
		Currently:
		
		(\acrfull{jaxa})
	} & \parbox{47pt}{\centering 
		Japan
	} & \parbox{42pt}{\centering
		1998-2005
	} & \parbox{193pt}{%\justify 
	\vskip 0.2cm
		$\circ$ The objective was to support future communications with high-speed links.
		\vskip 0.1cm
		
	\textcolor{black}{$\circ$ The project consisted of several airships positioned at an altitude of 20 km.}
		\vskip 0.1cm
			\textcolor{black}{$\circ$ Each airship would have about 200 m length and could operate for up to 3 years covering a radius  up to 100 km.}
		\vskip 0.1cm
		
		$\circ$ \textcolor{black}{Due to funding issues, the project was terminated \textcolor{black}{after the successful completion of several phases of the project.}} \vskip 0.2cm
	} \\
	\hline
	\parbox{54pt}{\centering 
		CAPANINA
		
		\cite{CAPANINA}
	} & \parbox{46pt}{\centering 
		Aerostatic- (Balloon)
	} & \parbox{61pt}{\centering 
Communications Research Group at the University of York
	} & \parbox{47pt}{\centering 
		United Kingdom
	} & \parbox{42pt}{\centering 
		2003-2006
	} & \parbox{193pt}{%\justify 
	\vskip 0.2cm
		\textcolor{black}{$\circ$ The goal of the project was to test the feasibility of HAPS for improving broadband access in Europe, particularly for rural communities.}
		\vskip 0.1cm
\textcolor{black}{$\circ$ \textcolor{black}{It was the first trial to use FSO links \textcolor{black}{for HAPS.}}}
		\vskip 0.1cm
		
		$\circ$ \textcolor{black}{It demonstrated successful communications \textcolor{black}{with a rate of} 1,25 Gbps at an altitude of 23 km \textcolor{black}{providing coverage over a radius} of 64 km.}
		\vskip 0.2cm
	} \\
	\hline
	\parbox{54pt}{\centering 
		X-station
		
		\cite{StratXX} 
}	& \parbox{46pt}{\centering 
		Aerostatic- (Airship)
	} & \parbox{61pt}{\centering 
		StratXX
	} & \parbox{47pt}{\centering 
		Switzerland
	} & \parbox{42pt}{\centering 
		2005-Now
	} & \parbox{193pt}{%\justify
	\vskip 0.2cm
$\circ$ \textcolor{black}{It supports different communication technologies such as TV, radio, mobile telephony, VoIP, and remote sensing.}
\vskip 0.1cm
\textcolor{black}{$\circ$ A set of 3 X-stations can be used to provide local GPS services \textcolor{black}{for an area of up to} $10^6$ $\text{km}^2$.}
\vskip 0.1cm
\textcolor{black}{$\circ$ Each \textcolor{black}{X-station can maintain an altitude of 21 km and cover up to 1,000 km diameter.}}
\vskip 0.1cm
\textcolor{black}{$\circ$ It  uses solar energy and batteries and \textcolor{black}{supports a payload of 100 kg and flight durations of up to one year.}}
\vskip 0.2cm
	} \\
	\hline
	\parbox{54pt}{\centering 
	Elevate
	
	\cite{Elevate} 
}	& \parbox{46pt}{\centering 
	Aerostatic- (Balloon)
} & \parbox{61pt}{\centering 
	Zero 2 Infinity
} & \parbox{47pt}{\centering 
	Spain
} & \parbox{42pt}{\centering 
	2009-Now
} & \parbox{193pt}{%\justify 
\vskip 0.2cm
	$\circ$ It is a transportation service  to lift payloads in the stratosphere for testing and validation of new HAPS technologies.
	\vskip 0.1cm
	$\circ$ \textcolor{black}{Its STRATOS vehicle can carry up to 100 kg for about 24 hours flight duration at an altitudes between 18-22 km.}
	
	\vskip 0.1cm
	 \textcolor{black}{$\circ$ The company provides different \textcolor{black}{options in terms of altitude, duration, and payload mass based on customer requirements.}} 
	\vskip 0.2cm
} \\
    \hline
	\parbox{54pt}{\centering 
		Loon
		
		\cite{Loon}
	} & \parbox{46pt}{\centering 
		Aerostatic- 
		(Balloon)
	} & \parbox{61pt}{\centering 
		Subsidiary of:
		(Alphabet Inc.)
		Previously: (Google X)
		
	} & \parbox{47pt}{\centering 
		United States
	} & \parbox{42pt}{\centering 
		2011-\textcolor{black}{2021}
	} & \parbox{193pt}{%\justify 
	\vskip 0.2cm
$\circ$ Its mission \textcolor{black}{was} to connect people everywhere using a network of HAPS.
\vskip 0.1cm

\textcolor{black}{$\circ$ It \textcolor{black}{was} the most mature \textcolor{black}{project whose fleet \textcolor{black}{constituted} a meshed network managed by Loon SDN, which provided service} to over 300,000 users.}
\vskip 0.1cm

\textcolor{black}{$\circ$ \textcolor{black}{Last design was able to} fly up to 312 days at an altitude around 18-23 km, with a 40 km coverage radius.}
\vskip 0.1cm

$\circ$ In 2019, Loon’s balloons accomplished over one million flight \textcolor{black}{hours, flying for a total of around 40 million km.}
\vskip 0.2cm
	} \\
	\hline
	\parbox{54pt}{\centering 
		Zephyr S
		
		\cite{Zephyr}
	} & \parbox{46pt}{\centering 
		Aerodynamic
	} & \parbox{61pt}{\centering 
		Airbus Defense and space
	} & \parbox{47pt}{\centering 
		United Kingdom
	} & \parbox{42pt}{\centering 
		2013-Now
	} & \parbox{193pt}{%\justify
	\vskip 0.2cm
		$\circ$ One of its goals is to connect isolated people \textcolor{black}{across} the globe.
		\vskip 0.1cm
		\textcolor{black}{$\circ$ \textcolor{black}{It has logged the longest continuous flight duration of any aerodynamic HAPS}
		 with a maiden flight of over 25 days.}
		\vskip 0.1cm
		\textcolor{black}{$\circ$ It \textcolor{black}{can broadcast at 100 Mbps with a payload of up to 12 kg and can fly continuously for 100 days.}}
		\vskip 0.1cm
		\textcolor{black}{$\circ$ \textcolor{black}{It has a 25 m wingspan, can maintain an altitude above 18 km, and} it is fully powered by solar energy, with secondary rechargeable batteries providing 250 W maximum payload power.}
		\vskip 0.2cm
	} \\
	\hline
	
\end{tabular}%}
\vspace{2pt}
\end{table*}
%========================================================
%COMPLETE THE TABLE

\begin{table*}	
\caption*{Table \ref{tab:CLASSIFICATION}: (Continued) Classification and description of popular HAPS examples}\label{tab:CLASSIFICATION_continue}
\vspace{3pt} \noindent
%\resizebox{\textwidth}{!}{
\begin{tabular}{|p{54pt}|p{46pt}|p{61pt}|p{47pt}|p{42pt}|p{193pt}|}
%\caption{Classification of popular HAPS examples}\label{tab:CLASSIFICATION}\\
	\hline
	\parbox{54pt}{\centering 
		\textbf{Project /Product}
	} & \parbox{46pt}{\centering 
		\textbf{Type}
	} & \parbox{70pt}{\centering 
		\textbf{Company/ Organization}
	} & \parbox{47pt}{\centering 
		\textbf{Country}
	} & \parbox{42pt}{\centering 
		\textbf{Project period}
	} & \parbox{193pt}{\centering 
		\textbf{\textcolor{black}{Description / Important features}}
	} \\
	\hline
		\parbox{54pt}{\centering 
		Aquila
		
		\cite{aquila}
	} & \parbox{46pt}{\centering 
		Aerodynamic
	} & \parbox{61pt}{\centering 
		Facebook
	} & \parbox{47pt}{\centering 
		United Kingdom
	} & \parbox{42pt}{\centering 
		2014-2018
	} & \parbox{193pt}{%\justify
	\vskip 0.2cm
		$\circ$ The aim of the project was to provide broadband coverage for remote areas \textcolor{black}{with an 80 km radius and 90-day flight duration.}
		
		\vskip 0.1cm
		\textcolor{black}{$\circ$ It was intended to fly at altitudes of 27 km during the day, dropping to 18 km at nights.}
		
		\vskip 0.1cm
		\textcolor{black}{$\circ$ After several successful tests, the project was ended to work with partners like Airbus.}
		\vskip 0.2cm
	} \\
	\hline
	\parbox{54pt}{\centering 
		Stratobus
		
		\cite{Stratobus}
	} & \parbox{46pt}{\centering 
		Aerostatic- (Airship)
	} & \parbox{61pt}{\centering 
		Thales Alenia Space
	} & \parbox{47pt}{\centering 
		France
	} & \parbox{42pt}{\centering 
		2014-Now
	} & \parbox{193pt}{%\justify
	\vskip 0.2cm 
		 $\circ$ One of its goals is to provide 5G telecommunications.
		\vskip 0.1cm
		\textcolor{black}{$\circ$ Its length and width are about (115 m x 34 m), and it can carry up to 450 kg payload for a 5-year mission with annual maintenance.}
		
		\vskip 0.1cm
		\textcolor{black}{$\circ$ It is positioned at an altitude of 20 km  and can cover up to 500 km in diameter.}
		
		\vskip 0.1cm
		\textcolor{black}{$\circ$ It is expected to be on the market in 2021.}
		\vskip 0.2cm
	} \\
	\hline
	\parbox{54pt}{\centering 
		HAWK30
		
		\cite{HAPSMobile}
	} & \parbox{46pt}{\centering 
		Aerodynamic
	} & \parbox{61pt}{\centering 
		HAPSMobile
	} & \parbox{47pt}{\centering 
		Japan
	} & \parbox{42pt}{\centering 
		2017-Now
	} & \parbox{193pt}{%\justify
	\vskip 0.2cm 
		$\circ$ Its objective is to connect mobiles, UAVs, and IoT nodes around the world.
		
		\vskip 0.1cm
		\textcolor{black}{$\circ$ \textcolor{black}{It has a wingspan of 78 m, deployed at an altitude of 20 km, and can provide a 100 km coverage radius for several months.}}
		\vskip 0.2cm
	} \\
	\hline
		\parbox{54pt}{\centering 
		PHASA-35
		
		\cite{prismatic,ABE}
	} & \parbox{46pt}{\centering 
		Aerodynamic
	} & \parbox{61pt}{\centering 
		BAE Systems and Prismatic
	} & \parbox{47pt}{\centering 
		United Kingdom
	} & \parbox{42pt}{\centering 
		2018-Now
	} & \parbox{193pt}{%\justify
		\vskip 0.2cm 
		$\circ$ It is designed for a variety of services including 5G communications.
		\vskip 0.1cm
		$\circ$ \textcolor{black}{It has a payload capacity of 15 kg and can remain airborne continuously for up to one year.}.
		
		\vskip 0.1cm
		\textcolor{black}{$\circ$ \textcolor{black}{It can maintain an altitude of 17-21 km with a payload power capacity of 300-1,000 w, and it can cover a radius of up to 200 km.}}
		\vskip 0.2cm
	} \\
	\hline
\end{tabular}
%}
\vspace{2pt}
\end{table*}

\subsection{Energy Management Subsystem} \label{energy}
\textcolor{black}{Managing the energy supplied to and consumed by a \acrshort{haps}} is an essential task and impacts  flight duration and deployment costs.  Since using a \acrshort{haps} system for    \textcolor{black}{communications is generally a prolonged operation,}  careful energy management is required in order to make \acrshort{haps}-based solutions feasible and cost-effective. 
% Generally, there are two sides to the energy management of the \acrshort{haps}. The first one is related to the platform’s supplied energy source, whereas the second one considers the energy consumption of the platform. Maximizing the obtained energy from the energy sources and minimizing the consumed energy by the \acrshort{haps} and its payload is a key enabler in utilizing \acrshort{haps} effectively for wireless communications.	

\subsubsection{HAPS Energy Sources}
% The sources of energy for a \acrshort{haps} is the most important design
% aspect of the \acrshort{haps} system, as a failure in this system results in the  loss of a \acrshort{haps} mission. 
There are three types of energy sources that have been used for \acrshort{haps} operations: conventional \textcolor{black}{energy sources (e.g., fuel tanks and electrical batteries), energy beams, and solar energy.}  \acrshort{haps} supplied by conventional energy sources have a very short flight duration of about 48 hours and require frequent landing for refueling \cite{grace2011broadband}.
%  When such a \acrshort{haps} is used for communications,  a seamless handoff between subsequent \acrshort{haps} is required. Thus, 
 The use of conventional energy sources is \textcolor{black}{suitable as a temporary solution} or in an emergency situation.
 Alternatively,  energy beams from the ground can be used to supply \textcolor{black}{a} \acrshort{haps} energy system. This idea was  proposed in the 1980s using microwave beams \cite{schlesak1988microwave,brown1986microwaver}. \textcolor{black}{An example of this is the SHARP project \cite{SHARP}, which
 %designed by  the Communications \acrfull{crc}
%  and built by the University of Toronto Institute for Aerospace Studies (UTIAS) 
consists} of a large ground antenna system for transmitting a large diameter of microwave beams. Such \acrshort{haps} energy systems use \textcolor{black}{collectors that consist of numerous rectifier antennas}  to convert the received energy to DC power.
 %, which powers the flight system as well as onboard control and communication circuits, and charge standby energy storage units.   
 Similarly, laser beams can also be used as %power transmission medium 
 an energy source. Several experiments were conducted in Japan and \textcolor{black}{the} USA using laser-beam powered \acrshort{haps}. However, due to the high power irradiation \textcolor{black}{risks posed by both microwave-powered and laser-powered platforms, these are not regarded as safe  solutions \cite{grace2011broadband}.}
%  Also, these types of  energy sources are not preferred due to low-efficiency transmission with high costs of the ground station \cite{karapantazis2005broadband}.

 Solar energy, on the other hand, is a renewable energy source and a safe option for powering  \acrshort{haps}, and it is the main energy source considered by most \acrshort{haps} projects. Solar energy is appropriate for \acrshort{haps} \textcolor{black}{for}  two basic reasons. First, \acrshort{haps}  \textcolor{black}{operate above the clouds}, where  natural solar  energy is abundant. Second, \acrshort{haps} are typically huge platforms that can have large solar panels to generate large amounts of energy. 
%  Although solar cells have low conversion efficiency \cite{aragon2008high},  advanced assisting technologies such as  photovoltaic-hydrogen systems  can collect sufficient energy for \acrshort{haps} operations \cite{knaupp2003photovoltaic}. 
 Solar-powered \acrshort{haps} \textcolor{black}{typically have secondary energy sources}  to power the \acrshort{haps} functions during nights or in winter. These secondary sources, \textcolor{black}{which may} include electrical batteries or hydrogen fuel cells, are recharged by the solar energy during the daytime. Accordingly, a control unit in the \acrshort{haps} energy subsystem is required to manage  operations between  primary and secondary energy sources. %Effective \acrshort{haps} solar system deigns, harvested solar energy analysis, efficient energy storage systems, and cooperation strategy between solar energy systems and secondary energy sources \cite{ gao2013energy, phillips1980some, alsahlani2017design, cestino2006design} are widely investigated in the literature. 
 \textcolor{black}{Effective \acrshort{haps} solar energy designs have been widely investigated in the literature along with analyses of harvested solar energy and cooperation strategies between solar energy systems and secondary energy sources \cite{ gao2013energy, phillips1980some, alsahlani2017design, cestino2006design}.}
 
 \textcolor{black}{Several recent studies have introduced} methods for improving the solar energy conversion efficiency \cite{yoshikawa2017exceeding, green2017energy, zheng2016photonic}. MicroLink Devices, a leading company in producing solar arrays for satellite and \acrshort{haps}, manufactures  high-efficiency solar sheets with powers exceeding 1.5 kW/kg.  These  ultra-thin and lightweight sheets  can achieve 37.75\% solar energy conversion efficiency, \textcolor{black}{which is highest record compared to any other solar cell technology}
  \cite{microLink}.  On the other hand, for the secondary sources, the  values of the energy density of  Lithium-ion  batteries in early \acrshort{haps} studies were around 100 Wh/kg \cite{karapantazis2005broadband}. However,  the current  commercial state-of-the-art  Lithium-ion batteries have 250 Wh/kg \cite{zubi2018lithium}.  
Fuel cells, which are the other alternative energy storage system, 
\textcolor{black}{have advanced greatly: in 2009,}
 the state-of-the-art fuel cells were around 400 Wh/kg \cite{thomas2009fuel}\textcolor{black}{; current}
 fuel cells have \textcolor{black}{an}  energy density of 1600 Wh/kg \cite{wilberforce2017developments}. 
 
% As a result of the extensive academic research efforts for the \acrshort{haps} energy system in the past two decades, and the recent  industry advancements  in the solar and storage systems,  \acrshort{haps} system attracts several network operators to be considered in the near-future deployment for tackling several wireless problems and providing  continuous and  uninterrupted services with long flight duration.

\subsubsection{HAPS Energy Consumption}
% The generated energy is mainly consumed by the two subsystems: the flight control and the communications payload subsystem. 
\textcolor{black}{Generally,  \acrshort{haps} energy is consumed by  the two subsystems: the flight control and the communications payload subsystem.}
The energy consumed by the flight subsystem includes the consumption for the stability and propulsion power, and the consumption caused by controlling the \acrshort{haps} altitude and direction.   
%, as discussed in Subsection \ref{sysComp}.
%associated mobility subsystems.
Also, the platform's type and its features, such as weight and size, impacts the  flight system energy consumption. Since aerodynamic platforms require \textcolor{black}{a} continuous circular movement, they have generally higher energy consumption \cite{karapantazis2005broadband,mohammed2011role, grace2011broadband}. In addition, as the size and the weight of the platform increases, more energy is required for the flight system. Since aerodynamic \acrshort{haps} also have a relatively smaller size than airships,
\textcolor{black}{the solar energy they generate is less than that of airships.}
 Consequently, their capacity for payload power is relatively small. 
 %energy consumption is related to the HAPS’s engineering design and its flight system. 
 \textcolor{black}{The remainder of the energy generated}
  is consumed by the payload for the communications operations\textcolor{black}{, which} chiefly depends on the type of  communications payload \textcolor{black}{and} the communication techniques. In general, as more active components and computation processes are included in the payload, %heavier payloads are required and more energy consumption is expected.
  \textcolor{black}{they get heavier and consume more energy.}
\subsection{Communications Payload Subsystem} \label{payload}
\subsubsection{Active Payload}
%  As explained in Section \ref{sysComp}, \acrshort{haps} can be as the sole station in a stand-alone infrastructure, or more likely in near-future deployments, it will  be integrated with terrestrial and non-terrestrial networks. Based on the network architecture and the mission type, the requirements of the communications payload can be different.  Also, the platform’s type and its weight and size impact the payload choice. Moreover, the availability of power for the communications payload system is an important factor for determining the features of  suitable payload options. 
 The active payload generally includes antennas, \textcolor{black}{transponders}, low-noise power amplifiers,  frequency converters, IF processors, and filters.
%  to obtain the best frequency response of each channel, maximize the gain, and limit the noise \cite{aragon2008high}.  
 The payload type is dependent on the intended application and use-cases. \textcolor{black}{Typically,} a \acrshort{haps} payload can be either as a relay station (HAPS-RS) or as a full base station (HAPS-BS) \cite{tozer2001high}. More active components and higher processing capabilities are associated with a HAPS-BS, which accordingly requires more energy \textcolor{black}{consumption and can necessitate} larger and heavier communications components. While a HAPS-BS can fully process signals and serve users directly, a HAPS-RS requires an intermediate station to process the users’ signals. Thus, a HAPS-RS involves an increase in the round-trip delay. \textcolor{black}{ On the other hand, as  discussed in Section \ref{sec:HAPS-use-cases20}, \acrshort{haps} can be equipped with a BS with superior caching, computing, and communication capabilities,  \textcolor{black}{capabilities (i.e., a HAPS-SMBS) for} serving dense urban environments. 
%  Due to its high processing ability, it will play an important role to facilitate several future wireless communications applications, as presented in Section II.
 However, \textcolor{black}{the} advanced and powerful capabilities %of the HAPS-SMBS comes with the costs of the payload weight and its consumed energy.
 \textcolor{black}{of a HAPS-SMBS come with the added costs of payload weight and the energy this consumes.}} 
 %In this case, the choice of the platform needs to be carefully considered. 
%  Due to some limitations of aerodynamic \acrshort{haps} for the available space, weight and power for the payload, they might not be suitable for such types of payload. 
%Airships, with their intrinsic features including  huge size and large payload capacity, have the potential to accommodate such heavyweight \textcolor{black}{and %provide for the 
%high energy consumption payload.} 
%  As a matter of fact, the specifications of some recent industry projects demonstrate the potential to support such advanced types of payloads. 
 %For instance, the Strtobus  airship, expected to be deployed late in 2021, can accommodate a payload of 450 kg with a power rating of 8 kW for a 5-year mission \tex\cite{Stratobus}. 
 \textcolor{black}{Therefore, careful analysis of the intended use-case with the chosen platform type and its payload size and weight capabilities as well as its energy consumption
are of paramount importance for a successful and cost-effective HAPS deployment. Sub-optimal mechanisms that trade  computational costs, and \textcolor{black}{therefore} consumed energy, for the performance in a controlled manner should be developed. Note also that as HAPS systems are expected to \textcolor{black}{supplement and coexist with terrestrial networks, one can also properly share required computational loads  across HAPS systems and terrestrial networks.} }
%  When a HAPS-BS is used, the payload energy consumption is a critical factor to be considered. A recent study proposed an energy consumption model for a HAPS-BS \cite{ arum2020energy}. By considering the energy consumption by both the flight system and the onboard BS, this study shows the feasibility of 15-24 h continuous service using aerodynamic HAPS-BS with a wingspan ranging between 25-35 m.  However, this result is based on the assumptions of the energy system and the selected type of the platform.    Indeed, current \acrshort{haps} projects validate continuous service of HAPS-BS for several months (e.g., Loon 223 days).  

\subsubsection{Passive payload- HAPS-RSS}
\acrshort{haps} deployment for wireless applications is profit-driven. The platform’s weight, energy efficiency, and flight duration play an important role in determining  \acrshort{haps} deployment costs.  
% Although it is possible to equip \acrshort{haps} with advanced communication, caching and computing functions, it might not be profitable in some scenarios, where the
% role of a \acrshort{haps} is restricted to relay signals, or it is only targeting a limited number of users in remote areas.
\textcolor{black}{Although it is possible to
equip a \acrshort{haps} with advanced communication, caching, and computing functions, it might not be profitable in some \textcolor{black}{scenarios: for instance, where the role} of a \acrshort{haps} \textcolor{black}{would be} limited to relay signals, \textcolor{black}{or if it only targeted a limited number} of users in remote areas. In such
situations, a}
 cost-effective \acrshort{haps} deployment \textcolor{black}{would require all of these features,} including low energy consumption, a light payload, and reliable communications for extended flight duration.  To this end, using passive \acrfull{rss} in \acrshort{haps} systems \textcolor{black}{would be beneficial.}
 \acrshort{rss} has recently gained a lot of \textcolor{black}{attention} in the research community \textcolor{black}{and has emerged} as a new technology---\textcolor{black}{indeed, as one of the}  driving technologies \textcolor{black}{of} 6G networks \cite{latva2019key, di2019smart, basar2019wireless}---to support wireless communications. It is shown in \cite{wu2019intelligent} that using \acrshort{rss}  can increase the received \acrshort{snr} and by doubling the number of \acrshort{rss} units, the \acrshort{snr} quadratically increases. Also, \textcolor{black}{as demonstrated in \cite{han2019large},  \acrshort{rss} can substantially improve data rates.}  
 
% An \acrshort{rss} consists of a thin layer of meta-surfaces, which can realize abrupt phase shifts and anomalous reflection and refraction beyond Snell’s law \cite{yu2011light}. 
% Through controlling the meta-surface units (meta-atoms),  customizable interactions of the electromagnetic waves can be acquired. 
\textcolor{black}{An} \acrshort{rss} \textcolor{black}{consists of a
thin layer of meta-surfaces that} can be used to deliberately manipulate the phases and  directions of incident waves in a \textcolor{black}{controlled manner and,} therefore, smartly reflect and refract the signals to the targeted directions. Due to its lightweight and flexible structure, it can be designed in thin films \textcolor{black}{to coat} different surfaces  \cite{liaskos2018new}. Moreover, using digital meta-materials \cite{cui2014coding}, the manipulation of the electromagnetic waves can be digitally controlled.
 To manipulate high-frequency signals, e.g., Gigahertz (GHz) and Terahertz (THz) bands, tunable materials such as liquid crystal \cite{perez2013design} and graphene \cite{carrasco2013tunable,carrasco2013reflectarray} are used. Experiments conducted in \cite{kowerdziej2014terahertz} confirmed that dynamic control of high-frequency signals is feasible. Also, \cite{liu2017ultra} showed that using graphene patches with meta-surfaces can effectively control the wavefront of THz signals with the advantages of adjusting phases up to $180^{\circ}$.
Due to the great potential of the \acrshort{rss} and \textcolor{black}{their} special features, industry efforts have started to utilize this technology for different wireless applications. For instance, NTT DOCOMO in collaboration with Metawave demonstrated \textcolor{black}{that by using \acrshort{rss} at 28 GHz, data rate can be increased by approximately ten times \cite{DOCOMO}.} %an increase in data rate by approximately ten times using \acrshort{rss} at 28 GHz \cite{DOCOMO}.
\textcolor{black}{The company Greenerwave} currently  manufactures an \acrshort{rss} applicable for \textcolor{black}{frequency ranges of up to 100 GHz \cite{greenerwave}}.

%  The reconfigurability feature of the \acrshort{rss} can be achieved by a control circuit with passive elements and switches. These elements include PIN diodes \cite{kamoda201160}, varactor or varicap diodes \cite{hum2007modeling} and radio frequency micro-electro-mechanical systems (RF-MEMS) \cite{hum2006integrated,perruisseau2008monolithic}. By tuning and controlling these elements, the resonant frequency of the reflectors is changed and desired wave manipulation and phase shift are achieved.

% Also, the VisoSurf project, funded by the European Union, is dedicated to designing and prototyping \acrshort{rss} \cite{visosurfweb}.

\begin{figure}[!htbp] 
	\begin{center}
	\centering
\includegraphics[scale=.35]{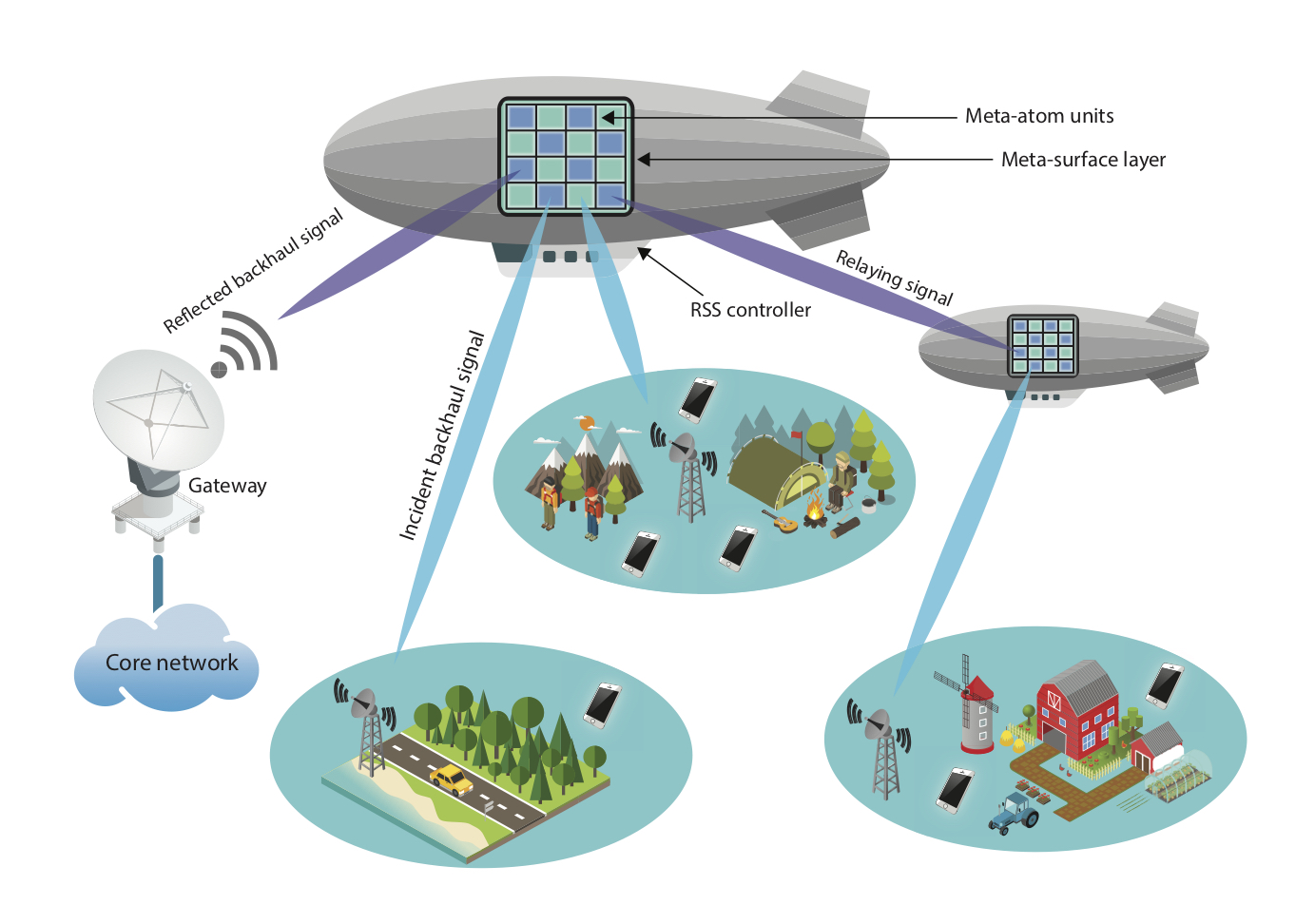} 
\caption{\textcolor{black}{HAPS-RSS for relaying signals and supporting backhaul from remote BSs.}} \label{Fig:HAPS_RSS}
\end{center}
\end{figure}

% Most of the research and industry efforts for utilizing \acrshort{rss} in wireless networks are restricted to terrestrial networks. 
%The integration of \acrshort{rss} in different aerial platforms are also focused in the literature
\textcolor{black}{More recently, researchers started investigating the integration of \acrshort{rss} in different aerial platforms}  \cite{long2020reflections,alfattani2020aerial}. \textcolor{black}{The idea of equipping or coating the HAPS with RSS (HAPS-RSS) was first introduced in \cite{alfattani2020aerial}.} Two possible scenarios for utilizing HAPS-RSS are illustrated in Fig. \ref{Fig:HAPS_RSS}:  Backhauling signals from rural areas to the gateway and inter-\acrshort{haps} \textcolor{black}{communication links.}  \textcolor{black}{In such scenarios, a} ground control station is responsible \textcolor{black}{for sending} the required configurations allowing the onboard \acrshort{rss} controller to configure the \acrshort{rss} to manipulate and direct the signals to the targeted direction.

%\textcolor{black}{HAPS-RSS resembles the function of HAPS-RS but with extra advantages, 

\textcolor{black}{The HAPS-RSS resembles the function of HAPS-RS, but the former has additional advantages,}
\textcolor{black}{as detailed in \cite{alfattani2020aerial}.} \textcolor{black}{For instance,} several recent studies \textcolor{black}{have indicated} that the capacity achieved through \acrshort{rss} is comparable to that achieved through radio relays \cite{huang2018energy,wu2019intelligent, ntontin2019reconfigurable}. As the performance of \acrshort{rss} supporting wireless communications is strongly dependent on the number of reflectors\footnote{The dimensions of each reflector unit are typically between $[\lambda/10, \lambda/5]$, where $\lambda$  is the wavelength \cite{liaskos2018new}.} and given that  \acrshort{haps} are typically  large platforms, it is possible to accommodate a large number of reflectors, which \textcolor{black}{will enhance} the spectral efficiency. Also, \textcolor{black}{an \acrshort{rss} can support} full-duplex communications without suffering from high noise or residual loop-back self-interference, which are the typical limitations of relays \cite{wu2019intelligent , ntontin2019reconfigurable }. 
On the other hand, using \acrshort{rss}, the communication payload energy consumption of \acrshort{haps} declines to the required power for the \acrshort{rss} controller. A recent experiment demonstrated that the configuration of each reflector unit consumed 0.33 mW \cite{tang2019wireless}. As a result of the reduction in the platform’s weight and consumed energy, %the \acrshort{haps}'s flight duration could be prolonged,
the flight duration of the \acrshort{haps} could be prolonged,
 which \textcolor{black}{led} to lower maintenance \textcolor{black}{and deployment}  costs. Studies show that an RSS-assisted communication system could \textcolor{black}{be} 40\% more energy efficient than a relay-assisted system  \cite{huang2018energy}. Thus, HAPS-RSS can be regarded as an energy-efficient and cost-effective solution.

% \subsection{\textcolor{black}{Supporting Use-Cases}}
\begin{comment}

\textcolor{black}{The energy consumption management of the \acrshort{haps}, regarding to communication functionalities, are mainly studied considering the \acrshort{haps} as a conventional BS. This is acceptable for the conventional use-cases of \acrshort{haps}, which includes \acrshort{haps} as a relaying node or a BS. Nevertheless, the future of \acrshort{haps} is more broader than such restricted functionalities. For example, if \acrshort{haps} is intended to be used for data centers, the payload type and energy consumption model can be entirely different. This is also true when it is used as a computation platform or a machine learning platform. In many cases, we expect the station has at least another functionality, as mentioned, besides the common BS/relay one. Therefore, more sophisticated investigations on the energy management and continuity of the service should be considered.  }
\end{comment}

\textcolor{black}{\textcolor{black}{Despite the advancements in HAPS systems discussed} in terms of various \textcolor{black}{platform types, features, onboard energy systems, and communications payloads, several requirements need to be addressed before the HAPS can be deployed to support the novel envisioned use cases.}
%platforms types and features, onboard energy system and communications payload; embracing the novel vision of the HAPS to support the presented future use-cases necessitate satisfying several requirements.
These requirements are dependent on the intended use-cases. In situations where the HAPS are used to tackle the connectivity demands of temporary large events or flash crowds, fast deployment \textcolor{black}{is necessary, and therefore an aerodynamic HAPS might be most suitable in such cases.} However, considering the huge data \textcolor{black}{demands of such scenarios, this would require the aerodynamic HAPS to have relatively high} payload capabilities. Fortunately, this is feasible with recent progress in HAPS projects.  \textcolor{black}{For} example, the recent aerodynamic HAPS ``PHASA-35’’ \textcolor{black}{has a payload power capacity of up to 1,000 W} \cite{prismatic,ABE}, which indicates the great potential for supporting a huge number of users with high data rate requirements. On the other hand, utilizing HAPS for IoT services, supporting ITS/CAV, or as data centers require HAPS \textcolor{black}{with capabilities for prolonged flight durations as well as higher power, communication, caching, and computation.} In these cases, the choice of the platform needs to be carefully considered. Airships, with their intrinsic \textcolor{black}{features, particularly their immense size and high payload capacities,} have the potential to accommodate such heavyweight and high energy consumption payloads. For instance,  the Stratobus airship, expected to be deployed late in 2021, can accommodate a payload of 450 kg with a power rating of 8 kW for a 5-year mission \cite{Stratobus}. In addition, %sine the envisioned use-cases require typically a constellation of HAPS 
%unlike the stand-alone HAPS, 
the envisioned use-cases \textcolor{black}{typically demand a constellation of HAPS.} Building such \textcolor{black}{constellations will require} cost-effective and energy-efficient inter-HAPS communication links. The HAPS-RSS with its passive nature can be utilized to support the required multi-hop links in a cost-effective manner. 
}

\section{Channel Models for HAPS Systems}\label{sec:channels}

\textcolor{black}{To understand the full potential that \acrshort{haps} networks can offer\textcolor{black}{,} a deep understanding of the channel model\textcolor{black}{s} is necessary. This  has been addressed by several studies. }\textcolor{black}{An extensive overview of  channel models, \textcolor{black}{including their extension from \acrshort{haps} to satellite models was} given in \cite{arapoglou2011land}. Although a recent survey \textcolor{black}{of} \acrshort{haps} channel\textcolor{black}{s} is not available, there are numerous studies for modeling the channels of lower altitude platforms, such as  \cite{khawaja2019survey}.  However, \textcolor{black}{since scatterers are more present on the ground than in the stratosphere, channel characteristics} show significant deviations, hence extra caution is \textcolor{black}{needed when} selecting the suitable channel model. An overview of the current channel models is given below. }

%\subsection{Measurement Campaigns}
\subsection{RF Channel Models}

\subsubsection{Empirical-Statistical Models}

\textcolor{black}{The initial studies of \acrshort{haps} channels are adaptations from  land-mobile satellite channels. A high-resolution time series provided by the data form German Aerospace Centre  (DLR) \textcolor{black}{was} first presented in 1991 in  \cite{lutz1991land}, where the authors use\textcolor{black}{d} the land mobile satellite channel to model the link via a digital two-state Gilbert-Elliott structure. This model \textcolor{black}{was} then used to assess the performance of the proposed \acrshort{haps} channel models. An overview of the satellite channel models \textcolor{black}{applicable to} \acrshort{haps} \textcolor{black}{systems were} first noted in \cite{ vazquez2002channel}. \textcolor{black}{Studies of channel modeling in stratospheric} telecommunication systems started \textcolor{black}{with the modeling of atmospheric effects on system performance} \cite{ oestges2001coverage}.}% I could not find this work -  J.-M. Park, B-J Ku, and D.-S. Ahn, “The simulation modeling and performance analysis of stratospheric communications system,” in Proc. 4th Int. Symp.Wireless Personal Multimedia Communications, Aalborg, Denmark, Sept. 9–12, 2001, pp. 1641–1644.}.

%\subsubsection{Deterministic Channels}
\subsubsection{Non-Geometric Stochastic Models}

\textcolor{black}{The first study that considered the impact of multi-path (i.e., small scale) fading \textcolor{black}{in} the presence of terrestrial scatterers in \acrshort{haps} channels \textcolor{black}{was} \cite{dovis2002small}, where the authors derive\textcolor{black}{d a}  channel model for \textcolor{black}{the} 2 GHz band. \textcolor{black}{Since attenuation due to rain} is negligible in this band, it \textcolor{black}{was} not considered in the analyses. The authors considered an ellipsoid channel model by placing the transmitter and the receiver as foci. The scatterers \textcolor{black}{were assumed to be uniformly positioned along the ellipsoid}. } 

\textcolor{black}{\textcolor{black}{A} statistical model for mixed propagation conditions for land mobile-satellite systems \textcolor{black}{was} presented in the  \acrshort{itu-r} Recommendation P.681-11 \cite{ITU-681}, along with the duration, state distributions and the transition probabilities. The combination of the line-of-sight, slight shadowing and the total obstruction conditions for the \acrshort{haps} channel models using a semi-Markov studied \textcolor{black}{was} proposed in \cite{cuevas2004channel} and extended to tapped-delay lines in \cite{cuevas2004statistical}. The impact of these state switches on the error performance \textcolor{black}{was} noted.}
\textcolor{black}{ \cite{king2005physical} present\textcolor{black}{ed} a statistical model for jointly estimating the statistical time-series and power spatial delay profile for HAPS-MIMO channels. A model comparison \textcolor{black}{was} also provided with the data provided by DLR  \cite{lutz1991land}, using the corresponding first order statistics. }

\subsubsection{Geometry-Based  Stochastic Models}

\textcolor{black}{The first study that proposes a geometry-based stochastic model (GBSM), specifically  a geometry-based single-bounce (GBSB) model, \textcolor{black}{was} introduced in \cite{michailidis2008spatially}, where the authors \textcolor{black}{presented} a 3D scattering model for \textcolor{black}{a} stratospheric multi-path fading channel for isotropic and non-isotropic scattering environments. The spiral and temporal correlation functions \textcolor{black}{were} provided and the required antenna separation distance \textcolor{black}{was} derived. The authors \textcolor{black}{in \cite{michailidis2010capacity} studied} the impact of the channel model on the capacity expressions that \textcolor{black}{could} be obtained in \acrshort{haps} communication channels. This work \textcolor{black}{was} then extended in \cite{michailidis2010three}, where  the elevation angle of the platform, the array orientation and configuration, the Doppler spread, and the distribution of the scatterers \textcolor{black}{were} considered for non-isotropic scattering environments using a 3D geometry-based single-bounce reference model for Ricean fading channels. It \textcolor{black}{was} observed that the  model parameters \textcolor{black}{had} a significant effect on the space-time correlation, \textcolor{black}{and that the corresponding impact needed} to be taken into account in the array designs.  \cite{ michailidis2012statistical} define\textcolor{black}{d} a 3D GBSB  sum-of-sinusoids (SoS) principle-based statistical simulation model for \textcolor{black}{HAPS}-MIMO \textcolor{black}{channels using the framework of the reference model} in \cite{michailidis2010three}. Its wideband extension \textcolor{black}{was} presented in \cite{michailidis2014wideband}. }

\textcolor{black}{Considering the relay channel use-case of \acrshort{haps} nodes between two terrestrial nodes, a geometry-based modeling of \acrshort{mimo} M-to-M relay-based channels \textcolor{black}{was} presented in \cite{michailidis2012three}.  An extension of the \cite{michailidis2010three} model for low altitude air-to-ground \acrshort{uav} communication channels \textcolor{black}{was} introduced in \cite{mendoza2019application}. The main difference between the low-altitude versus high altitude channel lies in the probability of the line-of-sight presence. Furthermore, at high altitudes, the air node encounters no local scatterers.}

\subsubsection{Non-Stationary Models}

\textcolor{black}{A birth-death process-based non-stationary \acrshort{los} component appearance and disappearance \textcolor{black}{was} presented in \cite{nikolaidis2016dual}, where the authors detailed the  derivation of the multi-user spatial correlation function \textcolor{black}{and} extended the use-cases to multi-user \acrshort{haps} environments.  A non-stationary 3D \acrshort{mimo} GBSM \textcolor{black}{was} investigated in  \cite{lian2018non}, where the authors model\textcolor{black}{ed} the appearance and  disappearance of  multi-path components using a two-state continuous-time Markov process. Closed form expressions of survival probabilities \textcolor{black}{were} derived. Long distance and small-scale time-variant parameters \textcolor{black}{were} also considered to model the non-stationary aspects of the \acrshort{mimo} channel. The Space-Time Correlation Function and the Doppler Power Spectral Density expressions \textcolor{black}{were} presented. \textcolor{black}{A dual polarized \acrshort{mimo} channel model for \acrshort{haps} systems was studied}   in \cite{yang2017statistical}, where spatial correlation and polarization correlation expressions \textcolor{black}{were} also provided.} 

\textcolor{black}{The dynamic evolution \textcolor{black}{of an \acrshort{los} component in 3D models was investigated} in \cite{lian20193} by a two-state continuous-time Markov chain. Closed-form expressions are derived for the survival probabilities of the \acrshort{los} components using Chapman-Kolmogorov equations along with the corresponding space-time correlation function. }

\subsubsection{Air-to-Air Channels}
\textcolor{black}{Air-to-air \acrshort{haps} scenarios \textcolor{black}{have} also \textcolor{black}{been} considered in the literature. The authors in \cite{ma2019wideband} introduced  a 3D non-stationary geometry-based scholastic model for air-to-air channels  using a 3D Markov mobility model where the \textcolor{black}{nodes could move both horizontally and vertically, and their velocities could change in time.} The authors derive\textcolor{black}{d} the time-frequency correlation function and the Doppler power spectrum. The results highlight\textcolor{black}{ed} the importance of considering  vertical movement. Yet the model about the scatterers needs to be carefully addressed for \acrshort{haps} scenarios. }

\subsection{Free-Space Optical Channels}

\textcolor{black}{In \acrshort{fso} communications, light signals that carry information are transmitted in free-space environments\textcolor{black}{,} such as \acrshort{los} links on the ground or \textcolor{black}{in the} vacuum \textcolor{black}{of} space. Due to \textcolor{black}{their cost-effective, license-free, and high-bandwidth nature, \acrshort{fso} communications are a leading}
technology solution, especially for the \acrshort{haps}-to-\acrshort{haps} connectivity and  backhaul transmission links. Hence, the usage of \acrshort{fso} for HAPS-based communication systems have been thoroughly investigated. \cite{giggenbach2002stratospheric} numerically demonstrate\textbf{d} the possibility of 9,000 km of inter-HAPS distances with a high  reliability. The main impairments encountered in the \acrshort{fso} channel links include\textcolor{black}{d} fluctuations in the  received signal due to turbulence, wind\textcolor{black}{,} and pressure fluctuations  \textcolor{black}{along with irregularities in} temperature \cite{andrews2005laser}. Various statistical models have been proposed to model the channel characteristics. Gamma-gamma distribution \textcolor{black}{was} proposed in \cite{al2001mathematical}. \cite{majumdar2005free} proposed a lognormal model to model weak fluctuations.  \acrshort{fso} channels have also been used as backhaul links in \acrshort{haps} nodes. An excellent overview \textcolor{black}{was} provided in \cite{kaushal2016optical}.}

\subsection{System Performance Analysis}
\textcolor{black}{
\textcolor{black}{The channel models discussed above feature prominently in studies of \acrshort{haps} system performance.}  The Shannon capacity and coverage probability are considered as the main performance metrics. Furthermore, to tackle the pitfalls of \acrshort{fso} \textcolor{black}{communications, including beam wandering and pointing errors and a sensitivity to atmospheric conditions, the use} of multi-hop \acrshort{haps} communications and relaying \textcolor{black}{was} suggested.} \textcolor{black}{In \cite{palma2010wimax}, the use of WiMAX HAPS-based for delivering data to fixed terrestrial users on the ground \textcolor{black}{was} investigated. The authors introduce\textcolor{black}{d} a channel model comprised of geometrical and statistical components to derive the \acrshort{ber} performance. The channel model \textcolor{black}{took} into account the \acrshort{los} occurrence prediction along with statistical shadowing. The authors use\textcolor{black}{d} satellite communication system records to corroborate their analysis and introduced channel models. This model, however, \textcolor{black}{had} limitations for correctly measuring the transition state of the channel.  \textcolor{black}{By contrast}, \cite{zakia2017capacity} used \acrshort{haps} to provide capacity for \textcolor{black}{a} high-speed train using \acrshort{mimo} communication in Ka band. It \textcolor{black}{was} shown that despite the strong \acrshort{los}, the channel \textcolor{black}{was} ill-conditioned due to high speed of the train, \textcolor{black}{and that} the multiplexing gain of the \acrshort{mimo} system reduced substantially. Suitable antenna distancing, up to couple of centimeters, at the receiver \textcolor{black}{appeared to be effective in curbing the degradation of the ill-conditioned}  \acrshort{mimo} channel, which, compared to typical handsets, \textcolor{black}{was} affordable. The authors of \cite{michailidis2010three} adopt\textcolor{black}{ed} a 3D geometric channel model along with \textcolor{black}{a} Rician fading channel to study the impact of antenna \textcolor{black}{placement} for achieving \textcolor{black}{an} uncorrelated response in HAPS-MIMO communications. It \textcolor{black}{was} shown that Doppler spread, array configuration, and the distribution of scatteres \textcolor{black}{had a} fundamental impact on the statistics of the channel. The theoretical results \textcolor{black}{could} be used to evaluate the performance of HAPS-MIMO channels. In \cite{Sudheesh2013CSIimperfectMIMO}, the authors \textcolor{black}{studied} the capacity of a HAPS-MIMO interference channel \textcolor{black}{comprised} of two \acrshort{haps} systems and two ground users. It \textcolor{black}{was} observed that to achieve the best performance, which corresponded to  independent channel power gains, the users \textcolor{black}{had to} be sufficiently separated spatially. The high spatial correlation \textcolor{black}{appeared to be due to the} angle-of-departure and angle-of-arrival at the transmitters and receivers, respectively. In \cite{grace2005improving}, the authors used mmWave communication to increase the capacity of a single \acrshort{haps} system and a \textcolor{black}{constellation of eight} \acrshort{haps} systems. Their analysis included the evaluation of Shannon capacity as well as the impact of modulation and coding by incorporating the \textcolor{black}{angular separation between ground users and a \acrshort{haps},}  the link length ratio, and the side-lobe of the antenna. Their analysis \textcolor{black}{indicated} that 4 km circular spacing is optimal. In \cite{thornton2003optimizing}, the authors \textcolor{black}{investigated} the coverage performance of \acrshort{haps} \textcolor{black}{systems} operating in 28 GHz and 48 GHz via approximating curve-fitting of the antenna pattern radiation. The analysis allowed the authors to shed some light on  \textcolor{black}{important} issues, \textcolor{black}{such as antenna beam type and frequency re-use, which affect cell planning} in \acrshort{haps} systems. 
}

% \textcolor{magenta}{The error performance of the channel models is investigated in \cite{palma2010wimax}. The channel capacity of MIMO-aided HAPS communications is investigated by Zakia in \cite{zakia2017capacity} for high-speed trains.  \cite{jia20193d} \cite{ma2019wideband} \cite{dovis2002small} \cite{michailidis2010three} \cite{vazquez2002channel}}

\textcolor{black}{The use of relays to extend the range of the communication systems due to  beam pointing errors and turbulence has been investigated in the literature in \textcolor{black}{the} presence of gamma-gamma atmospheric turbulence channels \cite{datsikas2010serial}.} \textcolor{black}{The impact of different types of relaying, such as amplify-and-forward, channel-state-information-assisted or fixed-gain relays, on the statistical properties of the \acrshort{sinr} \textcolor{black}{was} investigated  to derive the coverage probability via \textcolor{black}{a} moment-generating function. Adopting the composite channel gamma-gamma distribution\textcolor{black}{,} the closed-form expressions for the channel capacity and \acrshort{ber} \textcolor{black}{were} derived in  \cite{ sharma2016high} by adopting free-space optical links for multi-hop \acrshort{haps} communication. It \textcolor{black}{was} shown that the side effect of beam wandering and random pointing errors can be mitigated via multi-hop communications. The considered system \textcolor{black}{was} shown to be robust against fog, rain, snow, and other atmospheric turbulence. The conducted analysis \textcolor{black}{conducted by} \cite{salhab2016new} suggested that the worst relay channel \textcolor{black}{had a} profound impact on the reduction of \textcolor{black}{performance, such as capacity, outage probability, and \acrshort{ber}, of multi-hop}
 \acrshort{haps} systems. The authors then proposed the use of power allocation in order to minimize   negative effects.} \textcolor{black}{  \cite{michailidis2018outage} \textcolor{black}{investigated a triple-hop} triple-hop RF-FSO-RF communication system supported with an \acrshort{fso} link between two \acrshort{haps} nodes, which \textcolor{black}{were connected to a} terrestrial network via RF links.} \textcolor{black}{The performance of the system in term of outage probability \textcolor{black}{was} derived, which show\textcolor{black}{ed} that the \acrshort{nlos} occurrence and  vulnerability of the \acrshort{fso} link to atmospheric turbulence, pointing errors, and beam wandering render\textcolor{black}{ed} the growth of the outage probability. }

\section{\textcolor{black}{Radio Resource Management, Interference \textcolor{black}{Management,} and Waveform Design in HAPS Systems}}\label{resource}
\begin{figure}[t]
	\begin{center}
	\centering
\includegraphics[width=0.95\textwidth]{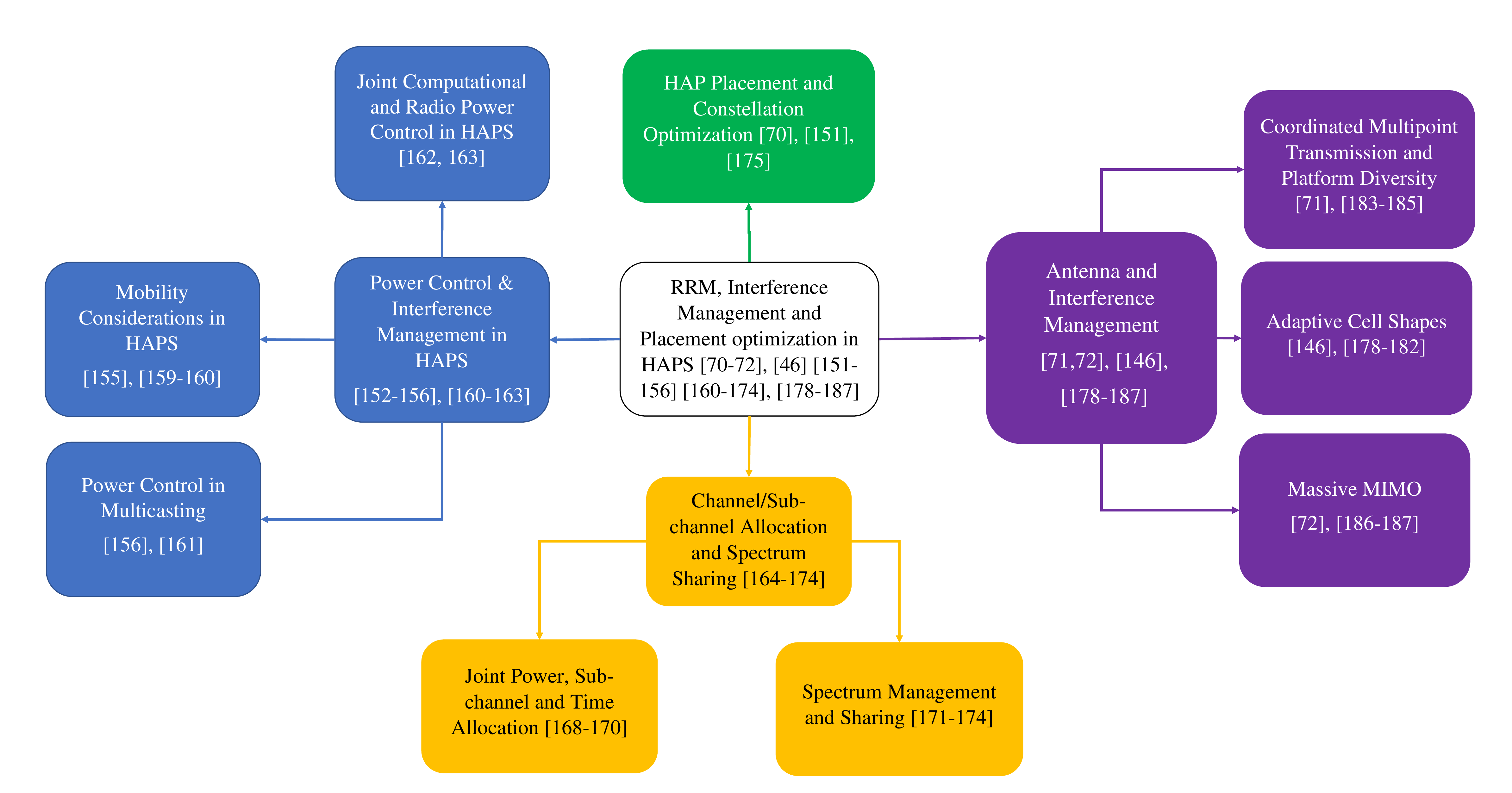}
\caption{\textcolor{black}{\textcolor{black}{Taxonomy diagram for the key \textcolor{black}{radio resource management, antenna and interference management, and} \acrshort{haps} placement strategies in the literature.} 
% \textcolor{black}{N.B:There is one additional reference to be added here. I will add it and renumber there reference numbers here in the figure at the very end, when we are sure no more changes in the order of the references could happen.} 
}}
\label{tax}
\end{center}
\end{figure}

\textcolor{black}{As in all wireless communication networks, radio resource management, interference management, \textcolor{black}{and waveform design} are crucial aspects for ensuring the performance of a \acrshort{haps} system. \textcolor{black}{From a service provider point of view, network performance} \textcolor{black}{can be measured by the spectral efficiency}, i.e., throughput (in bits/sec) per unit Hertz, which needs to be maximized. In addition, energy efficiency is one of the system design objectives that needs to be addressed, since the platform's energy \textcolor{black}{source mostly consists of rechargeable batteries and solar panels}, \textcolor{black}{i.e., a HAPS does not have the permanent directly connected power supply that the terrestrial BSs enjoy.} \textcolor{black}{By contrast, from a user point of view,} the performance of \textcolor{black}{a HAPS network can be assessed on the basis of certain} \acrfull{qos} \textcolor{black}{metrics, which largely depend} on the type of application. For example, in \acrshort{urllc} the end-to-end latency and packet error rate is crucial. %Another example is interactive applications such as online gaming and video calls, in which delay variation (also known as jitter) could become a more important key performance indicator while allowing higher tolerance to packet losses. 
\textcolor{black}{The way the HAPS on-board radio power and frequencies, antenna \textcolor{black}{beams, and} time scheduling are managed all play an important role in interference management determining the \acrshort{sinr} levels.} \textcolor{black}{This has a direct impact} on the outage probability, \acrshort{ber}, transmission delay, throughput and/or spectral efficiency at the system level. \textcolor{black}{In an integrated aerial/terrestrial/satellite system that is comprised of a HAPS constellation, \textcolor{black}{LEO-satellites,} and terrestrial BSs, the placement strategies of the HAPS nodes play a vital role as well. \textcolor{black}{Therefore, we highlight} some of the key radio resource and interference management schemes developed for \acrshort{haps} systems in the literature, focusing on power control, channel allocation, user association, beamforming, and the placement algorithms for \acrshort{haps} systems for which a related taxonomy diagram is provided in Fig. \ref{tax}.}}

\textcolor{black}{In connection \textcolor{black}{with the use cases discussed} in Section II-B, a HAPS-SMBS \textcolor{black}{can} provide coverage for unplanned events, \textcolor{black}{where terrestrial wireless capacity and/or coverage are insufficient for the users attending the event}. For this \textcolor{black}{use case, a} number of issues need to be taken into \textcolor{black}{consideration:}}
\begin{itemize}
\item \textcolor{black}{The largest number of users need to be admitted, \textcolor{black}{which is equivalent to the minimization of the user blocking probability.}}
\item \textcolor{black}{ Minimization of the service dropping probability. This is the probability a certain service QoS requirement (e.g. minimum rate requirement) is not satisfied. }
\item \textcolor{black}{The uses have heterogeneous service types because of different subscription plans. This means their uplink and downlink target rates are not all the same. The HAPS-SMBS's \textcolor{black}{resources need} to be managed such that the target rates are satisfied and fairness is achieved among users in the same service type.}
\end{itemize}
\textcolor{black}{\textcolor{black}{Some HAPS RRM} and admission control techniques that consider some of the aforementioned requirements \textcolor{black}{are discussed} in this section. Additionally, we believe that the HAPS-SMBS should also be `event' aware, i.e., \textcolor{black}{have} the ability to distinguish between unplanned event types and their requirements. For an emergency \textcolor{black}{event or disaster}, the HAPS-SMBS needs to \textcolor{black}{provide only voice call services}. On the other hand, in an event like outdoor musical festivals, it may be of little importance to spare much resources for audio calls due to the high musical noise in the surroundings which discourage users from making calls. It would rather need to prioritize hologram and augmented reality services, that could be part of the festival needs. If the unplanned \textcolor{black}{event} is a protest, perhaps video upstreaming could be of more significance than video \textcolor{black}{downstreaming.} }

\textcolor{black}{The use of HAPS-SMBS for backhauling of small-cell or isolated \textcolor{black}{BSs,} as discussed in Section II-B, requires joint backhaul link power allocation and associations as in \cite{alsharoa2019improvement} (discussed in Section VI-C). \textcolor{black}{More research is needed, however, especially for HAPS-SMBS mega-constellations} with possibly \textcolor{black}{no cooperation} from LEO satellites (as in \cite{alsharoa2019improvement}) for this particular use case. The design problem is expected to involve the association of small-cell (or isolated BS) to the HAPS-SMBSs as well as the power allocation jointly. \textcolor{black}{Moreover, using-point-to-multi-point narrow beams between the HAPS and the terrestrial BSs by exploiting the mMIMO beamforming technology, rather than using a wide coverage footprint is expected to come with merits for this use-case. This is expected to \textcolor{black}{considerably improve} the resulting SINR of the backhaul connections. That being said, the use of very high-order  modulation techniques will be possible in order to increase the spectral efficiency of the backhaul links without any degradation in BER. This is \textcolor{black}{especially crucial since the required data rates over backhaul links are expected to be tremendously large (Tbps)}. Although the positions of terrestrial BSs are fixed, beam pointing compensation schemes will need to be employed, especially in aerodynamic HAPS platforms that need to move in circles.} A more complex scenario that needs to be studied is when the backhaul traffic of a terrestrial BS is relayed over more than one HAPS, which also involves relay selection.}

\textcolor{black}{In addition \textcolor{black}{to system} level design considerations, link level design aspects are paramount in specifying how the communication will take place at the bit and waveform level through the channel. The link level (or PHY) design is concerned with a single wireless link between a transmitter and a \textcolor{black}{receiver and plays} a vital role in determining the achievable transmission reliability (measured in BER or outage probability) as well as the achievable spectral efficiency. 
% The trade-offs at this level are the ratio of bit energy to noise power spectral density, $\frac{E_{b}}{N_{o}}$, the spectral efficiency, BER and the complexity involved in the demodulation, detection, and channel decoding. 
In a \textcolor{black}{multi-user} system, the adopted waveform technology \textcolor{black}{would be a main factor} in determining/developing the multiple access scheme. % (e.g. OFDMA for OFDM). 
\textcolor{black}{Unfortunately, neither waveform technology nor the related multiple access scheme(s) have yet received much attention in the literature on HAPS systems.} In this section, we also discuss \textcolor{black}{some promising} waveforms for \acrshort{haps} systems.
% with focus on several prominent waveform technologies including \textit{faster-than-Nyquist} (FTN) signalling, \textit{spectrally efficient frequency division multiplexing} (SEFDM), \textit{non-orthogonal multiple access (NOMA), and multiple-input multiple-output NOMA (MIMO-NOMA).}
}

% \textcolor{black}{The aforementioned system-level and link-level (PHY/waveform) design aspects jointly impact the QoS of the target user/application and the overall system objectives like spectral efficiency, energy efficiency, and latency. Hence, cross-layer design becomes necessary to optimize the system performance where the inputs and outputs of the corresponding algorithms are possibly parameters in different layers,  such as power, queue status, sub-channel power gain, scheduling time, modulation and coding scheme (MCS), \acrshort{mimo} complex weights, subcarrier spacing, and signaling acceleration parameter. }

\subsection{\textcolor{black}{Power Control/Allocation and Interference Management in HAPS systems}}
\textcolor{black}{Power is the fundamental radio resource in any wireless system including \acrshort{haps} systems. %Unfortunately, the desired signal of one UE device is seen as an interference on another in the system, if they share the same frequency at the same time. 
In order to serve a large number of users/devices with different QoS requirements, the \acrshort{sinr} \textcolor{black}{level at the} receiver \textcolor{black}{is critical and requires} sophisticated power management schemes.
A significant portion of the proposed power control schemes \textcolor{black}{in the literature on HAPS} systems dates back to early \textcolor{black}{2000s, when most wireless communications research employed wide-band CDMA (WCDMA)} for their air interface technology \cite{abrardo2003centralized,foo2002centralized1,foo2002centralized2,foo2004call,araniti2007multicast}.  The frameworks of these power allocation schemes are more general and could be suitable (with minor modifications) for other potential radio \textcolor{black}{access technologies, like} multicarrier-CDMA \cite{adachi2005broadband} or power domain NOMA \cite{ding2017survey}.  While CDMA is primarily built \textcolor{black}{on the idea that} users are separated by exploiting the differences among their spreading codes,  NOMA   allows multiple users to employ exactly the same code and allocates more \textcolor{black}{power to UEs with lower channel gain}. The interference is removed \textcolor{black}{at the UE} using the \textit{successive interference cancellation} (SIC) scheme, where the UE  first decodes the interferer's message and then removes this message from its observation before decoding its own message.} \textcolor{black}{\textcolor{black}{By contrast, since we expect that HAPS will play a major role} in providing a worldwide network connectivity, the CDMA technology may emerge \textcolor{black}{as a possible solution} for a particular region. A summary of the power control schemes reported in the \acrshort{haps} literature is provided in Table \ref{TAb_RMM1}.} 

\textcolor{black}{The works \cite{abrardo2003centralized,foo2002centralized1,foo2002centralized2,foo2004call,araniti2007multicast} \textcolor{black}{studied power control} for the purpose of call admission control (CAC)\footnote{It \textcolor{black}{should} be noted that the term `call' here is equivalent to user admission request and is not limited to voice calls.}, where users \textcolor{black}{were} admitted into the system such that the Grade of Service (GoS) \textcolor{black}{was} maximized by minimizing a weighted sum of the dropping and blocking probabilities, while satisfying the \acrshort{sinr} and power constraints.}
\textcolor{black}{In \cite{foo2002centralized1}, the unique characteristic of \acrshort{haps} systems that all base stations are collocated on the same platform \textcolor{black}{was} exploited for uplink connections. Unlike terrestrial cellular networks, this feature allows the exchange of information on the interference conditions within the cells between base stations with no signaling overhead. 
\textcolor{black}{If the total power at any BS is less than or equal \textcolor{black}{to} a power outage threshold, the call gets \textcolor{black}{accepted; otherwise} it gets blocked. The central admission controller \textcolor{black}{in \cite{foo2002centralized1}} updates the BS total received power levels on a call-by-call basis so that the admission decision for new calls can be made more accurately.}
%Two centralized call CAC schemes are possible. The first one processes the calls in random order which could possibly admit a UE in a cell that already has high total received power leading to blockage of other subsequent calls in neighboring cells. The second one is  based  on  priority,  where  highest  priority  for  admission  is  given  to  a  call  request  in  the  cell  with  the  lowest  total  received power. 
} \textcolor{black}{The work in \cite{foo2002centralized2} extend\textcolor{black}{ed} the schemes in \cite{foo2002centralized1} and explore\textcolor{black}{d} a downlink CAC scheme. It studie\textcolor{black}{d} a \acrshort{haps} that centrally manage\textcolor{black}{d} the radio power at the platform level and allocate\textcolor{black}{d} it to the cells based on their demands.} \textcolor{black}{The basic idea for the BS-based downlink CAC \textcolor{black}{in \cite{foo2002centralized2}} is to manage incoming calls according to the increase in the interference levels of the target cell as well as adjacent cells. Hence, with the admission of the new call, the downlink powers for all UEs must be increased to satisfy all UEs' \acrshort{sinr} requirements. A call is blocked if admitting the call would cause the UE's target base station as well as other neighboring base stations to exceed the maximum allowable output powers and is admitted if the total platform power is not exceeded and the SIR thresholds are satisfied.}

\subsubsection{\textcolor{black}{{Mobility and Power Control}}}
\begin{figure}[!htbp]
	\begin{center}
	\centering
\includegraphics[width=1.0\columnwidth]{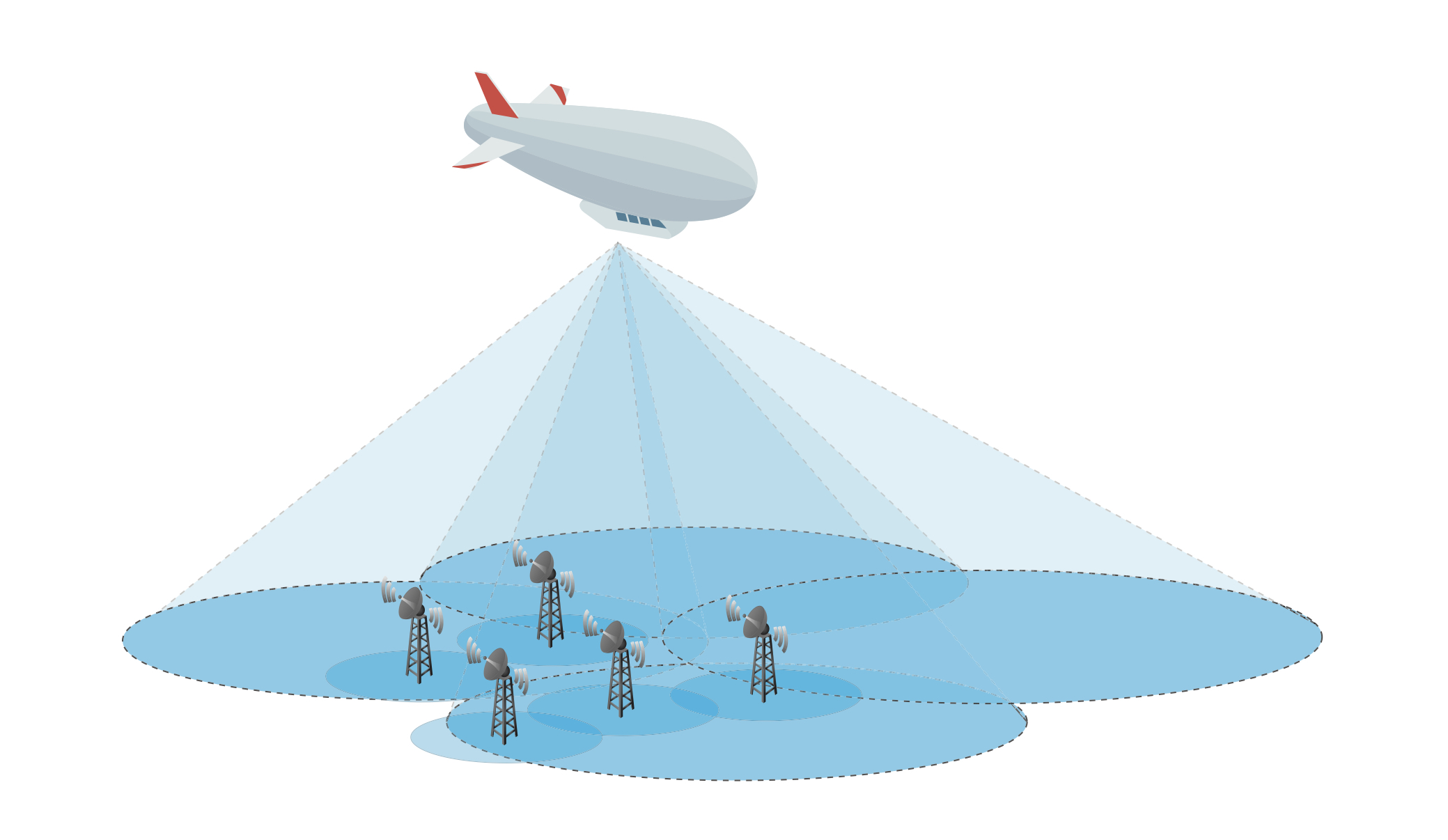}
\caption{\textcolor{black}{A hierarchical \textcolor{black}{HAPS-SMBS} and terrestrial cellular system.}}
\label{HAPandTerr}
\end{center}
\end{figure}

\textcolor{black}{\textcolor{black}{The mobility of UEs} in a HAPS service area has an impact on how power should be controlled such that UE admission is optimized. Only a \textcolor{black}{few publications} have considered the mobility of UEs in power control of \acrshort{haps} systems.  \cite{foo2004call} consider\textcolor{black}{ed} a hierarchical system in which a single \acrshort{haps} and a terrestrial cellular network \textcolor{black}{were} jointly \textcolor{black}{deployed, as} illustrated in Fig. \ref{HAPandTerr}. \textcolor{black}{In that system,} the \acrshort{haps} \textcolor{black}{would be} used to provide \textcolor{black}{SMBS} coverage and the terrestrial cellular towers \textcolor{black}{would be} used for \textcolor{black}{macro}-cell coverage at a different frequency band, therefore cross-layer interference is avoided. \textcolor{black}{The CAC scheme \textcolor{black}{that the authors' developed} uses a combination of overflow and speed sensitive strategies to direct calls arriving within overlapp\textcolor{black}{ing} service areas served by both \acrshort{haps} macro-cell and terrestrial micro-cell layers to the appropriate layer.} %A speed threshold is used such that UEs with speeds greater than the threshold get directed to the HAP layer, while UEs with speeds lower than the threshold get directed to the micro-cell layer for admission.
}

\textcolor{black}{In \cite{foo2005speed}, a speed and direction-based CAC scheme \textcolor{black}{was} developed for a  standalone \acrshort{haps} system with the objective of reducing the handoff call dropping probability as much as possible, as forced termination \textcolor{black}{is less desirable than the blocking of a new call.} For this scheme, the system continuously track\textcolor{black}{ed} the SIR received from the UE's serving BS's pilot channel and the next strongest SIR received from the UE's neighboring base stations' pilot signals. It \textcolor{black}{was} used to derive the speed and direction of the mobile UE relative to the rest of the UEs.
%Therefore, the new call admission thresholds in each cell are dynamically adjusted to ensure that all cells reserve sufficient resources for the handoff users that are predicted to enter their service areas from other neighboring cells. 
}
\textcolor{black}{ In \cite{ceran2017optimal}, \textcolor{black}{the authors} studied high altitude on-the-move flying wireless access points powered by renewable energy. The access point allocates its available energy to maximize the total utility (reward) provided to a sequentially observed set of users demanding service. %The problem is formulated as a 0/1 dynamic knapsack problem with incremental capacity over a finite time horizon. 
% The problem is addressed through deterministic and stochastic formulations followed by a model where the statistics of the underlying processes are not known and learned through rule-based and neural network approaches. \textcolor{black}{ For the deterministic problem,  online approximations including optimization via genetic algorithm and rule-based approach are proposed according to an instantaneous threshold that can adapt to short-time-scale dynamics. For the stochastic model, after showing the optimality of a threshold-based solution on a dynamic programming formulation, an approximate threshold-based policy is obtained.}
} 

\subsubsection{\textcolor{black}{{Power Control for Multicast Services}}}
\textcolor{black}{Only a few papers \textcolor{black}{have considered} power control for multicasting. In \cite{araniti2007multicast}, \textcolor{black}{an integrated terrestrial cellular and \acrshort{haps} system was} considered for Multicast Broadcast Multimedia Services (MBMS) applications with the aim of efficiently allocating transmission resources to multicast traffic streams by suitably selecting terrestrial and/or \textcolor{black}{HAPS} channels while still preserving the desired QoS \textcolor{black}{for} unicast traffic. 
% The authors consider two different  approaches, where one takes into account the number of users in each multicast group to drive the RRA policies; while the second  bases the  allocation behavior on the distribution of users belonging to each multicast group. %The \textit{Number of Multicast Users}  policy moves the multicast connections which belong to the largest multicast group, onto the HAP channel since doing the opposite is expected to degrade the performance due to congestion in the terrestrial network. \textit{Distribution of Multicast Users} achieve an improvement in performance in certain situations by assigning the forward access channel  of the \acrshort{haps} to the multicast group that is most scattered across the cells. This frees more FACH of the terrestrial system for other multicast groups. The performance improvement achieved is both in terms of cell resource utilization and QoS of unicast connections.
} 
\textcolor{black}{In \cite{raschella2009high}, a technique \textcolor{black}{was} proposed to improve the overall system capacity by selecting the most efficient multicast transport channel in terms of power consumption by defining the switching thresholds between point-to-point and point-to-multipoint connections \textcolor{black}{while taking into account the radio channel conditions, the cell coverage radius, and two sample MBMS application bit rates.} It was observed that for \acrshort{mbms} services, the choice of the most efficient transport channel \textcolor{black}{was} a key aspect \textcolor{black}{that \textcolor{black}{impacted} the overall system capacity.}
% since a wrong transport channel selection could adversely affect the overall capacity of the system. %The  RRM policy aims to identify the best `switching thresholds' among dedicated channels and forward access channels, by taking into account the radio channel conditions, the cell coverage radius, and two sample MBMS application bit rates. %The results that were obtained demonstrated that a smart selection of transport channels coupled with the RRM policy, leads to an efficient management of MBMS services in a HAP standalone scenario. 
}

\subsubsection{\textcolor{black}{{Joint Radio and Computational Power Management}}}
\textcolor{black}{\textcolor{black}{\textcolor{black}{HAPS-SMBSs are also envisioned for aerial edge computing, as discussed} in Section II-B. \textcolor{black}{In such cases, power and time consumed} in onboard computation should be managed jointly with radio resources. \textcolor{black}{Two recent works that have explored this use case are \cite{8766778, wang2020federated}}. In \cite{8766778} \textcolor{black}{the authors explored task offloading problem in a two-tier aerial network consisting of a low-altitude UAV tier and a HAPS tier}. The HAPS nodes \textcolor{black}{were} equipped with MEC servers to perform the computations of the tasks offloaded from the low-altitude UAVs. The design choices that were optimized to minimize \textcolor{black}{the} offloading delay \textcolor{black}{were} the offloading ratios of the tasks, which determine\textcolor{black}{d} the number of tasks to be processed locally and the number of tasks the low-altitude UAVs \textcolor{black}{were} to offload based on the available computational power of the servers onboard. The second optimization design choice \textcolor{black}{was} the uplink transmission power of the low-altitude UAVs. The authors in \cite{8766778} tackle\textcolor{black}{d} this problem by modeling it as a  multi-leader multi-follower Stackelberg game \textcolor{black}{and solving it} using the lower complexity equilibrium problem with \textcolor{black}{an} equilibrium constraints (EPEC) model.}  \textcolor{black}{The work in \cite{wang2020federated} explore\textcolor{black}{d} the multi-objective problem of minimizing energy and time consumption for task computation and transmission in a MEC-enabled balloon network}.
% In the considered network, each user needs to process a computational task in each time instant, where \acrshort{haps} acting as flying wireless base stations (i.e. HAPS-BS) can use their powerful computational abilities to process the tasks offloaded from their associated users. 
Since the data size of each user’s computational task varies over time, the HAPS-BSs must dynamically adjust the user association, service sequence, and task partition scheme, \textcolor{black}{which are the design choices that \cite{wang2020federated} considered, to meet \textcolor{black}{the needs of users.}} A support vector machine (SVM)-based federated learning (FL) algorithm determine\textcolor{black}{d} the user association proactively, \textcolor{black}{ before the service sequence and task allocation of each user \textcolor{black}{were} optimized so as to minimize the weighted sum of the energy and time consumption.}}

\subsection{\textcolor{black}{{Channel/\textcolor{black}{Sub-Channel} Allocation and Spectrum Sharing}}}

\textcolor{black}{Channel allocation is one of the principal functions in any multi-user wireless system. When it comes to multi-user/multi-HAPS systems, channel allocation schemes can exploit \textcolor{black}{the inherent diversity of such systems} through a dynamic allocation scheme. Basically, the spectrum gets divided into sub-channels, where one or more of these \textcolor{black}{sub-channels} can be allocated to \textcolor{black}{one or more UE}. The channel attenuation---due to path-loss, shadowing, and fast fading---seen by each user is \textcolor{black}{different, as} discussed in Section \ref{sec:channels}. 
%due to the following reasons:
% \begin{itemize}
% \item Users are at different distances from the \acrshort{haps}, and hence they experience different signal path losses,
% \item The shadowing losses due to buildings and other obstacles are different and independent for distant users,
% \item The multipath effect for each user is different as the number of multipath components, their amplitudes and their phases are different and independent for each user.
% \end{itemize}
}
\textcolor{black}{
\textcolor{black}{Therefore, each user is expected to experience different and independent attenuation on a given channel, which could be exploited for achieving multi-user diversity gain.}
Moreover, in integrated \acrshort{haps} systems, i.e., \acrshort{haps}/terrestrial/satellite systems, more than one of these layers may be operational within the same band. Hence spectrum sharing schemes as well as inter-layer interference management \textcolor{black}{will be crucial} to guarantee the performance of the system. In the rest of this sub-section, we outline some of the key channel/sub-channel approaches reported in the literature and summarize \textcolor{black}{them in Table \ref{TAb_RMM1}.}}

\textcolor{black}{One of the specific challenges of \textcolor{black}{a} \acrshort{haps} is its horizontal back and forth \textcolor{black}{movement caused by crosswinds in the stratosphere.} \cite{jiang2015channel}. \textcolor{black}{This movement poses a problem for ground users near cell edges who then need to handoff} between cells, even when they are stationary. To solve the problem, a channel assignment algorithm combining channel reservation and handoff queuing with priorities based \textcolor{black}{on the platform's horizontal movement was proposed}. 
% This algorithm takes full account of the needs of different types of user terminals on the service grade, prioritizes the user terminals, and introduces handoff queuing based on channel reservation from the point of reducing the handoff dropping rate. 
In \cite{guan2019intelligent}, an \acrshort{ai}-based wireless channel allocation algorithm based on \textcolor{black}{a} reinforcement learning algorithm for a 5G \acrshort{haps} massive \acrshort{mimo} communication system \textcolor{black}{was} proposed. A Q-learning algorithm combined with a back-propagation neural network enable\textcolor{black}{d} the system to learn independently according to the environment and intelligently according to the channel load and blocking conditions. 
% The network performance of the proposed algorithm is compared with that of a random channel allocation algorithm and the worst acceptable channel allocation algorithm. %The performance of the Q-learning intelligent channel allocation algorithm based on reinforcement learning algorithm turns out to be higher than the other two algorithms for various traffic.
}

\begin{figure}[!htbp]
	\begin{center}
	\centering
\includegraphics[width=1.0\columnwidth]{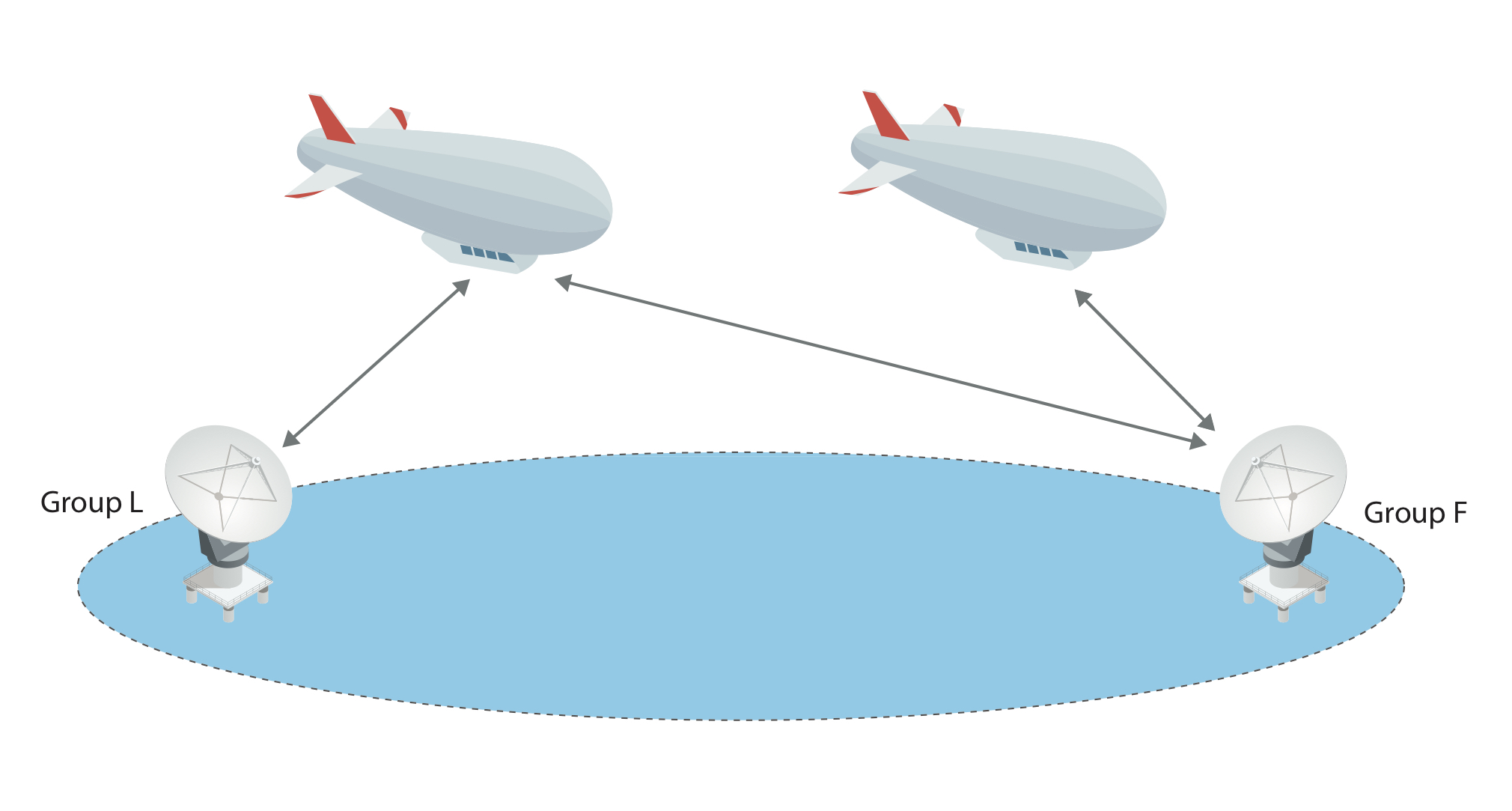}
\caption{A system of two \acrshort{haps} nodes and two groups of users with different access choices (adapted from \cite{liu2009exploiting}).}
\label{HAPCTMC}
\end{center}
\end{figure}

\textcolor{black}{In \cite{liu2009exploiting}, a heterogeneous network with two \acrshort{haps} \textcolor{black}{nodes (illustrated in Fig. \ref{HAPCTMC}) was} considered where UEs with a limited \acrshort{haps} choice (labeled Group L) and UEs with a full \acrshort{haps} choice (labeled Group F) coexiste\textcolor{black}{d} in the same system. Group F UEs \textcolor{black}{had} access to both HAPS1 and HAPS2 by smart or steerable antennas, while Group L UEs only \textcolor{black}{had} access to one of the \acrshort{haps} nodes due to some physical constraints such as fixed antennas. 
% The channel allocation process is modeled as a birth-death two dimensional \textit{Continuous Time Markov Chain}. 
In order to improve the potentially inferior QoS of Group L, a restriction \textcolor{black}{was} imposed to Group F to deliberately restrict its \textcolor{black}{channel} availability. %The restriction mechanism is for equalizing the blocking probability of Group F and Group L. The restriction blocks some Group F users to reserve more channels for Group L users which have more limited HAP availability.
Using this compensation effect, a balanced blocking probability \textcolor{black}{was achieved.} The study in \cite{guan2016channel} propose\textcolor{black}{d} a measure for deciding the minimum distance in mobile user access systems. Based on this measure, the channel allocation problem \textcolor{black}{for HAPS communications hot spot areas} \textcolor{black}{was dealt with on the basis of the prediction} of user number change and call volume change that could effectively solve the problem of insufficient or wasted channels caused by the lack of proactive cooperation in conventional channel allocation methods. %These approaches are validated through simulation, in which the proposed channel allocation approaches demonstrated the reasonable allocation of channels and avoidance of call blocking while improving channel utilization.
} 

\subsubsection{ \textcolor{black}{{Joint Power, Sub-\textcolor{black}{C}hannel and Time-Slot Allocation}}}
\textcolor{black}{In \cite{ibrahim2015using}, the authors \textcolor{black}{studied} radio resource allocation for multicasting in OFDMA-based \acrshort{haps} systems. An optimization problem \textcolor{black}{was} formulated and solved to provide the best allocation of \acrshort{haps} \textcolor{black}{resources, such} as radio power, sub-channels, and time slots. The problem also \textcolor{black}{determined} the best possible frequency \textcolor{black}{re-use} across the cells that constitute\textcolor{black}{d} the service area of the \acrshort{haps}. 
% The objective is to maximize the number of UEs that receive the requested multicast streams in the HAP service area in a given OFDMA frame. %A bounding subroutine in a branch and bound algorithm can be obtained by decomposing it into two relatively less complex subproblems, due to its high complexity, and solving them iteratively.  
\textcolor{black}{\cite {ibrahim2019optimizing} investigated multicast group users receiving multicast session transmissions from more than one antenna simultaneously.} This also allow\textcolor{black}{ed} the user\textcolor{black}{s} to receive multicast sessions transmitted \textcolor{black}{from} neighboring cells, not just those in the \textcolor{black}{cell in which users resided}. 
The \textcolor{black}{users then could have different priority levels} from the system's perspective, and the objective \textcolor{black}{was} to maximize the admission of highest priority users to the system rather than maximizing the number of admitted users \textcolor{black}{as in} \cite{ibrahim2015using}. The \textcolor{black}{solution was} based on branch and cut framework \textcolor{black}{(see also \cite{ibrahim2019optimization} for more details)}, in which linear outer approximation using McCormick underestimators \textcolor{black}{was} applied for the relaxation of the mixed binary \textcolor{black}{and} quadratically constrained problem.  %Various branching schemes are explored and a presolving reformulation linearization scheme for a specific set of quadratic constraints is proposed. 
% Ibrahim and Alfa published their book \cite{ibrahim2019optimization} in which they also discuss in depth several other aspects for the branch-and-cut acceleration  which include several cutting planes, domain propagation and primal heuristics. %To the best of our knowledge, this book is the only one that discusses  optimizing resource allocation for multicasting over a multi-antenna HAP in details.
}

\subsubsection{ \textcolor{black}{{Spectrum Management and Sharing}}}

\textcolor{black}{The authors in \cite{oodo2002sharing} highlight\textcolor{black}{ed} the impact of the minimum operational elevation angle, antenna radiation \textcolor{black}{patterns, and} the potential of dynamic channel assignment \textcolor{black}{\textcolor{black}{for} the purpose of sharing and compatibility between \textcolor{black}{a} fixed service using \textcolor{black}{a} HAPS and
other services in the 31/28 GHz bands.}
% These tools are essential to provide the inter-operability of the \acrshort{haps} nodes with the terrestrial networks.  
Moreover, \cite{likitthanasate2008spectrum}  investigate\textcolor{black}{d} the potential of cognitive radio-based dynamic spectrum management in integrated \acrshort{haps} terrestrial networks. The impact of the antenna beamwidths and adaptive modulation \textcolor{black}{were} considered. The authors in \cite{mokayef2013spectrum} investigate\textcolor{black}{d} the co-existence of \acrshort{haps} and fixed terrestrial networks in the \textcolor{black}{5,850-7,075 MHz} band \textcolor{black}{by considering both physical distance and frequency separation in terms of the \textcolor{black}{co-channel} and adjacent channel frequency through the use of spectrum emission masks as guided by the ITU recommendations.} 
In \cite {wang2019dynamic}, the authors consider\textcolor{black}{ed} a spectrum sharing problem in a LEO-HAPS cognitive system in \textcolor{black}{a} non-ideal spectrum sensing situation, in which a cognitive network model from \textcolor{black}{a} multi-beam \acrshort{leo} satellite and \acrshort{haps} access scenario \textcolor{black}{was} introduced. Aiming at dynamic spectrum and power allocation strategy for 
the non-ideal spectrum sensing case, a correction coefficient \textcolor{black}{was} accordingly included. \textcolor{black}{The reported simulation results \textcolor{black}{validated the capacity of \textcolor{black}{HAPS-to-ground} downlink connections are improved} with their proposed strategy in imperfect estimation scenario compared with the case in which estimation errors \textcolor{black}{were} ignored.}
% Simulation results corroborate the effectiveness of their approach. 
An overview of the potential of cognitive radio techniques \textcolor{black}{was also given in \cite{mohammed2011role}.}}

\begin{table*}[t]
\caption{\textcolor{black}{A summary of channel, power \textcolor{black}{allocation, and} spectrum sharing schemes reported in the literature about the \acrshort{haps} systems } }
\centering \footnotesize
\begin{tabular}{|p{4cm}|p{4cm}|p{3.5cm}|p{3.5cm}|p{.9cm}|} 
\hline
 Design Aspect(s)/Parameters	 & Objective& Technique & Network Type  &Reference\\ 
\hline 
 		
Power, \textcolor{black}{sub-channels}, user-selection, time scheduling		& Maximize the number of multicast admissions & Lagrangian relaxation		&  Stand-alone \acrshort{haps}		&	 \cite{ibrahim2015using}	\\	
\hline
	Associate users with multiple antennas $+$ design parameters in \cite{ibrahim2015using}	&Admit the highest priority users to the system&	McCormick outer approximation relaxations, branch and cut techniques,  cloud branching etc. 	& 	\textcolor{black}{Stand-alone} \acrshort{haps}	&	\cite {ibrahim2019optimizing}	\\
\hline
Complexity reduction for the problem in \cite{ibrahim2015using}	 by reformulations& 	Admit the highest priority users to the system&  Different types of cuts, acceleration heuristics, and variable domain propagation techniques 	& 	\textcolor{black}{Stand-alone} \textcolor{black}{\acrshort{haps}s}	&	\cite{ibrahim2019optimization}	\\
\hline
	Channel allocation	&  System capacity maximization	&Q-learning reinforcement learning and combines
back-propagation neural network	& System of	Multiple \acrshort{haps} &	\cite{guan2019intelligent}	\\
\hline
	Energy allocation	& Serving users with \textcolor{black}{the} highest priority  &Genetic algorithm, rule-based learning neural networks and dynamic programming	& Single	access point \acrshort{haps} moving on a trajectory	&	\cite{ceran2017optimal}	\\
\hline
	Channel allocation 	& Fair call blocking probability  &Continuous time Markov chain and restriction functions	& Two \textcolor{black}{\acrshort{haps}s} system	&	\cite{liu2009exploiting}	\\
\hline
Radio and computing powers & Minimizing energy and time consumption&  SVM-based FL algorithm& 	Mutliple HAPS-BSs	& \cite{wang2020federated}\\

\hline
\textcolor{black}{Offloading fraction and low-altitude UAV uplink powers} & \textcolor{black}{ Minimize the offloading delay}& \textcolor{black}{Multi-leader multi-follower Stackelberg game and EPEC} & \textcolor{black}{Two tier low-altitude and high-altitude UAVs} & \textcolor{black}{\cite{8766778}}
\\
\hline
Uplink transmit powers & Maximize system's uplink throughput while achieving local fairness&  SRA-LF algorithm& 	Standalone \acrshort{haps}	& \cite{abrardo2003centralized}\\
\hline
Downlink transmit powers&	CAC: Maximize System's Capacity	&	Centralized Transmit Power Based, Platform Power
Limited (CTP-PF) CAC heuristic	&	Standalone \acrshort{haps}	&\cite{foo2002centralized2}\\
\hline
Downlink transmit powers& Minimize dropping and blocking probabilities	& Overflow and speed sensitive strategies $+$ centralized resource reservation- random model and traffic selection	&	Two layer Terrestrial/\acrshort{haps} network	&\cite{foo2004call}\\
\hline
Spectrum sharing with conventional fixed service system& Interference mitigation	&  Dynamic Channel Allocation (DCA) $+$ increasing minimum elevation angle &	Two layer terrestrial/\acrshort{haps} network	&\cite{foo2005speed}\\
\hline
Power driven switching between  terrestrial and \textcolor{black}{HAPS} FACHs and DCHs for MBMS &Maximize GoS, i.e., minimize call blocking and dropping probabilities 	&  \textcolor{black}{number of multicast users policy $+$ distribution of multicast users policy} &	Two layer terrestrial/\acrshort{haps} network	&\cite{araniti2007multicast}\\
\hline
Power driven switching between  \textcolor{black}{HAPS} FACHs and DCHs for MPMS &	Maximize power utilization: system capacity at a given power level		&	Defines power switching thresholds to switch between \textcolor{black}{dedicated and common channels}		&	Standalone \textcolor{black}{HAPS}	&	\cite{raschella2009high}\\
\hline
Spectrum sharing for downlink coexistance of \acrshort{haps} and terrestrial fixed broadband systems&		Improve coexistence performance by reducing outage probability at the user	&	Interference-to-noise-based scheme $+$	CINR-based scheme	&	Two layer terrestrial/\acrshort{haps} network		& \cite{likitthanasate2008spectrum}\\
\hline
 Spectrum sharing of \textcolor{black}{5,850-7,075} MHz band with \textcolor{black}{fixed services} and effect of channel bandwidth&		Prevention of harmful interference	&  Co-channel, zero-guard-band and adjacent channel criteria and methods. Separation distance and frequency separation  used 	& terrestrial/\acrshort{haps} system &\cite{mokayef2013spectrum} \\
\hline
Spectrum and power allocation  &	Maximize sum rate of \textcolor{black}{HAPS}-ground downlinks in imperfect channel estimation	&	Decomposition, then \textcolor{black}{simplex algorithm} and convex optimization	&	LEO-\textcolor{black}{HAPS} cognitive system	& \cite {wang2019dynamic}\\
\hline
\end{tabular}
\label{TAb_RMM1} 
\vspace{0.5cm}
\end{table*}

\subsection{\textcolor{black}{HAPS Placement and Constellation Optimization}}
% \textcolor{black}{An aerial network providing cellular like coverage, consists of  {aerial flying platforms} (AFPs) of different types of UAVs such as drones, balloons, and high-altitude/medium-altitude/low-altitude platform stations (HAPS/MAPS/LAPS) to extend the cellular coverage and deliver broadband services to regions where terrestrial infrastructure is possibly not available and expensive to deploy.  }
One of the main challenges of \acrshort{haps} systems is the design and management of \textcolor{black}{self-organizing networks}. In conventional self-organizing networks, self-configuration and optimization as well as self-healing capabilities are among the basic functionalities. \textcolor{black}{This is particularly essential in an} aerial network \textcolor{black}{due to its more dynamic nature compared to a fixed cellular \textcolor{black}{network, because the position of the elements in the former may change over time.}} This may be due to changes in user requirements, atmospheric conditions, coverage necessities, \textcolor{black}{battery status, or abrupt} traffic changes in the network.

\textcolor{black}{In \cite{ahmadi2017novel}, the authors \textcolor{black}{explored} a layered architecture with aerial flying platforms (AFPs) of various types, flying in low/ medium/ high layers. In \textcolor{black}{their work}, the positions of LAPS nodes in the low layer \textcolor{black}{were} defined \textcolor{black}{centrally, and} the AFP \textcolor{black}{had} the ability to \textcolor{black}{reorganise} the layer to achieve its target, which \textcolor{black}{could} be maximizing the number of UEs served, maximizing the achievable \textcolor{black}{rate} and/or fairness among UEs. The optimum placement problem \textcolor{black}{was} formulated as a linear binary program. \textcolor{black}{The \textcolor{black}{results showed} show that an aerial self-organizing network \textcolor{black}{outperformed} a fixed placement. }
% For the architecture proposed in \cite{ahmadi2017novel}, \acrshort{haps} and MAPS are flying with fixed locations. %An airborne self-organizing-network (SON) that reorganizes itself with an NFP with fixed LL placement. 
%  which can be solved to optimally by any of the commercial solvers, such as CPLEX. Sub-optimal placements using meta heuristics such as ant colony, particle swarm, and genetic algorithm are possible with lower computational complexity.  
}
\textcolor{black} {A \textcolor{black}{cost-efficient} \acrshort{haps} constellation design with \textcolor{black}{a} QoS and user demand guarantee \textcolor{black}{was} investigated in \cite{dong2016constellation}. The QoS metrics \textcolor{black}{were} established by considering the \acrshort{sinr}, BER, throughput, and availability. For \acrshort{haps} broadband networks, availability is the percentage of coverage that the system can provide with a BER no worse than the desired target. %The user demand is modeled by production of the broadband size, the population distribution density, and a scale factor. 
Based on the network coverage model, the design vector of \acrshort{haps} system layout optimization, i.e., the number of \acrshort{haps} nodes, downlink antenna area, power of payload, longitude of \textcolor{black}{HAPS}, and latitude of \textcolor{black}{HAPS}, \textcolor{black}{was} devised. It was found that by applying the proposed constellation design methodology, the optimal cost-efficient broadband network can be realized. %The rest of the parameters were seen as a constant vector, and the network capacity model is derived and the HAP constellation optimization design framework gets decomposed into modules. 
% A nonlinear, non-convex, combinatorial optimization problem is formulated and solved using an improved artificial immune algorithm based on immune review. 
% For HAP placement over China, the authors give out an optimal placement in the sense of network capacity per cost with QoS and user demand guarantee.  %The constellation design might be treated as a benchmark guideline for the constellation of multiple \acrshort{haps} design as well as the HAP broadband network construction for decision makers.
}
% \subsubsection{\textcolor{black}{\underline{Joint Power Control and HAP Placement}}}
 \textcolor{black}{In \cite{alsharoa2019improvement}, power allocation with fronthauling  and backhauling  associations in order to improve the global connectivity using satellite, airborne, and terrestrial networks integration \textcolor{black}{was} investigated jointly with determining the \textcolor{black}{HAPS} \textcolor{black}{nodes placement}. It \textcolor{black}{was} shown that the satellite stations and \acrshort{haps} nodes \textcolor{black}{could} play a significant role in global connectivity when terrestrial BSs are overloaded or to support users with high throughput located outside terrestrial \textcolor{black}{BS} coverage areas (e.g., suburban and remote areas).}

\subsection{\textcolor{black}{Antenna and Interference Management}}
\textcolor{black}{Communication between a \acrshort{haps}s and UEs requires highly directive antennas to overcome the high attenuation caused by path loss and to prevent interference between UEs receiving transmission from antennas of different HAPS (especially those at the HAPS coverage edge). Additionally, given that the ITU-R allocates terrestrial cellular spectrum to \acrshort{haps} systems, the expected interference to terrestrial wireless systems requires that the antennas onboard be equipped with dynamic beam pointing to facilitate interference management. This \textcolor{black}{is} crucial given that a HAPS is \textcolor{black}{may} have hundreds of antennas onboard. Electronically steerable multi-beam antennas \textcolor{black}{were} used in early \textcolor{black}{HAPS} projects like CAPANINA and HELINET. Overlapping antenna main-lobes and side-lobes of the same frequencies introduces interference, which necessitates interference management in an intra-\acrshort{haps} system to mitigate interference from the serving and adjacent beams on \acrshort{haps} users. \textcolor{black}{A summary of the key HAPS antenna and interference management schemes reported in the literature is given in Table \ref{AAS}. In the rest of this subsection we discuss antenna and interference management schemes in terms of adaptive cell shaping, multipoint transmission and platform \textcolor{black}{diversity, and} massive MIMO for HAPS systems.}
% The performance of the system can be assessed considering carrier-to-interference-ratio (CIR) at the users, which largely depends on the HAP's beam/cell shaping technologies. %There is quite a number of published research in this area that investigates this aspect, which we discuss herein. 
}
\subsubsection{\textcolor{black}{Adaptive Cell Shaping}}
\textcolor{black}{\textcolor{black}{The design of \acrshort{haps} antenna beams \textcolor{black}{was} first discussed in \cite{djuknic1997establishing}}. Ideally an antenna beam illuminates its corresponding cell with equal power across the cell and with zero power outside the cell, \textcolor{black}{and in doing so acts} like a spatial filter. \textcolor{black}{However, for} practical antennas, the spot beams \textcolor{black}{do not} have this ideal pattern, particularly at millimeter-wave frequencies, where array beam synthesis techniques are challenging. \textcolor{black}{Aperture type antennas may be suitable for this purpose due to their well-studied radiation characteristics.} It is highly desirable to be able to construct beams with very low side-lobes and a steep
roll-off in the main-lobe. Side-lobe \textcolor{black}{levels} can be minimized with corrugated horn designs \cite{olver1994microwave}; however the roll-off rate is mainly affected by the main-lobe width and directivity. \textcolor{black}{There is a trade-off in that if a} highly directive main-lobe is used, the cell will experience excessive power roll-off at its edges leading to a low received power there. \textcolor{black}{By contrast,} if the directivity chosen is too low, excessive power will fall outside the cell leading to high level of inter-cell interference.}

\textcolor{black}{ The authors \textcolor{black}{proposed} a general formulation in \cite{thornton2003optimizing} for optimum directivity in order to maximize the received power at the cell edge. 
\textcolor{black}{This \textcolor{black}{was} in contrast to earlier works where the \textcolor{black}{HAPS cell was defined} as being within the footprint of the corresponding antenna's half-power \textcolor{black}{beamwidth}.} 
The impact of the beam patterns and the frequency re-use technique were further investigated for circular and elliptical beam patterns. \textcolor{black}{Elliptical} beams have been demonstrated to be superior in terms of optimized power at cell edges, which is crucial \textcolor{black}{when} RF link budgets are marginal. %By tailoring each antenna’s beam-widths to its corresponding cell's subtended angles an array of antennas can serve a chosen coverage area. For classical reuse patterns of four and seven channels, the elliptic antenna beams provide maximum power at cell edges. Specifically, for the case of four channels, cell edges suffer interference from neighboring beams and a reduction in side-lobe level turns out to produce a corresponding improvement in carrier-to-interference (CIR) at cell centers only. When seven channels are used, the benefit of reducing side-lobe levels is exploited across the majority of the coverage area.  
\cite{albagory2013flat} \textcolor{black}{investigated} the use of a vertical antenna array with windowing to change the cell shapes, specifically to obtain flat-top ring-shaped cells. The beam shape of the ring cells \textcolor{black}{was} improved by making use of a composite weighting functions for flattening the power pattern over the cell stripe as well as reducing the in-cell ripples and side-lobe levels. The analysis of this technique \textcolor{black}{showed} that a uniform power pattern \textcolor{black}{with an in-cell ripple of less than 0.25 dB was attainable} and this \textcolor{black}{reduced} \textcolor{black}{the power control needed} for the roll of conventional beam shapes towards the cell boundaries. As a result of the improved power pattern, the signal's CIR \textcolor{black}{was} improved both in value and distribution within the serving ring cells. %which increases the immunity to the propagation issues in cellular systems.
}

\textcolor{black}{Beam steering \textcolor{black}{was} considered in \cite{capstick2005high}. The authors \textcolor{black}{presented} two \textcolor{black}{steering \textcolor{black}{scenarios: one where all antennas were steered, and one involving a four actuator solution.}} It \textcolor{black}{was} concluded that from a complexity perspective, the four actuator solution \textcolor{black}{was} much simpler than having each aperture antenna on its own gimbals arrangement, especially when there might be more than 100 cells. %A four actuator solution has been reported to be preferable also from the perspective of weight and power considerations. In terms of the signal strength variation the performance for the four actuator solution is equal to or better than the all antennas individually steered solution and in the majority of situations the CIR performance was reported to be better as well. 
The main disadvantage of this solution \textcolor{black}{was} the high number of handoffs required. 
In \cite{thornton2005effect}, the authors \textcolor{black}{investigated} the impact of frequency re-use patterns and different antenna models in a multi-beam/multi-cell \acrshort{haps}-based communication system. \textcolor{black}{In so doing, they compared} different models for the antenna side-lobe region and \textcolor{black}{quantified} the CIR for a \textcolor{black}{three-channel} re-use plan for networks of 121 and 313 cells. \textcolor{black}{The results showed that the ITU recommended pattern} for the 47/48 GHz band can lead to poor results compared to an adapted pattern based on fitting the measured data for an \textcolor{black}{elliptical} beam lens antenna. } 
\textcolor{black}{\textcolor{black}{The authors in \cite{albagory2013smart} \textcolor{black}{considered}} irregular cell shapes  to obtain the target cell characteristics by grouping pixel \textcolor{black}{spots, which aimed} to limit co-channel interference. 
% The desired cell pattern is converted into a spatial mask that selects the desired pixel spots with a proper pre-weight smoothing function.
This cell design technique \textcolor{black}{optimized} the cell shape according to the user distribution and behavior in the \textcolor{black}{coverage area, and thus it was expected that this would reduce} the frequent handoff and signaling traffic of location updating from moving users. \textcolor{black}{The simulation results showed that a cell} with any irregular shape can be formed \textcolor{black}{with a side-lobe level as low as 40 dB} using \textcolor{black}{a} Gaussian concentric ring array.  \cite{arum2019beam} \textcolor{black}{introduced} a cell-pointing \textcolor{black}{approach that made use of HAPS-induced beams} to provide contiguous coverage delivery by \textcolor{black}{providing an overlap between beams} using a planar antenna array over an extended coverage area.}

\subsubsection{\textcolor{black}{Exploiting Multipoint Transmission and Platform Diversity}}
\textcolor{black}{ User-centric joint transmission coordinated multipoint (JT-CoMP) has proved to boost the capacity of terrestrial cellular systems by overcoming cell-edge interference. In \cite{zakaria2019exploiting}, the authors \textcolor{black}{investigated} how JT-CoMP \textcolor{black}{could} be extended to a \acrshort{haps} system architecture by exploiting a phased
array antenna, which \textcolor{black}{generated} multiple beams that \textcolor{black}{formed cells, each of which could be mapped} on to pooled virtual BS equipment, thereby replacing multiple terrestrial cell sites. CoMP \textcolor{black}{was} designed to enhance the user experience at the edge of the \acrshort{haps} cells. Methods to overcome the known trade-off for JT-CoMP between carrier-to-interference plus noise ratio (CINR) gain and loss of capacity accessible to the users were explored. 
% Two different methods of identifying non-CoMP and CoMP users are based upon the centralized CINR threshold and flexible CINR threshold approaches. For the bandwidth allocation technique, two approaches are used: full bandwidth  and half bandwidth. %These four approaches are combined, delivering the FBW,HBW, Flex FBW, and Flex HBW schemes that are used to control the JT-CoMP. It is shown that 57\% and 45\% of users gain benefit from the use of HBW and FBW, respectively. The schemes based on the flexible CINR threshold approach provide a balance between loss and gain of the user capacity, while the centralized CINR threshold-based schemes performed well, benefiting up to 57\% of the users, but with the drawbacks of a higher percentage of losing users.
}
% \textcolor{black}{\sout{The use of multiple \acrshort{haps} have also been considered to improve coverage and capacity by using the nodes as a virtual \acrshort{mimo} system through the use of platform diversity}}.
\textcolor{black}{In \cite{chen2005performance}, the performance of using  multiple \acrshort{haps} nodes by using antennas \textcolor{black}{was} investigated. The impact \textcolor{black}{of} the distance between \acrshort{haps} nodes on the CINR distribution \textcolor{black}{was} considered. The use of multiple \acrshort{haps} \textcolor{black}{as HAPS constellations was considered} in \cite{grace2005improving}, where the authors \textcolor{black}{considered} the millimeter-wave band transmission. \textcolor{black}{The potential gains in capacity that various \textcolor{black}{HAPS} constellations can deliver, both theoretically using the
Shannon equation and also while operating a number of practical modulation and coding schemes \textcolor{black}{were} quantified. An evaluation methodology consisting of minimum angular separation of HAPs as seen by the user, link length ratio, and sidelobe floor beamwidth was developed. For a $5^{\circ}$ beamwidth user antenna, the optimum HAP spacing radius \textcolor{black}{was shown to be approximately 4 km.}} The authors in \cite{dong2015diversity} \textcolor{black}{determined} the diversity order improvement that \textcolor{black}{could} be obtained by using multiple \acrshort{haps} nodes \textcolor{black}{via virtual MIMO transmission}. The ergodic capacity improvement \textcolor{black}{was} also quantified. }
\begin{table*}
\caption{\textcolor{black}{A summary of antenna and interference management related aspects and approaches in \acrshort{haps} systems } }\label{AAS}
\footnotesize
\begin{tabular}{|m{4cm}|m{4cm}|m{3.5cm}|m{3.5cm}|m{1.2cm}|} 
\hline
\textcolor{black}{HAPS Antenna Related Aspect}	 & \textcolor{black}{Investigated Parameters, Objectives and Approach}& \textcolor{black}{Important Findings} & \textcolor{black}{HAPS System Type}  &\textcolor{black}{Reference}\\ 
\hline 
\multirow{6}{*}{  Adaptive cell shaping} &$\circ$ \textcolor{black}{Optimized} directivity to maximize the received power at the cell edge. 
\vskip 0.1cm
$\circ$ Impact of the beam patterns and the frequency \textcolor{black}{re-use} technique \textcolor{black}{was} investigated for circular and elliptical beam patterns.  &\textcolor{black}{Elliptical} beams have been demonstrated to be superior in terms of optimized power at cell edges. & Mutli-cellular ($>$100 cells) \textcolor{black}{stand-alone} single \acrshort{haps}.	&	 \cite{thornton2003optimizing}\\ \cline{2-5}
&\textcolor{black}{Improved} beam shape of  ring cells using composite weighting functions for flattening the power pattern over the cell stripe as well as reducing the in-cell ripples and side-lobe levels.&Uniform power pattern with a lower in-cell ripple of less than 0.25 dB is attainable reducing \textcolor{black}{the power control needed} for the roll of conventional beam shapes at cell boundaries.	& 	System of multiple concentric antenna beam \acrshort{haps}.	&	\cite{albagory2013flat}	\\ \cline{2-5}
& $\circ$ Platform antenna adjustment mechanisms for horizontal and vertical position variation. 
\vskip 0.1cm
$\circ$  Two steering \textcolor{black}{scenarios: all antennas steered; and a four actuator scenario.} & The actuator solution \textcolor{black}{was} better in terms of complexity, weight, power and CIR.	& 	Single standalone \acrshort{haps}.	&  \cite{capstick2005high}.	\\\cline{2-5}
 & Impact of different antenna models.  &$\circ$ The ``mask" (ITU) approach \textcolor{black}{over-estimated} mean \textcolor{black}{side-lobe} level. 
\vskip 0.1cm
$\circ$ The flat \textcolor{black}{side-lobe} approximation \textcolor{black}{remained} very effective and computationally straightforward. 
\vskip 0.1cm
$\circ$ Regular hexagonal cell layouts \textcolor{black}{outperformed} the equivalent equiangular hexagonal cell layout in terms of CIR.  & Multi-beam/multi-cell HAPS-based system.	&\cite{thornton2005effect}	\\\cline{2-5}
 & \textcolor{black}{Optimized the cell shapes} according to the user distribution and behavior in the \textcolor{black}{coverage
area.}& A cell with any irregular shape \textcolor{black}{could be formed} \textcolor{black}{with a side-lobe level as low as 40 dB using a} Gaussian concentric ring array. & Single HAPS with pixel spot beams.	&\cite{albagory2013smart}	\\\cline{2-5}
 & Cell-pointing algorithm that \textcolor{black}{accounted for the broadening} of cells at low elevation angles. & Scheme significantly \textcolor{black}{improved} user CNR and CINR, achieving a CINR improvement of 5-15dB compared with the other schemes. &	Mutli-cellular stand-alone single \acrshort{haps}.	&	\cite{arum2019beam}\\\hline
\multirow{4}{*}{Coordinated multipoint transmission} & $\circ$ \textcolor{black}{Improved user experience at the edge of HAPS cells.}
\vskip 0.1cm
$\circ$ Two different methods of identifying non-CoMP and CoMP users are \textcolor{black}{based on} the centralized CINR threshold and flexible CINR threshold.  
\vskip 0.1cm
$\circ$ Two bandwidth allocation approaches: full bandwidth (FBW) and half bandwidth (HBW). & The schemes based on the flexible CINR threshold approach \textcolor{black}{provided} the best balance between loss and gain of the user capacity, while the centralized CINR threshold-based schemes performed \textcolor{black}{well and benefitted up to 57\% of users.} & HAPS system architecture with phased array antenna and pooled virtual eNodeBs mapped onto directional beams generated by the
phased array controller.	&\cite{zakaria2019exploiting}	\\\cline{2-5}
&Impact of the distance between HAPS nodes on the CINR distribution \textcolor{black}{was} considered. &  Locating HAPS nodes at a specific spacing radius outside the coverage area \textcolor{black}{was shown to improve performance.}& Multiple HAPS nodes system.	& \cite{chen2005performance}	\\\cline{2-5}
& \textcolor{black}{Maximized} CINR and spectral \textcolor{black}{efficiency by} determining the optimal  HAPS nodes spacing for given antenna beamwidths. & $\circ$ For a $5^{\circ}$ beamwidth user antenna, the optimum HAP spacing radius \textcolor{black}{was} approximately 4 km. 
\vskip 0.1cm
$\circ$ Capacity increases \textcolor{black}{were} commensurate with the increase in the number of platforms, up to
10 HAPs& Multiple HAPS nodes system. & \cite{grace2005improving}	\\\cline{2-5}
 & $\circ$ \textcolor{black}{Determined} the diversity order improvement that \textcolor{black}{could} be obtained by using multiple HAPS nodes via virtual MIMO transmission in wireless sensor networks.
\vskip 0.1cm
$\circ$ PDF and CDF of received SNR derived.& Virtual MIMO with multiple \textcolor{black}{HAPSs was shown to be a promising solution} for future high-data-rate and frequency-efficient smart wireless sensor networks. & Multiple HAPS network.	&	\cite{dong2015diversity}\\
\hline
\end{tabular}
\end{table*}
\begin{table*}
\caption*{\textcolor{black}{Table \ref{AAS}: (Continued) A summary of antenna and interference management related aspects and approaches in \acrshort{haps} systems. } }
\footnotesize
\begin{tabular}{|m{4cm}|m{4cm}|m{3.5cm}|m{3.5cm}|m{1.2cm}|} 
\hline
\textcolor{black}{HAPS \textcolor{black}{Antenna} Related Aspect}	 & \textcolor{black}{Investigated Parameters, Objectives and Approach}& \textcolor{black}{Important Findings} & \textcolor{black}{HAPS System Type}  &\textcolor{black}{Reference}\\ 
\hline 
\multirow{3}{*}{Massive MIMO for HAPS systems}&$\circ$ \textcolor{black}{Investigated} the interference caused by beamforming
technology  using massive MIMO. 
\vskip 0.1cm
$\circ$ Intelligent beamforming
algorithm based on game theory is proposed.	& 	The AI-based algorithm \textcolor{black}{had} better array gain than the traditional beamforming algorithm which \textcolor{black}{could} focus on the desired user and place spatial nulls in the
direction of undesired users. &	Single multi-antenna \textcolor{black}{stand-alone HAPS}.	&\cite{yue2019diversity}	   \\\cline{2-5}
&A robust multi-objective Pareto-optimal beamforming for sum rate maximization and total transmit power minimization. 	& A better \textcolor{black}{performance could be achieved in comparison to} the commonly used SCA-based scheme. 	& Integrated multibeam satellite and HAPS system sharing mmWave band.	& \cite{lin2019robust} \\\cline{2-5}
&User grouping and beamforming based on statistical-eigenmode.& The solution \textcolor{black}{was shown to outperform} existing schemes based on channel correlation matrix.	& A single HAPS mMIMO system with a uniform planar array.	& \cite{lian2019user}\\\hline
\end{tabular}
\end{table*}

\subsubsection{\textcolor{black}{Massive MIMO for HAPS Systems}}
\textcolor{black}{As in terrestrial wireless systems, massive \acrshort{mimo} takes advantage of a large number (hundreds) of antenna elements in an \textcolor{black}{array, which can be used for the following}}
\textcolor{black}{\begin{enumerate}
\item \textcolor{black}{To improve} the diversity gain in wireless fading channels, where the independence of the paths of each signal is exploited to ensure that an outage probability is minimized, or the receiver signal power is always above an acceptable \textcolor{black}{level;}
\item  \textcolor{black}{To improve} the \acrshort{sinr} in an interference limited system by controlling the direction and width of a beam's main lobe as well as the spatial nulls \textcolor{black}{(i.e., beamforming);}
\item spatial multiplexing to boost the system's throughput by feeding each antenna element with a different data stream.
\end{enumerate}}
\textcolor{black}{It \textcolor{black}{was} shown in \cite{yue2019diversity}  that a distributed sub-array architecture \textcolor{black}{yielded} a significantly better diversity performance than the co-located antenna architectures. This \textcolor{black}{means} that the diversity gain in \acrshort{haps} systems is achievable by the \acrshort{haps} mega-constellations, which adds another \textcolor{black}{benefit} to its envisioned advantages.}
\textcolor{black}{The problem of system interference caused by beamforming technology in a \acrshort{haps} communication system using massive \acrshort{mimo} \textcolor{black}{was} explored in  \cite{ guan2019efficiency}. An intelligent beamforming algorithm based on game theory \textcolor{black}{was} proposed, and a mathematical model of a \textcolor{black}{beamforming game algorithm was constructed}. 
% The transmit power of different users is described and controlled based on a multi-user game model. A particle swarm optimization (PSO) algorithm obtains the Nash equilibrium point, which can effectively suppress the interference between multiple users in \acrshort{haps} communication. The proposed AI-based algorithm focuses on the desired user and place spatial nulls in the direction of undesired users which can mitigate the interference between users and increase coverage area.
}
\textcolor{black}{A robust beamforming scheme \textcolor{black}{was} proposed in \cite{lin2019robust} for an integrated satellite and HAPS network, where a multi-beam satellite system \textcolor{black}{shared} the millimeter wave spectrum with a \acrshort{haps} system. A multi-objective optimization problem \textcolor{black}{was} formulated to obtain the Pareto optimal trade-off between two conflicting yet desirable objectives of the sum rate maximization and total transmit power minimization, while satisfying the QoS constraints of both earth stations and mobile terminals and per-antenna transmit power budget. 
% Then, by using the angular information-based imperfect channel state information, a low-complexity discretization method is used to transform the non-convex objective function and constraints to the convex ones. A monotonic optimization scheme combined with iterative penalty function algorithm is used to obtain the beamforming weight vectors with low computational complexity and fast convergence rate. 
Finally, motivated by the fact that signal power is mainly concentrated on the statistical eigen-mode, user grouping and beamforming is applied for \acrshort{haps} nodes in \cite{lian2019user}. Numerically it \textcolor{black}{was} shown that significant performance gains \textcolor{black}{could} be achieved through the use of massive \acrshort{mimo} \textcolor{black}{and that the technique proposed by the authors \textcolor{black}{outperformed} the existing schemes based on the channel correlation matrix. These works all contribute towards enabling the use of HAPS-SMBS \textcolor{black}{to fill coverage gaps}, as discussed in Section II-B.}}

\subsection{\textcolor{black}{The Waveform: Signaling and Multiple Access}}\label{wave}
%\subsubsection{Single Carrier}

\textcolor{black}{Signaling and multiple access formats, also referred to as the waveform design, have witnessed a long and radical evolution in wireless communications where they serve as its foundation. For example, WCDMA is the \textcolor{black}{technological} pillar of 3G, while OFDM/OFDMA and SC-FDMA are the main approaches of 4G. 
% For any wireless communication system, these techniques have always constituted an issue of paramount significance. 
% Due to the increasing signal bandwidths needed to support data applications, orthogonal frequency-division multiplexing (OFDM) was adopted by consensus for 4G along with scheduled FDMA/TDMA as the pros of orthogonality were looked at with appreciation. 
OFDM is also the main waveform technology in 5G New Radio (NR) \textcolor{black}{technology, which allows} dynamic sub-carrier \textcolor{black}{spacings in multiples} of \textcolor{black}{15 kHz to} support applications with different latency requirements. OFDMA is still the main multiple access scheme besides the optional Non-Orthogonal Multiple Access (NOMA) technology \cite{benjebbour2017overview}. \textcolor{black}{However, it is worth considering the issue} of candidate waveform structures for \acrshort{haps} systems. There is no standard yet that defines a specific waveform structure \textcolor{black}{for HAPS and, to the best of our knowledge, no active research on this has been reported in the literature so far}. However, wireless waveform solutions that are under investigation for 6G wireless communications technologies can be exploited for use in \textcolor{black}{HAPS} access link's radio interface or for inter-platform/backhaul links. }

\textcolor{black}{\textcolor{black}{Despite the prominence of OFDM, it presents a number of drawbacks, discussed in \cite{andrews2014will}, which need to be tackled:}
\begin{enumerate}
\item  High peak-to-average-power ratio due to the summation of uncorrelated inputs in the \textit{inverse-fast-Fourier transform}. This drawback can be mitigated by precoding the OFDM signals at the cost of a slightly higher equalization complexity at the receiver. To reduce the complexity at the \textcolor{black}{receiver, the} novel solutions based on advanced deep learning architectures seem promising. 
\item The need for higher \textcolor{black}{spectral efficiency, which} can potentially be improved by relaxing the orthogonality provided that the cyclic prefixes---included to circumvent inter-symbol interference---\textcolor{black}{are reduced or discarded outright}. \textcolor{black}{Technologies based on filter bank multi-carrier modulation (FBMC) could be adopted.}
\item 
Issues related to the applicability of OFDM to \textcolor{black}{the mmWave} spectrum given the enormous bandwidths therein and the difficulty of developing efficient power amplifiers at those frequencies. A single carrier with faster-than-Nyquist (FTN) signaling technology \cite{rusek2009constrained}, which can achieve higher spectral efficiencies for the same target BER compared to Nyquist signalling over a single carrier, could be a promising solution for \acrshort{haps} transmissions in the \textcolor{black}{mmWave} band.
\end{enumerate}}
\textcolor{black}{In the rest of this section, we  overview alternative \textcolor{black}{approaches that are actively being investigated in the PHY research community and that could be suitable waveform candidates for HAPSs.} \textcolor{black}{These approaches, which are also summarized in Table \ref{Table:Waveform}}, could be considered \textcolor{black}{as slight departures from OFDM/OFDMA-based designs} rather than eruptive changes. }
\subsubsection{\textcolor{black}{Filter Bank MultiCarrier (FBMC) Scheme}}
\textcolor{black}{FBMC is one of the \textcolor{black}{best} well known multi-carrier modulation formats in \textcolor{black}{the} wireless communications literature. It offers a great \textcolor{black}{advantage in being able to shape each sub-carrier} and enables a flexible utilization of the spectrum while satisfying different system
\textcolor{black}{requirements, such} as low latency and multiple access. It also offers \textcolor{black}{advantages by making} the transmitted signal immune to many channel impairments, such as dispersion in time and frequency domains. For example, rectangular filters are desirable for time dispersive \textcolor{black}{channels, while} raised cosine filters are more robust against frequency dispersion. Many other pulse shaping \textcolor{black}{filters have also been} investigated to cope \textcolor{black}{with the various} effects of the \textcolor{black}{channel, and they} provide a reliable system design based on different scenarios, such as the isotropic weighted Hermite pulse \cite{prakash2013efficient}. \textcolor{black}{By contrast, FBMC has drawbacks that} include long filter lengths resulting in enormously large symbol duration, which might not be suitable for low latency applications or short bursts of
machine type communications. Moreover, when used with \acrshort{mimo} technologies, the signal detection computational complexity is expected to be quite large as the channel coherence bandwidth would fall below the \textcolor{black}{sub-carrier} bandwidth \cite{ankarali2020enhanced}. }

\subsubsection{\textcolor{black}{Faster-than-Nyquist Signaling}}
\textcolor{black}{One of the promising ideas to increase spectral efficiency at the physical layer that has been under research for many years is \textcolor{black}{faster-than-Nyquist} (FTN) signalling \cite{anderson2013faster}. In this technique, \textcolor{black}{rather than making the duration of pulses shorter}, we increase their degree of overlap in time by transmitting them at a rate higher than the Nyquist’s signalling rate. In this way, we avoid occupying larger \textcolor{black}{bandwidths but introduce intentional yet controllable inter-symbolic interference (\acrshort{isi})} at the receiver sampling instants, which Nyquist signalling avoids under perfect synchronization, as Figure \ref{FTN} demonstrates. Despite the presence of ISI, in 1975 James Mazo showed that for a binary sinc-pulse, the minimum Euclidean distance of the signals at the receiver \textcolor{black}{experienced} no reduction for an acceleration parameter $\tau\geq 0.802$ \cite{mazo1975faster}. This means that if an optimal maximum likelihood sequence estimation (\acrshort{mlse}) detector is used, then the \textcolor{black}{performance measured} in BER is not compromised. The \textcolor{black}{gain is an increase of around 25\% in the data rate} within the same bandwidth and energy, but at the expense of a more complex receiver. It was shown that the same phenomenon occurred with root raised cosine  pulses \cite{liveris2003exploiting}, the ones most commonly used in different applications. An example is the binary 30\% excess bandwidth root raised cosine pulse, which ceases to have distance 2 at $\tau=0.703$ yielding a 42\% increase in the bit density. }

\textcolor{black}{\textcolor{black}{The spectral and energy efficiency of FTN comes with the cost of high complexity of the optimal detection. This has made low complexity detection a critical issue of investigation over the past few years.} The first symbol-by-symbol FTN sequence estimation was proposed for binary phase shift keying (BPSK) and quadrature phase shift keying (QPSK) in \cite{7886296}. Under these modulation schemes, \textcolor{black}{an 11\% increase in throughput is guaranteed without any penalties in terms of additional bandwidth, energy, or BER.} In \cite{7990502}, the authors proved the feasibility of  high order quadrature amplitude modulation (QAM) FTN signaling detection using a polynomial time complexity semi-definite relaxation-based detector. \textcolor{black}{Their simulation results showed that a data rate increase of up to 25\% could be achieved} at root raised cosine roll off factor of 0.3 without increasing the BER, the bandwidth, or the data symbols energy, when compared to Nyquist signaling; or up to 42.86\% increase in the data rate can be achieved at a roll off factor 0.5 with 0.7 dB penalty in the \acrshort{snr}.}
\begin{comment}
\begin{figure}[t]
	\begin{center}
	\centering
\includegraphics[width=0.65\textwidth]{HAP_RRA/FTN1.PNG}
\caption{\textcolor{black}{Illustration of Nyquist and FTN transmission \cite{fan2017faster}. 
(a)$\tau=1$ (Nyquist transmission), (b)
$\tau=0.82$ (FTN transmission).} (\textcolor{black}{\textbf{Important: This figure needs to be redrawn!)}}}
\label{FTN1}
\end{center}
\end{figure} 
\end{comment}

\textcolor{black}{In narrow-band (NB) IoT services that HAPS-SMBS is envisioned to support, uplink connections of IoT devices have limited transmission power of up to 23dBm and small transmission bandwidths of up to 180 kHz \cite{8641430}. In order to achieve higher transmission rates within the given bandwidth restrictions, higher SNR will be needed in order to transmit with higher modulation \textcolor{black}{orders than the NB-IoT's QPSK without} increasing the BER. A higher SNR cannot be achieved by simply increasing the power due to the low power and long battery \textcolor{black}{life} requirements. This is even more challenging in HAPS given the much longer distance \textcolor{black}{between IoT devices} and the HAPS-SMBS. FTN signaling could instead be used in the uplink connections to enable the sensors pack more bits/sec/Hertz, without any additional \textcolor{black}{SNR requirements} and without increasing the transmission errors. \textcolor{black}{The cost paid would be additional complexity, which HAPS receivers would easily endure because of the huge on-board computational power they are expected to have.} This is especially important since the bit rate requirements for many IoT applications could increase in the future for new applications. For example, if the IoT devices are artificial skin, nose or tasting devices \cite{EricssonIoS}, the bit rate requirements would be higher than existing NB-IoT could support. It is worth noting that further spectral efficiency improvements could be obtained by combining FTN transmission at the IoT device and mMIMO beamforming at the HAPS-SMBS to achieve the target data rate and BER. In such a system, two types of interference \textcolor{black}{would need to be eliminated}, the ISI, and the IoT inter-device interference in the uplink at the HAPS receiver. We believe this is a very important area that needs more research.}

\begin{table*}
\centering
\caption{\textcolor{black}{Summary of waveform candidates for HAPS systems: signaling and multiple access}}
\label{Table:Waveform}
\begin{tabular}{|m{3cm}|m{5cm} |m{5cm} |m{4cm} |}%{|l|l|l|}%{|p{6cm}|p{4cm}|p{6cm}|}
\hline
\textcolor{black}{\textbf{Waveform or Multiple Access Technology}} & \textcolor{black}{\textbf{Key Distinguishing Feature(s)}} &\textcolor{black}{\textbf{Advantages}} & \textcolor{black}{\textbf{Disadvantages}}\\ 
\hline 
\hline
FBMC &  Employs high quality pulse shape filtering for each \textcolor{black}{sub-carrier} separately.	&$\circ$ Addresses the need for large guard bands in OFDM. \vskip 0.1cm $\circ$ Permits a robust estimation of very large propagation delays and of arbitrarily high carrier frequency offsets \cite{andrews2014will}.& $\circ$  Falls short in handling MIMO channels. \vskip 0.1cm $\circ$ The design of wideband and high  dynamic range systems with FBMC results in significant RF development challenges.	\\
\hline
FTN & Accelerates the pulse rate beyond the Nyquist signaling rate, thereby introducing a controlled ISI.  &$\circ$ Packs more bits/second/Hertz by packing more pulses per unit time. This increases the spectral efficiency in a given single-carrier modulation type and order without requiring additional SNR and without degrading the BER, asymptotically. \vskip 0.1cm $\circ$ Substantial acceleration of the transmit pulses increases the \textit{constrained capacity}. & $\circ$ The required additional precoding and/or detection needed to mitigate the artificial ISI effect increases the transceiver complexity. \vskip 0.1cm $\circ$ Results in a higher peak-to-average power ratio. \vskip 0.1cm $\circ$ Complicates the synchronization problem.\\
\hline  
SEFDM & $\circ$ Reduces the sub-carrier spacing in a multi-carrier modulation below the minimum orthogonality spacing, hence, introducing controlled ICI. More data streams in a given frequency channel can hence be transmitted. \vskip 0.1cm $\circ$ The demodulator collects statistics of
the incoming signal by projecting it onto orthogonal bases while the detector estimates the transmit sequence based on the collected statistics.& Enhances the  spectral efficiency of a multicarrier modulation system requiring no additional SNR and no BER degradation. & $\circ$ The removal of the artificially introduced ICI increases the complexity of the system. \vskip 0.1cm $\circ$ Complicates the synchronization and carrier frequency-offset compensation.\\
\hline

NOMA		&	$\circ$	 Multiple users of different channel conditions are multiplexed in the power domain on the transmitter side.  \vskip 0.1cm  $\circ$	Superposition coding used at the \textcolor{black}{transmitter, and} SIC is employed at  the  receiver.& $\circ$  Enhances spectral efficiency by \textcolor{black}{assigning} multiple users \textcolor{black}{on the same} frequency resource.  \vskip 0.1cm $\circ$	Outperforms  orthogonal multiplexing and can achieve the capacity region of the downlink broadcast channel. \vskip 0.1cm $\circ$	Achieves a better trade-off between system capacity and user fairness in the uplink. \vskip 0.1cm $\circ$	NOMA along with MIMO delivers enhanced performance.&	$\circ$	Each user needs to decode information of all the other users even ones with poorest channel gains. This leads to higher complexity and energy consumption at the receiver. \vskip 0.1cm $\circ$ If an SIC decoding error occurs for a single user, decoding of \textcolor{black}{all other user information} will be erroneous. This limits the number of users.	\\
\hline
\end{tabular}
\end{table*} 

\subsubsection{\textcolor{black}{Spectrally Efficient Frequency Division Multiplexing} (\textcolor{black}{SEFDM})}
\begin{figure*}[t!]
    \centering
    \begin{subfigure}[b]{0.5\textwidth}\label{FTN1}
        \centering
      \includegraphics[width=0.5\textwidth]{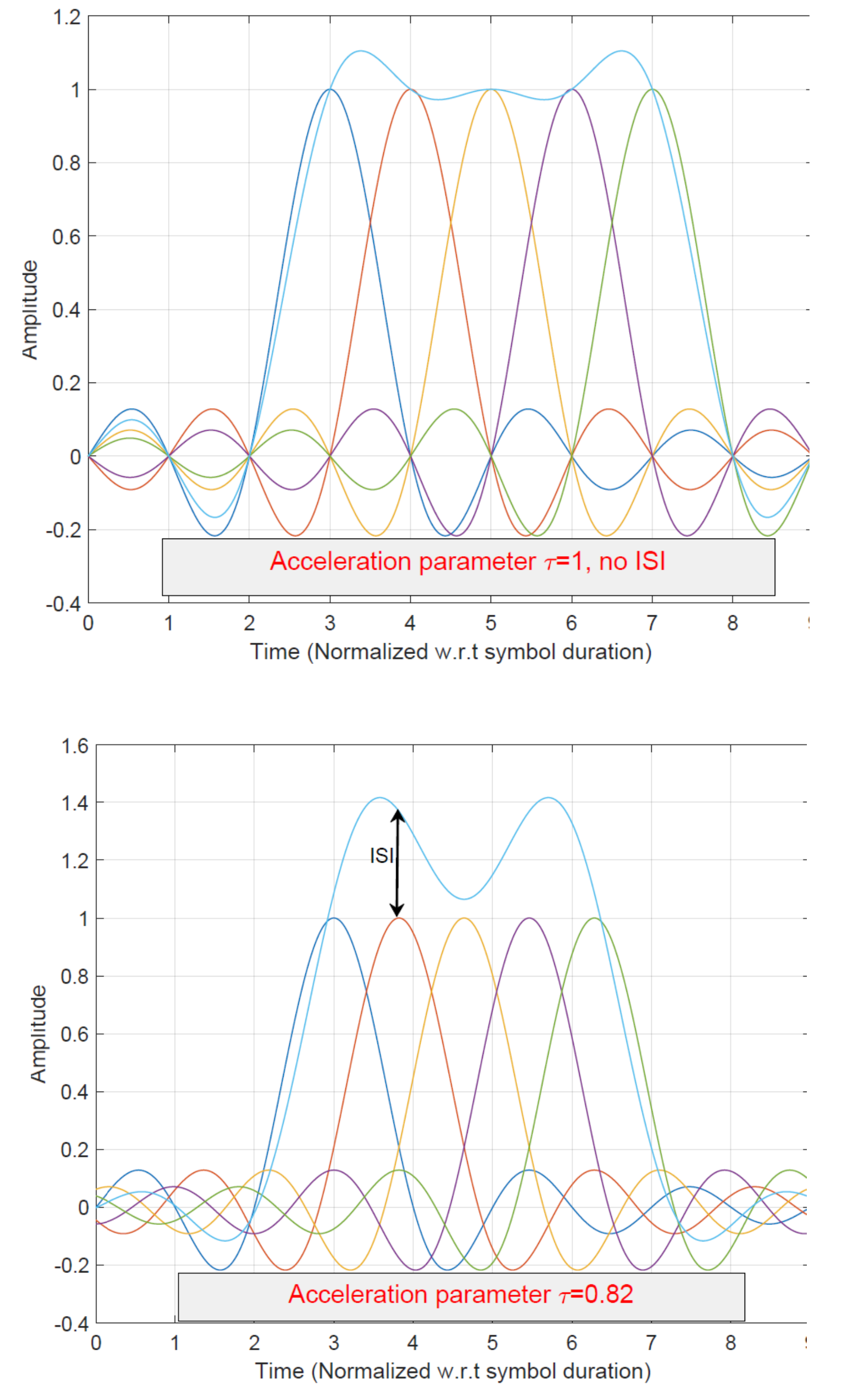}
        \caption{\textcolor{black}{ }}
    \end{subfigure}%
    ~ 
    \begin{subfigure}[b]{0.5\textwidth}\label{FTN2}
        \centering
        \includegraphics[width=0.99\textwidth]{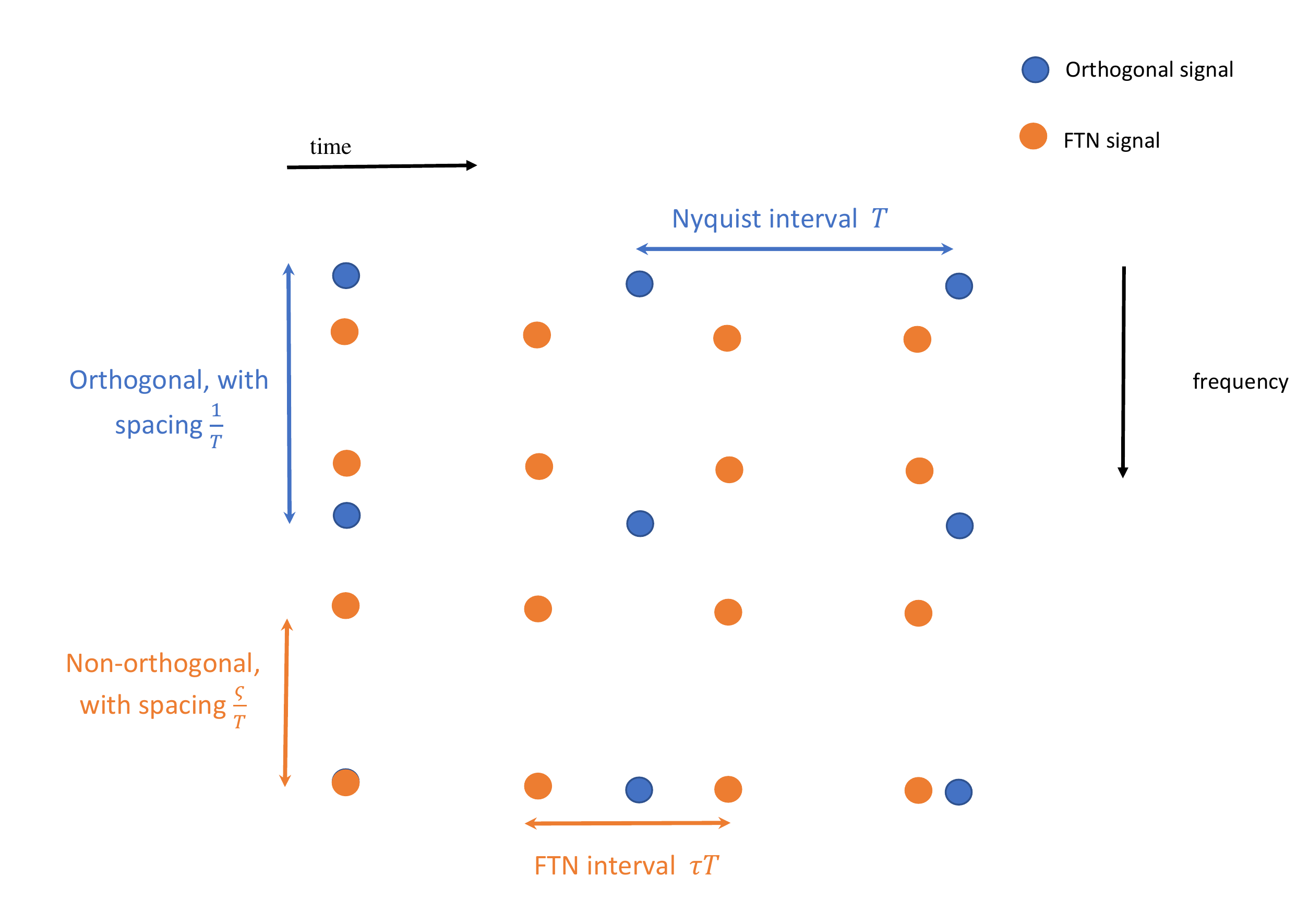}
        \caption{\textcolor{black}{ }}
    \end{subfigure}
    \caption{\textcolor{black}{FTN and SEFDM (\textcolor{black}{multi-carrier} FTN), (a) An illustration of Nyquist and FTN transmission. (i)$\tau=1$ (Nyquist transmission), (ii)$\tau=0.82$ (FTN transmission), (b) 2D orthogonal versus FTN symbols.}}
    \label{FTN}
\end{figure*}

\textcolor{black}{The idea of FTN signaling has been also studied in frequency domain, which is known as spectrally efficient frequency division multiplexing (SEFDM), extending it to \textcolor{black}{both time and frequency} domains, as \textcolor{black}{shown} in Figure \ref{FTN}. A squeezing factor $\varsigma $ allows the \textcolor{black}{sub-carriers} to be packed closer than in the orthogonal case, hence improving the spectrum utilization at the expense of a controlled  inter-carrier interference (\acrshort{ici}). The existence of Mazo limit in both time and frequency domains \textcolor{black}{was} proved in \cite{rusek2005two} and \cite{rusek2009multistream}.  SEFDM (multicarrier FTN) transceiver design, optimization, and performance \textcolor{black}{were} addressed in \cite{dasalukunte2010multicarrier,peng2018spectral,cai2020low,ma2019low}. 
% With the proposed transceiver architectures, FTN signaling can be combined with the current communication technologies, such as OFDM \cite{5755914}.
Besides the high spectral efficiency offered, SEFDM can further enhance the energy efficiency of communication systems because it improves the bandwidth efficiency without extra energy consumption. Since SEFDM is a spectrally efficient multicarrier modulation scheme, it is certainly a promising technology to be used for \acrshort{haps} access links, i.e., \textcolor{black}{between \acrshort{haps} and UEs in order} to achieve the ultra-high broadband data rates anticipated for 6G. The selection of the sub-carriers and their spacing for each user is an \textcolor{black}{open area of HAPS research, depending on the application type} and QoS requirements of each UE.} 

\subsubsection{\textcolor{black}{Non-Orthogonal Multiple Access}}
\textcolor{black}{One of the relatively novel multiple access approaches is the non-orthogonal multiple access (NOMA) scheme proposed by the mobile phone operator NTT DOCOMO in Japan \cite{benjebbour2017overview}. In this scheme, multiple users of different channel conditions are multiplexed in the power domain on the transmitter side, requiring multi-user signal separation on the receiver side. From an information-theoretic point of view, using superposition coding at the transmitter and successive interference cancellation (SIC) at the receiver, non-orthogonal user multiplexing not only outperforms orthogonal multiplexing, but can also achieve the capacity region of the downlink broadcast channel. NOMA can be also applied to the \textcolor{black}{uplink, and despite} the fact that under the orthogonal multiple access (OMA) one can achieve the capacity of the channel, NOMA achieves a better trade-off between system capacity and user fairness \cite{higuchi2015non}.}

\textcolor{black}{
By taking advantage of the spatial dimensions that multi-antenna \acrshort{haps} systems offer, \textit{spatial division multiple access} (SDMA) is facilitated allowing multiple users to communicate at the same time and frequency but in different spaces/beams. MIMO-NOMA overloads SDMA by allocating a cluster of UEs to each beam and using SC-SIC within each group \cite{vaezi2019non}. The interference between the clusters is managed by assigning a different beam to each of the clusters. MIMO-NOMA differs from \textcolor{black}{multi-user} \acrshort{mimo} \textcolor{black}{in that the former allows a cluster of users, not just one user, to share one beam.} Hence, \textcolor{black}{MIMO-NOMA can} serve a larger number of users and paves the way for massive connectivity.}
\section{Handoff Management in HAPS Networks}\label{sec:handoff}
\textcolor{black}{Terrestrial cellular networks \textcolor{black}{ support the mobility of \acrshort{ue}s across different \acrshort{bs}s. } When a \acrshort{ue} moves out of \textcolor{black}{ one \acrshort{bs} coverage area to another,} the communication will be handed over from the first \acrshort{bs} to the next without \textcolor{black}{a} discernible disruption to the \acrshort{ue}'s call or data session, which is carried out \textcolor{black}{ by a } handoff procedure\footnote{Handover is used within Europe, whereas handoff is the term used in North America.}. A properly designed handoff algorithm is essential \textcolor{black}{ for reducing } the overhead of the handoff process, while maintaining the desired \acrshort{qos} of the \acrshort{ue}\textcolor{black}{,} and reducing the probability of blocking new calls or sessions. Basically, the handoff process consists of three main phases: (1) handoff information gathering, (2) handoff decision, and (3) handoff execution. However, terrestrial handoff algorithms are designed to manage the mobility of the \acrshort{ue}s while assuming that \acrshort{bs}s are stationary. }

\textcolor{black}{\textcolor{black}{The stratosphere } is relatively stable \textcolor{black}{ but can be affected by short-term airflow. } Therefore, \textcolor{black}{a} \acrshort{haps} needs to maintain a quasi-stationary \textcolor{black}{position}. \textcolor{black}{ However, some \acrshort{haps} systems need to move in a certain pattern and area, as discussed in Section~\ref{sec:characteristics}.} According to the recommendations of the \acrshort{itu}, \textcolor{black}{ the position of a \acrshort{haps}}  \textcolor{black}{ should be maintained in a cylinder with a radius of 400 m and a height of approximately 700 m \cite{lTURF1500}. } \textcolor{black}{From this, \acrshort{haps} movements can be classified into four categories:} horizontal, vertical, rotation and swing.}

\textcolor{black}{Disturbances to the position of a \acrshort{haps} cause changes} in the size/position of \textcolor{black}{cell} coverage on Earth (i.e., \acrshort{haps} footprint). This can lead to the instability of  \textcolor{black}{user} communication links, which increases the probability and frequency of handoffs. Unfortunately, the handoff management algorithms of terrestrial networks are inadequate for handoff management in \acrshort{haps} systems. The main reason is that in \acrshort{haps} systems not only the \acrshort{ue}s are moving but also the \acrshort{haps} coverage \textcolor{black}{area} and position are changing. In addition, 
\textcolor{black}{the disturbance of \acrshort{haps} positions is irregular,}  which means that it is difficult to establish a clear relation between the speed and position information of a \acrshort{ue} and the cell coverage area. In \acrshort{haps} systems,  \textcolor{black}{handoffs} can be inter-\acrshort{haps} or intra-\acrshort{haps} (i.e., inter-cell handoff). There are two types of inter-cell handoff in \acrshort{haps} communication \textcolor{black}{systems: one is a user mobility} initiated handoff, which results
\textcolor{black}{from a user moving into a neighboring cell;} the other \textcolor{black}{ 
is triggered only by the instability of the platform} \cite{he2016handover}.

%This phenomenon puts forward a challenge for the design and application of handover algorithms.

%Therefore, the handover decision algorithm assisted by speed and position information is not applicable. While conventional RSS-based and its improved algorithms (such as RSSTH algorithm employing a hysteresis margin and an absolute threshold) mainly use a fixed threshold. When the threshold is too low, frequent handovers are likely to happen; in contrast, if the threshold is too high, it is supposed to increase the non-handover probability. However, the selection of the optimal threshold is related to the change rate of signal strength. Due to the constant change of the signal strength, the algorithms with a fixed threshold is not suitable for HAPS systems.

%\textcolor{orange}{Handoff is widely employed in cellular communications in order to maintain service continuity when mobile users move between several cells controlled by a fixed-base-station infrastructure.}
\textcolor{black}{In an intra-HAPS handoff, where the handoff occurs between cells served by the same HAPS, the centralized architecture and control can be exploited.} The important issue 
\textcolor{black}{ is controlling which cell to switch to at which time. } In \cite{grace2010low}, the handoff algorithm \textcolor{black}{ was } based on a time-reuse time-division multiple/time-division multiple access frame structure that is similar to that available with IEEE 802.16. A single-frequency variant \textcolor{black}{ was }
suggested, where the \acrshort{haps} transmits to/receives from different spot beams (i.e., cells) in different portions of the frame.
A multiple-frequency variant \textcolor{black}{was} also  suggested as a way of increasing the system's capacity. In this case, each cell transmits/receives using a sequence of frequencies in different parts of the frame, with each cell in a cluster starting at a different point in the sequence so as to create a hybrid time/frequency \textcolor{black}{re-use} plan.

\textcolor{black}{A radial based function neural network \textcolor{black}{was} used in \cite{alsamhi2015intelligent} to make intelligent handoff decisions, while considering the parameters of \acrfull{rssi}, direction of user mobility, \acrshort{haps} position, traffic intensity, steerable antenna, elevation angle of \acrshort{haps} systems and delay as inputs of the neural networks. By taking into account the curvature of the earth, the author in  \cite{albagory2015handover} analyzed the influence of the rotational movement on the user handoff probability in the equal beam-width coverage model. It was pointed out \textcolor{black}{that the outer layer cell was more susceptible} to the rotational movement.}

\textcolor{black}{In \cite{he2016handover}, after establishing an antenna beam coverage geometry model for \acrshort{haps} systems, the author used the Monte Carlo method to calculate the overlap area and analyze the handoff probability during the swing movement in the equal coverage area model. The average and maximum handoff probability of the different tier cellular \textcolor{black}{were} deduced. The simulation \textcolor{black}{results showed} that the handoff performance of the cells \textcolor{black}{was} severely affected by the swing state, especially for outer tier cells, \textcolor{black}{but that the effect could be reduced by increasing the cell coverage radius, to a certain extent.}}

\textcolor{black}{The traditional handoff algorithms that depends on fixed thresholds usually  deal effectively with the handoff issue caused by \textcolor{black}{UE} mobility. However, \textcolor{black}{given the quasi-stationary state of \acrshort{haps} systems,} these algorithms show poor handoff management. The quasi-stationary
state of a \acrshort{haps} causes the cell edge users to receive a variable signal strength, resulting \textcolor{black}{in frequent handoffs}  between cells or the ping pong effect \footnote{The ping-pong effect occurs when  a UE keeps performing handoffs between two adjacent cells due to fluctuation in received signal strength.}. In addition, representing the unbalanced cell load becomes inaccurate. The effects of the quasi-stationary
state on the inner and outer layers are different. Employing a handoff algorithm with a fixed threshold fails to provide efficient handoff in the entire communication system. When the threshold is too low, frequent handoffs are likely to happen; in contrast, if the threshold is too high, the handoff will be triggered too late causing long periods of communication disruption. In \cite{he2016improved}, by considering the received signal strength, the terminal of mobile speed, and the platform disturbance factors, the author proposed an adaptive handoff algorithm that \textcolor{black}{predicted} the received signal strength. In a similar approach, the author in \cite{ni2016handover} proposed a
prediction-based handoff decision algorithm with an adaptive threshold. The algorithm \textcolor{black}{predicted} the values of received signal strength using  \textcolor{black}{a} time series analysis model and dynamically \textcolor{black}{adjusting} the handoff initiation time according to the prediction. }

\textcolor{black}{However, in these studies a simple coverage model was used, such as circular or regular hexagonal cell coverage, to analyze the handoff probability. Although these models simplify the analyses of handoff probability, they are difficult to implement in practical engineering, which leads to fewer applications. The equal beam-width coverage model \textcolor{black}{was mainly proposed on the basis of the attenuation characteristics} of the antenna directional gain. However, 
\textcolor{black}{as \acrshort{haps} systems are located at high altitudes,} the difference in path loss at each location in the coverage area should also be considered.}

\textcolor{black}{In order to solve the problem of the high outage probability and longtime service interruption during the handoff process between \acrshort{haps} systems, 
\textcolor{black}{an} adaptive handoff scheme that uses cooperative transmission was proposed \cite{he2017adaptive}. 
\textcolor{black}{In this scheme, the HAPS with the higher channel gain was selected} for cooperative transmission to improve the system reliability, and the handoff 
\textcolor{black}{was determined} by the direction of terminal motion and channel gain 
\textcolor{black}{reduce service interruptions}  caused by frequent handoffs.}

\textcolor{black}{Under the influence of a stratospheric wind, \textcolor{black}{a}  \acrshort{haps} will inevitably move within a certain range. In \cite{wang2019effect}, the author \textcolor{black}{discussed both the vertical and swing movements of HAPS systems and the effect of these movements} on path loss. In addition, the author \textcolor{black}{analyzed the coverage on the basis of a derived ground coverage model} and calculated the handoff probability of the two movement modes. The \textcolor{black}{simulation results showed} that the \acrshort{haps} swing movement has greater influence on the handoff's probability than the vertical movement. This is because the swing movement of the \acrshort{haps} generates cell position drift and shape change.}

\textcolor{black}{%HAPS systems are not regarded as competitors of the satellite technology. On the contrary, 
Exploiting effective and seamless integration among heterogeneous aerospace segments (e.g., \acrshort{leo} and \acrshort{haps}) in order to globally extend broadband wireless connectivity \cite{mohammed2011role} seems promising for  improving handoff performance. In \textcolor{black}{such scenarios,} \acrshort{leo} satellites can provide the backhual link of a \acrshort{haps}. Under this assumption, the author in \cite{li2019handover}  \textcolor{black}{proposed} a dynamic handoff strategy to optimize the \textcolor{black}{moment of handoff} and resource allocation. %Lagrange duality is used and a sub-gradient algorithm is presented to solve this problem. 
The author considered factors of user priority, minimum rate requirement, delay requirement, channel gain\textcolor{black}{,} and the traffic of beams.}

\textcolor{black}{In \cite{li2010directional}, a directional traffic-aware intra-\textcolor{black}{HAPS} handoff scheme 
\textcolor{black}{was} proposed, where users
in overlapping areas of overloaded cells \textcolor{black}{would} be forced to
handoff earlier than their optimal handoff boundaries in
order to partially balance traffic among  adjacent
cells. A cooperative directional inter-cell handoff scheme for \acrshort{haps} systems  \textcolor{black}{was} studied in \cite{li2011cooperative}, where the handoff target cell and the two cells adjacent to it  \textcolor{black}{worked} cooperatively to exploit the traffic fluctuation to improve handoff performance. Basically, users in the overlap area of the overloaded handoff target cell \textcolor{black}{were} forced to handoff directionally before their optimal handoff boundary in order to free up resources for the handoff calls which would otherwise be dropped due to the shortage of resources and queue time out.}

\textcolor{black}{\cite{katzis2010inter} \textcolor{black}{investigated} the inter-\acrshort{haps} handoff process when \textcolor{black}{a} \acrshort{haps} operates with fixed or steerable
directional antennas in the case of a \acrshort{haps} replacement
either for maintenance or periodic replacement of short-endurance  \acrshort{haps} systems. Handoff performance was evaluated for both types of antennas based on a number of criteria, such as the \textcolor{black}{antenna beamwidth, platform} height, and the  \textcolor{black}{HAPS} position cylinder. Results showed that for users employing fixed antennas pointing towards the center of the position cylinder, \textcolor{black}{the} handoff can start as soon as the new platform enters the cylinder. Users from the center of the service area are required to employ the \textcolor{black}{widest beam} width antenna
(29°), whereas users at the edge of the service area \textcolor{black}{could}
employ a narrower beam width. Users employing steerable
antennas \textcolor{black}{could also use a narrower beam width.} \textcolor{black}{However, it was shown that connections would be dropped for any beam width of less than} 5°, unless users \textcolor{black}{employed} two 
\textcolor{black}{antennas or unless the new platform followed} a close flight path to the current serving platform
(±305 m vertical separation).}

\textcolor{black}{\cite{li2010novel}  \textcolor{black}{proposed} a connection admission
control scheme, referred to as the Rate Transition Area assisted Guaranteed
Handoff Scheme, which \textcolor{black}{utilized}   geographical
information, rate transition areas, and overlap areas to help \textcolor{black}{eliminate} both the inter-cell and the intra-cell
handoff failures through adaptive modulation and coding in
the physical layer.}

\textcolor{black}{\textcolor{black}{Using the \acrshort{rssi} values as the decision criterion will make the \acrshort{haps} with large \acrshort{rssi} overloaded, causing data congestion in these nodes.} \textcolor{black}{By contrast}, \acrshort{haps} systems with small \acrshort{rssi} \textcolor{black}{values} might remain idle, which \textcolor{black}{can lead} to insufficient utilization of network resources. Meanwhile, due to the limited energy of high altitude \textcolor{black}{platforms}, the energy of \acrshort{haps} \textcolor{black}{systems} with large \acrshort{rssi}  \textcolor{black}{quickly run out,} and the energy of a \acrshort{haps}  with small \acrshort{rssi} \textcolor{black}{values} remains excessive, resulting in \textcolor{black}{an}  imbalance of energy consumption among \acrshort{haps} systems. To address this issue, the author of  \cite{an2013load}  \textcolor{black}{proposed} a load balancing handoff algorithm based on \acrshort{rssi} \textcolor{black}{values} and energy-awareness in \acrshort{haps} networks. Table \ref{HandoffComparison} provides a comparison between available handoff schemes in terms of their proposed \textcolor{black}{solutions}, 
\textcolor{black}{(inter-/intra-HAPS), parameters considered in making handoff decisions, and type of movement causing a handoff.}}

\begin{figure}
	\begin{center}
	\centering
\includegraphics[width=1\columnwidth]{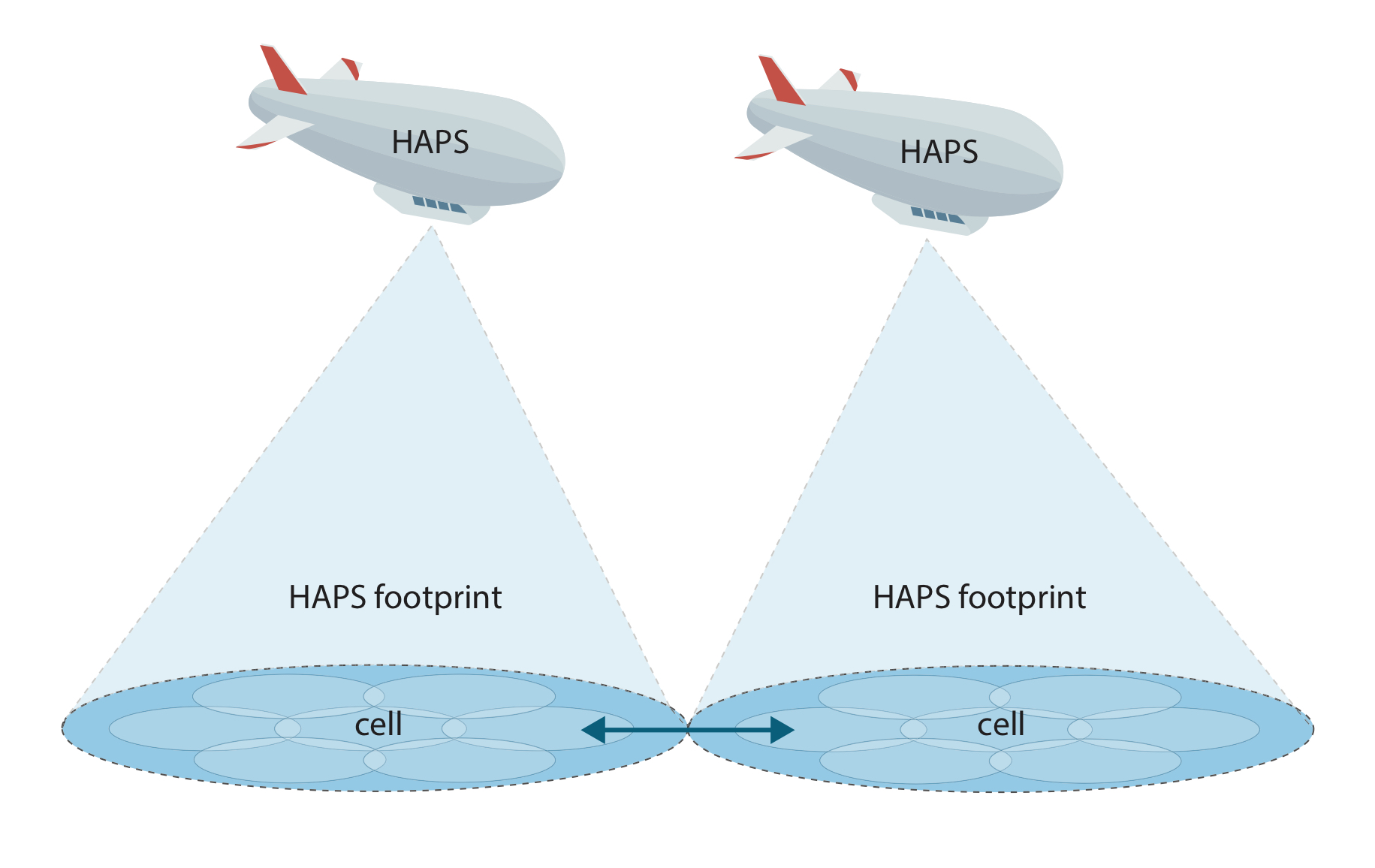}
\caption{\textcolor{black}{Inter-HAPS handoff.}}
\label{InterHAPS}
\end{center}
\end{figure} 

\begin{figure}
	\begin{center}
	\centering
\includegraphics[width=.6\columnwidth]{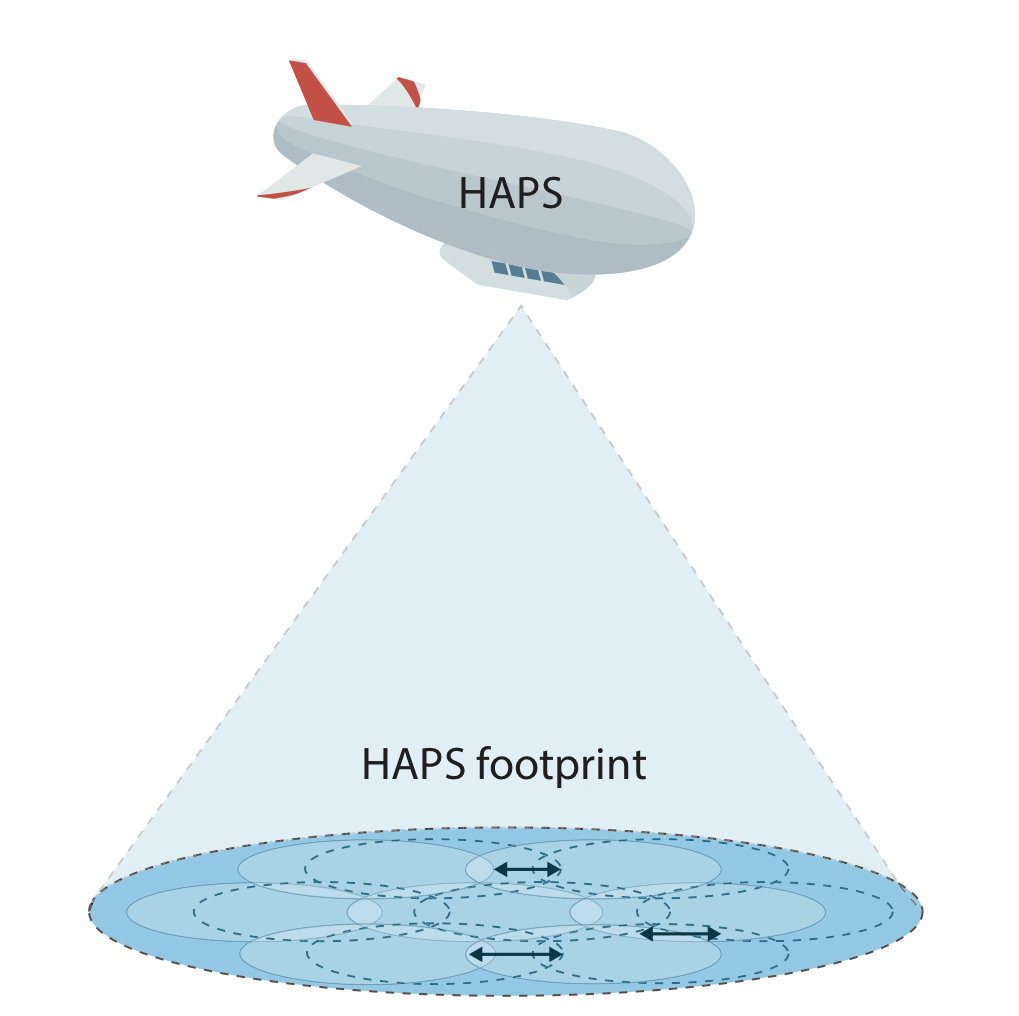}
\caption{\textcolor{black}{Intra-HAPS handoff.}}
\label{IntraHAPS}
\end{center}
\end{figure} 

\begin{table}[t]
\caption{A comparison of handoff management techniques in HAPS networks} 
\centering 
\begin{tabular}{|m{1cm}|m{6.8cm}|m{2cm}|m{3cm}|m{3cm}|}
\hline
 Reference	 & Proposed solution & \textcolor{black}{Inter-/Intra-HAPS} & \textcolor{black}{Parameters considered} & Movement type\\ 
\hline 
 \cite{grace2010low} & \textcolor{black}{Low-latency} MAC layer handoff& Intra-HAPS & RSSI, CIR, traffic load, and user position & Pitch movement\\
\hline

\cite{alsamhi2015intelligent} & \textcolor{black}{Proposed a radial-based neural network} to make intelligent handoff decisions & Intra-HAPS &  RSSI, direction of user mobility, HAPS position, traffic intensity, steerable antenna, elevation angle of HAPS systems, and delay & Vertical and horizontal movement  \\
\hline

\cite{he2016improved} & Proposed an adaptive handoff algorithm that predicts the received signal strength & Intra-HAPS & RSSI, the mobile terminal speed, and the platform disturbance factors & \textcolor{black}{Swing} movement\\
\hline

\cite{ni2016handover} & Proposed a prediction-based handoff decision algorithm with an adaptive threshold  to dynamically adjust the handoff initiation time & Intra-HAPS & RSSI values set & Horizontal movement \\
\hline

\cite{he2017adaptive} & Proposed an adaptive handoff scheme that \textcolor{black}{used} cooperative transmission to improve system reliability & Inter-HAPS & The direction of terminal motion and channel gain & Mobile terminal movement \\
\hline

\cite{li2019handover} & Proposed a dynamic handoff strategy to \textcolor{black}{optimize the moment of handoff and} resource allocation & \textcolor{black}{HAPS with LEO satellites} & User priority, minimum rate requirement, delay requirement, channel gain, and the traffic of beams & Mobile terminal movement\\

\hline

 \cite{katzis2010inter} & Studied the effect of antennas type and beamwidth on handoff during HAPS replacement & Inter-HAPS & \textcolor{black}{Antenna beamwidth, platform height, and user antenna direction} & Replacement movement\\
\hline

\cite{li2010novel} & Proposed \textcolor{black}{using} adaptive modulation and coding in the physical layer to ensure that  ongoing calls are not disrupted \textcolor{black}{by} the platform movement & Intra-HAPS & the geographical information, rate transition areas and \textcolor{black}{overlapping} areas & Replacement movement\\
\hline

\cite{li2011cooperative} & Proposed neighbouring cells cooperatively force users in the \textcolor{black}{overlapping} area of the overloaded handoff target cell to handoff before their optimal handoff boundary in order to free up resources for the handoff calls & Intra-HAPS & \textcolor{black}{Position of users}, HAPS movement direction, and HAPS capacity load & Rotational movement\\
\hline

\cite{an2013load} & Proposed a load balancing handoff algorithm based on RSSI and energy-awareness in HAPS networks & Inter-HAPS & RSSI, HAPS residual energy, and mobile terminal movement direction & Mobile terminal movement\\
\hline

\cite{li2010directional}  & \textcolor{black}{Proposed forcing} UEs in overlapping areas to handoff \textcolor{black}{in order to} balance load among cells		& 	Intra-HAPS	& HAPS traffic load and UE position	& \textcolor{black}{Horizontal} displacement and rotation \\	
\hline

		\hline
\end{tabular}
\label{HandoffComparison} 
\end{table}

% \subsection{\textcolor{black}{Use Cases Requirements}}

\textcolor{black}{Overall, it needs to be mentioned that an efficient handoff management is extremely important to support the HAPS-SMBS role in the \textcolor{black}{previously mentioned} use cases of delivering IoT services, covering unplanned user events, and supporting intelligent transportation systems. Three main handoff related points should be considered to fulfil the HAPS use cases requirements. First, the mobility \textcolor{black}{patterns are random in IoT and unplanned events use cases,} whereas in intelligent transportation use case mobility \textcolor{black}{patterns are} more predictable. Second, mobility speed should be considered in order to perform fast handoff or normal handoff. Obviously, intelligent transportation and aerial network support requires fast handoff as users tend to move \textcolor{black}{at} high speeds e.g., cars, trains, and aerial vehicles (see use cases 6 and 7 in Section \ref{sec:USECases}), while IoT devices (sensors) or users in unplanned events may tolerate normal handoff delays. Thus, future handoff management solutions need to consider time sensitive applications for rapidly moving network entities. Third, user applications requirements  of each use case should be considered in handoff algorithms as some applications are time-sensitive (e.g., use case 6 and 7) and require fast handoff algorithms with low packet loss \textcolor{black}{rates}. Note that under 5G specifications, handoff protocols rely on uplink synchronization which may require the random access procedure. However, the 3/4-way handshaking for initial access could result in unacceptable propagation  \textcolor{black}{delays}, especially in use case 6 and 7 (i.e., HAPS-SMBS to support and manage aerial network and HAPS-SMBS to support intelligent transportation systems).  \textcolor{black}{It is not entirely clear yet whether it will be possible with conventional solutions to adhere to the latency requirements} of 5G and beyond.}

\section{\textcolor{black}{HAPS Network Management and Computational role}}\label{sec:management}

%\textbf{Note: I am not sure if computation should be in this section. (This section is not complete)} 

\textcolor{black}{To operate a \acrshort{haps}, two to four ground-based crew members are required to oversee various aspects of mission planning, flight control, sensor operation\textcolor{black}{,} and data assessment \cite{hehtke2017autonomous}.  Operational complexity and cost will likely scale up in scenarios where multiple \acrshort{haps} systems are deployed and need to coordinate and work together as a swarm. To overcome the technical and economical problems of deploying a network of \acrshort{haps} systems, \acrshort{haps} systems control and 
\textcolor{black}{coordination will require some level of autonomy.} Autonomy will eliminate the need for direct human intervention on many operational levels and allow \acrshort{haps} systems to make intelligent decisions in a collaborative manner.} \textcolor{black}{In effect, \acrshort{haps}s can play important roles in aerial network management and network slicing, as described in the sixth use case in Section \ref{sec:USECases}. This is due to \textcolor{black}{the HAPS} higher position which enables it to collect data and network status information from a large part of the aerial network. Another advantage is that \textcolor{black}{a HAPS} can be equipped with computational devices, which enables full or partial computations to be accomplished in \textcolor{black}{the} air without congesting the communication links towards the terrestrial data \textcolor{black}{centers}. However, this approach requires strong and reliable collaboration between \acrshort{haps} systems to fully utilize \textcolor{black}{their}  distributed computational resources. We should also mention that as privacy is one of the main concerns in data collection and analysis, federated learning may offer learning without moving data from devices to a centralized server, thus preserving user privacy. One recommended solution might be the utilization of federated learning in future \acrshort{haps} networks.}

\textcolor{black}{The implementation
\textcolor{black}{of a semi-autonomous high-altitude} platform swarm with self-organizing capabilities was investigated to maximize communications area coverage \textcolor{black}{in} \cite{8797881}.  The author compared the application of \acrfull{rl} and \acrfull{si} based methods for resolving the problem of coordinating multiple
\acrshort{haps} to maximize communications area coverage.
It was observed that the \acrshort{si} algorithm showed faster convergence and \textcolor{black}{a} more stable user coverage profile due to the simple rule-based logic. However, the \acrshort{rl} algorithm %(applying dynamic epsilon-greedy technique and decaying learning rate) 
achieved higher overall peak user coverage rates but with some coverage dips due to individual \acrshort{haps} exploration strategy. \acrshort{rl}-based techniques \textcolor{black}{demonstrated} inherent coordination resilience due to independence from feedback loops and cross-agent communications. Therefore in \acrshort{haps} systems coordination, swarm intelligence-based approaches may be more efficient and reliable but with less optimal coverage  \textcolor{black}{results. By contrast, RL algorithms may} achieve better coverage peaks but at the risk of occasional dips.} \textcolor{black}{This conclusion should be considered in the design of the fifth use case, HAPS-SMBS \textcolor{black}{to fill coverage gaps,} which was described in Section \ref{sec:USECases}. \textcolor{black}{As it is very difficult to predict/detect coverage gaps using conventional mathematical models, \acrshort{rl} based solutions have high values as they circumvent the need for a tractable mathematical models, to some extents.} Nevertheless, one should note that the \acrshort{rl} approaches are generally sensitive to the design of a proper reward function, otherwise there is no guarantee that the algorithm could converge to a suitable solution. }

\textcolor{black}{The author of \cite{anicho2019autonomously} investigated the coordination among a swarm of four autonomous \acrshort{haps} \textcolor{black}{systems} in a volcanic ash cloud emergency scenario for aerial communications coverage, where terrestrial or satellite infrastructure \textcolor{black}{was} degraded or \textcolor{black}{non-existent.} Due to the \textcolor{black}{extreme nature of the environment, it was shown that} a \acrshort{haps} platform may fail and require replacement. The swarm of \acrshort{haps} 
\textcolor{black}{were autonomously coordinated and used their} self-organizational capabilities to react to the failure of one or more \acrshort{haps} and to autonomously adapt to the addition of a spare \acrshort{haps}. Autonomy in this regard refers to the ability of \textcolor{black}{a} \acrshort{haps} to make local decisions with limited or no global knowledge and still achieve network-wide objectives cooperatively. In \textcolor{black}{a} \acrshort{haps} swarm, self-organization and  coordination is crucial to provide   communications coverage in volcanic cloud emergency conditions, \textcolor{black}{and the author in \cite{anicho2019autonomously} developed} a swarm intelligence algorithm for this scenario. The participating \acrshort{haps} \textcolor{black}{systems} in the swarm \textcolor{black}{exchanged} essential data as they \textcolor{black}{explored} the environment. The \textcolor{black}{algorithm developed had four phases:} scouting mode, exploitation mode, decision making loop, and exploration mode. The simulation results showed that through self-organization and swarm coordination, within  \textcolor{black}{one} hour the spare \acrshort{haps} provided the needed boost in the global coverage performance. }

%\textcolor{orange}{cross layer management.}
%\textcolor{orange}{existing studies consider one or a small group of HAPS. In near future the whole Earth will be covered with a network of HAPS or large HAPS constellations. }

\textcolor{black}{\textcolor{black}{Promising solutions have also been introduced in the literature in the matter of network control and management softwarization.} For example, \textcolor{black}{in \cite{kaur2018edge},  \acrshort{sdn} decoupled } data and control planes from each other in order to reduce network control complexity. \textcolor{black}{In \cite{han2015network}, \acrshort{nfv} decoupled network functions} from physical devices, 
\textcolor{black}{which demonstrated the potential of facilitating} the deployment of new services with increased agility and \textcolor{black}{a} faster time-to-value. 
\textcolor{black}{Also in \cite{NS9003208}, network slicing enabled} connectivity for devices with diverse requirements via multiple logical networks built on top of the shared physical infrastructure. To enable the availability of the networks-as-a-service according to user demands, network slicing \textcolor{black}{employed} \acrshort{nfv}, \acrshort{sdn}, cloud computing, and edge computing. Enabling network slicing requires \textcolor{black}{the} successful interaction among these \textcolor{black}{technologies} which is a challenging task \cite{NS9003208}. Softwarization combined with intelligent algorithms is expected \textcolor{black}{to make network control and management automated and self-organized.} For example, software defined aerial networks components can be reprogrammed automatically and dynamically on the basis of intelligent decisions to adapt to changes in the communications environment. To \textcolor{black}{address} the challenges of dynamic network traffic, multiple service providers, and mobility, dynamic network slicing needs to be considered \cite{NS9003208}. To enable dynamic network slicing, it is necessary to accurately estimate the user demands and dynamically allocate resources accordingly. Several learning theory schemes\textcolor{black}{,} such as deep learning and reinforcement learning\textcolor{black}{,}  can be used for the prediction of user traffic. After \textcolor{black}{an} accurate prediction, effective resource allocation schemes can be used for enabling dynamic network slicing.}

\textcolor{black}{In \cite{8610425}, \acrshort{haps} systems \textcolor{black}{were} used as the control plane of \textcolor{black}{a} software defined aerial network. The controllers \textcolor{black}{were} deployed in \acrshort{haps} systems to take advantage of their wide coverage and relative \textcolor{black}{stability, which was shown to potentially reduce} the configuration updating time in the data plane of the aerial network caused by the length and connectivity variation of links among aerial network \textcolor{black}{components} due to their high mobility.} \textcolor{black}{This study can be quite useful in realizing and developing the sixth use case ``HAPS-SMBS to support and manage aerial network," \textcolor{black}{discussed in Section}  \ref{sec:USECases}.} \textcolor{black}{ In  \cite{chen2019segment}, a \acrfull{sd-abn}  \textcolor{black}{was} proposed to maintain coverage and provide reach-back to military units and \textcolor{black}{to} ensure network flexibility, openness, interoperability, and evolvability. To meet the challenges of traffic management in \acrshort{sd-abn}, segment routing \textcolor{black}{was} applied. Moreover, a network traffic scheduling algorithm  \textcolor{black}{was designed on the basis of} \acrshort{sd-abn} to improve the transmission reliability and bandwidth utilization by balancing network traffic \textcolor{black}{among} multiple reliable transmission paths.}

%%%%Note: write about network slicing in HAPS.

%%%%%Note: write about computing. orchestration is necessary to manage distributed resources. Computing enables intelligent network management. 

\textcolor{black}{In future \acrshort{haps} networks, network management functions \textcolor{black}{will} be distributed and automated to best meet multi-dimensional (cost, latency, availability, throughput, massive connectivity, etc.) service requirements. This process will be self-organizing and self-optimizing across administrative boundaries, either within a single operator or between operators with autonomous re-arrangement of network
partitions. When multiple providers/operators are involved, managing the whole network using a single management unit (orchestrator) increases complexity and delay. Such \textcolor{black}{an} increase in delay will be more prominent for massive machine-type communication in 5G and massive ultra-reliable \textcolor{black}{low-latency} communication in the upcoming 6G wireless systems. To cope with these issues, multiple distributed orchestrators can be used to reduce complexity. \textcolor{black}{In such a scheme, every orchestrator is designed} to control particular network segments. These multiple orchestrators are then controlled by another entity,  \textcolor{black}{called a hyperstrator,} whose job is to control the overall network resource allocation \cite{NS9003208}. Although this management model was not originally proposed for \acrshort{haps} systems, it \textcolor{black}{it is worth investigating more deeply}.} \textcolor{black}{ This high level of control and orchestration is a necessary  \textcolor{black}{requirement for realizing} many of the envisioned HAPS-SMBS use cases, such as creating an aerial data center, \textcolor{black}{filling coverage gaps}, and supporting and managing aerial networks.}  \textcolor{black}{ As \acrshort{haps} systems can form \textcolor{black}{an} \acrshort{mec} cluster or an aerial data centre for processing offloaded data from aerial or satellite \textcolor{black}{networks} (refer to the fourth use case, Section \ref{sec:USECases}), intelligent task scheduling schemes are required, which take into consideration  \acrshort{haps} energy consumption, computational capabilities, and processing loads. Intelligent decision-making algorithms are required to decide on when it is more efficient to process data in \acrshort{haps} \acrshort{mec} clusters than \textcolor{black}{to send} the raw data to terrestrial data centers. } 

\textcolor{black}{ Ultra-reliable low latency applications \textcolor{black}{that have emerged}  from the confluence of 5G, SDN/NFV, and \acrshort{ai}/\acrshort{ml} (e.g., autonomous driving, emergency response systems, remote medical, etc.) \textcolor{black}{require control and processing functions to be distributed} toward the point of data collection and consumption. In this regard, \acrshort{haps} systems can provide the services of a huge network \textcolor{black}{edge above aerial networks and below satellite networks.}}
\textcolor{black}{ \acrshort{haps} systems are expected to play the role of floating aerial data centre, as described \textcolor{black}{in the fourth use case, mentioned} in Section \ref{sec:USECases}. This is due to their wide coverage above \textcolor{black}{low-altitude} aerial network components (e.g., \acrshort{uav}s), which makes them ideal \textcolor{black}{for collecting} large \textcolor{black}{amounts} of data about aerial \textcolor{black}{network statuses and for using such data}  in network management. However, the critical issue is managing and scheduling the computational and communication resources in \acrshort{haps} systems to serve the speedy and dynamic environment of \acrshort{uav}s and satellites.}

\textcolor{black}{Recently, several studies 
\textcolor{black}{have proposed using multi-\acrshort{uav}s} to form a \acrfull{mec} cluster. In \cite{yang2020multi}, a multi-\acrshort{uav} aided \acrshort{mec} system \textcolor{black}{was} proposed, where ground \acrshort{iot} nodes \textcolor{black}{could} offload the computational tasks that \textcolor{black}{could not} be processed \textcolor{black}{with} their limited capabilities. The author introduced a load balancing algorithm to balance computational loads among \acrshort{uav}s and used deep reinforcement learning for computational task scheduling. In \textcolor{black}{an} \acrshort{mec} network formed by multiple \acrshort{uav}s, the sum power minimization problem was considered in \cite{yang2019energy}. The author minimized the power \textcolor{black}{by} jointly optimizing user association, power control, computation capacity allocation\textcolor{black}{,} and \acrshort{uav} location planning. In \cite{zhou2020mobile}, the author introduced two architectures where a \acrshort{uav} \textcolor{black}{could} work as either a node in a distributed \acrshort{mec} cluster or \textcolor{black}{or as a relay node that assisted in computational} offloading from \acrshort{iot} devices to a far terrestrial edge computing node. A game-theoretic and reinforcement learning framework \textcolor{black}{was} introduced in \cite{asheralieva2019hierarchical} for computational offloading in \textcolor{black}{an} \acrshort{mec} network operated by multiple service providers. The network \textcolor{black}{was} formed by \acrshort{mec} servers installed at stationary \acrshort{bs}s and \acrshort{uav}s which \textcolor{black}{were}  quasi-stationary. Although these studies  \textcolor{black}{proposed using} multiple \acrshort{uav}s to form \textcolor{black}{an} \acrshort{mec} cluster and process the offloaded computational tasks, the same ideas can be implemented in \acrshort{haps} networks. In fact, a network of \acrshort{haps} \textcolor{black}{systems} can provide a more stable \acrshort{mec} cluster with stronger computational capabilities in comparison to \textcolor{black}{a} \acrshort{uav}-base \acrshort{mec}. This is because of the quasi-stationary status of \acrshort{haps} systems \textcolor{black}{with} their ability to carry more advanced computational servers and their longer flight \textcolor{black}{durations}. \textcolor{black}{A} \acrshort{haps}-based \acrshort{mec} can not only serve terrestrial \acrshort{ue}s but also aerial \acrshort{ue}s and satellites. \textcolor{black}{However, to achieve the vision of creating \textcolor{black}{a HAPS-based} MEC, reliable communications among \acrshort{haps} systems, intelligent task scheduling, and advanced resource allocation techniques are necessary.}}

\textcolor{black}{Regarding support for the use cases mentioned in Section \ref{sec:USECases}, HAPS network management should be automated to enable \textcolor{black}{a network of HAPS systems} to be a self-evolving network \cite{Tasneem2020}. As HAPS systems are envisioned to provide \textcolor{black}{aerial data centers} (the fourth use case), several requirements should be considered\textcolor{black}{,} including reliable collaboration among HAPS \textcolor{black}{systems} and efficient computational and storage resource management among distributed HAPS systems. In addition, intelligent task scheduling, which \textcolor{black}{considers the capabilities, energy consumption, and processing loads of each HAPS}. Moreover, it is very important to have intelligent decision-making algorithms for data offloading to decide on where to process the data (e.g., in \textcolor{black}{a} HAPS or terrestrial network), which is necessary to support the second, third, fourth, sixth, and seventh use cases mentioned in Section \ref{sec:USECases}. Another crucial requirement for all use cases is that HAPS networks \textcolor{black}{will need to} support the SDN and NFV paradigms. In fact, HAPS systems are a potential candidate for SDN controller palcement in an SDN-based VHetNet architecture.}

\section{The Role of AI in HAPS Systems}\label{sec:AI}
\textcolor{black}{ Active research is currently being carried out to enable \acrshort{ml} in highly resource-scarce \acrfull{mcu}s and Field-Programmable Gate Arrays \cite{Branco2019}. In 2017, Microchip \textcolor{black}{manufactured} the first \acrshort{mcu} \textcolor{black}{with} a high-performance 2D GPU, the PIC32MZ DA, \textcolor{black}{to handle parallel calculations}  \cite{Microchip2019}.   Implementing low-power tiny \acrshort{mcu}s with embedded GPU capabilities is a step  toward implementing advanced \acrshort{ml} algorithms in \acrshort{haps} systems. In 2019, ARM launched its Helium technology, which will be present in the next generation of ARMv8.1 \acrshort{mcu}s (Cortex-M). This technology is intended to provide high digital signal processing and machine learning capabilities  \cite{ARM2019}. STM Microelectronics \textcolor{black}{sells sensors with incorporated machine learning cores that have embedded classifiers} \cite{STMicroelectronics2019}.  As companies race to provide high digital signal processing and \acrshort{ml} capabilities to their \acrshort{mcu}s, we will witness a tight combination between electronic devices and novel \acrshort{ml} algorithms designed to be executed \textcolor{black}{with} limited resources.}

\textcolor{black}{Advances in \acrshort{ml} \textcolor{black}{have} produced a number of emerging powerful \acrshort{ml}  \textcolor{black}{algorithms, such as deep neural networks consisting of two stages (i.e., offline training and online execution).} In the on-line execution stage,  deep neural  \textcolor{black}{networks make}   decisions \textcolor{black}{using the environment states as input}, even when \textcolor{black}{some of the environment states have} not been experienced in the offline training phase \cite{zhang20196g}. 
Another  powerful \acrshort{ml} approach is reinforcement learning, which \textcolor{black}{resembles the trial and error process of the human brain.} The decision-making entity of the reinforcement learning framework interacts with the environment continuously through iterative \textcolor{black}{observations} of the environment state. The reinforcement learning framework \textcolor{black}{then} selects the actions that affect the environment and obtain immediate rewards, before observing new environment states. Basically, the decision-making entity tends to select the best action with the greatest long-term reward for each environmental state \cite{guan2019intelligent}. In recent publications, reinforcement learning \textcolor{black}{has been} adopted to address decision-making problems in communications \textcolor{black}{environments}, such as access radio technology handoffs \cite{nguyen2017reinforcement}, spectrum sharing \cite{raj2018spectrum}, and user scheduling \cite{qiao2018topology}.  However, in a large state-action space, reinforcement learning performance drops since many state-action pairs may not be explored. \textcolor{black}{More recently, a} new version of reinforcement learning, “deep reinforcement learning,” has emerged, which applies the intelligent data representation of the deep neural network in the reinforcement learning \cite{luong2019applications}. Although merging deep neural networks and reinforcement learning shows \textcolor{black}{promise} capabilities in adapting to complicated and dynamic communications environments with extensive state-action spaces, the scalability of such solutions \textcolor{black}{needs} to be considered. For resource-limited equipment, some simplified novel ML algorithms (e.g., compressed deep neural network learning) have been proposed. FastGRNN and FastRNN are algorithms to implement Recurrent Neural Networks (RNNs), and gated \acrshort{rnn}s in  tiny devices \cite{kusupati2018fastgrnn}. }

\textcolor{black}{In future networks, AI will play an essential role \textcolor{black}{in the orchestration and management of HAPS systems.} On the other hand, \acrshort{haps} systems will be a great enabler for AI and computing in aerial and space networks, \textcolor{black}{as HAPS systems} can carry an aerial data centre and perform edge computing functionality. In fact, \acrshort{haps} systems are physically 
\textcolor{black}{located at high altitudes between satellite and terrestrial networks.}  Due to their high \textcolor{black}{altitudes} and wide coverage, \textcolor{black}{HAPS systems can collect a massive volume of data.} In the \acrfull{ioe} era, data is the precious fuel of data analytics and \acrshort{ml} algorithms. Such data can be used to reveal trends, hidden patterns, unseen correlations, and achieve automated decision making. It can also be used to continuously learn about wireless network \textcolor{black}{environment and} user behaviour and enable the network to proactively adapt to changes. \textcolor{black}{This will allow HAPS systems to achieve optimal performance.}  \acrshort{haps} systems are \textcolor{black}{a} potential candidate for collaborative computing and distributed \acrshort{ml}. In big data centers, complex ML \textcolor{black}{tasks are divided into smaller ones} that are executed in parallel on multiple virtual or physical machines. This makes the idea of collaborative computing \cite{Hochul2019} feasible by distributing the tasks of \acrshort{ml} among a group of collaborating \acrshort{haps} systems forming an aerial data \textcolor{black}{center}, as described in the fourth use case in Section \ref{sec:USECases}. As a leading alternative to centralized \acrshort{ml} algorithms, federated learning techniques can provide a platform to achieve distributed  \acrshort{ml} with high prediction accuracy in a privacy-preserving manner.  However, to support \textcolor{black}{artificial} intelligence in future \acrshort{haps} systems through collaborative \acrshort{ml} execution, reliable communications among \acrshort{haps} are required. }

\textcolor{black}{ The current trends demonstrate that \acrshort{ai} algorithms started to gain more interest among researchers to optimize the functionality of \acrshort{haps} systems, reduce the operational cost, and adapt to changes in communication environment. Current studies on \acrshort{haps} systems consider the deployment of a single or a small number of \acrshort{haps}. However, in future networks, it is expected that a \acrshort{haps} network will consist of several \acrshort{haps} \textcolor{black}{systems} of different types and characteristics. Managing, controlling, and operating such systems in conventional ways will not be efficient and might be impossible. Therefore, there is a \textcolor{black}{vital} need to introduce automation in \acrshort{haps} systems \textcolor{black}{by} exploiting the power of AI algorithms. For example, when an unexpected change happens in the density distribution of \acrshort{ue}s, an intelligent  \acrshort{haps} system can detect such a change \textcolor{black}{by observing the movement of UEs}. Afterward, \textcolor{black}{by analyzing} the collected data, the \acrshort{haps} system can make an intelligent decision to redirect or form a beam towards the newly emerging \acrshort{ue}s groups. \textcolor{black}{In such a situation,} the required characteristic of the formed beam (e.g., capacity and coverage area) can be predicted using a machine learning algorithm.}

\textcolor{black}{Recently, some 
\textcolor{black}{studies have investigated AI to address certain complex optimization problems in HAPS systems.} To maximize the network capacity per cost via optimizing a \acrshort{haps} network constellation, an artificial immune algorithm was used in \cite{dong2016constellation}. The author considered the constraints of \acrshort{qos} (e.g., signal to noise ratio, bit error rate, bits per second coverage) and user demand metrics. In a different scenario, neural networks \textcolor{black}{were} used to 
\textcolor{black}{handle the issue of frequent handoffs} that users at cell edge may experience due to \acrshort{haps} movement. The author in \cite{alsamhi2015intelligent} used radial-based function neural network to make intelligent handoff decisions. The \acrshort{rssi}, direction of user mobility, position of \acrshort{haps}, traffic intensity, steerable antenna, elevation angle of \acrshort{haps} and delay \textcolor{black}{were} the inputs of the neural networks.}

\textcolor{black}{In a wireless communication
network operated by \acrshort{haps} systems, the key factor \textcolor{black}{for the improvement of the \acrfull{cir}} \textcolor{black}{is a reduction in the antenna \acrfull{sll}.} In \cite{ismaiel2018performance}, the author \textcolor{black}{optimized} the beamforming parameters using a comprehensive learning particle swarm optimizer to reduce the \acrshort{sll}. The antenna array configuration \textcolor{black}{was} chosen as \textcolor{black}{a} concentric circular antenna array and the \acrshort{haps} cellular system \textcolor{black}{consisted} of 169 cells. The proposed method significantly suppressed the \acrshort{sll} which led to a significant improvement in \acrshort{cir}.}

\textcolor{black}{To address the limited power and poor computational capabilities of \acrshort{uav}s, the author in  \cite{liu2019minimization} \textcolor{black}{studied} mobile edge computing services through \acrshort{haps} systems, where \acrshort{uav}s \textcolor{black}{could} offload their computing tasks. The author proposed a multi-leader multi-follower Stackelberg game to formulate the offloading problem. As the leaders of the game, the \acrshort{haps} systems \textcolor{black}{optimized} their pricing by considering the behavior of their competitors to maximize their revenue. Each \acrshort{uav} \textcolor{black}{selected} the best computing tasks offload strategy to minimize latency. From this perspective, the stochastic equilibrium problem of equilibrium program with equilibrium constraints model was proposed to develop the optimal supply strategies for \acrshort{haps} to maximize their profits and minimize \acrshort{uav}s’ cost. Computational task planning in \acrshort{haps} systems is essential to optimize \acrshort{haps} computational services and resource utilization. }

\textcolor{black}{A hierarchical task planning structure is favorable for its capability to accommodate constraints at different abstraction levels. This structure is adopted for task planning among multiple \acrshort{haps} systems. As the combinatorial search problem grows with the presence of multiple agents, the author in \cite{kiam2019hierarchical} \textcolor{black}{proposed} a genetic algorithm-based method that \textcolor{black}{guided} the decomposition of the tasks in order to find quality plans within limited time.}

\textcolor{black}{To prove the feasibility of executing AI algorithms using \acrshort{haps} resources,
\cite{clark2019testing} \textcolor{black}{described} a successful test of a commercial off-the-shelf neural network accelerator on a \acrshort{haps}. Various advances in hardware acceleration for specific algorithms and approaches (e.g. neuromorphic processors) can offer advantages when compared to general-purpose CPUs that would otherwise be necessary to accomplish an equivalent task. These improvements have led to a marked interest in the idea of running nontrivial  \textcolor{black}{computing} tasks directly on \textcolor{black}{a} \acrshort{haps} before data passes through the link. \textcolor{black}{In so doing, the amount of data transmitted to terrestrial networks can be significantly reduced}  while simultaneously improving the speed at which a system can analyze and react to a dynamic environment. \textcolor{black}{This \textcolor{black}{raised} the motivation of incorporating \textcolor{black}{artificial} intelligence with \acrshort{haps} systems and utilizing \acrshort{haps} systems to form an aerial data \textcolor{black}{center}.}}

\section{Open Issues} \label{sec:open}
\textcolor{black}{The challenges and open issues related to \acrshort{haps} system can be categorized into two groups, one that mainly covers the next-generation (up to 10 years) challenges and the other the next-next-generation (10-20 years) challenges. The \textcolor{black}{ former will require} intensive research but in a more incremental fashion with regard to the current technologies \textcolor{black}{for} communications systems. \textcolor{black}{For} example, the use of massive \acrshort{mimo} and mmWave communications for \acrshort{haps} system\textcolor{black}{s} can be categorized as \textcolor{black}{a} next-generation challenge as the required theory and practice \textcolor{black}{has been well investigated}  for the terrestrial networks. However, the use of \textcolor{black}{these technologies} for \acrshort{haps} \textcolor{black}{systems will require additional investigations}. Corresponding research investigations could be related to new communications techniques to compensate for the lack of enough \acrfull{dof} in \acrshort{haps} system channels, the restricted transmission energy, or the detection without availability of the channel statistics/model. }

\textcolor{black}{For next-next-generation wireless networks, there needs to be a disruptive shift about how we design and configure these networks.} 
\textcolor{black}{An example of challenges for the next-next generation might be how \acrshort{haps} mega-constellations will interact with satellite mega-constellations.}
 \textcolor{black}{As the potential and pitfalls of satellite mega-constellations are not yet known, since} participating companies tend to  \textcolor{black}{keep certain technologies secret,} the design of  \acrshort{haps} \textcolor{black}{mega-constellations will be} dramatically more challenging and highly speculative in the initial phases.  Should the former be more of a complementary technology compensating for the latter's shortcomings? Or, should \textcolor{black}{satellite and HAPS mega-constellations} be considered as competing technologies targeting \textcolor{black}{potentially}  separate use cases? \textcolor{black}{With} this categorization in mind, in this section we enumerate many important challenges and open issues. We do our best to provide suitable solutions and\textcolor{black}{, where possible,} road maps to tackle the challenges.

\textcolor{black}{For the list of challenges and where these are discussed in the paper refer to Table \ref{Table:challenges}. Note that although some of the challenges  \textcolor{black}{may be discussed in several sections,} for simplicity we only provide the relation of the challenge to the section that overwhelmingly influence it. \textcolor{black}{For} example, \textcolor{black}{h}andoff management \textcolor{black}{challenges are mainly discussed in Section} \ref{sec:handoff} while network management Section \ref{sec:management} and Section \ref{sec:AI} have some influence on it.}

\begin{table*}
\centering
\caption{\textcolor{black}{Section-wise Classification of Challenges}}
\label{Table:challenges}
\begin{tabular}{|m{10cm}|m{6cm} |}%{|l|l|l|}%{|p{6cm}|p{4cm}|p{6cm}|}
\hline
Section of the Paper & Challenge(s)\\ 
\hline 
The Role of AI in HAPS Systems (Section \ref{sec:AI}) &  $\circ$ Efficient Network Re-configuration \\ & $\circ$ Support for Edge Intelligence \\  & $\circ$ Efficient Network Re-configurations  \\ \hline
Regulatory Aspects (Section \ref{sec:regulations}) &  $\circ$ Regulatory Aspects  \\ &  $\circ$ Integration with Satellite Network \\ \hline
The \acrshort{haps} Subsystems (Section \ref{sec:characteristics}) &  $\circ$ System Issues  \\ \hline

Channel Models for \acrshort{haps} Systems (Section \ref{sec:channels}) & $\circ$ Channel Model and Performance Evaluation \\ \hline

Radio Resource Management, Interference Management and Waveform Design in \acrshort{haps} (Section \ref{resource}) & $\circ$ PHY and Related Cross Layer Design \\ & $\circ$ Radio Resource Management \\ & $\circ$ Massive MIMO Communications \\ & $\circ$ Beam  Tracking \\ \hline

Handoff Management in \acrshort{haps} Networks (Section \ref{sec:handoff}) & $\circ$ Handoff Management in HAPS Networks \\ \hline

\acrshort{haps} Network Management and Computational Role (Section \ref{sec:management}) & $\circ$ Networks  Management  of  HAPS  Systems \\ & $\circ$ Computational Roles \\ & $\circ$ Privacy  and  Security  Concerns \\ & $\circ$ HAPS Mega-Constellation \\ \hline

\end{tabular}
\end{table*} 

\subsection{Next 10 Years: On the Use of HAPS in Next-Generation Networks}

\subsubsection{\textcolor{black}{Regulatory Aspects}}
\textcolor{black}{The spectrum provided by \acrshort{itu} for dedicated \acrshort{haps} usage is critical. Unlicensed bands are specifically designated bands worldwide that are intended for industrial\textcolor{black}{,} scientific and medical (ISM) applications. Although WiFi-based systems highlight the successful usage of communication purposes in ISM bands, this is not their main functionality, as the name ISM also implies. In fact, the use of unlicensed ISM bands may have a significant effect on radio astronomy due to electromagnetic interference. This matter \textcolor{black}{was} substantiated in Google’s Project Loon tests in Oceania \cite{AMSAT-UK}. Hence the use of these bands in \acrshort{haps} nodes have to be carefully planned in order to protect radio-astronomy research from unintended interference. To this end, dynamic frequency allocation techniques with cognitive radio capabilities seem promising to manage interference. }

\subsubsection{System Issues}
\textcolor{black}{Since different types of \acrshort{haps} \textcolor{black}{have different payload capacities and different energy consumption specifications,} understanding  the \textcolor{black}{trade-offs between platform type, cost, performance,} and flight endurance is necessary. In general, the energy consumption of a  \acrshort{haps} related to its conventional communication functionalities \textcolor{black}{has been discussed} in the literature. Nevertheless, the future of \acrshort{haps} networks is broader than current functionalities. For example, if \acrshort{haps} are intended to be used for data centers, the payload type and energy consumption requirements need to be discussed. This is also true when HAPS are used as computation or machine learning platforms. In many cases, we expect the station has
at least another functionality besides the common BS/relay one. Therefore, more investigations into energy management and continuity of  service is needed. This might require supporting a HAPS partially with other sources of energy sources such as remote charging or nuclear energy, to ensure sustainable and continuous operations.
}

\textcolor{black}{On the other hand, the use of \acrshort{rss}, while beneficial for reducing  payload weights and energy consumption, results in a smaller usable surface area for the installation of solar panels, thus reducing the amount of solar energy absorbed in the long run. In effect, as we mentioned, to increase the directionality of the reflected signal, and thus the spectral efficiency, more surface area needs to be dedicated to the \acrshort{rss}. Therefore, a balance between the necessity of solar panels for  energy absorption and \acrshort{rss} for reducing energy consumption seems important. One of the potential solutions to this is to utilize the upper surface of the platform for the solar energy, while the bottom surface would be dedicated to the RSS functions. In addition, since different types of \acrshort{haps} nodes have different surface shapes, (e.g., flat or curved), the effects of different surface \textcolor{black}{shapes} on  \acrshort{rss} performance need to be studied.}

\subsubsection{\textcolor{black}{PHY and Related Cross Layer Design}} %\label{sec:open}
\textcolor{black}{
As discussed in \textcolor{black}{S}ection \ref{resource}, there is not yet a suggested PHY waveform specified for \acrshort{haps}. Indeed it is desirable to exploit the promising technologies that are currently under active research in the literature for terrestrial wireless \textcolor{black}{systems; however}, simulations and careful system performance analysis \textcolor{black}{will be required}. The analysis will assess the suitability of candidate waveforms as well as pulse shaping filters for \textcolor{black}{the}  mm-Wave band taking into account the unique propagation
and channel fading nature of a channel propagation of \acrshort{haps}. After developing or \textcolor{black}{determining}  the most suitable waveform, a rigorous design for the corresponding detectors \textcolor{black}{that accounts for computational complexity}  as well as performance could be established. This should \textcolor{black}{take two important goals into account:}   1) the integration with massive \acrshort{mimo}, and 2) \textcolor{black}{the} lack of model-based detection due to lack of complete/reliable knowledge of the underlying channel model. 
% However, the \acrshort{haps} channel models are still not mature for that purpose and the  conventional model-based detection methods are not suitable, as they would have to sacrifice accuracy (and hence performance) with models that are still under development.  Moreover, conventional channel-model-based techniques require the instantaneous channel state information to be estimated, this process entails overhead, which decreases the data transmission rate. 
% Another problem that arises is that inaccurate state information estimation typically degrades the detection performance. 
Using advanced AI/ML techniques offer\textcolor{black}{s} benefits over traditional model-based approaches \cite{farsad2018neural}. First, ML methods are independent of the underlying stochastic model and thus can operate efficiently in scenarios where \textcolor{black}{a} model is \textcolor{black}{unknown or} its parameters cannot be accurately estimated. Second, when the underlying model is extremely complex, ML algorithms have demonstrated the ability to extract meaningful features from the observed data,  which is difficult to carry out using traditional model-based approaches. Finally, ML techniques often lead to faster convergence compared to iterative model-based approaches, even when the model is known \cite{shlezinger2020viterbinet}.}

\textcolor{black}{Another challenge that needs to be considered is the non-linearity \textcolor{black}{of FSO communications,}  which are used in inter-platform link communications, \acrshort{leo}/\acrshort{haps} communications\textcolor{black}{,}
 and back-\textcolor{black}{haul}
 link communications. These non-linearities \textcolor{black}{make using high order modulations} quite challenging\textcolor{black}{. H}ence, there should be ways to increase the spectral efficiency\textcolor{black}{,} as the most commonly used modulation scheme is On-Off keying, or binary amplitude shift keying (BASK). One possibility could be to use single carrier FTN for that purpose and design the suitable detectors for a BASK-based FTN signaling while taking into account the FSO pulse shaping filters that are commonly used.}
\textcolor{black}{This could be very promising to support the use of HAPS-SMBSs in providing Tbps FSO link backhauling for small-cell (or isolated) BSs, which is one of the use cases discussed in Section II-B.}

\textcolor{black}{When new waveform technologies are introduced, cross layer design challenges appear. \textcolor{black}{For} FTN and SEFDM technologies, power and  channel allocation as well as acceleration and/or squeezing parameters can be considered jointly \textcolor{black}{as} a single design problem. For instance, if we consider a single carrier FTN system, then as discussed in Section \ref{resource}, decreasing the acceleration parameter would increase the spectral efficiency of a particular user but would degrade the performance due to larger ISI. A channel allocation and parameter selection scheme can assign the channels and select acceleration parameters for \acrshort{haps} users depending on the channel fading each user experiences, hence taking advantage of multi\textcolor{black}{-}user diversity. This can be done such that the overall spectral efficiency of the system is maximized. This becomes more complex in SEFDM multicarrier HAPS access links, where we expect time and frequency squeezing parameters to be jointly optimized with power and sub-channel allocation\textcolor{black}{s} for each \textcolor{black}{HAPS} user, where each sub-channel could even have a different number of sub-carriers depending on the frequency-squeezing parameter.}

\subsubsection{Radio Resource Management \textcolor{black}{(RRM)}}\label{RRM_Challanges}
\textcolor{black}{Despite the fact that research on the RRM and interference management for \acrshort{haps} systems dates back to \textcolor{black}{the} early 2000s, there are still many open issues and challenges. Developing techniques that yield acceptable performance \textcolor{black}{with} low computational overhead \textcolor{black}{is a balance that designers and researchers try to strike}. This is \textcolor{black}{partly}  why we expect that AI and ML techniques for \textcolor{black}{the design and optimization of } \acrshort{haps} systems will prosper. AI/ML schemes\textcolor{black}{,} such as reinforcement learning for channel allocation in a 5G massive \acrshort{mimo} \acrshort{haps} \textcolor{black}{have begun to} appear in the \acrshort{haps} RRM  literature. There is a gap actually, when it comes to the desired performance that is possible using model-based mathematical optimization and the \textcolor{black}{real-time} implementation requirements in terms \textcolor{black}{of the computational overhead required.}
   In \cite {sun2017learning}, an AI deep neural network (DNN) was proposed to \textcolor{black}{address this,}  where the input and output of an RRA algorithm \textcolor{black}{was treated} as an unknown non-linear mapping \textcolor{black}{problem}  to approximate the algorithm. It was demonstrated that DNNs can achieve orders of magnitude speedup in computational time \textcolor{black}{when}  compared to state-of-the-art power allocation algorithms based on optimization. The role of AI/ML for system design aspects (e.g. RRM, interference management\textcolor{black}{,}  etc.) is even more emphasized for integrated networks of \acrshort{haps}, \acrshort{leo} satellites, terrestrial networks\textcolor{black}{,} and UAVs\textcolor{black}{,} where the mathematical models are still not mature and \textcolor{black}{are}  expected to be quite complex, possibly making model\textcolor{black}{-}based optimization techniques \textcolor{black}{unsuitable}. Little work has been done on RRM and interference management \textcolor{black}{in}  integrated \acrshort{haps}/ \acrshort{leo} systems.} %As discussed in \cite{zappone2019wireless} a mobile AI paradigm envisions both a cloud intelligence, which every node of the network can access by connecting to the cloud, and a device intelligence specific to each network device. 

\textcolor{black}{\textcolor{black}{There is a gap in the literature  that}    heterogeneous types of services (data traffic) \textcolor{black}{ have not been considered for}
 HAPS. For example, URLLC with massive broadband and/or massive machine type communications (mMTC) have almost completely different QoS requirements which must be satisfied and hence need to be considered in the mathematical formulations. This requires \textcolor{black}{the} development of multi-objective schemes that can be executed in real-time, and \textcolor{black}{therefore} an ultra-low complexity is needed. For example, the \acrshort{haps} system might want to maximize throughput for massive broadband UEs while minimizing \textcolor{black}{end-to-end} delay and packet loss for URLLCs.} Additionally, technology \textcolor{black}{has thus far interacted}  primarily with only two senses, sight and sound. Interestingly, in \cite{EricssonIoS}, Ericsson Research \textcolor{black}{envisioned} enabling an Internet of  \textcolor{black}{the} five senses \textcolor{black}{in} around 2025, \textcolor{black}{which would require} the establishment of new QoS metrics that reflect the user's convenience or satisfaction for these new types of services. This will need to be considered in \acrshort{haps} RRM, interference management\textcolor{black}{,} and placement schemes as they are being developed.  Also, more research needs to be conducted on emerging technologies like hologram streaming,  its QoS metrics\textcolor{black}{,} and the development of related novel \acrshort{haps} RRM schemes.

\textcolor{black}{\acrshort{haps} systems are expected to be powered by rechargeable batteries and solar panels. This was not sufficiently taken into consideration in the vast majority of \acrshort{haps} RRM, interference management\textcolor{black}{,} and placement research papers. We believe that for optimum energy utilization, we need to take this into account while developing suitable model-based or AI-based schemes for power management in the \textcolor{black}{HAPS} access downlink, inter-platform links and backhaul links, possibly jointly. In addition,   \textcolor{black}{as we anticipate deployments of}
  \textcolor{black}{ \acrshort{haps}  mega-constellations}, it is important to consider relaying between platforms over multihops to facilitate communication between two devices associated \textcolor{black}{with} different \acrshort{haps} for certain applications (e.g. URLLC), possibly in different cities, rather than pushing the communication through the core network. For such scenarios, inter-platform link power control together with relay selection for inter-platform routing over multiple hops will be necessary. As \acrshort{haps} systems are expected to be a major part of 6G communication systems, \textcolor{black}{more research is needed} into resource allocation and interference management in \acrshort{haps} systems \textcolor{black}{for new} use cases, such as mission-critical robotics, self-driving cars, high-capacity AR/VR applications, and high-stakes cargo drones. \textcolor{black}{Specifically, cargo drones (in huge numbers) are part of the intelligent transportation system that a HAPS-SMBS is envisioned to support. The channel models \textcolor{black}{to capture 3D mobility effects between HAPS-SMBSs and drones have yet to be developed}. These \textcolor{black}{will be crucial for analyzing SINR and outage probabilities that drones could experience and whose insights will be of great importance} for developing RRM schemes. Moreover, \textcolor{black}{associating cargo drones with terrestrial BSs}, a medium altitude platform or a \textcolor{black}{HAPS-SMBS needs} to be addressed. Under what \textcolor{black}{conditions should a drone connect} to a HAPS or a terrestrial BS? Or should we advocate for double connectivity in \acrshort{haps} systems?  This needs to \textcolor{black}{take distance into account as well as available radio resources}, such as power. It is worth noting that \textcolor{black}{the rechargeable nature of power supplies for aerial platforms} must be taken into account as well.} It is worth keeping in mind that about 80$\%$ of the traffic demand is media-driven, \textcolor{black}{which calls for more research into video streaming}, caching, and \acrshort{qoe} in \acrshort{haps} systems. \textcolor{black}{Finally}, \textcolor{black}{in the context of joint control-communication system design}, new performance \textcolor{black}{metrics have emerged, such as} age of information (AoI) and information freshness, \textcolor{black}{which have thus far been completely overlooked} in the \acrshort{haps} \textcolor{black}{literature, and} therefore new studies need to consider \textcolor{black}{these}.}

\subsubsection{\textcolor{black}{Channel Model and Performance Evaluation}}
\textcolor{black}{\textcolor{black}{Channel models, especially for HAPS-to-LAPS and HAPS-to-satellite links,} need further elaboration as this \textcolor{black}{is still an open}   area in the current literature.} Despite its vital importance and many unresolved issues, the performance evaluation of \acrshort{haps} systems \textcolor{black}{has also been} overlooked. One reason might be due to the lack of a universally agreed-upon, easy-to-use, and practically substantiated channel model facilitating the performance evaluation. On the other hand, the \textcolor{black}{analyses found in the literature have typically considered a single} \acrshort{haps} along with a limited number of users on the ground. \textcolor{black}{Moveover}, interaction\textcolor{black}{s} between terrestrial, \acrshort{haps}, and satellite networks have not been considered. Due \textcolor{black}{to the complex structure of} \acrshort{haps} and its role in \acrshort{vhetnet} more sophisticated tools should be adopted for the performance evaluation. 

\textcolor{black}{To study the coverage and capacity of a} HAPS or cluster of HAPS nodes, we recommend tools from stochastic geometry for modeling the spatial locations of UEs and UAVs. \textcolor{black}{These tools have been widely adopted} for investigating various aspects of terrestrial networks as well as \acrshort{uav} systems. In general, these tools are able to exploit some measures regarding the average behavior of the network to anticipate easy-to-use performance bounds of the network, which can be used to better understand \textcolor{black}{the} large-scale impact of various system parameters. A powerful aspect \textcolor{black}{of these tools} lie in \textcolor{black}{their} abilities to incorporate a mathematically amenable formulation of the inter-cell interference in the analysis of the network, which \textcolor{black}{is difficult with}  other approaches.  \textcolor{black}{These tools also seem} promising for evaluating the performance of a large-scale \acrshort{haps} systems by modeling the location of \acrshort{haps} nodes via sophisticated point processes, such as \textcolor{black}{the} Determinantal Point process \cite{Li2019Detrimental} and Ginibre Point Process \cite{Na2019Ginibre}, as these mathematical models allow \textcolor{black}{for}  the inclusion of the (deliberate) repulsion that exists between the stations. Accordingly, \textcolor{black}{an}  accurate account of the inter-cell interference between \acrshort{haps} cells \textcolor{black}{and} terrestrial cells can be included in the analysis. In \acrshort{haps} systems, the effect of inter-cell interference is far more \textcolor{black}{severe} than that of the \acrshort{uav} and terrestrial networks due mainly to highly dominated \acrshort{los} air-to-ground/ground-to-air channel component\textcolor{black}{s} \cite{khosh2019HetNet}. In effect, even stations hundreds of kilometers away  can still \textcolor{black}{cause} severe interference \textcolor{black}{for} ground users, even merely due to side-lobe antenna gain \cite{khosh2019NonIdeal}. This \textcolor{black}{means} that more advanced resource allocation along with sophisticated antenna techniques should be adopted at the stations, \textcolor{black}{ideally without creating} high computational burdens \textcolor{black}{for user terminals or}  IoT devices. Note, on the other hand, that since each platform may be a collection of several macro BSs, the typical assumptions regarding the independency of large-scale path-loss attenuation (including the \acrshort{los} occurrence) and shadowing, which are the main assumptions for deriving the coverage and capacity performance of the terrestrial networks, need to be revisited. This makes the performance evaluation of \acrshort{haps} systems  \textcolor{black}{particularly} challenging in comparison to its counterparts, which calls for new investigations. Finally, we expect that  \textcolor{black}{stochastic geometry will play} a key role for understanding the performance of a \acrshort{haps} system for robotic applications and edge intelligence---via analysis information freshness or the age of information---which deviates from the conventional techniques based on the average analysis of the performance metrics. This matter can be addressed via the meta distribution analysis of Poisson networks \cite{Kalamkar2019meta}.  

\subsubsection{Massive MIMO Communications}
\textcolor{black}{Massive \acrshort{mimo} communication is among the disruptive technologies \textcolor{black}{that has made the 1,000x capacity growth of \acrshort{5g} possible}. Nevertheless, the promise of the technology, as it is developed i\textcolor{black}{in the context of} the ground terrestrial networks, \textcolor{black}{may or may not be the right fit for} \acrshort{haps} applications. This is simply because air-to-ground and ground-to-air channels are highly \acrshort{los} dominant and also suffer from low scattering profile\textcolor{black}{;} therefore the exploitable \acrshort{dof} can be limited. In effect, pilot contamination, \textcolor{black}{which is a principal} performance-degrading phenomenon in massive \acrshort{mimo} communications, \textcolor{black}{can be even more detrimental} in the \acrshort{haps} systems. The pilots can be contaminated from stations \textcolor{black}{located} hundreds of kilometers away, \textcolor{black}{which can lead to a} small frequency re-use factor. Furthermore, the possibility of the antenna array to properly disjoint signals in the spatial domain can \textcolor{black}{be} less effective, perhaps regardless of the number of antennas \textcolor{black}{or} the processing power of the stations, as the received signals from large areas might be highly correlated. One solution to tackle this issue might \textcolor{black}{be to intelligently cluster the users in the spatial domain} in order to minimize \textcolor{black}{the} effect of the correlated received signals. However, given the size of the coverage area and massive number of users in each coverage zone, \textcolor{black}{attention should be given so as not to deplete computation and energy resources.} In general, we expect novel breeds of \acrshort{mimo} techniques that are tuned for highly correlated signals, allowing to exploit the signal correlations for jointly encode/decode signals for the best possible performance.}

\subsubsection{Beam Tracking}
\textcolor{black}{In \acrshort{5g} and beyond, mmWave communication \textcolor{black}{is} among the key enablers for \acrshort{5g} \acrfull{nr} developed by the 3GPP. Due to very high antenna gain and narrow beams\textcolor{black}{,} it is possible to substantially increase the data rate and reduce the latency. \textcolor{black}{However}, it is more efficient to spatially disjoint multiple users with different radios within the coverage area, and thus serve them simultaneously. This \textcolor{black}{would lead to a} much higher capacity per coverage area, which can be useful for serving swarm\textcolor{black}{s} of \acrshort{uav}s via \acrshort{haps} system\textcolor{black}{s} and also serving \textcolor{black}{a} massive number of devices on the ground. Nevertheless, accurate beam steering/alignment and diligent beam tracking\textcolor{black}{,} communication should be taken care of. For example, \acrshort{uav}s are able to maneuver very fast or could be blocked by large objects \textcolor{black}{or}  buildings temporarily. Without accurate, low-cost, and fast beam tracking the communication can be jeopardized---or worse, lost---which is not acceptable for many mission-critical applications. Conventional solutions, which only rely on the radio signals for estimation and adjustment of the beams, may not be suitable any more. Novel solutions utilizing machine learning in order to predict the mobility of  device\textcolor{black}{s} seems to be crucial. \textcolor{black}{T}he use of computer vision in order to extract valuable information regarding the existence of blockages could \textcolor{black}{also}  improve the overall performance of mmWave communication.}

%\subsubsection{Note yet Decided or Written}
%\begin{itemize}
%\item \textcolor{magenta}{Security  in integrated networks \cite{jiang2015security} provides the key aspects of the authentication process, including secure handoff, secure transmission control, key management, and secure routing.}

%\item\textcolor{magenta}{Regulation of the international near-space: The aviation rules concerning the HAPS deployments are determined by the national civil aviation authorities of the corresponding country, as noted in \cite{liu2019regulating}. The lack of a cohesive set of  rules and regulations to monitor and license the use of  18km to 100 km near-space region may hamper the deployment of large scale HAPS networks to address the high capacity demand in urban areas. This may ensure the safety, security, and privacy of the users. This open issue needs effort from cross disciplines including the avionics and the international law. } 
  %  \item Envisioned Physical Layer Signaling Design
    %\item Computational offloading 
    
   % \item Machine learning applications
  %  \item Energy harvesting / SWIPT options

%\end{itemize}
\subsubsection{Networks Management of HAPS Systems}
\textcolor{black}{\acrshort{haps}-enabled wireless systems are relatively fast to deploy and to some extent re-configurable, which is important for ever-changing demand\textcolor{black}{s}. Nevertheless, the need for the 3D systems \textcolor{black}{in}  accordance with \textcolor{black}{onboard} energy limitation\textcolor{black}{s} and permissible payload weight \textcolor{black}{presents} unprecedented challenges \textcolor{black}{for} network management. This implies the essential role that the optimal deployment of \acrshort{haps} has  for coverage extension as well as capacity improvement, while energy and computation flows are also crucially important. Furthermore, usually the deployment of \acrshort{haps} systems could be short-term---compared with terrestrial networks that are long-term---where the functionalities and responsibilities are subject to change, modification, or augmentation. For example, a station might be initially deployed for the purposes of communications as a flying BS or a relay node, but with possible upgrades and sufficient provisions will be promoted to a computation platform. Hence, there is a need to develop intelligent self-organizing control algorithms to optimize the network resources and deployment of \acrshort{haps} with respect to the functionalities or responsibilities. \acrshort{ai} will play a critical role in designing and optimizing HAPS architectures, protocols, and operations accordingly.}

\textcolor{black}{\textcolor{black}{In} future networks, multiple \acrshort{haps} networks \textcolor{black}{will}  be deployed and instead of working in isolation, they will form a network. Coordinating \acrshort{haps} network\textcolor{black}{s} through ground stations \textcolor{black}{will not be} efficient due to response \textcolor{black}{delays, and ground stations with their limited footprints} cannot have communication coverage to all the \acrshort{haps} network. Therefore, it is envisioned that \acrshort{haps} networks \textcolor{black}{will} be self-organizing with either centralized or distributed control and management system\textcolor{black}{s}. In the centralized approach, one \acrshort{haps} \textcolor{black}{will be designated as a manager} while the \textcolor{black}{others will be designated as followers}. In the distributed approach, the available \acrshort{haps} nodes in a network \textcolor{black}{will} need to negotiate and coordinate in distributing the communication tasks in order to avoid interference, wasting resources, overlapping footprints or beamforming, etc. In this regard, intelligent control and management, based on data analysis and predictions, \textcolor{black}{will} be super valuable. }

    % As future networks are expected to be an integration of satellite, aerial, and terrestrial networks, \acrshort{haps} systems need to be involved in the overall integrated network orchestration. With the use of dynamic beamforming and steering capabilities, it is very important to have continuous coordination between \acrshort{haps} systems to 
    % prevent coverage areas overlapping and interference.  
 
%   In many types of terrestrial networks (e.g., wireless sensor networks), network clustering is implemented to enhance energy consumption efficiency, routing performance, and distributed computational resources management. As \acrshort{haps} networks will consist of clusters of \acrshort{haps} systems that form a sphere around the earth, clustering techniques should be explored to optimize \acrshort{haps} computation resource management, routing among \acrshort{haps} systems, and efficient use of energy. \acrshort{haps} network slicing should consider dynamic spectrum slicing to avoid underutilization or overutilization.

\subsubsection{Handoff Management in HAPS Networks}

\textcolor{black}{
The existing studies 
\textcolor{black}{on handoff management in \acrshort{haps}  systems consider} simple scenarios that might occur in the early deployment stage. However, such scenarios might not be realistic for future \acrshort{haps} systems \textcolor{black}{whose networks will span the globe} (i.e, \acrshort{haps} mega constellations). The future 
\textcolor{black}{  \acrshort{haps} networks are expected to have multiple layers} with several hundred  \acrshort{haps} components. Managing handoff in such a complicated network cannot be efficiently achieved using conventional approaches. There are a number of issues that need to be considered to manage handoffs in an efficient way in future \acrshort{haps} networks.}

%\begin{itemize}
    %\item 
    \textcolor{black}{It is expected that \acrshort{haps} systems will be part of the all-IP network. Thus, handoff management solutions should consider both Layer 2 (i.e., scanning and selecting a new radio channel then associating to a new cell) and Layer 3 (i.e., configuring a new IPv6 address, registering the new IPv6 address using the mobility management protocol, rerouting packets) handoff management.}
   
    \textcolor{black}{\acrshort{haps} systems will provide coverage for not only \textcolor{black}{smartphone holders but also for network entities} moving \textcolor{black}{at} high speeds (e.g., cars, trains, and aerial vehicles). Thus, future handoff management solutions need to consider time sensitive applications for rapidly moving network entities. }
%    \item 

    \textcolor{black}{As handoff management in future networks need to consider many parameters that change in a very dynamic way, intelligent and self-adaptive handoff management solutions are required for both inter- and intra-\acrshort{haps} handoff management.}  \textcolor{black}{Dynamic beamforming techniques should be utilized to reduce the handoff frequency for the largest number of users.} \textcolor{black}{\textcolor{black}{T}o minimize the transmitted power or to maximize the capacity, needs to be revised accordingly. This is because such solutions may render ping-pong effect, which is very undesirable form handoff perspective. Consequently, a more holistic \textcolor{black}{solution for beamforming that addresses the requirements of handoffs seems necessary.}}
    %\item
    
    \textcolor{black}{In the 5G handoff protocol, the random access  procedure plays an important role in uplink synchronization. However, the three-way handshake will result in unacceptable propagation delay\textcolor{black}{s}. }\textcolor{black}{In particular, it is not yet clear if under the conventional solutions it is possible to adhere to the latency requirements of 5G and beyond.}
    \textcolor{black}{\textcolor{black}{Furthermore}, in 5G networks, \acrshort{haps} systems will use the mmWave communication frequencies, which can be absorbed by the atmosphere and affected by weather conditions (e.g., rain, fog, moisture in the air). Thus, mmWave signals might have high attenuation resulting in reduced signal strength. As most of the handoff algorithms depend on signal strength as a main indication to establish handoff, the characteristics of mmWave signals might result in unnecessary handoffs.} As a remedy, \textcolor{black}{a double connectivity solution that would allow} connectivity via microwave for handoff \textcolor{black}{and} beam management and payload communication via mmWave link\textcolor{black}{s}
 should be investigated for \acrshort{haps} systems.

\subsubsection{ Computational Roles}
A \acrshort{haps} can play a role in the aerial network management and network slicing. This is \textcolor{black}{due to the higher position of a \acrshort{haps} that enables it to collect data} and network status information from a large part of the aerial network. Another advantage is that it can be equipped with computational devices, which enables full or partial computations to be accomplished in \textcolor{black}{the} air without congesting the communication \textcolor{black}{links with terrestrial data centers.} In fact, \textcolor{black}{due to their quasi-stationary positions and large footprints} (no frequent handoff\textcolor{black}{s} required), \acrshort{haps} systems are ideal for computational offloading either form satellite networks or from aerial networks (e.g. UAVs). In comparison to offloading computations from satellites and aerial networks to terrestrial networks, offloading to \acrshort{haps} systems can reduce the response delays, reduce the interruption\textcolor{black}{s} during offloading due to \textcolor{black}{the} mobility of satellites or UAVs, and free the terrestrial networks links for terrestrial-aerial or terrestrial-satellite communications.

However, this approach requires strong and reliable collaboration between \acrshort{haps} systems to fully utilize the distributed computational resources . As privacy is one of the main concerns in data collection and analysis tasks, the federated learning may offer learning without moving data from devices to a centralized server, thus preserving user privacy. Therefore, it is recommended to utilize federated learning in future \acrshort{haps} networks.
%  Due to its low altitude when compared to satellites, \acrshort{haps} can support reliable and low latency computational offloading for some delay sensitive applications that are running in areas that do not have terrestrial coverage. However, this issue still needs further investigation and a feasibility analysis through simulations. 
\textcolor{black}{However, in the near future,} it is envisioned that there will be groups of \acrshort{haps} systems surrounding the \textcolor{black}{E}arth. Thus, collaboration and coordination between \acrshort{haps} systems is  essential to achiev\textcolor{black}{ing} optimized  \acrshort{haps} resource management, load balancing, and \acrshort{ue} mobility management. 
    
Note also that as \acrshort{haps} systems can form a\textcolor{black}{n} \acrshort{mec} cluster for processing offloaded data from aerial or satellites, intelligent task scheduling schemes are required, which take into consideration the \acrshort{haps} energy consumption, computational capabilities, and processing loads. Intelligent decision-making algorithms are required to decide on when it is more efficient to process data in \acrshort{haps} \acrshort{mec} clusters rather than sending the raw data to terrestrial data centers.

\subsubsection{Privacy and Security Concerns}
\textcolor{black}{The security of \acrshort{haps} systems can be challenging due to \textcolor{black}{their unique characteristics} and also the integrated ground-\acrshort{haps}-satellite communication paradigm. On the one hand, if by any means a \acrshort{haps} node is compromised, the integrity of any communication passing through the node will be questionable. This \textcolor{black}{could be catastrophic given the enormous footprint of a HAPS, which could include numerous devices and users} with different level\textcolor{black}{s} of security and privacy vulnerabilities. Cautions must be practiced regarding the applicability of the current techniques that merely rely upon detection and localization of malicious devices by exploiting conventional signal sensing and ranging techniques, \textcolor{black}{as these can fall short of effectively combating passive} eavesdropping. Large scale radar surveillance and computer vision techniques can be helpful\textcolor{black}{; however, this can increase} the payload and energy consumption of the \acrshort{haps}. }

\textcolor{black}{From a physical layer communication perspective, one can guarantee highly directional signals with great resolutions via mmWave link or, if possible, \acrshort{fso} communications. Apart from the immediate benefits \textcolor{black}{of lowering interference and increasing the} data rate, mmWave communications can enhance the security of the communication channel as well as protect the ground users against passive eavesdropping and active jamming \cite{bao2017mmWave}, \cite{Xiao2017mmWave}. However, due to imperfect beam alignment and also leaky antenna patterns due to side lobes, the communication links may \textcolor{black}{be} vulnerable. \textcolor{black}{This calls for more effective techniques rely} on information theoretic security \cite{Zhang2019mmWave}, \cite{Liu2017mmWave} and also covert communications \cite{Hu2017covert} for \acrshort{haps} communications. }\textcolor{black}{In addition, the feasibility of HAPS systems for quantum key distribution  is also under investigation to help improve the security and promising results are obtained according to the presented link budget analysis \cite{QKDDD}.}

\textcolor{black}{\textcolor{black}{Given} the diverse roles of a station (e.g., as a communication platform, a data center platform, and a computation platform) the magnitude of security and privacy issues \textcolor{black}{is} even greater. This \textcolor{black}{means} that simply protecting the communication link may not be sufficient. Furthermore, analogous \textcolor{black}{the vulnerability of autonomous vehicles to hijacking and the dangers this poses, the hijacking of a HAPS also poses great dangers.} For example, \textcolor{black}{this could put airplanes traveling in the vicinity at risk of colliding with a HAPS.} One shall designate dedicated \acrshort{haps} nodes with the only responsibility of security monitoring \textcolor{black}{and also} preservation. \textcolor{black}{These dedicated} stations must be equipped with advanced radar, computer vision, and jamming functionalities in order to detect any possible threats. Also, we ought to allow such a station to practice preemptive rights for freezing (with respect to functionalities) and towing (or take over the responsibilities \textcolor{black}{and} functionalities) compromised stations if deemed necessary.}

%\textcolor{magenta}{Security  in integrated networks \cite{jiang2015security} provides the key aspects of the authentication process, including secure handoff, secure transmission control, key management, and secure routing.}

\subsection{Next 20 Years: On the Use of HAPS in Next-Next-Generation Networks}

\subsubsection{Integration with Satellite Network}
\textcolor{black}{Vertical integration of \acrshort{haps} system\textcolor{black}{s} with the satellite networks, known also as multilevel satellite/\acrshort{haps} architecture \cite{SatelliteHAPSsIoT2016}, \cite{WaterQuality2016}, seems attractive and is deemed imperative to attain super connectivity. Such a reliance on the satellite communications \textcolor{black}{will be effective}, noting that many projects, e.g., SpaceX's Starlink, OneWeb, Amazon's Project Kuiper, Telesat, are geared toward providing worldwide 5G/6G coverage via satellite mega-constellations. For instance, Starlink \textcolor{black}{is considering} launching up to 42,000 satellites for occupying different orbital shells. Meanwhile, SpaceX is targeting \textcolor{black}{deployment of}  up to 12,000 satellites through low Earth and very low Earth orbit (500 km to 2000 km, roughly speaking). Smaller players\textcolor{black}{,} such as the satellite startup OneWeb \textcolor{black}{is planing to launch} 900 small satellites into orbit in order to provide broadband internet connections to remote areas. 
\textcolor{black}{D}espite many indispensable advantages, such \textcolor{black}{integrations present} vulnerability issues for \acrshort{haps} system\textcolor{black}{s}. In effect, \textcolor{black}{any partial collapse of satellite communications,} \textcolor{black}{such as a collision} between satellites, \textcolor{black}{could degrade a} \acrshort{haps} network's performance. Accordingly, such challenges \textcolor{black}{make the design of a robust \acrshort{haps} network even more complex.} \textcolor{black}{S}hould the \acrshort{haps} network compensate for the resulted coverage holes? If yes, what regulations \textcolor{black}{should be put in place} and what extra functionalities \textcolor{black}{should be} provisioned for \acrshort{haps} \textcolor{black}{nodes}? Equally important, \textcolor{black}{one should also discuss how transparent the satellite networks should be to \acrshort{haps} networks and vice versa.}}

\subsubsection{HAPS Mega-Constellation}
\textcolor{black}{The emergence of satellite mega-constellations to provide broadband Internet access across the globe could \textcolor{black}{present} a major shift in the future of telecommunication systems. For \textcolor{black}{a} satellite mega-constellation to be \textcolor{black}{economically feasible}, it must provide Internet access faster \textcolor{black}{than what is  already available through fiber-optics}. Provided that satellites are equipped with laser-link communication capabilities\textcolor{black}{,} this goal is indeed reachable \cite{Handley2018mega}. However, this technology \textcolor{black}{is only in its infancy, and recently} launched satellites are not equipped with it. Accordingly, \textcolor{black}{a} stand\textcolor{black}{-}alone satellite mega-constellation \textcolor{black}{may} not guarantee fast, long-distance Internet access as long as fast and cost-effective \textcolor{black}{technologies} for inter-satellite communications \textcolor{black}{are} missing. Current practices advocate reducing the altitude of the satellites \textcolor{black}{and} installing millions of ground-based relay nodes. It is speculated that with these adjustments fast communications across satellites \textcolor{black}{will still be} possible without the existence of laser-link communication \cite{Handley2019mega}. However, such a solution \textcolor{black}{will be} costly and may not be \textcolor{black}{attainable for whole world}. A better solution might be the use of a mega-constellation \textcolor{black}{that allows} multi-hop FSO communications, hence the traverse of the data up to thousands of kilometers becomes feasible, eliminating the requirement of frequent satellite-relay zigzag data exchanges. In this way, the number of hops that \textcolor{black}{would be needed from one satellite to another would be substantially}  decreased. Note also that the installation/monitoring/protection of ground-based nodes in remote/coastal areas could be costly, which is less problematic in the case of \acrshort{haps} \textcolor{black}{systems}. }

\textcolor{black}{\textcolor{black}{In addition,} by using \acrshort{haps} \textcolor{black}{systems,} the coverage zone of each satellite can be considerably extended due to much higher computation/communication \textcolor{black}{capabilities} of \acrshort{haps} \textcolor{black}{systems}. For example, as satellite signals might get too weak on the boundary cells and due to excessive interference form neighboring satellites, \textcolor{black}{a} \acrshort{haps} can boost, combine, and transmit satellite signals via joint transmission techniques. We should further point out that although we mention \acrshort{haps} mega-constellations as a solution for enhancing the coverage performance of satellite mega-constellation, we also advocate for the stand\textcolor{black}{-}alone \acrshort{haps} mega-constellations as a robust solution for fast internet access across \textcolor{black}{borders}. In effect, a large cluster of \acrshort{haps} systems for communications, relaying, routing, data centers\textcolor{black}{,} servers, computation platforms, and security/cyber-policing, can provide the backbone infrastructure of a mobile Internet.  }

\subsubsection{Efficient Network Reconfigurations}
\textcolor{black}{\acrshort{haps} network\textcolor{black}{s are} highly dynamic and heterogeneous. For example, some stations may disappear intermittently for a while to recharge their energy resources. If \textcolor{black}{a} network is primarily configured \textcolor{black}{to provide} maximum capacity, \textcolor{black}{a} new configuration may require \textcolor{black}{adjustments for preserving the coverage requirements.} Given the vast geographic \textcolor{black}{area} that each station is capable of serving, such a consequential alternation in functionality \textcolor{black}{would be} unprecedented. The complexity of network reconfiguration can augment as each station may have distinctive functionalities and given that different types of \acrshort{haps}, such as aerostatic and aerodynamic platforms, have distinctive traits---some stations are quasi-stationary while the others must be keep moving. In effect, the continuous coordination among \textcolor{black}{a} diverse array of stations and simultaneously with the ground stations or satellite mega-constellation to preserve \textcolor{black}{coverage would be} daunting. \textcolor{black}{C}oordinated action across heterogeneous stations \textcolor{black}{would require a tremendous} amount of data and extensive optimization routines. From \textcolor{black}{a} computational perspective, \textcolor{black}{caution is advised to avoid exhausting HAPS energy and computational resources.} It appears that common approach\textcolor{black}{es} of coordination, resource allocation, and networking, \textcolor{black}{which rely on the selection of actions based on a given network structure and specified task, are impractical.}}

\textcolor{black}{One way to smoothly cope with this issue is \textcolor{black}{by using} meta learning\footnote{\textcolor{black}{In machine learning\textcolor{black}{,} the meta-learning is also known as “learning to learn” \cite{Chelsea2017metalearning}. In a nutshell, meta-learning attempts to design models that can learn new skills or adapt to new environments rapidly with a few training examples. \textcolor{black}{By contrast,} meta-reinforcement learning (meta-RL) is meta-learning on reinforcement learning tasks. Here, after training the agents over a distribution of tasks, the agent is able to solve a new task by developing a new reinforcement learning algorithm with its internal activity dynamics. For example, instead of solving a particular graph problem, e.g., minimum cut problem, meta-RL intends to learn a whole set of algorithms on the graph such as shortest path, graph coloring, minimum spanning tree, and the like.}}; Instead of \textcolor{black}{having to constantly solving} the optimization problems (for routing, coverage, backhauling, resource allocation, computation offloading, and the like) to derive action parameters \textcolor{black}{on the basis of new network configurations,} the network \textcolor{black}{could} learn the underlying optimization structures. For example, one can train the network for several prominent tasks, such as coverage preservation, capacity enhancement, energy consumption minimization, computation offloading, and latency reduction, and then \textcolor{black}{with} meta learning an emerging task/environment can be quickly recognized and dealt with.  }

\subsubsection{Support for Edge Intelligence}
Under 6G, \textcolor{black}{we are expected to achieve}
\begin{enumerate}
    \item very high data rate\textcolor{black}{s}, up to 1 Tbps, in order to facilitate the broad uses of \acrfull{vr} and large-scale machine learning applications,
    \item secure connected globe,
    \item extreme reliability and relatively low latency communication\footnote{\textcolor{black}{Under 5G, \acrshort{urllc} requires the 5-nine (99.999$\%$) reliability and 1 ms latency targets. With the emerge of new mission-critical applications such as self-driving cars and high-precision robots, 6G needs to address extreme \acrshort{urllc} with 9-nine (99.9999999$\%$) reliability and at least 0.1 ms latency targets.}}, for control and monitoring of massive number of intelligent, mission-critical high-stake robots, \acrshort{uav}s, and devices,
    \item over-the-air connected intelligence allowing widespread use of machine learning and data analytics tools on the edge.
\end{enumerate}

In effect, the concept of edge computing is already under investigation and is part of 5G wireless communications networking. \textcolor{black}{In addition}, a \textcolor{black}{coherent integration of} edge computing and machine learning is under development, known as edge ML\textcolor{black}{. This} is expected to be a crucial \textcolor{black}{element} of 6G as an enabling computation-communication paradigm for omnipresent machine intelligence. %The goal of edge ML. This is expected to circumvent the reliance on dedicated servers for training ML models and solving ML tasks. In effect, 
\textcolor{black}{T}he evolution of telecommunication infrastructures
towards 6G will call for dispersing \textcolor{black}{artificial} intelligence \textcolor{black}{utilizing} edge computing resources. Edge devices such as AI-enabled \acrshort{uav}s, self-driving cars, robots, and the like are expected to locally train sub-models and share the trained models instead of sharing data, which has important consequences form \textcolor{black}{a} privacy perspective \textcolor{black}{as well}. For this large-scale, distributed edge ML, \acrshort{haps} networks can provide a universal intelligence blanket. In effect, for a given application, e.g., self-driving cars, flying taxi\textcolor{black}{s}, or cargo deliver\textcolor{black}{ies}, a station can collect hundreds of thousands of sub-models from AI-enabled devices in \textcolor{black}{a} synchronous or asynchronous manner. The station \textcolor{black}{can then} apply training routine\textcolor{black}{s} by including its own collected data/intelligence.  

\textcolor{black}{Nevertheless, to stand as an effective universal edge intelligence provider, one must ensure the timely, secure, and efficient communication links to the diverse array of devices including, robots, drones, street lights, servers, driving cars, and the like. In effect, one should optimize the shared communication resource\textcolor{black}{s} for two disjoint purposes: communication for data communication and communication for intelligence. The former is well understood\textcolor{black}{, being the main focus telecommunication system design.} Nevertheless, new applications such as \textcolor{black}{mission-critical} robotics requires a join\textcolor{black}{t} communication-control resource allocation, which is largely unprecedented in the design of communication networks. \textcolor{black}{A joint communication-control resource allocation framework could be used} to facilitate large-scale distributed machine learning tasks. Therefore, common approaches in resource allocation, scheduling, and computational offloading should be geared \textcolor{black}{toward} the required communication and computation media that such algorithms require. \textcolor{black}{However,}
 resource sharing between these \textcolor{black}{two communications paradigms} is inevitable. Such resource sharing in large-scale \acrshort{haps} network should be discussed. }

\section{Conclusions}\label{sec:conclusion}
This article \textcolor{black}{has aimed} to highlight the unexplored potential of the \acrfull{haps} systems. \textcolor{black}{With} the potential to address the ubiquitous connectivity target of  6G networks,  \acrshort{haps} \textcolor{black}{systems seem}  indispensable in  future deployments. Several prospective use-cases for the near future and beyond \textcolor{black}{have been described above}, along with the technological synergies that will enable the dense use of \acrshort{haps} \textcolor{black}{systems} in terms of mega-constellations. 

\textcolor{black}{Along with the diverse set of applications addressed \textcolor{black}{in these} use-cases, \textcolor{black}{a}  {\acrshort{haps}-mounted \acrfull{smbs}} paradigm \textcolor{black}{was} introduced as a promising and cost-effective solution for addressing the traffic demands of  future networks. \textcolor{black}{As we showed, this}  platform is also capable of supporting computing, caching, and processing in a plethora of application domains\textcolor{black}{} including sensing, machine type communications, UAV communications, and various IoT applications.} 

\textcolor{black}{A wide spectrum of topics \textcolor{black}{were} discussed with a forward looking perspective. The evolution of  HAPS network architecture \textcolor{black}{was} highlighted with a focus on HAPS energy subsystems and the latest technologies introduced for communications payload\textcolor{black}{s}.   The promising technology of passive payloads offered by the \acrfull{rss} \textcolor{black}{was} introduced.  A  detailed review and discussion of the \acrfull{rrm}  and interference management schemes \textcolor{black}{were} reported. Suitable waveform designs and multiple access techniques \textcolor{black}{were} elaborated. The mobility management \textcolor{black}{was} also studied by discussing both inter-\acrshort{haps} and intra-\acrshort{haps} handoff algorithms. The interaction between existing softwarized techniques\textcolor{black}{,} such as network slicing, software defined networks\textcolor{black}{,} and network function virtualization techniques\textcolor{black}{,}  and \acrshort{haps} networks \textcolor{black}{were} detailed. The necessary \acrfull{ai} enablers in future HAPS systems \textcolor{black}{were} also introduced.}

The current literature is expected to evolve, targeting the realization of the proposed visionary framework by addressing the \textcolor{black}{open issues discussed}. The challenges and open issues related to VHetNets and HAPS systems can be categorized into two groups\textcolor{black}{:} one \textcolor{black}{that} mainly covers  next-generation (up to 10 years) challenges\textcolor{black}{;} and the other  next-next-generation (10-20 years) challenges.  The former \textcolor{black}{will require} intensive research but in a more incremental fashion with regards to the current technologies of the communications systems.  As an example, the use of massive MIMO and mmWave communications for  HAPS system\textcolor{black}{s} can be categorized as \textcolor{black}{a} next-generation challenge\textcolor{black}{,} as the required theory and practice  \textcolor{black}{are jointly} investigated for  terrestrial networks.  However, \textcolor{black}{the use of these technologies for HAPS systems will require further investigation.} \textcolor{black}{Future research activities should target} the lack of enough HAPS system channels, the restricted transmission  energy,  or  the  detection  without  availability  of the  channel  statistics/model.

%\begin{figure*}[!t]
%\centering
%\subfloat[Case I]{\includegraphics[width=2.5in]{box}%
%\hfil
%\subfloat[Case II]{\includegraphics[width=2.5in]{box}%
%\label{fig_second_case}}
%\caption{Simulation results for the network.}
%\label{fig_sim}
%\end{figure*}

\section*{Acknowledgment}

 This work was supported by Huawei Canada Co., Ltd.

\ifCLASSOPTIONcaptionsoff
  \newpage
\fi



%\bibliographystyle{IEEEtran}

\bibliographystyle{IEEEtran}
\bibliography{main_final_submission}

\begin{thebibliography}{100}
\providecommand{\url}[1]{#1}
\csname url@samestyle\endcsname
\providecommand{\newblock}{\relax}
\providecommand{\bibinfo}[2]{#2}
\providecommand{\BIBentrySTDinterwordspacing}{\spaceskip=0pt\relax}
\providecommand{\BIBentryALTinterwordstretchfactor}{4}
\providecommand{\BIBentryALTinterwordspacing}{\spaceskip=\fontdimen2\font plus
\BIBentryALTinterwordstretchfactor\fontdimen3\font minus
  \fontdimen4\font\relax}
\providecommand{\BIBforeignlanguage}[2]{{%
\expandafter\ifx\csname l@#1\endcsname\relax
\typeout{** WARNING: IEEEtran.bst: No hyphenation pattern has been}%
\typeout{** loaded for the language `#1'. Using the pattern for}%
\typeout{** the default language instead.}%
\else
\language=\csname l@#1\endcsname
\fi
#2}}
\providecommand{\BIBdecl}{\relax}
\BIBdecl

\bibitem{T338811}
3GPP, ``Study on {New Radio (NR)} to support non-terrestrial networks {V15.4.0
  (Rel. 15)},'' Tech. Rep., 2020.

\bibitem{ITU-RR}
ITU, ``Radio regulations articles,''
  \url{http://www.itu.int/pub/R-REG-RR-2016}, 2016.

\bibitem{cao2018airborne}
X.~Cao, P.~Yang, M.~Alzenad, X.~Xi, D.~Wu, and H.~Yanikomeroglu, ``Airborne
  communication networks: A survey,'' \emph{IEEE Journal on Selected Areas in
  Communications}, vol.~36, no.~9, pp. 1907--1926, Sep. 2018.

\bibitem{tozer2001high}
T.~Tozer and D.~Grace, ``High-altitude platforms for wireless communications,''
  \emph{Electronics \& Communication Engineering Journal}, vol.~13, no.~3, pp.
  127--137, Jun. 2001.

\bibitem{qiu2019air}
J.~Qiu, D.~Grace, G.~Ding, M.~D. Zakaria, and Q.~Wu, ``Air-ground heterogeneous
  networks for {5G} and beyond via integrating high and low altitude
  platforms,'' \emph{IEEE Wireless Communications}, vol.~26, no.~6, pp.
  140--148, Dec. 2019.

\bibitem{hoshino2019study}
K.~Hoshino, S.~Sudo, and Y.~Ohta, ``A study on antenna beamforming method
  considering movement of solar plane in {HAPS} system,'' in \emph{IEEE 90th
  Vehicular Technology Conference (VTC2019-Fall)}, 2019, pp. 1--5.

\bibitem{lTURF1500}
ITU-R, ``Preferred characteristics of systems in the fixed service using high
  altitude platforms operating in the bands 47.2-47.5 {GHz} and 47.9-48.2
  {GHz}.'' International Telecommunication Union, Geneva, Recommendation
  F.1500, Jan. 2000.

\bibitem{milas2003interference}
V.~Milas, M.~Koletta, and P.~Constantinou, ``Interference and compatibility
  studies between satellite service systems and systems using high altitude
  platform stations,'' in \emph{ESA Special Publication}, vol. 541, Jul. 2003.

\bibitem{miura2001wireless}
R.~Miura and M.~Oodo, ``Wireless communications system using stratospheric
  platforms: {R} and {D} program on telecom and broadcasting system using high
  altitude platform stations,'' \emph{JCRL}, vol.~48, no.~4, pp. 33--48, Nov.
  2001.

\bibitem{Handley2019mega}
M.~Handley, ``Using ground relays for low-latency wide-area routing in mega
  constellations,'' in \emph{ACM Workshop on Hot Topics in Networks (HotNets)},
  2019, pp. 125--132.

\bibitem{karapantazis2005broadband}
S.~Karapantazis and F.~Pavlidou, ``Broadband communications via high-altitude
  platforms: A survey,'' \emph{IEEE Communications Surveys \& Tutorials},
  vol.~7, no.~1, pp. 2--31, firstquarter 2005.

\bibitem{widiawan2007high}
A.~K. Widiawan and R.~Tafazolli, ``High altitude platform station ({HAPS}): A
  review of new infrastructure development for future wireless
  communications,'' \emph{Wireless Personal Communications}, vol.~42, no.~3,
  pp. 387--404, Aug. 2007.

\bibitem{gavan2009concepts}
J.~Gavan, S.~Tapuchi, and D.~Grace, ``Concepts and main applications of
  high-altitude-platform radio relays,'' \emph{URSI Radio Science Bulletin},
  vol. 2009, no. 330, pp. 20--31, Sep. 2009.

\bibitem{fidler2010optical}
F.~Fidler, M.~Knapek, J.~Horwath, and W.~R. Leeb, ``Optical communications for
  high-altitude platforms,'' \emph{IEEE Journal of Selected Topics in Quantum
  Electronics}, vol.~16, no.~5, pp. 1058--1070, Aug. 2010.

\bibitem{aubineau2010itu}
P.~Aubineau, S.~Buonomo, W.~Frank, and K.~Hughes, ``{ITU}'s regulatory
  framework, technical studies in {ITU-R}, and future activities in relation to
  high-altitude-platform stations {(HAPS)},'' \emph{URSI Radio Science
  Bulletin}, vol. 2010, no. 333, pp. 67--74, Mar. 2010.

\bibitem{mohammed2011role}
A.~Mohammed, A.~Mehmood, F.-N. Pavlidou, and M.~Mohorcic, ``The role of
  high-altitude platforms ({HAPs}) in the global wireless connectivity,''
  \emph{Proceedings of the IEEE}, vol.~99, no.~11, pp. 1939--1953, Nov. 2011.

\bibitem{d2016high}
F.~A. d’Oliveira, F.~C. L.~d. Melo, and T.~C. Devezas, ``High-altitude
  platforms—present situation and technology trends,'' \emph{Journal of
  Aerospace Technology and Management}, vol.~8, no.~3, pp. 249--262, Jul./Sep.
  2016.

\bibitem{aragon2008high}
A.~Aragon-Zavala, J.~L. Cuevas-Ru{\'\i}z, and J.~A. Delgado-Pen{\'\i}n,
  \emph{{High-Altitude Platforms for Wireless Communications}}.\hskip 1em plus
  0.5em minus 0.4em\relax Wiley Online Library, 2008.

\bibitem{grace2011broadband}
D.~Grace and M.~Mohorcic, \emph{{Broadband Communications via High Altitude
  Platforms}}.\hskip 1em plus 0.5em minus 0.4em\relax John Wiley \& Sons, 2011.

\bibitem{cianca2005integrated}
E.~Cianca, R.~Prasad, M.~De~Sanctis, A.~De~Luise, M.~Antonini, D.~Teotino, and
  M.~Ruggieri, ``Integrated satellite-{HAP} systems,'' \emph{IEEE
  Communications Magazine}, vol.~43, no.~12, pp. 33--39, Dec. 2005.

\bibitem{thornton2001broadband}
J.~Thornton, D.~Grace, C.~Spillard, T.~Konefal, and T.~Tozer, ``Broadband
  communications from a high-altitude platform: The european {HeliNet}
  programme,'' \emph{Electronics \& Communication Engineering Journal},
  vol.~13, no.~3, pp. 138--144, Jun. 2001.

\bibitem{wang2014high}
W.-Q. Wang and H.~Shao, ``High altitude platform multichannel {SAR} for
  wide-area and staring imaging,'' \emph{IEEE Aerospace and Electronic Systems
  Magazine}, vol.~29, no.~5, pp. 12--17, May 2014.

\bibitem{arapoglou2011land}
P.-D. Arapoglou, E.~T. Michailidis, A.~D. Panagopoulos, A.~G. Kanatas, and
  R.~Prieto-Cerdeira, ``The land mobile earth-space channel,'' \emph{IEEE
  Vehicular Technology Magazine}, vol.~6, no.~2, pp. 44--53, Jun. 2011.

\bibitem{kanatas2016radio}
A.~G. Kanatas and A.~D. Panagopoulos, \emph{Radio Wave Propagation and Channel
  Modeling for Earth--Space Systems}.\hskip 1em plus 0.5em minus 0.4em\relax
  CRC Press, 2016.

\bibitem{Mozaffari2019SurveyUAV}
M.~Mozaffari, W.~Saad, M.~Bennis, Y.-H. Nam, and M.~Debbah, ``A tutorial on
  {UAVs} for wireless networks: Applications, challenges, and open problems,''
  \emph{IEEE Communications Surveys \& Tutorials}, vol.~21, no.~3, pp.
  2334--2360, Thirdquarter 2019.

\bibitem{Zhang2019SurveymmWaveUAV}
L.~Zhang, H.~Zhao, S.~Hou, Z.~Zhao, H.~Xu, X.~Wu, Q.~Wu, and R.~Zhang, ``A
  survey on {5G} millimeter wave communications for {UAV}-assisted wireless
  networks,'' \emph{IEEE Access}, vol.~7, pp. 117\,460--117\,504, 2019.

\bibitem{Zeng2019proceedingUAV}
Y.~Zeng, Q.~Wu, and R.~Zhang, ``Accessing from the sky: A tutorial on {UAV}
  communications for {5G} and beyond,'' \emph{Proceedings of the IEEE}, vol.
  107, no.~12, pp. 2327--2375, Dec. 2019.

\bibitem{Haijun2019SurveyCyberUAV}
H.~Wang, H.~Zhao, J.~Zhang, D.~Ma, J.~Li, and J.~Wei, ``Survey on unmanned
  aerial vehicle networks: A cyber physical system perspective,'' \emph{IEEE
  Communications Surveys \& Tutorials}, vol.~22, no.~2, pp. 1027--1070,
  Thirdquarter 2019.

\bibitem{arum2020review}
S.~C. Arum, D.~Grace, and P.~D. Mitchell, ``A review of wireless communication
  using high-altitude platforms for extended coverage and capacity,''
  \emph{Computer Communications}, vol. 157, no.~1, pp. 232--256, May 2020.

\bibitem{bhushan2014network}
N.~Bhushan, J.~Li, D.~Malladi, R.~Gilmore, D.~Brenner, A.~Damnjanovic, R.~T.
  Sukhavasi, C.~Patel, and S.~Geirhofer, ``Network densification: the dominant
  theme for wireless evolution into {5G},'' \emph{IEEE Communications
  Magazine}, vol.~52, no.~2, pp. 82--89, Feb. 2014.

\bibitem{andrews2016we}
J.~G. Andrews, X.~Zhang, G.~D. Durgin, and A.~K. Gupta, ``Are we approaching
  the fundamental limits of wireless network densification?'' \emph{IEEE
  Communications Magazine}, vol.~54, no.~10, pp. 184--190, Oct. 2016.

\bibitem{alam2020high}
M.~S. Alam, G.~K. Kurt, H.~Yanikomeroglu, P.~Zhu, and N.~D{\`a}o, ``High
  altitude platform station based super macro base station {(HAPS-SMBS)}
  constellations,'' \emph{IEEE Communications Magazine}, vol.~59, no.~1, pp.
  103--109, Jan. 2021.

\bibitem{sibiya2019reliable}
S.~Sibiya and O.~O. Olugbara, ``Reliable internet of things network
  architecture based on high altitude platforms,'' in \emph{IEEE Conference on
  Information Communications Technology and Society (ICTAS)}, 2019, pp. 1--4.

\bibitem{gineste2017narrowband}
M.~Gineste, T.~Deleu, M.~Cohen, N.~Chuberre, V.~Saravanan, V.~Frascolla,
  M.~Mueck, E.~C. Strinati, and E.~Dutkiewicz, ``Narrowband {IoT} service
  provision to {5G} user equipment via a satellite component,'' in \emph{IEEE
  Globecom Workshops (GC Wkshps)}, 2017, pp. 1--4.

\bibitem{abu2019performance}
A.~A. Abu-Arabia, R.~Hakimi \emph{et~al.}, ``Performance of {5G} services
  deployed via {HAPS} system,'' in \emph{IEEE 13th International Conference on
  Telecommunication Systems, Services, and Applications (TSSA)}, 2019, pp.
  168--172.

\bibitem{dahrouj2015cost}
H.~Dahrouj, A.~Douik, F.~Rayal, T.~Y. Al-Naffouri, and M.-S. Alouini,
  ``Cost-effective hybrid {RF/FSO} backhaul solution for next generation
  wireless systems,'' \emph{IEEE Wireless Communications}, vol.~22, no.~5, pp.
  98--104, Oct. 2015.

\bibitem{taori2015point}
R.~Taori and A.~Sridharan, ``Point-to-multipoint in-band {mmWave} backhaul for
  {5G} networks,'' \emph{IEEE Communications Magazine}, vol.~53, no.~1, pp.
  195--201, Jan. 2015.

\bibitem{Yuan19TropicalLoss}
F.~Yuan, Y.~H. Lee, Y.~S. Meng, S.~Manandhar, and J.~T. Ong, ``High-resolution
  {ITU-R} cloud attenuation model for satellite communications in tropical
  region,'' \emph{IEEE Transactions on Antennas and Propagation}, vol.~67,
  no.~9, pp. 6115--6122, Sep. 2019.

\bibitem{alzenad2018fso}
M.~Alzenad, M.~Z. Shakir, H.~Yanikomeroglu, and M.-S. Alouini, ``{FSO}-based
  vertical backhaul/fronthaul framework for {5G+} wireless networks,''
  \emph{IEEE Communications Magazine}, vol.~56, no.~1, pp. 218--224, Jan. 2018.

\bibitem{kaushal2016optical}
H.~Kaushal and G.~Kaddoum, ``Optical communication in space: Challenges and
  mitigation techniques,'' \emph{IEEE Communications Surveys \& Tutorials},
  vol.~19, no.~1, pp. 57--96, Firstquarter 2016.

\bibitem{alsharoa2019facilitating}
A.~Alsharoa and M.-S. Alouini, ``Facilitating satellite-airborne-terrestrial
  integration for dynamic and infrastructure-less networks,'' \emph{arXiv
  preprint arXiv:1912.03819}, 2019.

\bibitem{pham2015hybrid}
A.~T. Pham, P.~V. Trinh, V.~V. Mai, N.~T. Dang, and C.-T. Truong, ``Hybrid
  free-space optics/millimeter-wave architecture for {5G} cellular backhaul
  networks,'' in \emph{IEEE Opto-Electronics and Communications Conference
  (OECC)}, 2015, pp. 1--3.

\bibitem{bor2016new}
I.~Bor-Yaliniz and H.~Yanikomeroglu, ``The new frontier in {RAN} heterogeneity:
  Multi-tier drone-cells,'' \emph{IEEE Communications Magazine}, vol.~54,
  no.~11, pp. 48--55, Nov. 2016.

\bibitem{mirahsan2017hethetnets}
M.~{Mirahsan}, R.~{Schoenen}, and H.~{Yanikomeroglu}, ``{HetHetNets:
  Heterogeneous} traffic distribution in heterogeneous wireless cellular
  networks,'' \emph{IEEE Journal on Selected Areas in Communications}, vol.~33,
  no.~10, pp. 2252--2265, Oct. 2015.

\bibitem{alzenad2018coverage}
M.~Alzenad and H.~Yanikomeroglu, ``Coverage and rate analysis for unmanned
  aerial vehicle base stations with {LoS/NLoS} propagation,'' in \emph{IEEE
  Globecom Workshops (GC Wkshps)}, 2018, pp. 1--7.

\bibitem{kumar2013survey}
K.~Kumar, J.~Liu, Y.-H. Lu, and B.~Bhargava, ``A survey of computation
  offloading for mobile systems,'' \emph{Mobile networks and Applications},
  vol.~18, no.~1, pp. 129--140, Apr. 2013.

\bibitem{wang2017computation}
C.~Wang, C.~Liang, F.~R. Yu, Q.~Chen, and L.~Tang, ``Computation offloading and
  resource allocation in wireless cellular networks with mobile edge
  computing,'' \emph{IEEE Transactions on Wireless Communications}, vol.~16,
  no.~8, pp. 4924--4938, Aug. 2017.

\bibitem{NWS}
\BIBentryALTinterwordspacing
``{NWS} jetstream - layers of the atmosphere.'' [Online]. Available:
  \url{www.weather.gov}
\BIBentrySTDinterwordspacing

\bibitem{bai2014analysis}
T.~Bai, R.~Vaze, and R.~W. Heath, ``Analysis of blockage effects on urban
  cellular networks,'' \emph{IEEE Transactions on Wireless Communications},
  vol.~13, no.~9, pp. 5070--5083, Sep. 2014.

\bibitem{halbauer20133d}
H.~Halbauer, S.~Saur, J.~Koppenborg, and C.~Hoek, ``{3D} beamforming:
  Performance improvement for cellular networks,'' \emph{Bell Labs Technical
  Journal}, vol.~18, no.~2, pp. 37--56, Sep. 2013.

\bibitem{cao2018mobile}
X.~Cao, J.~Xu, and R.~Zhangt, ``Mobile edge computing for cellular-connected
  {UAV}: Computation offloading and trajectory optimization,'' in \emph{IEEE
  19th International Workshop on Signal Processing Advances in Wireless
  Communications (SPAWC)}, 2018, pp. 1--5.

\bibitem{messous2017computation}
M.-A. Messous, H.~Sedjelmaci, N.~Houari, and S.-M. Senouci, ``Computation
  offloading game for an {UAV} network in mobile edge computing,'' in
  \emph{IEEE International Conference on Communications (ICC)}, 2017, pp. 1--6.

\bibitem{nikitas2019examining}
A.~Nikitas, E.~T. Njoya, and S.~Dani, ``Examining the myths of connected and
  autonomous vehicles: analysing the pathway to a driverless mobility
  paradigm,'' \emph{International Journal of Automotive Technology and
  Management}, vol.~19, no. 1-2, pp. 10--30, Mar. 2019.

\bibitem{talebpour2016influence}
A.~Talebpour and H.~S. Mahmassani, ``Influence of connected and autonomous
  vehicles on traffic flow stability and throughput,'' \emph{Transportation
  Research Part C: Emerging Technologies}, vol.~71, pp. 143--163, Oct. 2016.

\bibitem{zhang2011data}
J.~Zhang, F.-Y. Wang, K.~Wang, W.-H. Lin, X.~Xu, and C.~Chen, ``Data-driven
  intelligent transportation systems: A survey,'' \emph{IEEE Transactions on
  Intelligent Transportation Systems}, vol.~12, no.~4, pp. 1624--1639, Dec.
  2011.

\bibitem{wang2011real}
J.~Wang, J.~Cho, S.~Lee, and T.~Ma, ``Real time services for future cloud
  computing enabled vehicle networks,'' in \emph{IEEE International Conference
  on Wireless Communications and Signal Processing (WCSP)}, 2011, pp. 1--5.

\bibitem{stojmenovic2014fog}
I.~Stojmenovic and S.~Wen, ``The fog computing paradigm: Scenarios and security
  issues,'' in \emph{IEEE Federated Conference on Computer Science and
  Information Systems}, 2014, pp. 1--4.

\bibitem{stanczak2018enhanced}
J.~Stanczak, D.~Kozio{\l}, I.~Z. Kov{\'a}cs, J.~Wigard, M.~Wimmer, and
  R.~Amorim, ``Enhanced unmanned aerial vehicle communication support in
  {LTE}-advanced,'' in \emph{IEEE Conference on Standards for Communications
  and Networking (CSCN)}, 2018, pp. 1--6.

\bibitem{bamburry2015drones}
D.~Bamburry, ``Drones: Designed for product delivery,'' \emph{Design Management
  Review}, vol.~26, no.~1, pp. 40--48, Jul. 2015.

\bibitem{sarddar2011handover}
D.~Sarddar, J.~Banerjee, S.~Chatterjee, P.~Ghosh, S.~Chakraborty, K.~Hui, and
  M.~K. Naskar, ``A handover management in {LEO} satellite network using
  angular and distance based algorithm,'' \emph{International Journal of
  Computer Applications}, vol. 975, p. 8887, Oct. 2011.

\bibitem{ITU-1569}
ITU, ``Recommendation {ITU-R F.1569},'' 2002.

\bibitem{yuniarti2018regulatory}
D.~Yuniarti, ``Regulatory challenges of broadband communication services from
  high altitude platforms ({HAPs}),'' in \emph{IEEE International Conference on
  Information and Communications Technology (ICOIACT)}, 2018, pp. 919--922.

\bibitem{ICAO-UA}
ICAO, ``Draft regulatory guidance for unmanned aircraft operations – general
  regulations, {APUAS/TF/3 – WP/08},'' 2019.

\bibitem{liu2019regulating}
H.~Liu and F.~Tronchetti, ``Regulating near-space activities: Using the
  precedent of the exclusive economic zone as a model?'' \emph{Ocean
  Development \& International Law}, vol.~50, no. 2-3, pp. 91--116, Feb. 2019.

\bibitem{ITU-2471}
ITU, ``Report {ITU-R F.2471-0},'' 2019.

\bibitem{ITU-2472}
------, ``Report {ITU-R F.2472-0},'' 2019.

\bibitem{ITU-2475}
------, ``Report {ITU-R F.2475-0},'' 2019.

\bibitem{colella2000halo}
M.~Colella, J.~N. Martin, and F.~Akyildiz, ``The {HALO} network/sup {TM},''
  \emph{IEEE Communications Magazine}, vol.~38, no.~6, pp. 142--148, Jun. 2000.

\bibitem{ITU-WRC19}
ITU, ``{World Radio communication Conference 2019 ({WRC}-19) Final Acts},''
  2019.

\bibitem{dong2016constellation}
F.~Dong, H.~Han, X.~Gong, J.~Wang, and H.~Li, ``A constellation design
  methodology based on {QoS} and user demand in high-altitude platform
  broadband networks,'' \emph{IEEE Transactions on Multimedia}, vol.~18,
  no.~12, pp. 2384--2397, Dec. 2016.

\bibitem{grace2005improving}
D.~Grace, J.~Thornton, G.~Chen, G.~P. White, and T.~C. Tozer, ``Improving the
  system capacity of broadband services using multiple high-altitude
  platforms,'' \emph{IEEE Transactions on Wireless Communications}, vol.~4,
  no.~2, pp. 700--709, Mar. 2005.

\bibitem{lin2019robust}
Z.~Lin, M.~Lin, Y.~Huang, T.~de~Cola, and W.-P. Zhu, ``Robust multi-objective
  beamforming for integrated satellite and high altitude platform network with
  imperfect channel state information,'' \emph{IEEE Transactions on Signal
  Processing}, vol.~67, no.~24, pp. 6384--6396, Dec. 2019.

\bibitem{ehrenfried2014stratonauts}
M.~Ehrenfried \emph{et~al.}, \emph{Stratonauts: Pioneers venturing into the
  stratosphere}.\hskip 1em plus 0.5em minus 0.4em\relax Springer, 2014.

\bibitem{Loon}
\BIBentryALTinterwordspacing
``{Connecting People Everywhere}.'' [Online]. Available:
  \url{https://www.loon.com/}
\BIBentrySTDinterwordspacing

\bibitem{SHARP}
\BIBentryALTinterwordspacing
``{SHARP (Stationary High Altitude Relay Platform)}.'' [Online]. Available:
  \url{http://www.friendsofcrc.ca/Projects/SHARP/sharp.html}
\BIBentrySTDinterwordspacing

\bibitem{AV}
\BIBentryALTinterwordspacing
``Aerovironment, inc.'' [Online]. Available:
  \url{https://www.avinc.com/resources/press-releases/view/solar-high-altitude-long-endurance-uas}
\BIBentrySTDinterwordspacing

\bibitem{CAPANINA}
\BIBentryALTinterwordspacing
``{CAPANINA Test Results Summary Report}.'' [Online]. Available:
  \url{https://www.capanina.org/documents/CAP-D22a-WP44-CGS-PUB-01.pdf}
\BIBentrySTDinterwordspacing

\bibitem{StratXX}
\BIBentryALTinterwordspacing
``{StratXX}.'' [Online]. Available: \url{http://www.stratxx.com/}
\BIBentrySTDinterwordspacing

\bibitem{Elevate}
\BIBentryALTinterwordspacing
``Elevate.'' [Online]. Available: \url{http://www.zero2infinity.space/elevate/}
\BIBentrySTDinterwordspacing

\bibitem{Zephyr}
\BIBentryALTinterwordspacing
``{Zephyr Pioneering the Stratosphere}.'' [Online]. Available:
  \url{https://www.airbus.com/defence/uav/zephyr.html}
\BIBentrySTDinterwordspacing

\bibitem{aquila}
\BIBentryALTinterwordspacing
``{Flying Aquila: Early lessons from the first full-scale test flight and the
  path ahead}.'' [Online]. Available:
  \url{https://engineering.fb.com/connectivity/flying-aquila-early-lessons-from-the-first-full-scale-test-flight-and-the-path-ahead/}
\BIBentrySTDinterwordspacing

\bibitem{Stratobus}
\BIBentryALTinterwordspacing
``What's up with stratobus?'' [Online]. Available:
  \url{https://www.thalesgroup.com/en/worldwide/space/news/whats-stratobus}
\BIBentrySTDinterwordspacing

\bibitem{HAPSMobile}
\BIBentryALTinterwordspacing
``{HAPSMobile}.'' [Online]. Available: \url{https://www.hapsmobile.com/}
\BIBentrySTDinterwordspacing

\bibitem{prismatic}
\BIBentryALTinterwordspacing
``phasa-35.'' [Online]. Available:
  \url{http://prismaticltd.co.uk/products/phasa-35/}
\BIBentrySTDinterwordspacing

\bibitem{ABE}
\BIBentryALTinterwordspacing
``phasa-35 first flight.'' [Online]. Available:
  \url{https://www.baesystems.com/en/article/ground-breaking-solar-powered-unmanned-aircraft-makes-first-flight}
\BIBentrySTDinterwordspacing

\bibitem{schlesak1988microwave}
J.~J. Schlesak, A.~Alden, and T.~Ohno, ``A microwave powered high altitude
  platform,'' in \emph{IEEE MTT-S International Microwave Symposium Digest},
  1988, pp. 283--286.

\bibitem{brown1986microwaver}
W.~C. Brown, ``A microwaver powered, long duration, high altitude platform,''
  in \emph{IEEE MTT-S International Microwave Symposium Digest}, 1986, pp.
  507--510.

\bibitem{gao2013energy}
X.-Z. Gao, Z.-X. Hou, Z.~Guo, J.-X. Liu, and X.-Q. Chen, ``Energy management
  strategy for solar-powered high-altitude long-endurance aircraft,''
  \emph{Energy Conversion and Management}, vol.~70, pp. 20--30, Jun. 2013.

\bibitem{phillips1980some}
W.~Phillips, ``Some design considerations for solar-powered aircraft,''
  National Aeronautics and Space Administration (NASA), Hampton, VA (USA),
  Tech. Rep., 1980.

\bibitem{alsahlani2017design}
A.~Alsahlani, L.~Johnston, P.~Atcliffe \emph{et~al.}, ``Design of a high
  altitude long endurance flying-wing solar-powered unmanned air vehicle,''
  \emph{Progress in Flight Physics}, vol.~9, pp. 3--24, Oct. 2017.

\bibitem{cestino2006design}
E.~Cestino, ``Design of solar high altitude long endurance aircraft for multi
  payload \& operations,'' \emph{Aerospace Science and Technology}, vol.~10,
  no.~6, pp. 541--550, Sep. 2006.

\bibitem{yoshikawa2017exceeding}
K.~Yoshikawa, W.~Yoshida, T.~Irie, H.~Kawasaki, K.~Konishi, H.~Ishibashi,
  T.~Asatani, D.~Adachi, M.~Kanematsu, H.~Uzu \emph{et~al.}, ``Exceeding
  conversion efficiency of 26\% by heterojunction interdigitated back contact
  solar cell with thin film {Si} technology,'' \emph{Solar Energy Materials and
  Solar Cells}, vol. 173, pp. 37--42, Dec. 2017.

\bibitem{green2017energy}
M.~A. Green and S.~P. Bremner, ``Energy conversion approaches and materials for
  high-efficiency photovoltaics,'' \emph{Nature Materials}, vol.~16, no.~1, pp.
  23--34, 2017.

\bibitem{zheng2016photonic}
X.~Zheng and L.~Zhang, ``Photonic nanostructures for solar energy conversion,''
  \emph{Energy \& Environmental Science}, vol.~9, no.~8, pp. 2511--2532, Sep.
  2016.

\bibitem{microLink}
\BIBentryALTinterwordspacing
``{MicroLink Devices}.'' [Online]. Available:
  \url{http://www.mldevices.com/index.php/news}
\BIBentrySTDinterwordspacing

\bibitem{zubi2018lithium}
G.~Zubi, R.~Dufo-L{\'o}pez, M.~Carvalho, and G.~Pasaoglu, ``The lithium-ion
  battery: State of the art and future perspectives,'' \emph{Renewable and
  Sustainable Energy Reviews}, vol.~89, pp. 292--308, Jun. 2018.

\bibitem{thomas2009fuel}
C.~Thomas, ``Fuel cell and battery electric vehicles compared,''
  \emph{International Journal of Hydrogen Energy}, vol.~34, no.~15, pp.
  6005--6020, Aug. 2009.

\bibitem{wilberforce2017developments}
T.~Wilberforce, Z.~El-Hassan, F.~Khatib, A.~Al~Makky, A.~Baroutaji, J.~G.
  Carton, and A.~G. Olabi, ``Developments of electric cars and fuel cell
  hydrogen electric cars,'' \emph{International Journal of Hydrogen Energy},
  vol.~42, no.~40, pp. 25\,695--25\,734, Oct. 2017.

\bibitem{latva2019key}
M.~Latva-aho and K.~Lepp{\"a}nen, ``Key drivers and research challenges for
  {6G} ubiquitous wireless intelligence,'' in \emph{White Paper}.\hskip 1em
  plus 0.5em minus 0.4em\relax University of Oulu, Sep. 2019.

\bibitem{di2019smart}
M.~Di~Renzo, M.~Debbah, D.-T. Phan-Huy, A.~Zappone, M.-S. Alouini, C.~Yuen,
  V.~Sciancalepore, G.~C. Alexandropoulos, J.~Hoydis, H.~Gacanin \emph{et~al.},
  ``Smart radio environments empowered by reconfigurable {AI} meta-surfaces: An
  idea whose time has come,'' \emph{EURASIP Journal on Wireless Communications
  and Networking}, vol. 2019, no.~1, pp. 1--20, May 2019.

\bibitem{basar2019wireless}
E.~Basar, M.~Di~Renzo, J.~De~Rosny, M.~Debbah, M.-S. Alouini, and R.~Zhang,
  ``Wireless communications through reconfigurable intelligent surfaces,''
  \emph{IEEE Access}, vol.~7, pp. 116\,753--116\,773, Aug. 2019.

\bibitem{wu2019intelligent}
Q.~Wu and R.~Zhang, ``Intelligent reflecting surface enhanced wireless network
  via joint active and passive beamforming,'' \emph{IEEE Transactions on
  Wireless Communications}, vol.~18, no.~11, pp. 5394--5409, Nov. 2019.

\bibitem{han2019large}
Y.~Han, W.~Tang, S.~Jin, C.-K. Wen, and X.~Ma, ``Large intelligent
  surface-assisted wireless communication exploiting statistical {CSI},''
  \emph{IEEE Transactions on Vehicular Technology}, vol.~68, no.~8, pp.
  8238--8242, Aug. 2019.

\bibitem{liaskos2018new}
C.~Liaskos, S.~Nie, A.~Tsioliaridou, A.~Pitsillides, S.~Ioannidis, and
  I.~Akyildiz, ``A new wireless communication paradigm through
  software-controlled metasurfaces,'' \emph{IEEE Communications Magazine},
  vol.~56, no.~9, pp. 162--169, Sep. 2018.

\bibitem{cui2014coding}
T.~J. Cui, M.~Q. Qi, X.~Wan, J.~Zhao, and Q.~Cheng, ``Coding metamaterials,
  digital metamaterials and programmable metamaterials,'' \emph{Light: Science
  \& Applications}, vol.~3, no.~10, pp. e218--e218, Oct. 2014.

\bibitem{perez2013design}
G.~Perez-Palomino, P.~Baine, R.~Dickie, M.~Bain, J.~A. Encinar, R.~Cahill,
  M.~Barba, and G.~Toso, ``Design and experimental validation of liquid
  crystal-based reconfigurable reflectarray elements with improved bandwidth in
  {F}-band,'' \emph{IEEE Transactions on Antennas and Propagation}, vol.~61,
  no.~4, pp. 1704--1713, Apr. 2013.

\bibitem{carrasco2013tunable}
E.~Carrasco, M.~Tamagnone, and J.~Perruisseau-Carrier, ``Tunable graphene
  reflective cells for {THz} reflectarrays and generalized law of reflection,''
  \emph{Applied Physics Letters}, vol. 102, no.~10, pp. 1--5, 2013.

\bibitem{carrasco2013reflectarray}
E.~Carrasco and J.~Perruisseau-Carrier, ``Reflectarray antenna at terahertz
  using graphene,'' \emph{IEEE Antennas and Wireless Propagation Letters},
  vol.~12, pp. 253--256, Feb. 2013.

\bibitem{kowerdziej2014terahertz}
R.~Kowerdziej, M.~Olifierczuk, J.~Parka, and J.~Wrobel, ``Terahertz
  characterization of tunable metamaterial based on electrically controlled
  nematic liquid crystal,'' \emph{Applied Physics Letters}, vol. 105, no.~2,
  pp. 1--5, 2014.

\bibitem{liu2017ultra}
Z.~Liu and B.~Bai, ``Ultra-thin and high-efficiency graphene metasurface for
  tunable terahertz wave manipulation,'' \emph{Optics Express}, vol.~25, no.~8,
  pp. 8584--8592, Apr. 2017.

\bibitem{DOCOMO}
\BIBentryALTinterwordspacing
``{NTT DOCOMO} and metawave announce successful demonstration of {28GHz}-band
  {5G} using world's first meta-structure technology.'' [Online]. Available:
  \url{https://bwnews.pr/2IhH8cc}
\BIBentrySTDinterwordspacing

\bibitem{greenerwave}
\BIBentryALTinterwordspacing
``Greenerwave.'' [Online]. Available: \url{http://greenerwave.com/}
\BIBentrySTDinterwordspacing

\bibitem{long2020reflections}
H.~Long, M.~Chen, Z.~Yang, B.~Wang, Z.~Li, X.~Yun, and M.~Shikh-Bahaei,
  ``Reflections in the sky: Joint trajectory and passive beamforming design for
  secure {UAV} networks with reconfigurable intelligent surface,'' \emph{arXiv
  preprint arXiv:2005.10559}, 2020.

\bibitem{alfattani2020aerial}
S.~Alfattani, W.~Jaafar, Y.~Hmamouche, H.~Yanikomeroglu, A.~Yonga{\c{c}}oglu,
  N.~{\DJ}{\`a}o, and P.~Zhu, ``Aerial platforms with reconfigurable smart
  surfaces for {5G} and beyond,'' \emph{IEEE Communications Magazine}, vol.~59,
  no.~1, pp. 96--102, Jan. 2021.

\bibitem{huang2018energy}
C.~Huang, G.~C. Alexandropoulos, A.~Zappone, M.~Debbah, and C.~Yuen, ``Energy
  efficient multi-user {MISO} communication using low resolution large
  intelligent surfaces,'' in \emph{IEEE Globecom Workshops (GC Wkshps)}, 2018,
  pp. 1--6.

\bibitem{ntontin2019reconfigurable}
K.~Ntontin, M.~Di~Renzo, J.~Song, F.~Lazarakis, J.~de~Rosny, D.-T. Phan-Huy,
  O.~Simeone, R.~Zhang, M.~Debbah, G.~Lerosey \emph{et~al.}, ``Reconfigurable
  intelligent surfaces vs. relaying: {Differences,} similarities, and
  performance comparison,'' \emph{IEEE Open Journal of the Communications
  Society}, vol.~1, pp. 798--807, 2020.

\bibitem{tang2019wireless}
W.~Tang, M.~Z. Chen, X.~Chen, J.~Y. Dai, Y.~Han, M.~Di~Renzo, Y.~Zeng, S.~Jin,
  Q.~Cheng, and T.~J. Cui, ``Wireless communications with reconfigurable
  intelligent surface: {Path} loss modeling and experimental measurement,''
  \emph{IEEE Trans. on Wireless Commu.}, vol.~20, no.~1, pp. 421--439, Jan.
  2021.

\bibitem{khawaja2019survey}
W.~Khawaja, I.~Guvenc, D.~W. Matolak, U.-C. Fiebig, and N.~Schneckenburger, ``A
  survey of air-to-ground propagation channel modeling for unmanned aerial
  vehicles,'' \emph{IEEE Communications Surveys \& Tutorials}, vol.~21, no.~3,
  pp. 2361--2391, Thirdquarter 2019.

\bibitem{lutz1991land}
E.~Lutz, D.~Cygan, M.~Dippold, F.~Dolainsky, and W.~Papke, ``The land mobile
  satellite communication channel-recording, statistics, and channel model,''
  \emph{IEEE Transactions on Vehicular Technology}, vol.~40, no.~2, pp.
  375--386, May 1991.

\bibitem{vazquez2002channel}
M.~V{\'a}zquez-Castro, F.~P{\'e}rez-Font{\'a}n, and B.~Arbesser-Rastburg,
  ``Channel modeling for satellite and {HAPS} system design,'' \emph{Wireless
  Communications and Mobile Computing}, vol.~2, no.~3, pp. 285--300, May 2002.

\bibitem{oestges2001coverage}
C.~Oestges and D.~Vanhoenacker-Janvier, ``Coverage modelling of high-altitude
  platforms communication systems,'' \emph{Electronics Letters}, vol.~37,
  no.~2, pp. 119--121, Jan. 2001.

\bibitem{dovis2002small}
F.~Dovis, R.~Fantini, M.~Mondin, and P.~Savi, ``Small-scale fading for
  high-altitude platform {(HAP)} propagation channels,'' \emph{IEEE Journal on
  Selected Areas in Communications}, vol.~20, no.~3, pp. 641--647, Apr. 2002.

\bibitem{ITU-681}
ITU, ``Recommendation {ITU-R P.681-11},'' 2019.

\bibitem{cuevas2004channel}
J.~L. Cuevas-Ru{\'\i}z and J.~A. Delgado-Penin, ``Channel model based on
  semi-{Markovian} processes. an approach for {HAPS} systems,'' in \emph{IEEE
  14th International Conference on Electronics, Communications and Computers
  (CONIELECOMP)}, 2004, pp. 52--56.

\bibitem{cuevas2004statistical}
J.~L. Cuevas-Ruiz and J.~A. Delgado-Penin, ``A statistical switched broadband
  channel model for {HAPS} links,'' in \emph{IEEE Wireless Communications and
  Networking Conference}, 2004, pp. 290--294.

\bibitem{king2005physical}
P.~R. King, B.~G. Evans, and S.~Stavrou, ``Physical-statistical model for the
  land mobile-satellite channel applied to satellite/{HAP-MIMO},'' in
  \emph{11th VDE European Wireless Conference--Next Generation wireless and
  Mobile Communications and Services}, 2005, pp. 1--5.

\bibitem{michailidis2008spatially}
E.~T. Michailidis, G.~Efthymoglou, and A.~G. Kanatas, ``Spatially correlated
  {3-D HAP-MIMO} fading channels,'' in \emph{IEEE Globecom Workshops}, 2008,
  pp. 1--7.

\bibitem{michailidis2010capacity}
E.~T. Michailidis and A.~G. Kanatas, ``On the capacity of {3-D} space-time
  correlated {HAP-MIMO} channels,'' in \emph{IEEE Second International
  Conference on Advances in Satellite and Space Communications}, 2010, pp.
  87--92.

\bibitem{michailidis2010three}
------, ``Three-dimensional {HAP-MIMO} channels: Modeling and analysis of
  space-time correlation,'' \emph{IEEE Transactions on Vehicular Technology},
  vol.~59, no.~5, pp. 2232--2242, Jun. 2010.

\bibitem{michailidis2012statistical}
------, ``Statistical simulation modeling of {3-D HAP-MIMO} channels,''
  \emph{Wireless Personal Communications}, vol.~65, no.~4, pp. 833--841, Apr.
  2012.

\bibitem{michailidis2014wideband}
------, ``Wideband {HAP-MIMO} channels: A {3-D} modeling and simulation
  approach,'' \emph{Wireless Personal Communications}, vol.~74, no.~2, pp.
  639--664, Jul. 2014.

\bibitem{michailidis2012three}
E.~T. Michailidis, P.~Theofilakos, and A.~G. Kanatas, ``Three-dimensional
  modeling and simulation of {MIMO} mobile-to-mobile via stratospheric relay
  fading channels,'' \emph{IEEE Transactions on Vehicular Technology}, vol.~62,
  no.~5, pp. 2014--2030, Jun. 2012.

\bibitem{mendoza2019application}
H.~A. Mendoza and G.~Corral-Briones, ``On the application of three dimensional
  {HAP} {MIMO} model in {UAV} environment,'' \emph{Vehicular Communications},
  vol.~16, pp. 72--84, Apr. 2019.

\bibitem{nikolaidis2016dual}
V.~Nikolaidis, N.~Moraitis, and A.~G. Kanatas, ``Dual-polarized narrowband
  {MIMO LMS} channel measurements in urban environments,'' \emph{IEEE
  Transactions on Antennas and Propagation}, vol.~65, no.~2, pp. 763--774, Feb.
  2016.

\bibitem{lian2018non}
Z.~Lian, L.~Jiang, C.~He, and D.~He, ``A non-stationary {3-D} wideband {GBSM}
  for {HAP-MIMO} communication systems,'' \emph{IEEE Transactions on Vehicular
  Technology}, vol.~68, no.~2, pp. 1128--1139, Feb. 2019.

\bibitem{yang2017statistical}
M.~Yang, S.~Zhang, X.~Shao, Q.~Guo, and W.~Tang, ``Statistical modeling of the
  high altitude platform dual-polarized {MIMO} propagation channel,''
  \emph{China Communications}, vol.~14, no.~3, pp. 43--54, Mar. 2017.

\bibitem{lian20193}
Z.~Lian, L.~Jiang, C.~He, and D.~He, ``A {3-D} multiuser {HAP-MIMO} channel
  model based on dynamic evolution of {LOS} components,'' \emph{Chinese Journal
  of Electronics}, vol.~28, no.~3, pp. 465--450, May 2019.

\bibitem{ma2019wideband}
Z.~Ma, B.~Ai, R.~He, G.~Wang, Y.~Niu, and Z.~Zhong, ``A wideband non-stationary
  air-to-air channel model for {UAV} communications,'' \emph{IEEE Transactions
  on Vehicular Technology}, vol.~69, no.~2, pp. 1214--1226, Feb. 2019.

\bibitem{giggenbach2002stratospheric}
D.~Giggenbach, R.~Purvinskis, M.~Werner, and M.~Holzbock, ``Stratospheric
  optical inter-platform links for high altitude platforms,'' in \emph{20th
  AIAA International Communication Satellite Systems Conference and Exhibit},
  2002, p. 1910.

\bibitem{andrews2005laser}
L.~C. Andrews and R.~L. Phillips, \emph{{Laser Beam Propagation Through Random
  Media}}.\hskip 1em plus 0.5em minus 0.4em\relax SPIE Press, 2005.

\bibitem{al2001mathematical}
A.~Al-Habash, L.~C. Andrews, and R.~L. Phillips, ``Mathematical model for the
  irradiance probability density function of a laser beam propagating through
  turbulent media,'' \emph{Optical Engineering}, vol.~40, no.~8, pp.
  1554--1563, Aug. 2001.

\bibitem{majumdar2005free}
A.~K. Majumdar, ``Free-space laser communication performance in the atmospheric
  channel,'' \emph{Journal of Optical and Fiber Communications Reports},
  vol.~2, no.~4, pp. 345--396, Nov. 2005.

\bibitem{palma2010wimax}
I.~R. Palma-L{\'a}zgare and J.~A. Delgado-Pen{\'\i}n, ``{WiMAX HAPS}-based
  downlink performance employing geometrical and statistical
  propagation-channel characteristics,'' \emph{URSI Radio Science Bulletin},
  vol. 2010, no. 333, pp. 50--66, Mar. 2010.

\bibitem{zakia2017capacity}
I.~Zakia, ``Capacity of {HAP-MIMO} channels for high-speed train
  communications,'' in \emph{IEEE 3rd International Conference on Wireless and
  Telematics (ICWT)}, 2017, pp. 26--30.

\bibitem{Sudheesh2013CSIimperfectMIMO}
P.~G. Sudheesh, N.~Sharma, M.~Magarini, and P.~Muthuchidambaranathan, ``Effect
  of imperfect {CSI} on interference alignment in multiple-high altitude
  platforms based communication,'' \emph{Physical Communication}, vol.~29, pp.
  336--342, Aug. 2017.

\bibitem{thornton2003optimizing}
J.~Thornton, D.~Grace, M.~H. Capstick, and T.~C. Tozer, ``Optimizing an array
  of antennas for cellular coverage from a high altitude platform,'' \emph{IEEE
  Transactions on Wireless Communications}, vol.~2, no.~3, pp. 484--492, May
  2003.

\bibitem{datsikas2010serial}
C.~K. Datsikas, K.~P. Peppas, N.~C. Sagias, and G.~S. Tombras, ``Serial
  free-space optical relaying communications over gamma-gamma atmospheric
  turbulence channels,'' \emph{IEEE/OSA Journal of Optical Communications and
  Networking}, vol.~2, no.~8, pp. 576--586, Aug. 2010.

\bibitem{sharma2016high}
M.~Sharma, D.~Chadha, and V.~Chandra, ``High-altitude platform for free-space
  optical communication: Performance evaluation and reliability analysis,''
  \emph{IEEE/OSA Journal of Optical Communications and Networking}, vol.~8,
  no.~8, pp. 600--609, Aug. 2016.

\bibitem{salhab2016new}
A.~M. Salhab, ``A new scenario of triple-hop mixed {RF/FSO/RF} relay network
  with generalized order user scheduling and power allocation,'' \emph{EURASIP
  Journal on Wireless Communications and Networking}, vol. 2016, no.~1, pp.
  1--20, Oct. 2016.

\bibitem{michailidis2018outage}
E.~T. Michailidis, N.~Nomikos, P.~Bithas, D.~Vouyioukas, and A.~G. Kanatas,
  ``Outage probability of triple-hop mixed {RF/FSO/RF} stratospheric
  communication systems,'' in \emph{IEEE 10th International Conference on
  Advances in Satellite and Space Communications (SPACOMM)}, 2018, pp. 1--6.

\bibitem{alsharoa2019improvement}
A.~{Alsharoa} and M.~S. {Alouini}, ``Improvement of the global connectivity
  using integrated satellite-airborne-terrestrial networks with resource
  optimization,'' \emph{IEEE Transactions on Wireless Communications}, vol.~19,
  no.~8, pp. 5088--5100, 2020.

\bibitem{abrardo2003centralized}
A.~Abrardo and D.~Sennati, ``Centralized radio resource management strategies
  with heterogeneous traffics in {HAPS WCMA} cellular systems,'' \emph{IEICE
  Transactions on Communications}, vol.~86, no.~3, pp. 1040--1049, Mar. 2003.

\bibitem{foo2002centralized1}
Y.~C. Foo, W.~L. Lim, and R.~Tafazolli, ``Centralized total received power
  based call admission control for high altitude platform station {UMTS},'' in
  \emph{13th IEEE International Symposium on Personal, Indoor and Mobile Radio
  Communications}, 2002, pp. 1596--1600.

\bibitem{foo2002centralized2}
------, ``Centralized downlink call admission control for high altitude
  platform station {UMTS} with onboard power resource sharing,'' in \emph{IEEE
  56th Vehicular Technology Conference}, 2002, pp. 549--553.

\bibitem{foo2004call}
------, ``Call admission control schemes for high altitude platform station and
  terrestrial tower-based hierarchical {UMTS},'' in \emph{IEEE 9th
  International Conference on Communications Systems (ICCS)}, 2004, pp.
  531--536.

\bibitem{araniti2007multicast}
G.~Araniti, A.~Molinaro, and A.~Iera, ``Multicast in terrestrial-{HAP} systems:
  User number vs. user distribution oriented {RRM} policies,'' in \emph{IEEE
  66th Vehicular Technology Conference}, 2007, pp. 154--158.

\bibitem{adachi2005broadband}
F.~Adachi, D.~Garg, S.~Takaoka, and K.~Takeda, ``Broadband {CDMA} techniques,''
  \emph{IEEE Wireless communications}, vol.~12, no.~2, pp. 8--18, Apr. 2005.

\bibitem{ding2017survey}
Z.~Ding, X.~Lei, G.~K. Karagiannidis, R.~Schober, J.~Yuan, and V.~K. Bhargava,
  ``A survey on non-orthogonal multiple access for {5G} networks: Research
  challenges and future trends,'' \emph{IEEE Journal on Selected Areas in
  Communications}, vol.~35, no.~10, pp. 2181--2195, Oct. 2017.

\bibitem{foo2005speed}
Y.~C. Foo and W.~L. Lim, ``Speed and direction adaptive call admission control
  for high altitude platform station {(HAPS) UMTS},'' in \emph{IEEE Military
  Communications Conference (MILCOM)}, 2005, pp. 2182--2188.

\bibitem{ceran2017optimal}
E.~T. Ceran, T.~Erkilic, E.~Uysal-Biyikoglu, T.~Girici, and K.~Leblebicioglu,
  ``Optimal energy allocation policies for a high altitude flying wireless
  access point,'' \emph{Transactions on Emerging Telecommunications
  Technologies}, vol.~28, no.~4, p. e3034, Mar. 2016.

\bibitem{raschella2009high}
A.~Raschell{\`a}, G.~Araniti, A.~Iera, and A.~Molinaro, ``High altitude
  platforms: Radio resource management policy for {MBMS} applications,'' in
  \emph{International Conference on Personal Satellite Services}.\hskip 1em
  plus 0.5em minus 0.4em\relax Springer, 2009, pp. 85--93.

\bibitem{8766778}
J.~{Liu}, L.~{Li}, F.~{Yang}, X.~{Liu}, X.~{Li}, X.~{Tang}, and Z.~{Han},
  ``Minimization of offloading delay for two-tier {UAV} with mobile edge
  computing,'' in \emph{Proc. International Wireless Communications Mobile
  Computing Conference (IWCMC)}, 2019, pp. 1534--1538.

\bibitem{wang2020federated}
S.~Wang, M.~Chen, C.~Yin, W.~Saad, C.~S. Hong, S.~Cui, and H.~V. Poor,
  ``Federated learning for task and resource allocation in wireless high
  altitude balloon networks,'' \emph{arXiv preprint arXiv:2003.09375}, 2020.

\bibitem{jiang2015channel}
J.~Jiang, B.~Zhang, D.~Guo, and Z.~Ye, ``A channel assigning algorithm for
  platform displacement model in {HAPS} communication system,''
  \emph{Telecommunication Engineering}, vol.~55, no.~8, pp. 906--912, Aug.
  2015.

\bibitem{guan2019intelligent}
M.~Guan, Z.~Wu, Y.~Cui, X.~Cao, L.~Wang, J.~Ye, and B.~Peng, ``An intelligent
  wireless channel allocation in {HAPS 5G} communication system based on
  reinforcement learning,'' \emph{EURASIP Journal on Wireless Communications
  and Networking}, vol. 2019, no. 138, pp. 1--10, May 2019.

\bibitem{liu2009exploiting}
Y.~Liu, D.~Grace, and P.~D. Mitchell, ``Exploiting platform diversity for {GoS}
  improvement for users with different high altitude platform availability,''
  \emph{IEEE Transactions on Wireless Communications}, vol.~8, no.~1, pp.
  196--203, Feb. 2009.

\bibitem{guan2016channel}
M.~Guan, L.~Wang, and L.~Chen, ``Channel allocation for hot spot areas in
  {HAPS} communication based on the prediction of mobile user
  characteristics,'' \emph{Intelligent Automation \& Soft Computing}, vol.~22,
  no.~4, pp. 613--620, Apr. 2016.

\bibitem{ibrahim2015using}
A.~Ibrahim and A.~S. Alfa, ``Using {Lagrangian} relaxation for radio resource
  allocation in high altitude platforms,'' \emph{IEEE Transactions on Wireless
  Communications}, vol.~14, no.~10, pp. 5823--5835, Oct. 2015.

\bibitem{ibrahim2019optimizing}
------, ``Optimizing radio resources for multicasting on high-altitude
  platforms,'' \emph{EURASIP Journal on Wireless Communications and
  Networking}, vol. 2019, no.~1, p. 213, Aug. 2019.

\bibitem{ibrahim2019optimization}
A.~Ibrahim and A.~Alfa, \emph{Optimization Methods for User Admissions and
  Radio Resource Allocation for Multicasting over High Altitude
  Platforms}.\hskip 1em plus 0.5em minus 0.4em\relax River Publishers, 2019.

\bibitem{oodo2002sharing}
M.~Oodo, R.~Miura, T.~Hori, T.~Morisaki, K.~Kashiki, and M.~Suzuki, ``Sharing
  and compatibility study between fixed service using high altitude platform
  stations {(HAPS)} and other services in the 31/28 {GHz} bands,''
  \emph{Wireless Personal Communications}, vol.~23, no.~1, pp. 3--14, Oct.
  2002.

\bibitem{likitthanasate2008spectrum}
P.~Likitthanasate, D.~Grace, and P.~D. Mitchell, ``Spectrum etiquettes for
  terrestrial and high-altitude platform-based cognitive radio systems,''
  \emph{IET communications}, vol.~2, no.~6, pp. 846--855, Jul. 2008.

\bibitem{mokayef2013spectrum}
M.~Mokayef, T.~A. Rahman, R.~Ngah, and M.~Y. Ahmed, ``Spectrum sharing model
  for coexistence between high altitude platform system and fixed services at
  5.8 {GHz},'' \emph{International Journal of Multimedia and Ubiquitous
  Engineering}, vol.~8, no.~5, pp. 265--275, 2013.

\bibitem{wang2019dynamic}
S.~Wang, Y.~Li, Q.~Wang, M.~Su, and W.~Zhou, ``Dynamic downlink resource
  allocation based on imperfect estimation in {LEO-HAP} cognitive system,'' in
  \emph{IEEE 11th International Conference on Wireless Communications and
  Signal Processing (WCSP)}, 2019, pp. 1--6.

\bibitem{ahmadi2017novel}
H.~Ahmadi, K.~Katzis, and M.~Z. Shakir, ``A novel airborne self-organising
  architecture for {5G+} networks,'' in \emph{IEEE 86th Vehicular Technology
  Conference (VTC-Fall)}, 2017, pp. 1--5.

\bibitem{djuknic1997establishing}
G.~M. Djuknic, J.~Freidenfelds, and Y.~Okunev, ``Establishing wireless
  communications services via high-altitude aeronautical platforms: A concept
  whose time has come?'' \emph{IEEE Communications Magazine}, vol.~35, no.~9,
  pp. 128--135, Sep. 1997.

\bibitem{olver1994microwave}
A.~D. Olver, P.~Clarricoats, L.~Shafai, and A.~Kishk, \emph{Microwave {H}orns
  and {F}eeds}.\hskip 1em plus 0.5em minus 0.4em\relax IET, 1994, no.~39.

\bibitem{albagory2013flat}
Y.~Albagory, ``Flat-top ring-shaped cell design for high-altitude platform
  communications,'' \emph{International Journal of Computer Network and
  Information Security}, vol.~5, no.~7, p.~51, Jun. 2013.

\bibitem{capstick2005high}
M.~H. Capstick and D.~Grace, ``High altitude platform {mm-Wave} aperture
  antenna steering solutions,'' \emph{Wireless Personal Communications},
  vol.~32, no. 3-4, pp. 215--236, Feb. 2005.

\bibitem{thornton2005effect}
J.~Thornton, D.~A. Pearce, D.~Grace, M.~Oodo, K.~Katzis, and T.~C. Tozer,
  ``Effect of antenna beam pattern and layout on cellular performance in high
  altitude platform communications,'' \emph{Wireless Personal Communications},
  vol.~35, no. 1-2, pp. 35--51, Oct. 2005.

\bibitem{albagory2013smart}
Y.~Albagory and A.~E. Abbas, ``Smart cell design for high altitude platforms
  communication,'' \emph{AEU-International Journal of Electronics and
  Communications}, vol.~67, no.~9, pp. 780--786, Sep. 2013.

\bibitem{arum2019beam}
S.~C. Arum, D.~Grace, P.~D. Mitchell, and M.~D. Zakaria, ``Beam-pointing
  algorithm for contiguous high-altitude platform cell formation for extended
  coverage,'' in \emph{IEEE 90th Vehicular Technology Conference
  (VTC2019-Fall)}, 2019, pp. 1--5.

\bibitem{zakaria2019exploiting}
M.~D. Zakaria, D.~Grace, P.~D. Mitchell, T.~M. Shami, and N.~Morozs,
  ``Exploiting user-centric joint transmission--coordinated multipoint with a
  high altitude platform system architecture,'' \emph{IEEE Access}, vol.~7, pp.
  38\,957--38\,972, Mar. 2019.

\bibitem{chen2005performance}
G.~Chen, D.~Grace, and T.~C. Tozer, ``Performance of multiple high altitude
  platforms using directive {HAP} and user antennas,'' \emph{Wireless Personal
  Communications}, vol.~32, no. 3-4, pp. 275--299, Feb. 2005.

\bibitem{dong2015diversity}
F.~Dong, M.~Li, X.~Gong, H.~Li, and F.~Gao, ``Diversity performance analysis on
  multiple {HAP} networks,'' \emph{Sensors}, vol.~15, no.~7, pp.
  15\,398--15\,418, Jun. 2015.

\bibitem{yue2019diversity}
D.-W. Yue, S.~Xu, and H.~H. Nguyen, ``Diversity gain of millimeter-wave massive
  {MIMO} systems with distributed antenna arrays,'' \emph{EURASIP Journal on
  Wireless Communications and Networking}, vol. 2019, no.~1, p.~54, Mar. 2019.

\bibitem{lian2019user}
Z.~Lian, L.~Jiang, C.~He, and D.~He, ``User grouping and beamforming for {HAP}
  massive {MIMO} systems based on statistical-eigenmode,'' \emph{IEEE Wireless
  Communications Letters}, vol.~8, no.~3, pp. 961--964, Jun. 2019.

\bibitem{guan2019efficiency}
M.~Guan, Z.~Wu, Y.~Cui, X.~Cao, L.~Wang, J.~Ye, and B.~Peng, ``Efficiency
  evaluations based on artificial intelligence for {5G} massive {MIMO}
  communication systems on high-altitude platform stations,'' \emph{IEEE
  Transactions on Industrial Informatics}, vol.~16, no.~10, pp. 6632--6640,
  Oct. 2019.

\bibitem{benjebbour2017overview}
A.~Benjebbour, ``An overview of non-orthogonal multiple access,'' \emph{ZTE
  Communications}, vol.~15, no.~S1, pp. 21--30, June 2017.

\bibitem{andrews2014will}
J.~G. Andrews, S.~Buzzi, W.~Choi, S.~V. Hanly, A.~Lozano, A.~C. Soong, and
  J.~C. Zhang, ``What will {5G} be?'' \emph{IEEE Journal on Selected Areas in
  Communications}, vol.~32, no.~6, pp. 1065--1082, Jun. 2014.

\bibitem{rusek2009constrained}
F.~Rusek and J.~B. Anderson, ``Constrained capacities for faster-than-{Nyquist}
  signaling,'' \emph{IEEE Transactions on Information Theory}, vol.~55, no.~2,
  pp. 764--775, Feb. 2009.

\bibitem{prakash2013efficient}
J.~A. Prakash and G.~R. Reddy, ``Efficient prototype filter design for filter
  bank multicarrier ({FBMC}) system based on ambiguity function analysis of
  hermite polynomials,'' in \emph{IEEE International Mutli-Conference on
  Automation, Computing, Communication, Control and Compressed Sensing
  (iMac4s)}, 2013, pp. 580--585.

\bibitem{ankarali2020enhanced}
Z.~E. Ankaral{\i}, B.~Pek{\"o}z, and H.~Arslan, ``Enhanced {OFDM} for {5G}
  {RAN},'' \emph{ZTE Communications}, vol.~15, no.~S1, pp. 11--20, Jun. 2020.

\bibitem{anderson2013faster}
J.~B. Anderson, F.~Rusek, and V.~{\"O}wall, ``Faster-than-{Nyquist}
  signaling,'' \emph{Proceedings of the IEEE}, vol. 101, no.~8, pp. 1817--1830,
  Mar. 2013.

\bibitem{mazo1975faster}
J.~E. Mazo, ``Faster-than-{Nyquist} signaling,'' \emph{The Bell System
  Technical Journal}, vol.~54, no.~8, pp. 1451--1462, Oct. 1975.

\bibitem{liveris2003exploiting}
A.~D. Liveris and C.~N. Georghiades, ``Exploiting faster-than-{Nyquist}
  signaling,'' \emph{IEEE Transactions on Communications}, vol.~51, no.~9, pp.
  1502--1511, Sep. 2003.

\bibitem{7886296}
E.~{Bedeer}, M.~H. {Ahmed}, and H.~{Yanikomeroglu}, ``A very low complexity
  successive symbol-by-symbol sequence estimator for faster-than-{Nyquist}
  signaling,'' \emph{IEEE Access}, vol.~5, pp. 7414--7422, Mar. 2017.

\bibitem{7990502}
------, ``Low-complexity detection of high-order {QAM} faster-than-{Nyquist}
  signaling,'' \emph{IEEE Access}, vol.~5, pp. 14\,579--14\,588, Jun. 2017.

\bibitem{8641430}
L.~Feltrin, G.~Tsoukaneri, M.~Condoluci, C.~Buratti, T.~Mahmoodi, M.~Dohler,
  and R.~Verdone, ``Narrowband {IoT}: {A} survey on downlink and uplink
  perspectives,'' \emph{IEEE Wireless Communications}, vol.~26, no.~1, pp.
  78--86, Feb. 2019.

\bibitem{EricssonIoS}
\BIBentryALTinterwordspacing
E.~ConsumerLab. 10 hot consumertrends 2030: The internet of senses. [Online].
  Available:
  \url{https://www.ericsson.com/4ae13b/assets/local/reports-papers/consumerlab/reports/2019/10hctreport2030.pdf}
\BIBentrySTDinterwordspacing

\bibitem{rusek2005two}
F.~Rusek and J.~B. Anderson, ``The two dimensional {Mazo} limit,'' in
  \emph{IEEE International Symposium on Information Theory (ISIT)}, 2005, pp.
  970--974.

\bibitem{rusek2009multistream}
------, ``Multistream faster than {Nyquist} signaling,'' \emph{IEEE
  Transactions on Communications}, vol.~57, no.~5, pp. 1329--1340, May 2009.

\bibitem{dasalukunte2010multicarrier}
D.~Dasalukunte, F.~Rusek, and V.~Owall, ``Multicarrier faster-than-{Nyquist}
  transceivers: Hardware architecture and performance analysis,'' \emph{IEEE
  Transactions on Circuits and Systems I: Regular Papers}, vol.~58, no.~4, pp.
  827--838, May 2010.

\bibitem{peng2018spectral}
S.~Peng, A.~Liu, L.~Song, I.~Memon, and H.~Wang, ``Spectral efficiency
  maximization for deliberate clipping-based multicarrier faster-than-{Nyquist}
  signaling,'' \emph{IEEE Access}, vol.~6, pp. 13\,617--13\,623, Mar. 2018.

\bibitem{cai2020low}
B.~Cai, A.~Liu, and X.~Liang, ``Low-complexity selective mapping methods for
  multicarrier faster-than-{Nyquist} signaling,'' \emph{IEEE Access}, vol.~8,
  pp. 31\,420--31\,431, Feb. 2020.

\bibitem{ma2019low}
Y.~Ma, F.~Tian, N.~Wu, B.~Li, and X.~Ma, ``A low-complexity receiver for
  multicarrier faster-than-{Nyquist} signaling over frequency selective
  channels,'' \emph{IEEE Communications Letters}, vol.~24, no.~1, pp. 81--85,
  Jan. 2019.

\bibitem{higuchi2015non}
K.~Higuchi and A.~Benjebbour, ``Non-orthogonal multiple access ({NOMA}) with
  successive interference cancellation for future radio access,'' \emph{IEICE
  Transactions on Communications}, vol.~98, no.~3, pp. 403--414, Mar. 2015.

\bibitem{vaezi2019non}
M.~Vaezi, R.~Schober, Z.~Ding, and H.~V. Poor, ``Non-orthogonal multiple
  access: Common myths and critical questions,'' \emph{IEEE Wireless
  Communications}, vol.~26, no.~5, pp. 174--180, Oct. 2019.

\bibitem{he2016handover}
P.~He, N.~Cheng, and J.~Cui, ``Handover performance analysis of cellular
  communication system from high altitude platform in the swing state,'' in
  \emph{IEEE International Conference on Signal and Image Processing (ICSIP)},
  2016, pp. 407--411.

\bibitem{grace2010low}
D.~Grace, K.~Katzis, D.~Pearce, and P.~Mitchell, ``Low-latency {MAC}-layer
  handoff for a high-altitude platform delivering broadband communications,''
  \emph{URSI Radio Science Bulletin}, vol. 2010, no. 333, pp. 39--49, Mar.
  2010.

\bibitem{alsamhi2015intelligent}
S.~Alsamhi and N.~Rajput, ``An intelligent hand-off algorithm to enhance
  quality of service in high altitude platforms using neural network,''
  \emph{Wireless Personal Communications}, vol.~82, no.~4, pp. 2059--2073, Jan.
  2015.

\bibitem{albagory2015handover}
Y.~Albagory, M.~Nofal, and A.~Ghoneim, ``Handover performance of unstable-yaw
  stratospheric high-altitude stations,'' \emph{Wireless Personal
  Communications}, vol.~84, no.~4, pp. 2651--2663, May 2015.

\bibitem{he2016improved}
P.~He, N.~Cheng, and S.~Ni, ``Improved {LMS} predictive link triggering for
  handover in {HAPS} communication system,'' in \emph{IEEE 8th International
  Conference on Wireless Communications \& Signal Processing (WCSP)}, 2016, pp.
  1--5.

\bibitem{ni2016handover}
S.-y. Ni, S.~Jin, and H.-l. Hong, ``A handover decision algorithm with an
  adaptive threshold applied in {HAPS} communication system,'' in \emph{Theory,
  Methodology, Tools and Applications for Modeling and Simulation of Complex
  Systems}.\hskip 1em plus 0.5em minus 0.4em\relax Springer, 2016, pp. 38--47.

\bibitem{he2017adaptive}
P.~He, N.~Cheng, S.~Ni, and C.~Li, ``An adaptive handover scheme based on
  cooperative transmission from high altitude platform stations,'' in
  \emph{IEEE 2nd Advanced Information Technology, Electronic and Automation
  Control Conference (IAEAC)}, 2017, pp. 1306--1310.

\bibitem{wang2019effect}
X.~Wang, L.~Li, and W.~Zhou, ``The effect of {HAPS} unstable movement on
  handover performance,'' in \emph{IEEE 28th Wireless and Optical
  Communications Conference (WOCC)}, 2019, pp. 1--5.

\bibitem{li2019handover}
K.~Li, Y.~Li, Z.~Qiu, Q.~Wang, J.~Lu, and W.~Zhou, ``Handover procedure design
  and performance optimization strategy in {LEO-HAP} system,'' in \emph{IEEE
  11th International Conference on Wireless Communications and Signal
  Processing (WCSP)}, 2019, pp. 1--7.

\bibitem{li2010directional}
S.~Li, D.~Grace, J.~Wei, and D.~Ma, ``Directional traffic-aware intra-{HAP}
  handoff scheme for {HAP} communications systems,'' \emph{URSI Radio Science
  Bulletin}, vol. 2010, no. 334, pp. 11--18, Sep. 2010.

\bibitem{li2011cooperative}
S.~Li, L.~Wang, G.~David, and D.~Ma, ``Cooperative directional inter-cell
  handover scheme in high altitude platform communications systems,''
  \emph{Journal of Electronics (China)}, vol.~28, no.~2, p. 249, Oct. 2011.

\bibitem{katzis2010inter}
K.~Katzis and D.~Grace, ``Inter-high-altitude-platform handoff for
  communications systems with directional antennas,'' \emph{URSI Radio Science
  Bulletin}, vol. 2010, no. 333, pp. 29--38, Dec. 2010.

\bibitem{li2010novel}
S.~Li, D.~Grace, J.~Wei, and D.~Ma, ``A novel guaranteed handover scheme for
  {HAP} communications systems with adaptive modulation and coding,'' in
  \emph{IEEE Vehicular Technology Conference (VTC-Fall)}, 2010, pp. 1--5.

\bibitem{an2013load}
F.~An, Y.~Wang, and F.~Meng, ``A load balancing handoff algorithm based on
  {RSSI} and energy-aware in {HAPs} network,'' in \emph{IEEE International
  Conference on Signal Processing, Communication and Computing (ICSPCC 2013)},
  2013, pp. 1--6.

\bibitem{hehtke2017autonomous}
V.~Hehtke, J.~Kiam, and A.~Schulte, \emph{An Autonomous Mission Management
  System to Assist Decision Making of a {HALE} Operator}.\hskip 1em plus 0.5em
  minus 0.4em\relax Deutsche Gesellschaft f{\"u}r Luft-und
  Raumfahrt-Lilienthal-Oberth eV, 2017.

\bibitem{8797881}
O.~{Anicho}, P.~B. {Charlesworth}, G.~S. {Baicher}, A.~{Nagar}, and
  N.~{Buckley}, ``Comparative study for coordinating multiple unmanned {HAPS}
  for communications area coverage,'' in \emph{International Conference on
  Unmanned Aircraft Systems (ICUAS)}, 2019, pp. 467--474.

\bibitem{anicho2019autonomously}
O.~Anicho, P.~B. Charlesworth, G.~S. Baicher, and A.~Nagar, ``Autonomously
  coordinated multi-{HAPS} communications network: Failure mitigation in
  volcanic incidence area coverage,'' in \emph{IEEE International Conference on
  Communication, Networks and Satellite (Comnetsat)}, 2019, pp. 79--84.

\bibitem{kaur2018edge}
K.~Kaur, S.~Garg, G.~S. Aujla, N.~Kumar, J.~J. Rodrigues, and M.~Guizani,
  ``Edge computing in the industrial internet of things environment:
  Software-defined-networks-based edge-cloud interplay,'' \emph{IEEE
  Communications Magazine}, vol.~56, no.~2, pp. 44--51, Feb. 2018.

\bibitem{han2015network}
B.~Han, V.~Gopalakrishnan, L.~Ji, and S.~Lee, ``Network function
  virtualization: Challenges and opportunities for innovations,'' \emph{IEEE
  Communications Magazine}, vol.~53, no.~2, pp. 90--97, Feb. 2015.

\bibitem{NS9003208}
L.~U. {Khan}, I.~{Yaqoob}, N.~H. {Tran}, Z.~{Han}, and C.~S. {Hong}, ``Network
  slicing: Recent advances, taxonomy, requirements, and open research
  challenges,'' \emph{IEEE Access}, vol.~8, pp. 36\,009--36\,028, Feb. 2020.

\bibitem{8610425}
Y.~{Shi}, Y.~{Cao}, J.~{Liu}, and N.~{Kato}, ``A cross-domain {SDN}
  architecture for multi-layered space-terrestrial integrated networks,''
  \emph{IEEE Network}, vol.~33, no.~1, pp. 29--35, Jan./Feb. 2019.

\bibitem{chen2019segment}
K.~Chen, S.~Zhao, N.~Lv, W.~Gao, X.~Wang, and X.~Zou, ``Segment routing based
  traffic scheduling for the software-defined airborne backbone network,''
  \emph{IEEE Access}, vol.~7, pp. 106\,162--106\,178, 2019.

\bibitem{yang2020multi}
L.~Yang, H.~Yao, J.~Wang, C.~Jiang, A.~Benslimane, and Y.~Liu, ``Multi-{UAV}
  enabled load-balance mobile edge computing for {IoT} networks,'' \emph{IEEE
  Internet of Things Journal}, 2020.

\bibitem{yang2019energy}
Z.~Yang, C.~Pan, K.~Wang, and M.~Shikh-Bahaei, ``Energy efficient resource
  allocation in {UAV}-enabled mobile edge computing networks,'' \emph{IEEE
  Transactions on Wireless Communications}, vol.~18, no.~9, pp. 4576--4589,
  Sep. 2019.

\bibitem{zhou2020mobile}
F.~Zhou, R.~Q. Hu, Z.~Li, and Y.~Wang, ``Mobile edge computing in unmanned
  aerial vehicle networks,'' \emph{IEEE Wireless Communications}, vol.~27,
  no.~1, pp. 140--146, Feb. 2020.

\bibitem{asheralieva2019hierarchical}
A.~Asheralieva and D.~Niyato, ``Hierarchical game-theoretic and reinforcement
  learning framework for computational offloading in {UAV}-enabled mobile edge
  computing networks with multiple service providers,'' \emph{IEEE Internet of
  Things Journal}, vol.~6, no.~5, pp. 8753--8769, Oct. 2019.

\bibitem{Tasneem2020}
T.~Darwish, G.~K. Kurt, H.~Yanikomeroglu, G.~Senarath, and P.~Zhu, ``A vision
  of self-evolving network management for future intelligent vertical
  {HetNet},'' \emph{arXiv preprint arXiv:2009.02771}, 2020.

\bibitem{Branco2019}
S.~Branco, A.~G. Ferreira, and J.~Cabral, ``Machine learning in resource-scarce
  embedded systems, {FPGAs}, and end-devices: A survey,'' \emph{Electronics},
  vol.~8, no.~11, p. 1289, Nov. 2019.

\bibitem{Microchip2019}
A.~Chandler, ``Microchip introduces the industry’s first {MCU} with
  integrated {2D} {GPU} and integrated {DDR2} memory for groundbreaking
  graphics capabilities,'' \url{ https://cutt.ly/LaieMgt}, accessed:
  2020-01-20.

\bibitem{ARM2019}
R.~Dirvin, ``Next-generation {Armv8.1-M} architecture: Delivering enhanced
  machine learning and signal processing for the smallest embedded devices,''
  \url{https://www.arm.com/company/news/2019/02/next-generation-armv8-1-m-architecture},
  accessed: 2020-01-20.

\bibitem{STMicroelectronics2019}
STMicroelectronics, ``Ism330dhcx: Machine learning core,''
  \url{https://www.st.com/content/ccc/resource/technical/document/application_note/group1/60/c8/a2/6b/35/ab/49/6a/DM00651838/files/DM00651838.pdf/jcr:content/translations/en.DM00651838.pdf},
  accessed: 2020-01-20.

\bibitem{zhang20196g}
L.~Zhang, Y.-C. Liang, and D.~Niyato, ``{6G} visions: Mobile ultra-broadband,
  super {Internet}-of-things, and artificial intelligence,'' \emph{China
  Communication}, vol.~16, no.~8, pp. 1--14, Aug. 2019.

\bibitem{nguyen2017reinforcement}
D.~D. Nguyen, H.~X. Nguyen, and L.~B. White, ``Reinforcement learning with
  network-assisted feedback for heterogeneous {RAT} selection,'' \emph{IEEE
  Transactions on Wireless Communications}, vol.~16, no.~9, pp. 6062--6076,
  Sep. 2017.

\bibitem{raj2018spectrum}
V.~Raj, I.~Dias, T.~Tholeti, and S.~Kalyani, ``Spectrum access in cognitive
  radio using a two-stage reinforcement learning approach,'' \emph{IEEE Journal
  on Selected Topics in Signal Processing}, vol.~12, no.~1, pp. 20--34, Feb.
  2018.

\bibitem{qiao2018topology}
M.~Qiao, H.~Zhao, L.~Zhou, C.~Zhu, and S.~Huang, ``Topology-transparent
  scheduling based on reinforcement learning in self-organized wireless
  networks,'' \emph{IEEE Access}, vol.~6, pp. 20\,221--20\,230, 2018.

\bibitem{luong2019applications}
N.~C. Luong, D.~T. Hoang, S.~Gong, D.~Niyato, P.~Wang, Y.-C. Liang, and D.~I.
  Kim, ``Applications of deep reinforcement learning in communications and
  networking: A survey,'' \emph{IEEE Communications Surveys \& Tutorials},
  vol.~21, no.~4, pp. 3133--3174, Fourthquarter 2019.

\bibitem{kusupati2018fastgrnn}
A.~Kusupati, M.~Singh, K.~Bhatia, A.~Kumar, P.~Jain, and M.~Varma,
  ``{FastGRNN}: A fast, accurate, stable and tiny kilobyte sized gated
  recurrent neural network,'' in \emph{Advances in Neural Information
  Processing Systems (NIPS)}, 2018, pp. 9017--9028.

\bibitem{Hochul2019}
L.~Hochul, L.~Jaehun, C.~L. Young, H.~Hyuck, and K.~Sooyong, ``Mobile
  collaborative computing on the fly,'' \emph{IEEE Access}, vol.~58, pp.
  1574--1192, Aug. 2019.

\bibitem{ismaiel2018performance}
A.~M. Ismaiel, E.~Elsaidy, Y.~Albagory, H.~A. Atallah, A.~B. Abdel-Rahman, and
  T.~Sallam, ``Performance improvement of high altitude platform using
  concentric circular antenna array based on particle swarm optimization,''
  \emph{AEU-International Journal of Electronics and Communications}, vol.~91,
  pp. 85--90, Jul. 2018.

\bibitem{liu2019minimization}
J.~Liu, L.~Li, F.~Yang, X.~Liu, X.~Li, X.~Tang, and Z.~Han, ``Minimization of
  offloading delay for two-tier {UAV} with mobile edge computing,'' in
  \emph{IEEE 15th International Wireless Communications \& Mobile Computing
  Conference (IWCMC)}, 2019, pp. 1534--1538.

\bibitem{kiam2019hierarchical}
J.~J. Kiam, V.~Hehtke, E.~Besada-Portas, and A.~Schulte, ``Hierarchical
  planning guided by genetic algorithms for multiple {HAPS} in a time-varying
  environment,'' in \emph{International Conference on Intelligent Human Systems
  Integration}.\hskip 1em plus 0.5em minus 0.4em\relax Springer, 2019, pp.
  719--724.

\bibitem{clark2019testing}
G.~Clark, G.~Landis, E.~Barnes, B.~LaFuente, and K.~Collins, ``Testing a neural
  network accelerator on a high-altitude balloon,'' in \emph{IEEE Cognitive
  Communications for Aerospace Applications Workshop (CCAAW)}, 2019, pp. 1--8.

\bibitem{AMSAT-UK}
AMSAT-UK,
  \url{https://amsat-uk.org/2013/07/04/google-project-loon-interference-concerns/}.

\bibitem{farsad2018neural}
N.~Farsad and A.~Goldsmith, ``Neural network detection of data sequences in
  communication systems,'' \emph{IEEE Transactions on Signal Processing},
  vol.~66, no.~21, pp. 5663--5678, Sep. 2018.

\bibitem{shlezinger2020viterbinet}
N.~Shlezinger, N.~Farsad, Y.~C. Eldar, and A.~J. Goldsmith, ``Viterbinet: A
  deep learning based {Viterbi} algorithm for symbol detection,'' \emph{IEEE
  Transactions on Wireless Communications}, vol.~19, no.~5, pp. 3319--3331,
  Feb. 2020.

\bibitem{sun2017learning}
H.~Sun, X.~Chen, Q.~Shi, M.~Hong, X.~Fu, and N.~D. Sidiropoulos, ``Learning to
  optimize: Training deep neural networks for wireless resource management,''
  in \emph{IEEE International Workshop on Signal Processing Advances in
  Wireless Communications (SPAWC)}, 2017, pp. 1--6.

\bibitem{Li2019Detrimental}
Y.~Li, F.~Baccelli, H.~S. Dhillon, and J.~G. Andrews, ``Fitting determinantal
  point processes to macro base station deployments,'' in \emph{IEEE Global
  Communications Conference (GLOBECOM)}, 2014, pp. 1--6.

\bibitem{Na2019Ginibre}
N.~Deng, W.~Zhou, and M.~Haenggi, ``The ginibre point process as a model for
  wireless networks with repulsion,'' \emph{IEEE Transactions on Wireless
  Communications}, vol.~14, no.~1, pp. 107--121, Jan. 2015.

\bibitem{khosh2019HetNet}
M.~G. Khoshkholgh, K.~Navaie, H.~Yanikomeroglu, V.~C.~M. Leung, and K.~G. Shin,
  ``Coverage performance of aerial-terrestrial {HetNets},'' in \emph{IEEE
  Vehicular Technology Conference (VTC)}, 2019, pp. 1--5.

\bibitem{khosh2019NonIdeal}
------, ``How do non-ideal {UAV} antennas affect air-to-ground
  communications?'' in \emph{IEEE International Conference on Communications
  (ICC)}, 2019, pp. 1--7.

\bibitem{Kalamkar2019meta}
S.~Kalamkar and M.~Haenggi, ``Simple approximations of the {SIR} meta
  distribution in general cellular networks,'' \emph{IEEE Transactions on
  Communications}, vol.~67, no.~6, pp. 4393--4406, Jun. 2019.

\bibitem{bao2017mmWave}
J.~Bao, D.~Sprinz, and H.~Li, ``Blockage of millimeter wave communications on
  rotor {UAVs}: Demonstration and mitigation,'' in \emph{IEEE Military
  Communications Conference (MILCOM)}, 2017, pp. 768--774.

\bibitem{Xiao2017mmWave}
M.~Xiao, S.~Mumtaz, Y.~Huang, L.~Dai, Y.~Li, M.~Matthaiou, G.~K. Karagiannidis,
  E.~Bjornson, K.~Yang, I.~Chih-Lin, and A.~Ghosh, ``Millimeter wave
  communications for future mobile networks,'' \emph{IEEE Journal on Selected
  Areas in Communications}, vol.~35, no.~9, pp. 1909--1935, Sep. 2017.

\bibitem{Zhang2019mmWave}
G.~Zhang, Q.~Wu, M.~Cui, and R.~Zhang, ``Securing {UAV} communications via
  joint trajectory and power control,'' \emph{IEEE Transactions on Wireless
  Communications}, vol.~18, no.~2, pp. 1376--1389, Feb. 2019.

\bibitem{Liu2017mmWave}
C.~Liu, T.~Q.~S. Quek, and J.~Lee, ``Secure {UAV} communication in the presence
  of active eavesdropper,'' in \emph{IEEE International Conference on Wireless
  Communications and Signal Processing (WCSP)}, 2017, pp. 1--5.

\bibitem{Hu2017covert}
J.~Hu, S.~Yan, X.~Zhou, F.~Shu, and J.~Wang, ``Secure {UAV} communication in
  the presence of active eavesdropper,'' in \emph{IEEE Global Communications
  Conference (GLOBECOM)}, 2017, pp. 1--6.

\bibitem{QKDDD}
Y.~Chu, R.~Donaldson, R.~Kumar, and D.~Grace, ``Feasibility of quantum key
  distribution from high altitude platforms,'' \emph{arXiv preprint
  arXiv:2012.07479}, 2020.

\bibitem{SatelliteHAPSsIoT2016}
O.~Said and A.~Tolba, ``Performance evaluation of a dual coverage system for
  internet of things environments,'' \emph{Mobile Information Systems}, vol.
  2016, no. 3464392, pp. 1--20, Dec. 2016.

\bibitem{WaterQuality2016}
S.~O. Olatinwo and T.-H. Joubert, ``Enabling communication networks for water
  quality monitoring applications: A survey,'' \emph{IEEE Access}, vol.~7, pp.
  100\,332--100\,362, 2019.

\bibitem{Handley2018mega}
M.~Handley, ``Delay is not an option: Low latency routing in space,'' in
  \emph{ACM Workshop on Hot Topics in Networks (HotNets)}, 2018, pp. 85--91.

\bibitem{Chelsea2017metalearning}
C.~Finn, P.~Abbeel, and S.~Levine, ``Model-agnostic meta-learning for fast
  adaptation of deep networks,'' in \emph{International Conference on Machine
  Learning (ICML)}, 2017, pp. 1126--1135.

\end{thebibliography}


%\end{IEEEbiography}


%\newpage



%\vfill


\end{document}